\documentclass[12pt]{article}
\usepackage[letterpaper, left=1in, right=1in, top=1in, bottom=1in]{geometry}
\usepackage{ragged2e}
\usepackage{amsmath, amssymb, graphicx}
\usepackage{authblk}
\usepackage{cite}
\usepackage{hyperref}
\usepackage{cleveref}
\usepackage{bm}
\usepackage{indentfirst}
\usepackage{subfig}
\usepackage{xcolor}

\crefname{emptytype}{}{}

\title{Identification of MHD equilibrium $\beta$ limits for CFQS plasmas}

\author{Jian~Zhang$^1$}
\author{Ping~Zhu$^{1,2,*}$}
\author{Haifeng~Liu$^{3,*}$}
\author{Yasuhiro~Suzuki$^4$}
\author{Chris~C.~Hegna$^2$}
\author{Yuhong~Xu$^3$}

\affil{\justifying\small\noindent$^1$State Key Laboratory of Advanced Electromagnetic Technology, International Joint Research Laboratory of Magnetic Confinement Fusion and Plasma Physics, School of Electrical and Electronic Engineering, Huazhong University of Science and Technology, Wuhan, 430074, China}
\affil{\justifying\small\noindent$^2$Department of Nuclear Engineering and Engineering Physics, University of Wisconsin-Madison, Madison, WI 53706, United States of America}
\affil{\justifying\small\noindent$^3$Institute of Fusion Science, School of Physical Science and Technology, Southwest Jiaotong University, Chengdu 610031, China}
\affil{\justifying\small\noindent$^4$Graduate School of Advanced Science and Engineering, Hiroshima University, Higashi-Hiroshima, 739-8527, Japan}
\affil{\justifying\small\noindent$^*$Author to whom any correspondence should be addressed.}
\affil{\justifying\small\noindent\textbf{E-mail}: zhup@hust.edu.cn, hfliu@swjtu.edu.cn}
\date{}

\begin{document}
\maketitle
\vspace{-2em}

\begin{abstract}
The magnetohydrodynamic (MHD) equilibrium $\beta$ limits in the Chinese First Quasi-Axisymmetric Stellarator (CFQS) 
are investigated using the NTEC code, for both the standard and the magnetic island configurations. 
The equilibrium $\beta$ limit is identified upon the onset of the rapid destruction of nested flux surfaces by evaluating several numerical metrics, 
including the fractal dimension, weighted Birkhoff average, and effective volume of parallel diffusion. In the standard configuration, 
the net-current-free and the bootstrap-current-carrying equilibria can sustain well-ordered magnetic surfaces up to $\langle\beta\rangle\approx1.5\%$. 
The proliferation of stochastic field lines starts after the critical overlap between the internal major islands 
and the high-order island chains. 
Two types of divertor island configurations are studied based on net-current-free equilibria. 
It is found that the edge islands may transition into open field lines at low $\langle\beta\rangle$ values and lead to 
a drastic shrinkage of the last closed flux surface. Meanwhile, 
the threshold $\langle\beta\rangle$ value of the degradation of inner flux surfaces is similar to the standard configuration. 
\end{abstract}

\section{Introduction}

The performance of toroidal magnetic confinement fusion strongly scales with the attainable plasma pressure, 
making $\beta$ (defined as the ratio of plasma pressure to magnetic pressure) a critical figure of merit for potential reactor designs. 
High $\beta$ can reduce device size and the cost associated with magnets and power supplies, 
hence the device designs routinely aim for volume-averaged $\langle\beta\rangle$ of a few percent. 
In tokamaks, since the presence of nested flux surfaces can be theoretically guaranteed by the axisymmetry, 
the maximum $\beta$ is often constrained by pressure-driven MHD instabilities \cite{Gryaznevich_2008}. 
In contrast, in fully three-dimensional systems, $\beta$ limits are primarily imposed first 
by the degradation of equilibrium itself \cite{Helander_2012}. In stellarators, 
the magnetic configuration of the vacuum field is designed with a dense set of nested flux surfaces. As $\beta$ increases, however, 
a portion of these nested flux surfaces may give way to magnetic islands and chaotic regions, 
which tend to enhance the radial transport of both particles and energy \cite{Hudson_2010}. Furthermore, 
modern stellarators are highly optimized with advanced concepts such as omnigenity \cite{Landreman_2012} and quasisymmetry \cite{Boozer_1983} 
which rely on the assumption of nested flux surfaces. Therefore, it is particularly important to assess the robustness of 
well-ordered nested flux surfaces in high-$\beta$ stellarator equilibria.

Experimental observations and theoretical studies have demonstrated that distinct magnetic configurations
of stellarators give rise to different main causes for $\beta$ limits.
In Large Helical Device (LHD) \cite{Iiyoshi_1990}, the operation with a low magnetic field $B_{t}=0.41\mathrm{T}$ raised $\langle\beta\rangle$ up to 
$4.8\%$ \cite{Sakakibara_2008}, whereas another high-field operation achieved $\langle\beta\rangle=4.1\%$ with $B_{t}=0.41\mathrm{T}$ \cite{Sakakibara_2017}. 
The plasma periphery was located on the magnetic hill, which induced edge MHD instabilities observed in the corresponding experiments. 
But there was no clear evidence to confirm whether the $\beta$ limit primarily resulted from the stochastization of flux surfaces or
edge MHD instabilities. 
MHD equilibrium calculations implied that a soft $\beta$ limit, which violates the linearity between $\langle\beta\rangle$ and $\beta_\text{axis}$, 
should be higher than the experimentally achieved value \cite{Suzuki_2008, Suzuki_2020}. Unlike a conventional stellarator, 
Wendelstein 7-AS (W7-AS) \cite{Hirsch_2008}, which has an optimized configuration where the entire plasma resides in a magnetic well, 
has achieved MHD-quiescent discharge levels with $\langle\beta\rangle=3.4\%$ at a field of $0.9-1.25\mathrm{T}$ \cite{Weller_2003}. 
MHD equilibrium calculations found a large region of stochastic magnetic field lines at the plasma edge, 
which is believed to account for the confinement degradation \cite{Reiman_2007}. 
Nonlinear MHD simulations have also indicated that only mild MHD activities are observed 
during the plasma evolution terminated at a soft $\beta$ limit \cite{Ramasamy_2024}. 
Wendelstein 7-X (W7-X) \cite{Grieger_1992}, the successor to W7-AS, features more advanced design elements and is designed with a $5\%$ $\beta$ limit, 
even though the highest achieved $\langle\beta\rangle$ is to date $2.8\%$ with $B=1.5\mathrm{T}$ \cite{Grulke_2025}.
Relevant numerical studies suggested the W7-X standard configuration is fairly immune to major pressure-driven MHD instabilities 
and thus may only exhibit a soft $\beta$ limit \cite{Zhou_2024}.

CFQS is a modular-coil-type quasi-axisymmetric stellarator with a toroidal period number $N_\text{fp}=2$, major radius $R_{0}=1\mathrm{m}$, 
and an aspect ratio $A_\text{p}=4$ \cite{Liu_2018}. The standard magnetic configuration is produced with $16$ non-planar modular coils (MCs), 
and can be flexibly adjusted using $12$ toroidal field coils (TFCs) and $4$ poloidal field coils (PFCs) \cite{Liu_2025}. 
A series of simulation studies have been conducted to assess the MHD equilibrium characteristics 
\cite{Liu_2020, Liu_2023, Wang_2021, Su_2024, Fu_2024} in CFQS. 
The main findings show that various ideal MHD instabilities may occur only in high-$\beta$ equilibria. For net-current-free plasmas \cite{Liu_2020}, 
the interchange mode is stable up to $\langle\beta\rangle=2.0\%$, 
and ballooning instabilities emerge within a narrow gap of $\langle\beta\rangle=1.1\sim2.0\%$. 
When the effects of bootstrap currents are considered \cite{Liu_2023}, the external kink modes become destabilized at $\langle\beta\rangle\approx1.3\%$, 
and the quasi-axisymmetry property can be well preserved up to $\langle\beta\rangle=1.5\%$. 
Noted that these calculations are based on the nested flux surface assumption that might be broken due to the finite-$\beta$ effect. 
Other researches \cite{Wang_2021, Su_2024, Fu_2024} that have relaxed this assumption focus on how bootstrap 
and external currents modify equilibrium magnetic islands, 
with bootstrap currents typically obtained from semi-analytical formulas \cite{Shaing_1989}. 

In order to estimate the resilience of good flux surfaces in finite-$\beta$ CFQS plasmas, we use the NTEC code \cite{Zhang_2026} to
compute 3D MHD equilibria for various $\beta$ levels without the necessity of presuming nested magnetic flux surfaces. When the magnetic topology is significantly altered by 
the break of nested flux surfaces, we conclude that the equilibrium $\beta$ limit has been reached. The metrics used to evaluate 
the stochastization of magnetic field lines are described in the next section. 
We consider the standard and two magnetic island configurations. 
Since both bootstrap-current-carrying and net-current-free discharge operations will be implemented for future CFQS experiments \cite{Liu_2020}, 
in this paper, we consider both types of equilibria for the standard configuration. 
For the two island configurations, we only focus on net-current-free equilibria since the major islands already exist in their vacuum solutions. 

The rest of this paper is organized as follows. \Cref{sec:methods} elaborates on the numerical methods for equilibrium calculation and magnetic topology identification. 
\Cref{sec:standard} and \Cref{sec:island} present the results for the standard and the island configurations, respectively. 
Finally, \Cref{sec:summary} summarizes the main findings and discusses about future work. 

\section{Numerical and analysis methods}\label{sec:methods}

\subsection{3D MHD equilibrium solver: NTEC}

The NTEC code is a 3D MHD equilibrium solver capable of handling magnetic islands and chaotic regions with free or fixed plasma boundary. 
In order to properly tackle a highly shaped stellarator geometry, all calculations are performed within a computational domain in the coordinates $(x,y,\zeta)$ 
for the equivalent physical domain in the cylindrical coordinates $(R,Z,\phi)$ for example. 
A pseudo flux mapping is then established to connect the two coordinate systems. 
The numerical iteration adopts a non-variational method for plasma relaxation, evolving from an initial state toward a force-balanced equilibrium, 
following the numerically adapted single-fluid MHD equations briefly described next.

To seek the solutions where the pressure field is subject to the constriant $\bm{B}\cdot\nabla p = 0$, 
the parallel diffusion equation is extended into a second-order hyperbolic equation for the pressure field update
\begin{equation}\label{eq:pres}
\frac{\partial^2 p}{\partial t^2} + \frac{1}{\tau} \frac{\partial p}{\partial t} = \mathcal{F} (p) = \nabla \cdot (\nabla_{\parallel} p), 
\end{equation}
which permits the adoption of a larger time step $\delta t \sim O(h)$ ($h$ is the grid size) in an explicit time-advancing scheme. 
The parameter $\tau$ controls the numerical stability and convergence rate, and is set to be proportional to $\sqrt{p/\|\mathcal{F}\|}$. 

The relaxation process of the magnetic field is governed by the magnetic induction equation
\begin{equation}\label{eq:mag}
\frac{\partial \bm{B}}{\partial t} = \nabla \times[\bm{u}\times\bm{B}-\eta(\bm{J}-\bm{J}_\text{net})], 
\end{equation}
where the artificial resistivity allows the breaking of nested flux surfaces during the relaxation process. 
Here $\bm{J_\text{net}=\bm{B} \langle \bm{J} \cdot \bm{B} \rangle / \langle B^2 \rangle$} represents the field-line-averaged current density, such as the Ohmic, bootstrap and ECCD currents. 
A penalty term and an artificial diffusion term $\kappa_{div}\nabla\nabla\cdot \bm{B}$ are added to maintain the divergence-free condition for magnetic field. 
The flow relaxation is driven by the local residual force
\begin{equation}\label{eq:vel}
\rho\frac{\partial \bm{u}}{\partial t} = \bm{J} \times \bm{B} - \nabla p, 
\end{equation}
which should vanish upon the arrival at the equilibrium. 
The plasma density $\rho$ is set to unity and has little influence on the calculations. 
The advection term is ignored to address the static force balance problem, thereby, 
both velocity and resistivity will be dropped to zero when a static equilibrium state with $\partial\bm{B}/\partial t = 0$ and 
$\partial\bm{u}/\partial t = 0$ is found. 
The convergence properties and numerical setup of equilibria used in this paper are discussed in \nameref{App:A}. 

\subsection{Identification of magnetic topology change}

\subsubsection{Box-counting dimension}

Fractal measures provide a compact and quantitative way to determine whether a set of magnetic-field-line intersections, namely a Poincaré map, 
resembles a smooth curve, an island chain, a thin chaotic layer, or a space-filling chaotic region. 
The box-counting dimension $D_\text{bc}$ is one of the most intuitive and practically computable fractal dimensions \cite{Falconer_2013, Loizu_2017}. 
This method can be described as follows: cover the point set on the Poincaré section with a grid of boxes of side length $\epsilon$, 
and count how many boxes $N(\epsilon)$ contain at least one point of the set. As $\epsilon$ gradually decreases, the number of boxes increases. 
The box-counting dimension is defined as the ratio of their logarithms 
\begin{equation}\label{eq:BC}
D_\text{bc} = \lim_{\epsilon \to 0} \frac{\text{log}N(\epsilon)}{\text{log}(1/\epsilon)}. 
\end{equation}
If the field line tracing depicts a flux surface, $D_\text{bc}$ is expected to be $1$. 
For chaotic regions that fill a finite area, $D_\text{bc}$ is undoubtedly greater than $1$. 
For magnetic islands, $D_\text{bc}$ depends on whether the tracing position is near the X point, the O point, or somewhere in between. For some structures, 
such as those where a high-order perturbation slightly deforms the rational surface into a dotted-line-like chain of islands, 
$D_\text{bc}$ is sensitive to the numerical accuracy. Here, we focus on the regions of nested flux surface with $D_\text{bc}=1$.

\subsubsection{Weighted Birkhoff average}

The usual ergodic average of an observable $f$ along orbit points $\{x_{i}\}$ is $\overline{f}_{N} = \sum_{i = 1}^{N}f(x_{i})/N$. 
The weighted Birkhoff average (WBA) replaces the uniform weights $1/N$ by a smooth sequence of weights $\{w_{i}\}$ \cite{Tong_2024, Tong_2026, Gui_2026}: 
\begin{equation}\label{eq:WBA1}
WBA(f,w) = \sum_{i = 1}^{N}w_{i}f(x_{i})\big/\sum_{i = 1}^{N}w_{i}. 
\end{equation}
Through properly choosing $\{w_{i}\}$ that smoothly ramp up and down, the endpoint discontinuities can be suppressed for quasi-periodic orbits. 
We practically employ a $C^{\infty}$-continuous bump function $w(t) = \text{exp}\{-{[t(1-t)]}^{-1}\}$ where $t\in(0,1)$, 
and the rotational transform squared $\iota^2$ as the observable. 
The WBA method gives much faster convergence in regular orbits, such as flux surfaces and magnetic islands, than in chaotic orbits. 
As a result, the bigger the absolute difference of the WBAs between the middle and the end, the more chaotic the orbit. 
We utilize the quantity 
\begin{equation}\label{eq:WBA2}
|WBA_{[1/2,3/4]}(\iota^2) - WBA_{[3/4,1]}(\iota^2)| 
\end{equation}
as the evaluation of the stochasticity of magnetic field lines. 
The subscript $[a,b]$ represents a certain segment along a field line tracing, where $a$ and $b$ are the tracing point 
indices for both ends of that segment normalized to the total number of tracing points on that particular trajectory of field line tracing. 
For example, $[1/2,3/4]$ denotes the segment between the middle to three-quarters of the tracing points. 
The latter half of the tracing points, rather than the entire set are selected to enhance the convergence of regular trajectories.

\subsubsection{Effective volume of parallel diffusion}

In chaotic regions with open field lines connecting the vessel boundary, 
the force balance equation $\nabla p = \bm{J} \times \bm{B}$ could be pathological for numerical calculations \cite{Hudson_2010}. 
A straightforward way is to set the trivial solution $p=const$ in these areas so that $\bm{B}\cdot\nabla p = 0$, which however results in 
a pressure discontinuity between the integrable and the non-integrable field lines near the plasma-vacuum separatrix \cite{Suzuki_2024}. 
Instead of requiring an extremely high grid resolution to resolve the steep pressure gradient near the separatrix, a small numerical residue of the parallel gradient $\nabla_{\parallel}p$ is tolerated to maintain a smooth pressure profile, 
and consequently, can be used to identify the region whose field line connection length is not long enough for complete ergodicity. It should be noted that, in our numerical experiments, 
the pressure field can remain flat along stochastic field lines that are locally ergodic within a bounded layer of chaos. 
Theoretical study of anisotropic heat diffusion equation in a steady state similarly shows that the effective volume of parallel diffusion  
\begin{equation}\label{eq:VPD}
V_{PD} = \frac{1}{V} \int_{V} \mathcal{H} (\kappa_{\parallel}{|\nabla_{\parallel}T|}^2 - \kappa_{\perp}{|\nabla_{\perp}T|}^2) \,dV, 
\end{equation}
where $V$ is the total plasma volume and $\mathcal{H}$ the Heaviside function, 
can be used to assess the impact of non-integrability on the heat transport in stellarator equilibria \cite{Paul_2022, Baillod_2023}. 
We here replace the temperature by the pressure field since the plasma density is normalized to $1$. 

\section{Equilibrium \texorpdfstring{$\beta$}{beta} limits of standard configurations}\label{sec:standard}

\subsection{Net-current-free equilibria}\label{sec:standard_ncf}

\begin{figure}[htbp]
    % \vspace{-7em}
    \centering
    \subfloat{\includegraphics[width=0.35\linewidth]{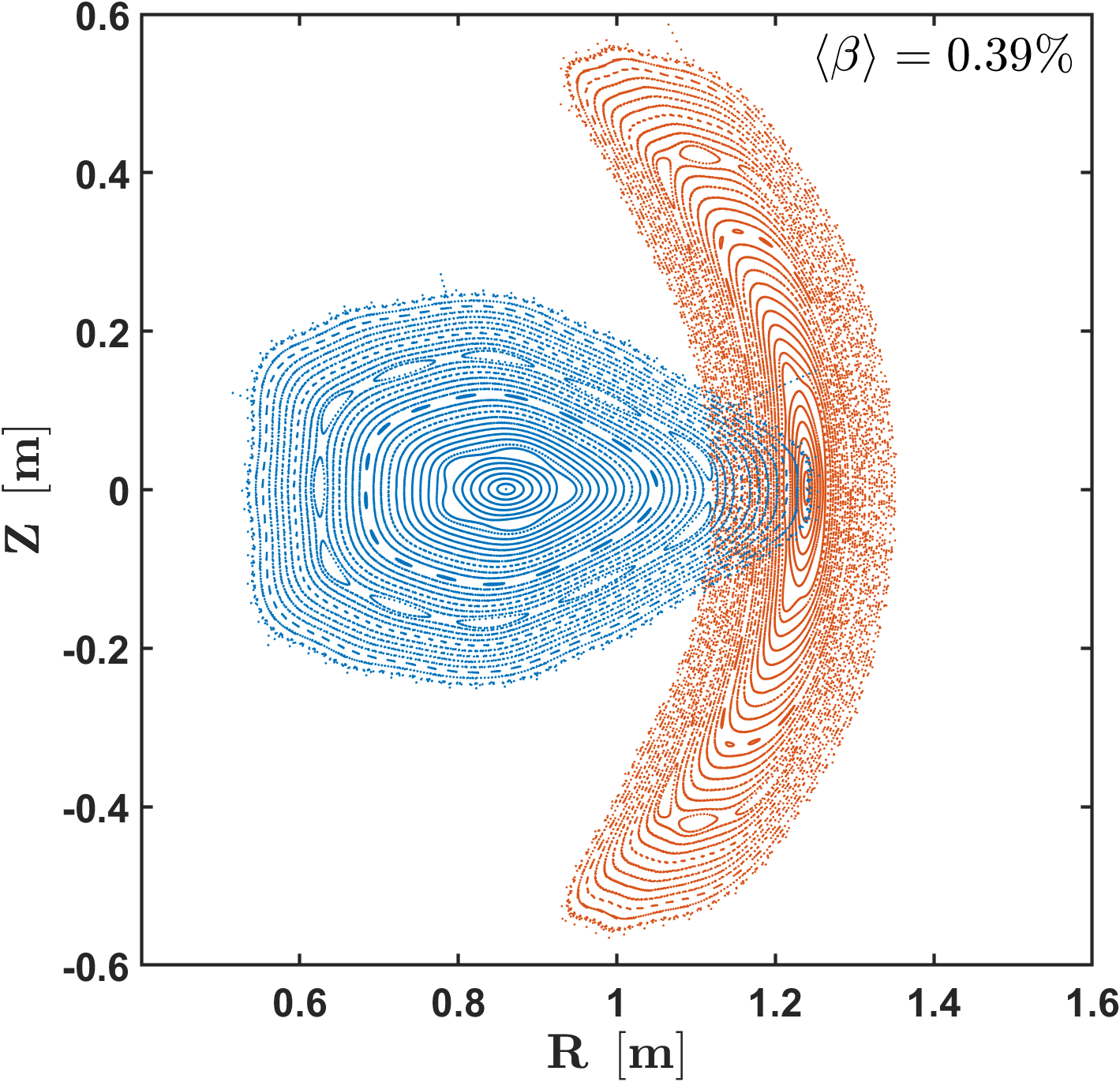}}
    \subfloat{\includegraphics[width=0.35\linewidth]{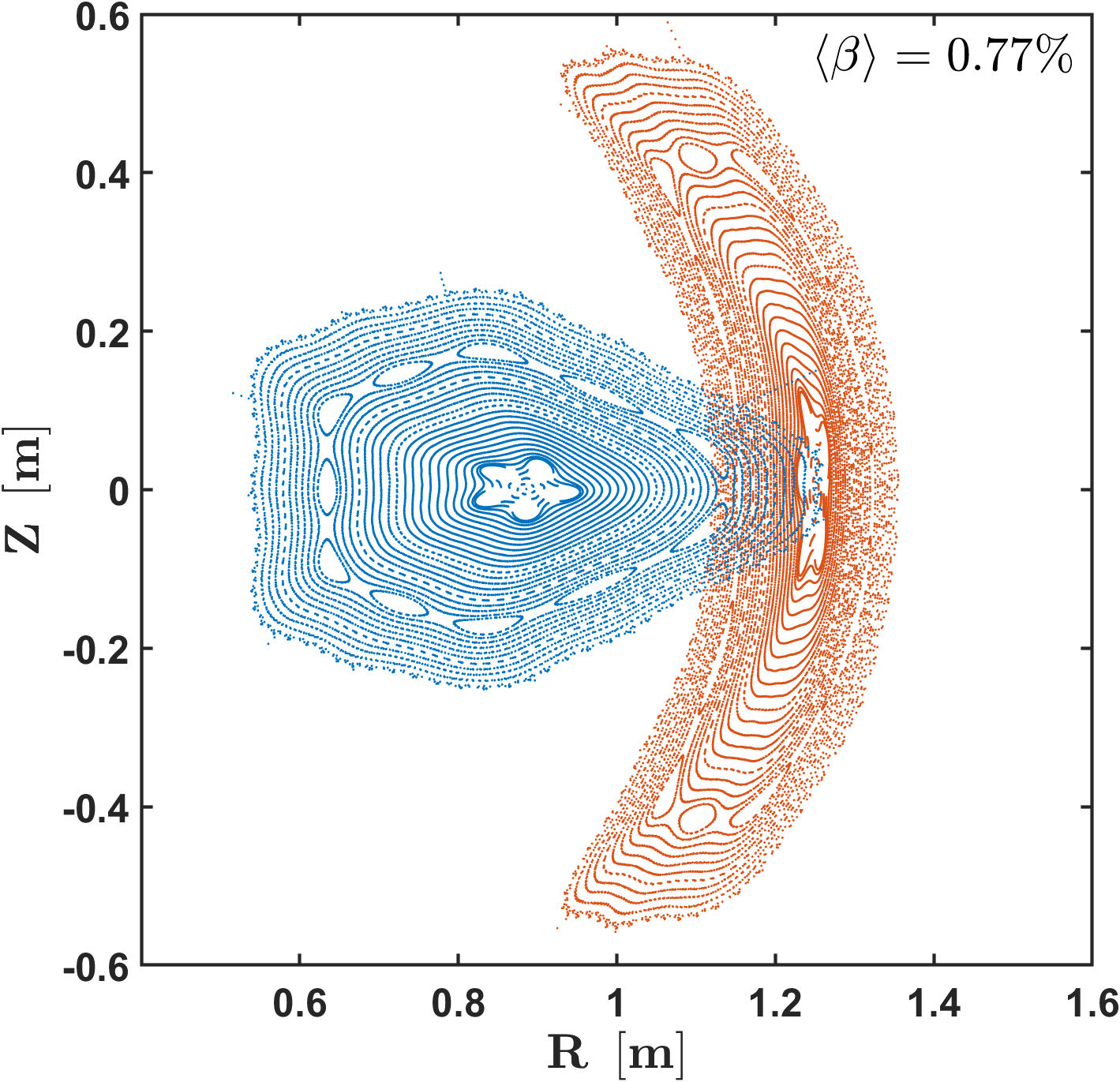}}
    \subfloat{\includegraphics[width=0.35\linewidth]{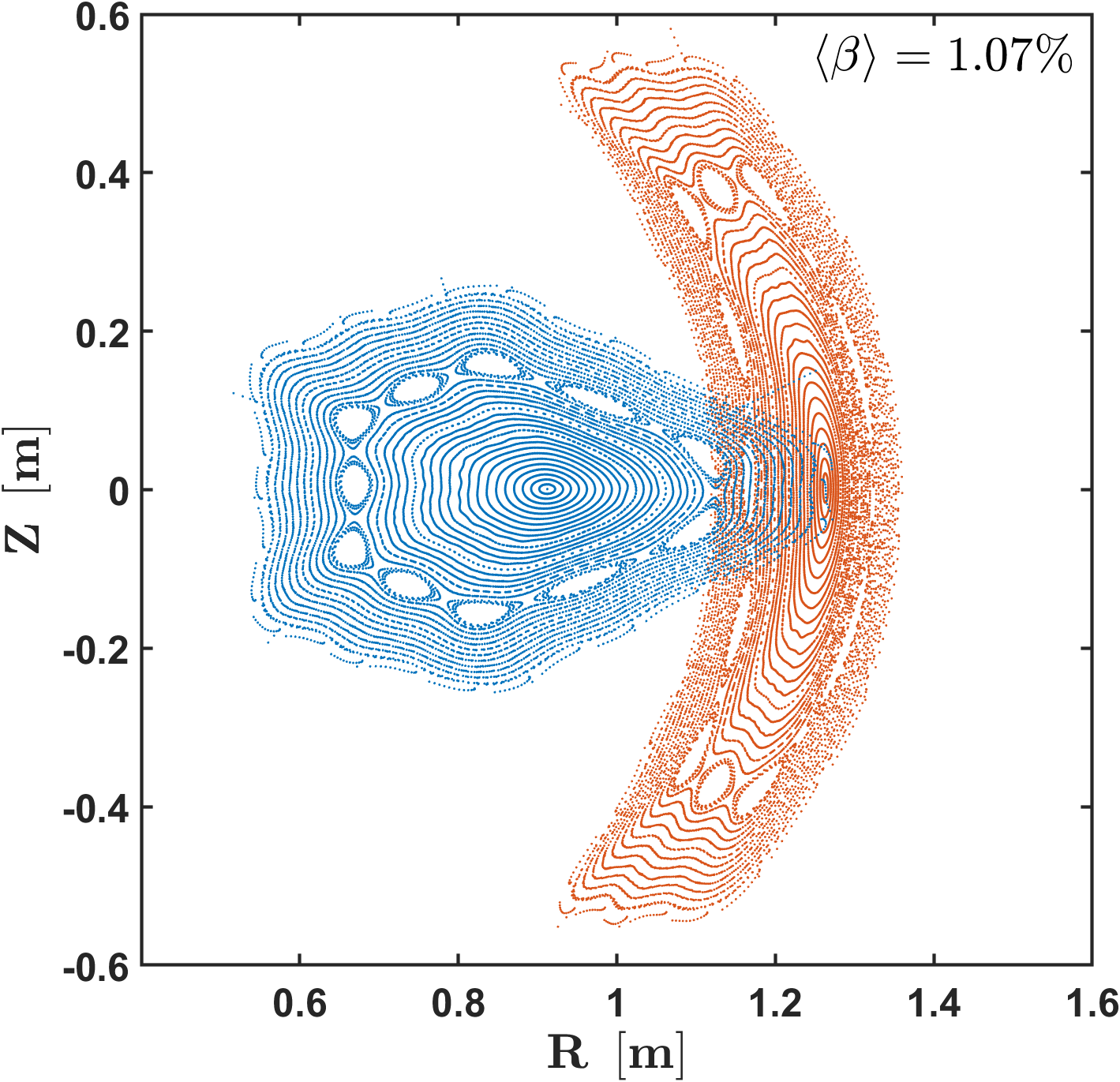}}\\ \vspace{-0.75em}
    \subfloat{\includegraphics[width=0.35\linewidth]{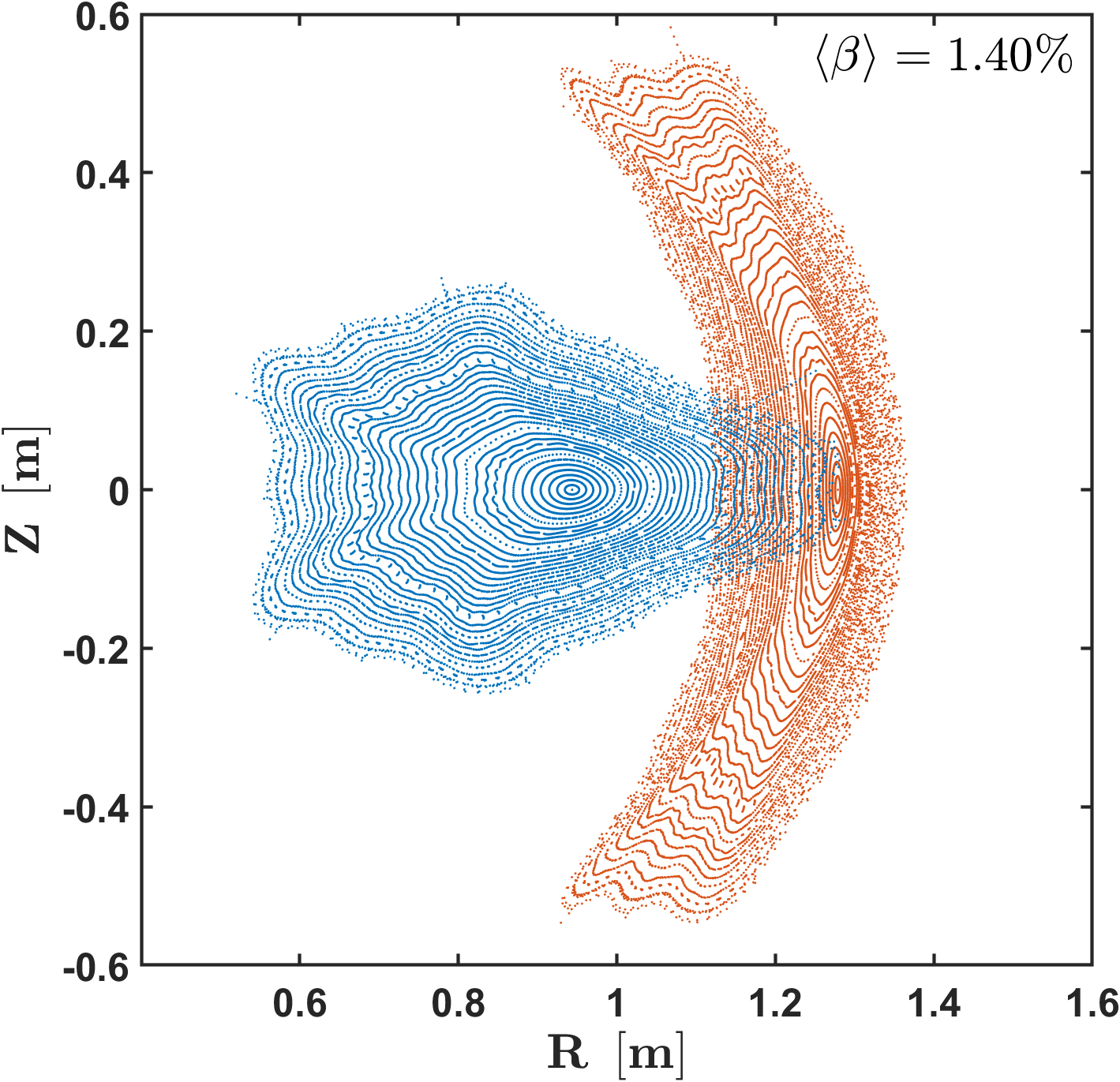}}
    \subfloat{\includegraphics[width=0.35\linewidth]{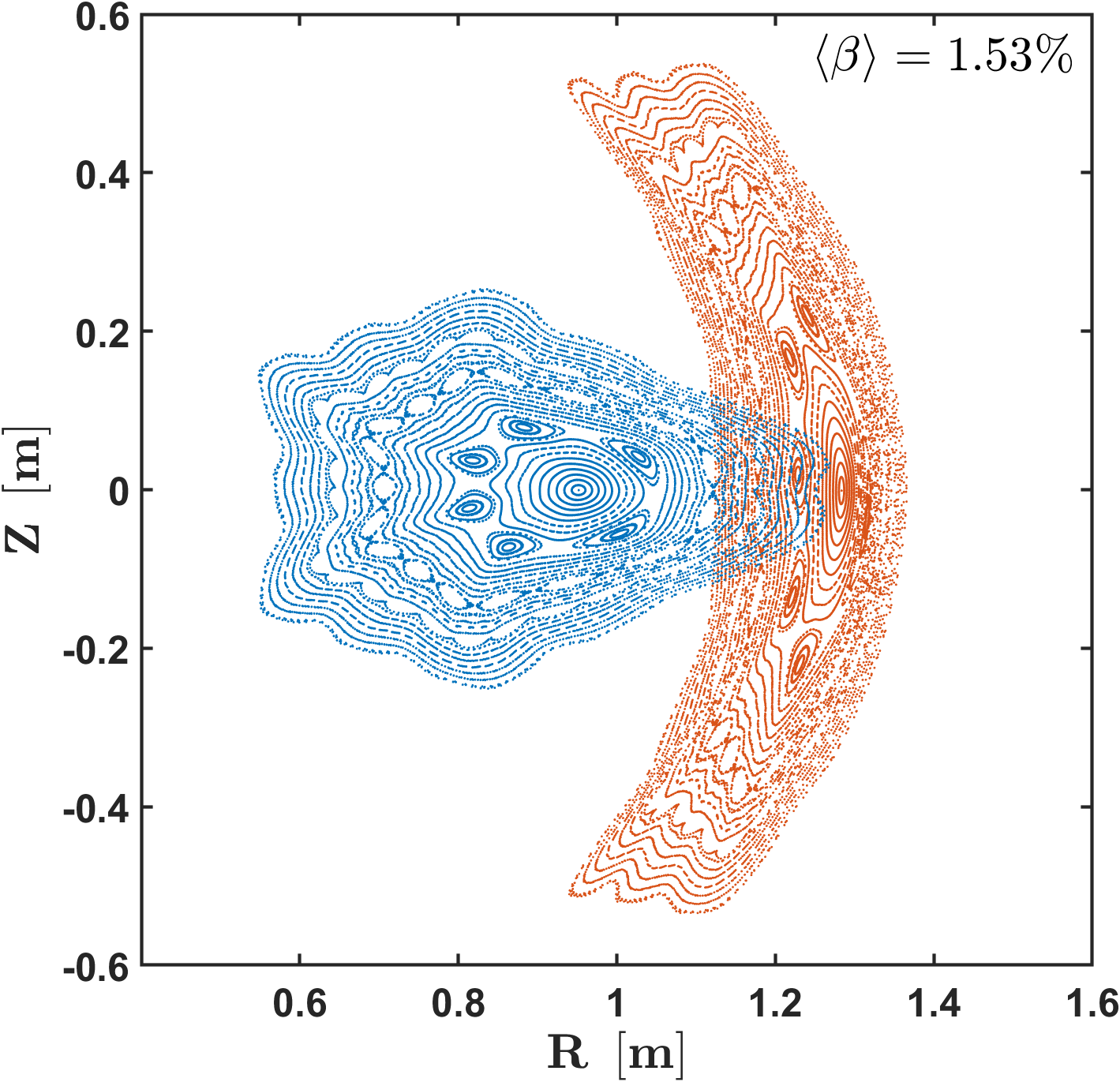}}
    \subfloat{\includegraphics[width=0.35\linewidth]{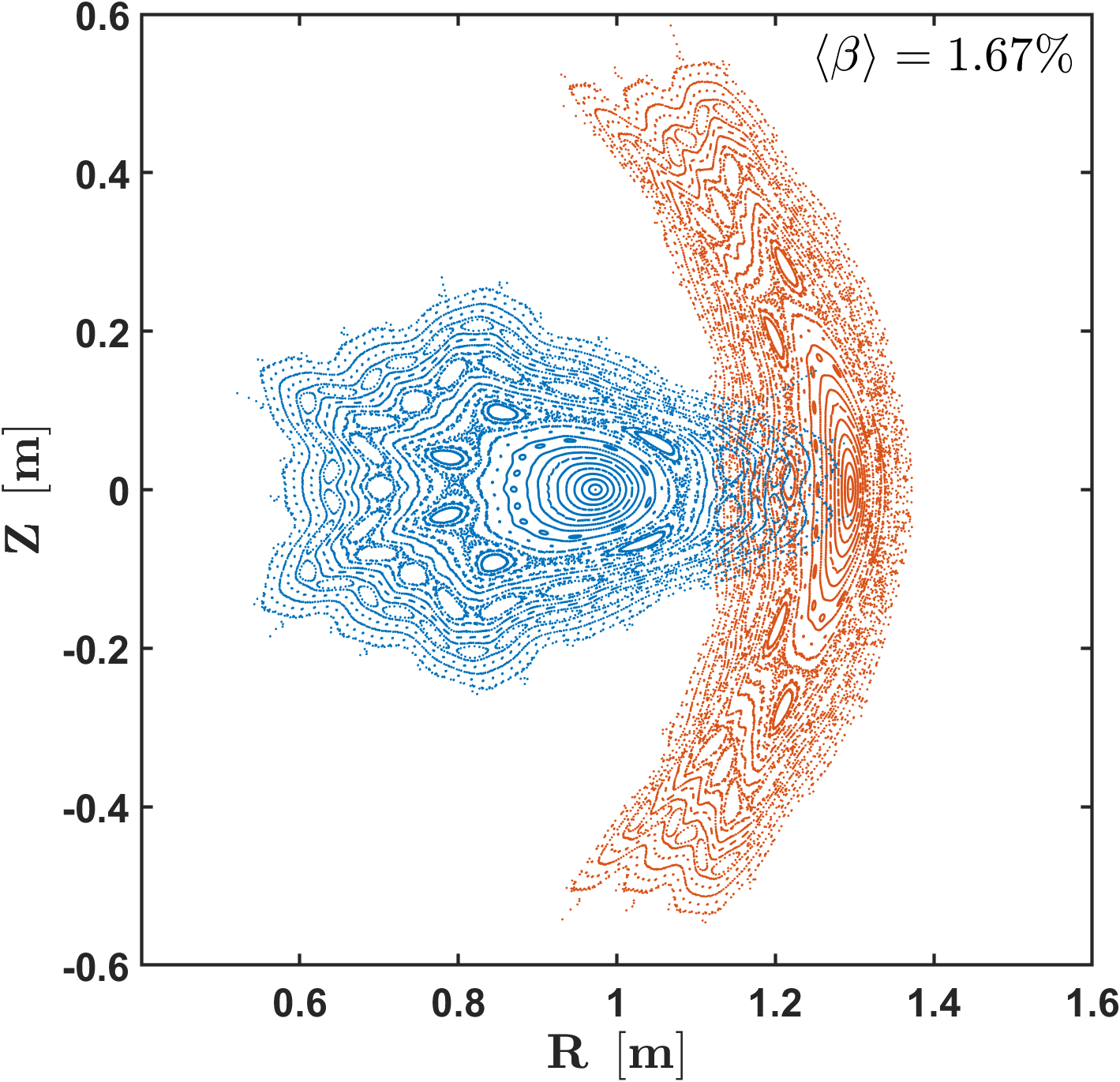}}\\ \vspace{-0.75em}
    \subfloat{\includegraphics[width=0.35\linewidth]{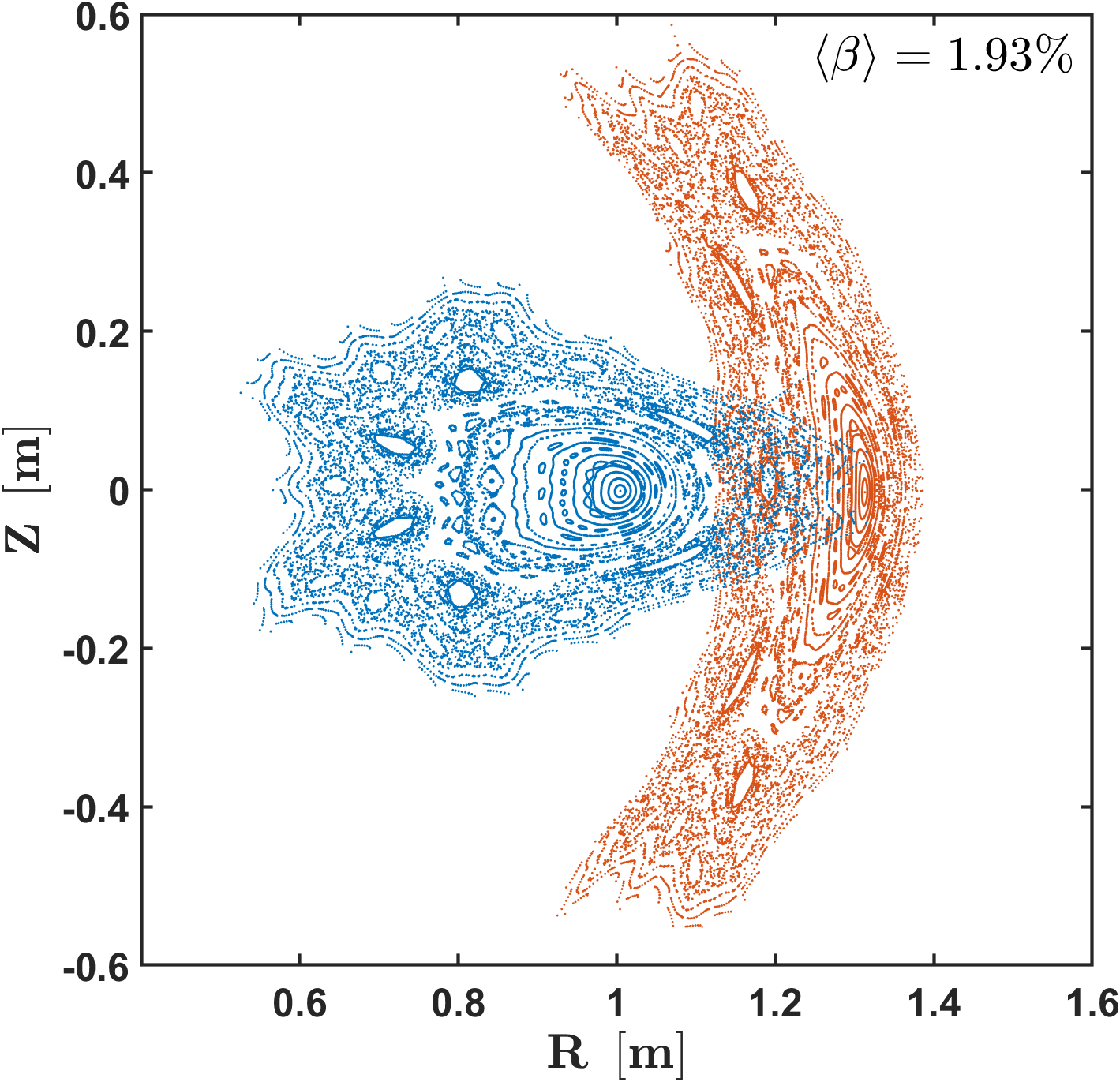}}
    \subfloat{\includegraphics[width=0.35\linewidth]{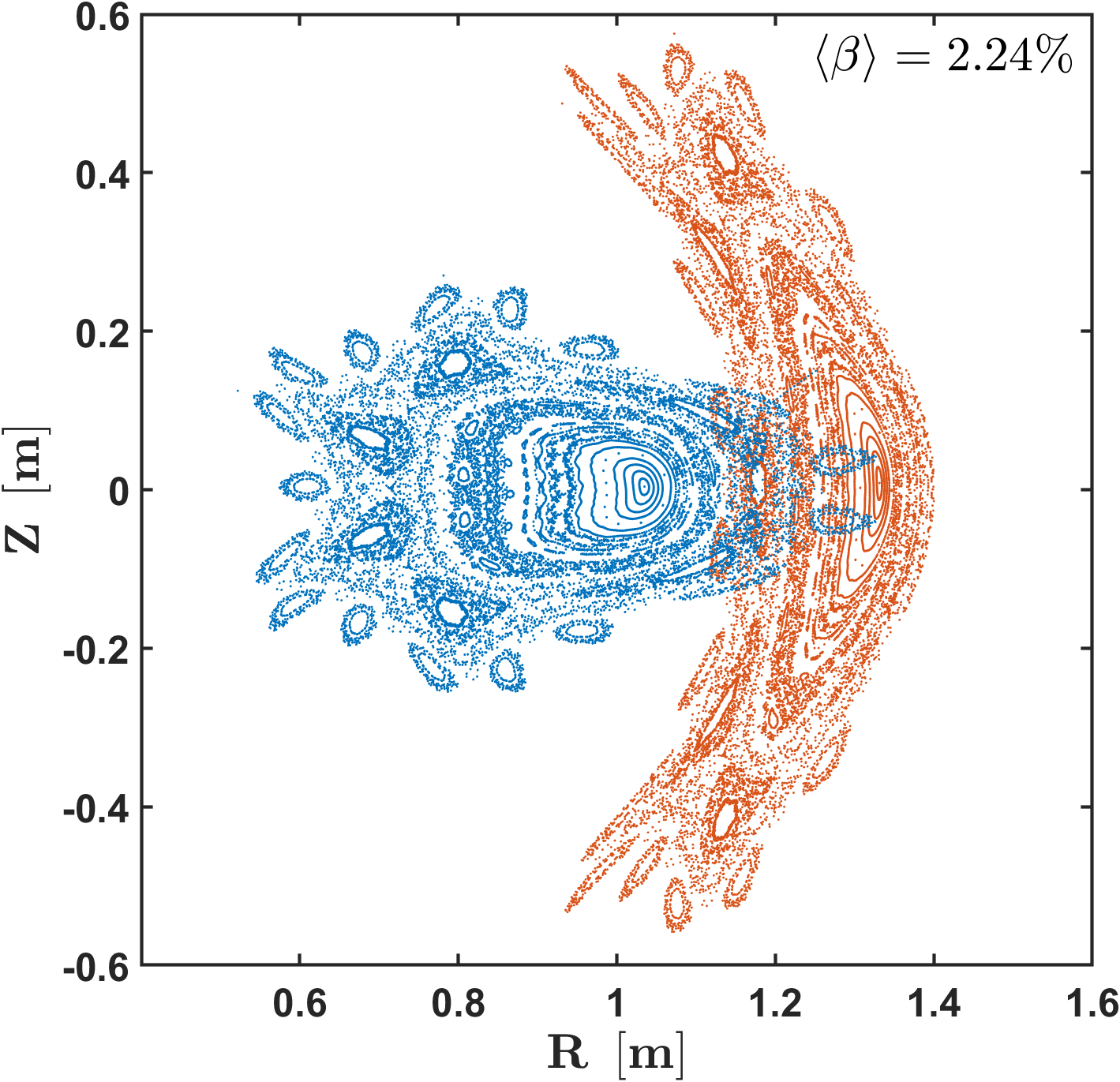}}
    \subfloat{\includegraphics[width=0.35\linewidth]{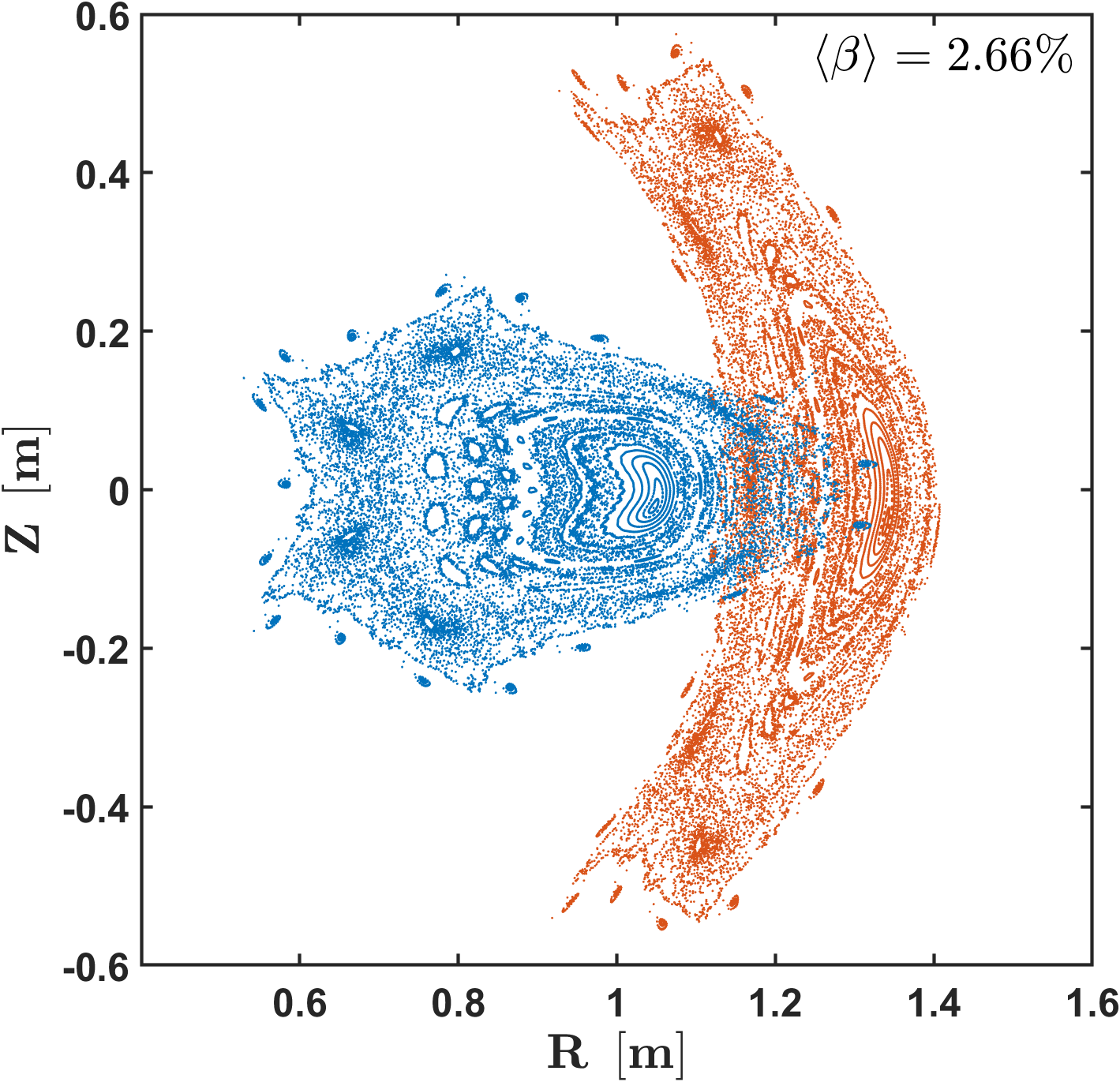}}
    \caption{Poincaré plots of the net-current-free finite-$\beta$ equilibria in the standard configuration at two cross sections 
             $\phi=0\:\text{(red)}\:\text{and}\:\pi/2\:\text{(blue)}$ with various volume-averaged $\langle\beta\rangle$.}
    \label{fig:poincare_QA}
\end{figure}

\begin{figure}[htbp]
    \centering
    \subfloat{\includegraphics[width=1\linewidth]{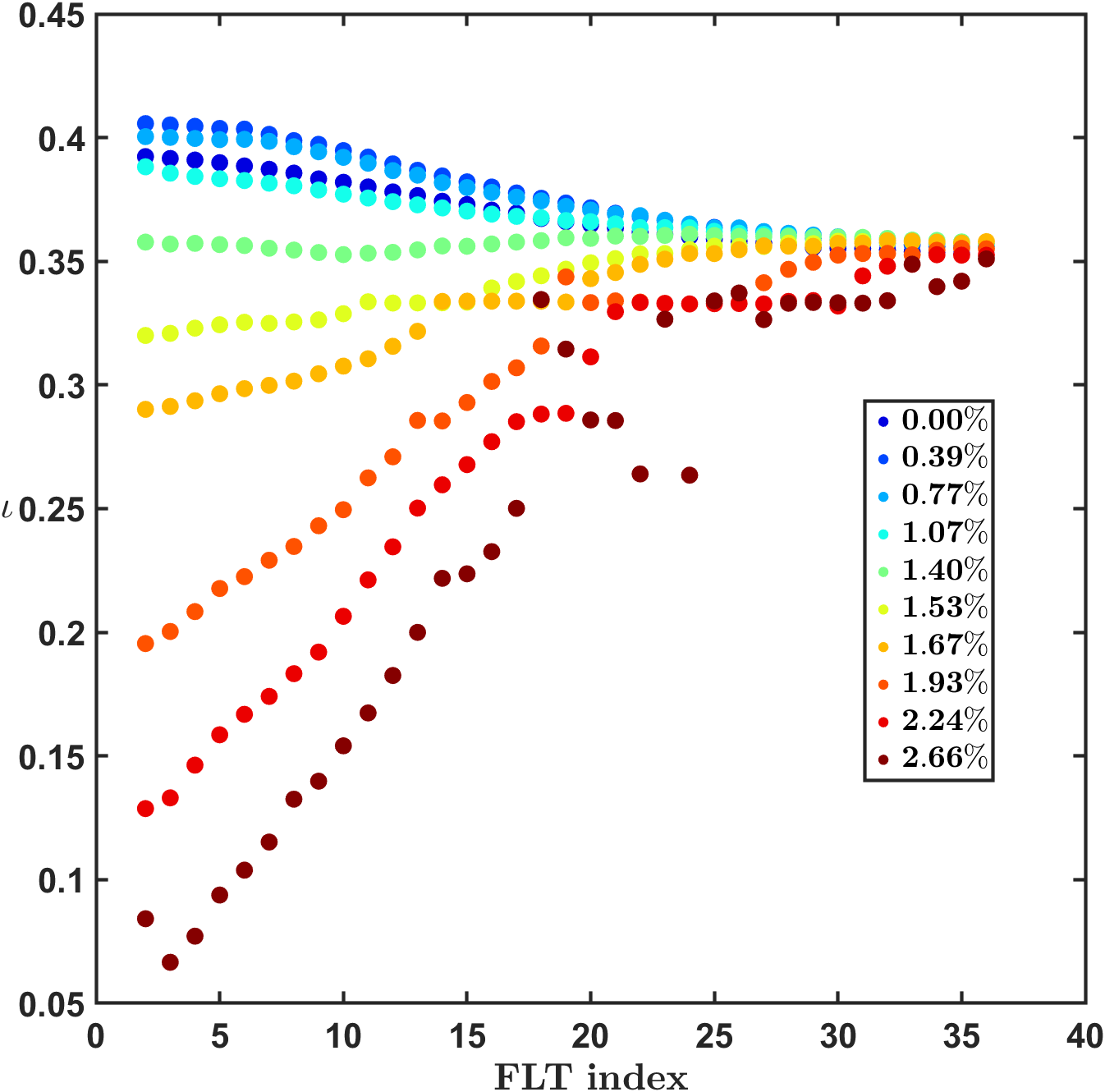}}
    \caption{Profiles of rotational transform $\iota$ with respect to the equilibria in \Cref{fig:poincare_QA} and the vacuum field. 
             Without flux surfaces, the rotational transform $\iota$ is calculated from field line tracing using 
             $\lim\limits_{\phi \to \infty} (\Delta\theta/\Delta\phi)$ where $\theta$ is poloidal angle corresponding to the toroidal angle $\phi$. 
             The abscissa represents the index of field line tracing from the magnetic axis to the plasma boundary.}
    \label{fig:beta_vs_iota}
\end{figure}

In this subsection, we examine the standard configuration, in which the vacuum field is produced by the MCs, 
for a set of net-current-free equilibria ranging from low to high $\langle\beta\rangle$. 
\Cref{fig:poincare_QA} shows Poincaré plots for the relevant equilibria solved using NTEC. 
Evidently, in cases with $\langle\beta\rangle<1.53\%$, nested flux surfaces persist and dominate the plasma regions 
in the presence of only thin layers of island chains. As $\langle\beta\rangle$ goes up, 
the majority of the regular regions become chaotic due to island overlap. 
The rotational transform $\iota$ profiles of the equilibria, displayed in \Cref{fig:beta_vs_iota}, 
help reveal the magnetic topology changes. The $\iota$ profile in low-$\beta$ equilibria does not change significantly compared to the vacuum field. 
The minimum value of $\iota$ keeps to be above $1/3$. In the equilibrium of $\langle\beta\rangle=0.39\%$, 
the rotational transform near the magnetic axis slightly passes through $2/5$, resulting in a small $n/m=2/5$ island chain. 
For the case with $\langle\beta\rangle=0.77\%$, the innermost plasma region has a nearly flat $\iota$ close to the rational value $2/5$, 
which partially relaxes the nested flux surfaces. Therefore, by and large, low-$\beta$ equilibria retain the design feature 
where $\iota$ varies within $[1/3,2/5]$ to avoid low-order rational surfaces. However, after $\langle\beta\rangle$ reaches $1.53\%$, 
$\iota$ in the core area drastically decreases, and in the intermediate zone $\iota$ is flattened to the value of $1/3$. 
Near axis, islands are suppressed by large magnetic shear and thus flux surfaces still survive. 
From the intermediate zone to the edge, high-order islands overlap with the main $n/m=2/6$ islands to bring about the stochastization of field lines. 
For instance, in the case with $\langle\beta\rangle=1.67\%$, the overlap of the $n/m=2/6$ island chain and another $n/m=6/17$ 
renders stochastic field lines near the X points of the $n/m=2/6$ islands. Additionally, previous equilibrium calculations using
HINT also found that large $n/m=2/6$ islands come into existence within the plasma region at $\langle\beta\rangle=1.58\%$ \cite{Wang_2021, Zhang_2026}. 

\begin{figure}[htbp]
    % \vspace{-7em}
    \centering
    \subfloat{\includegraphics[width=0.35\linewidth]{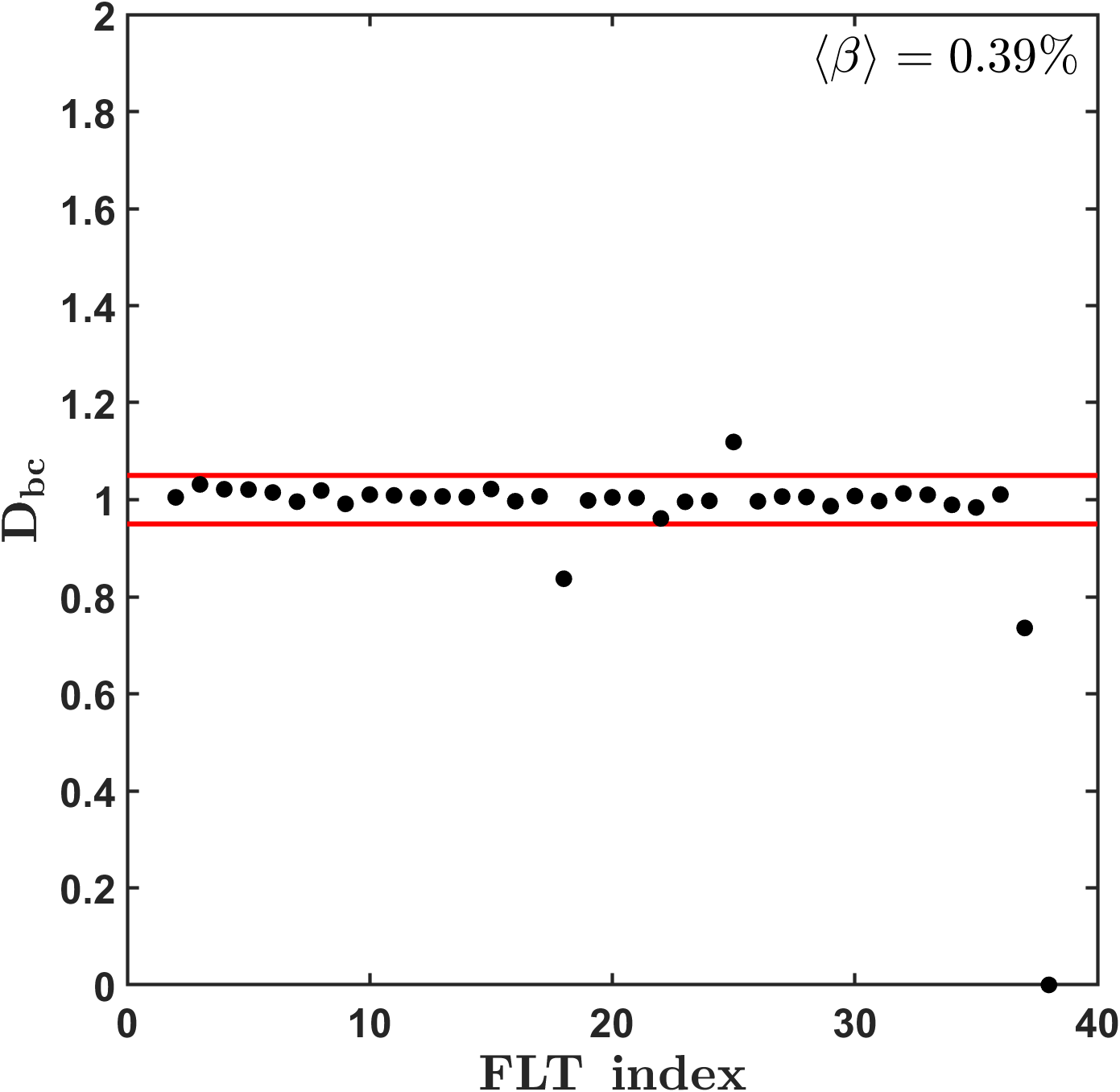}} \qquad
    \subfloat{\includegraphics[width=0.35\linewidth]{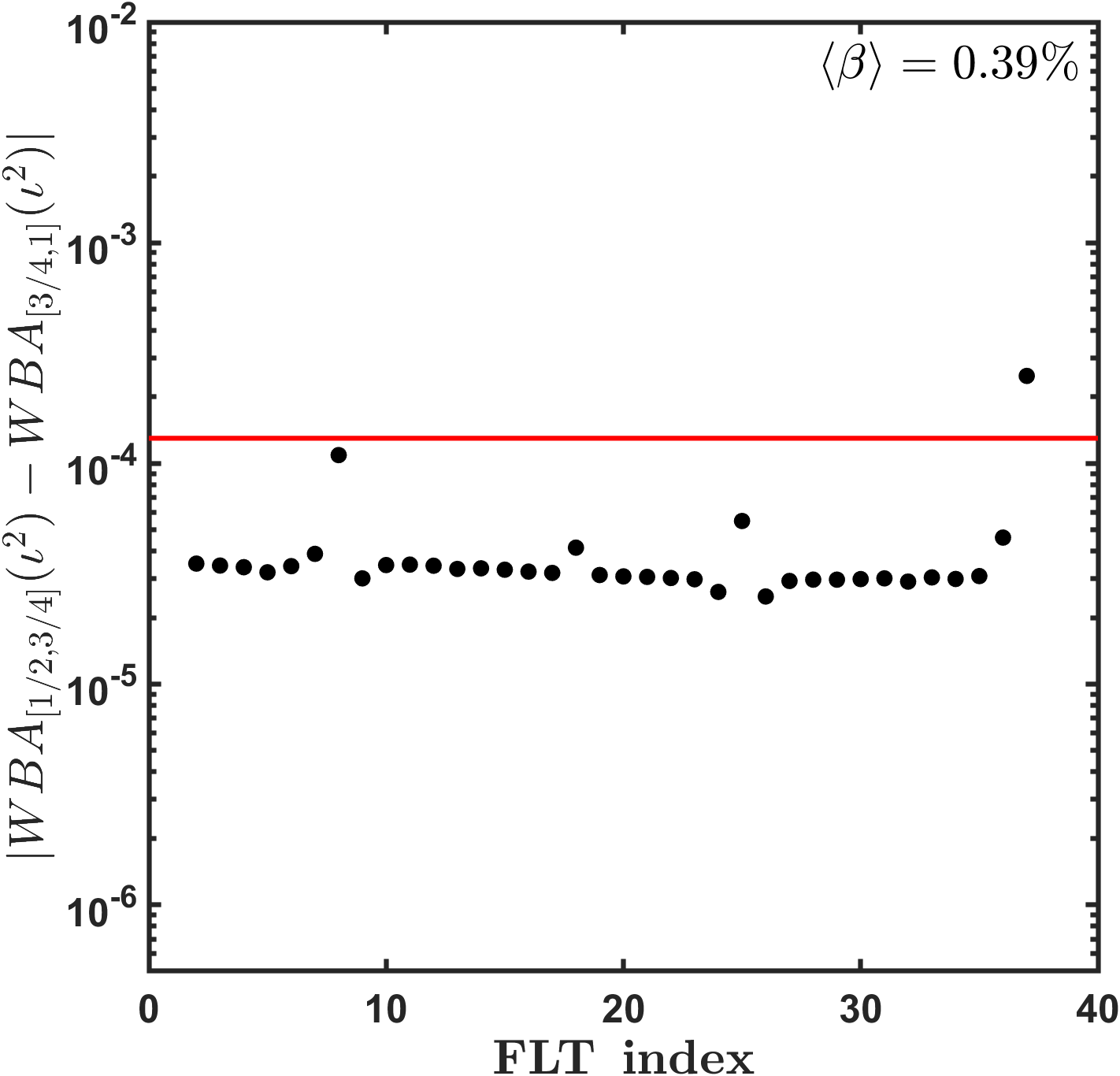}}\\ \vspace{-0.75em}
    \subfloat{\includegraphics[width=0.35\linewidth]{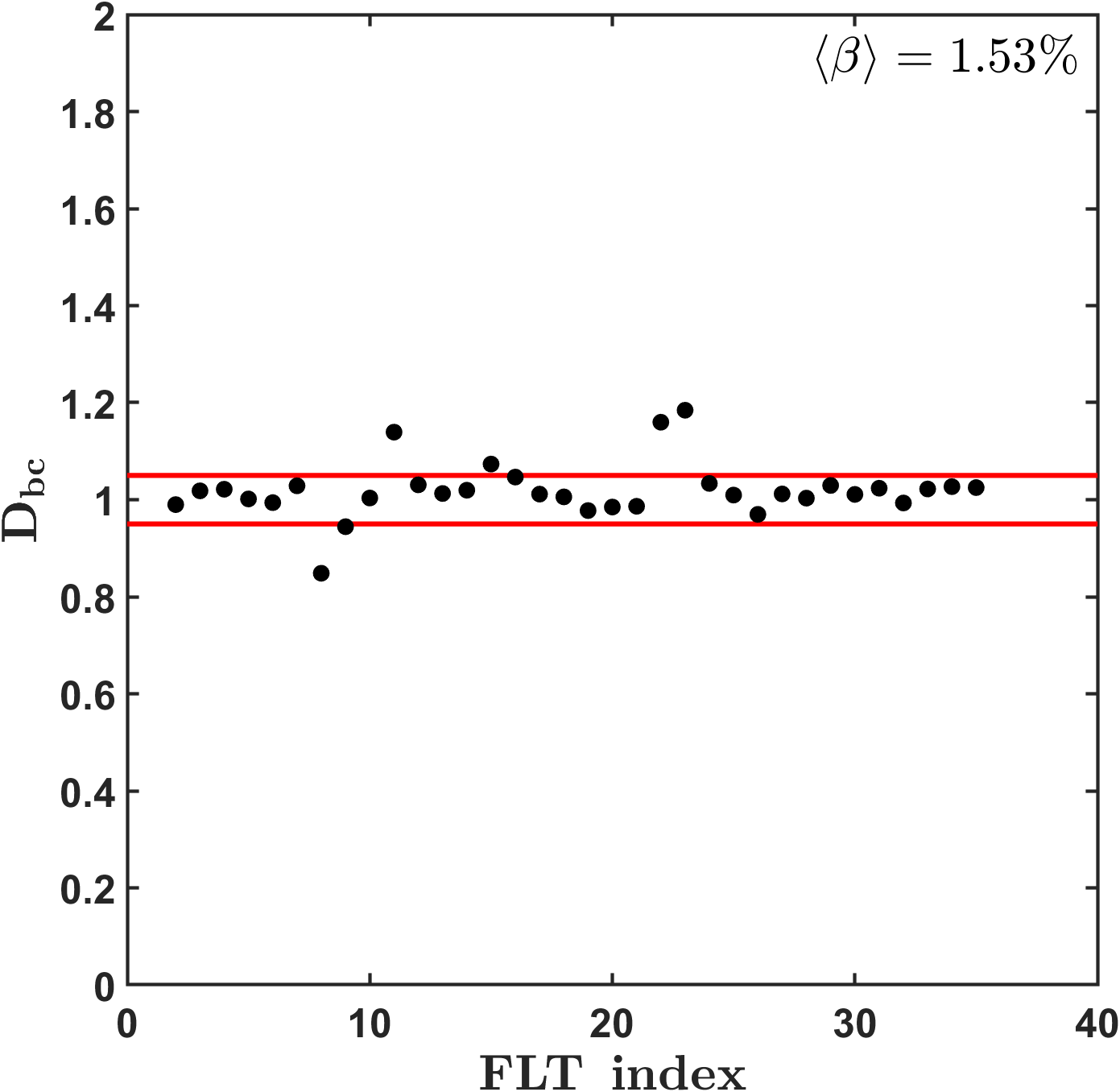}} \qquad
    \subfloat{\includegraphics[width=0.35\linewidth]{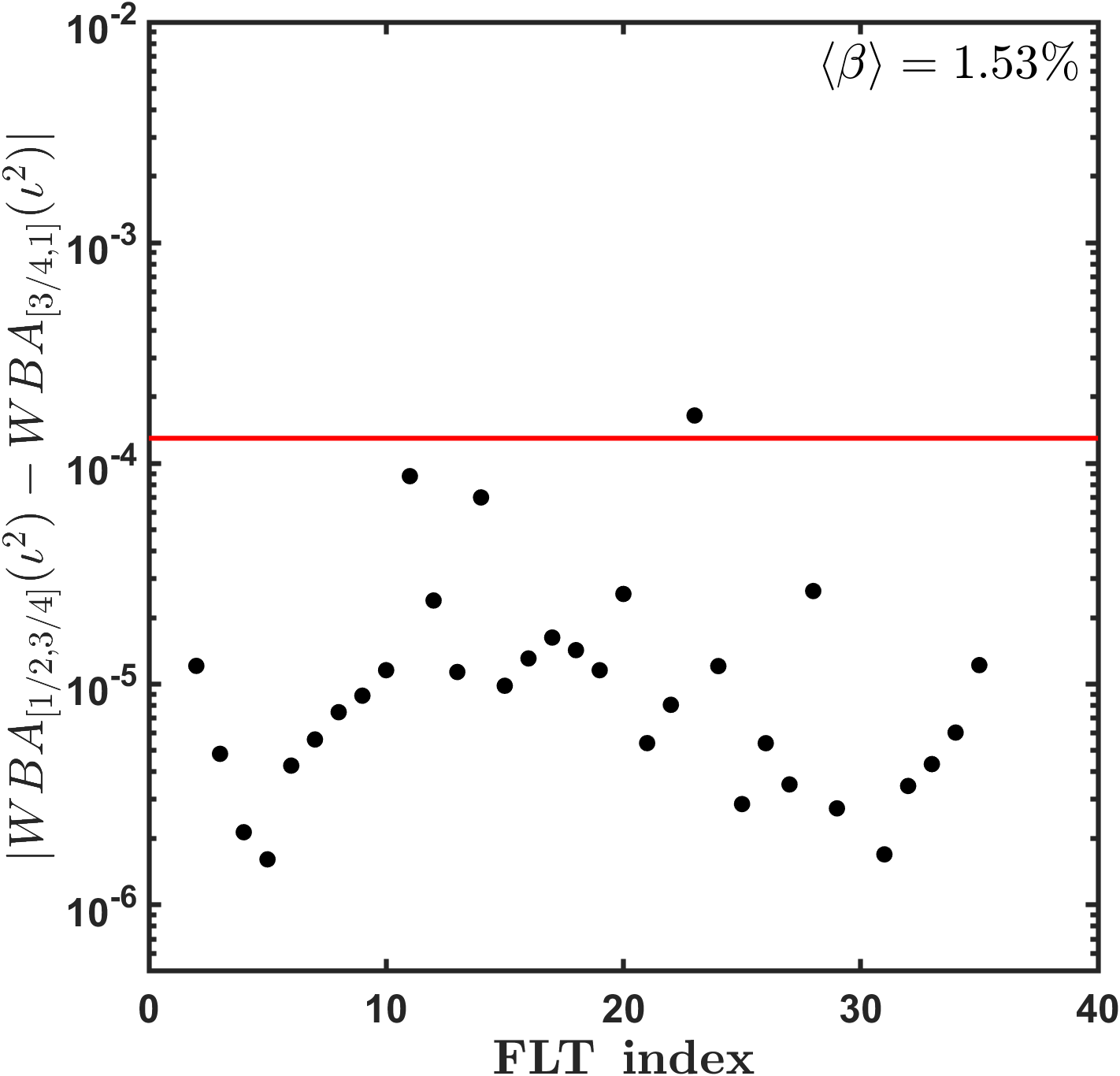}}\\ \vspace{-0.75em}
    \subfloat{\includegraphics[width=0.35\linewidth]{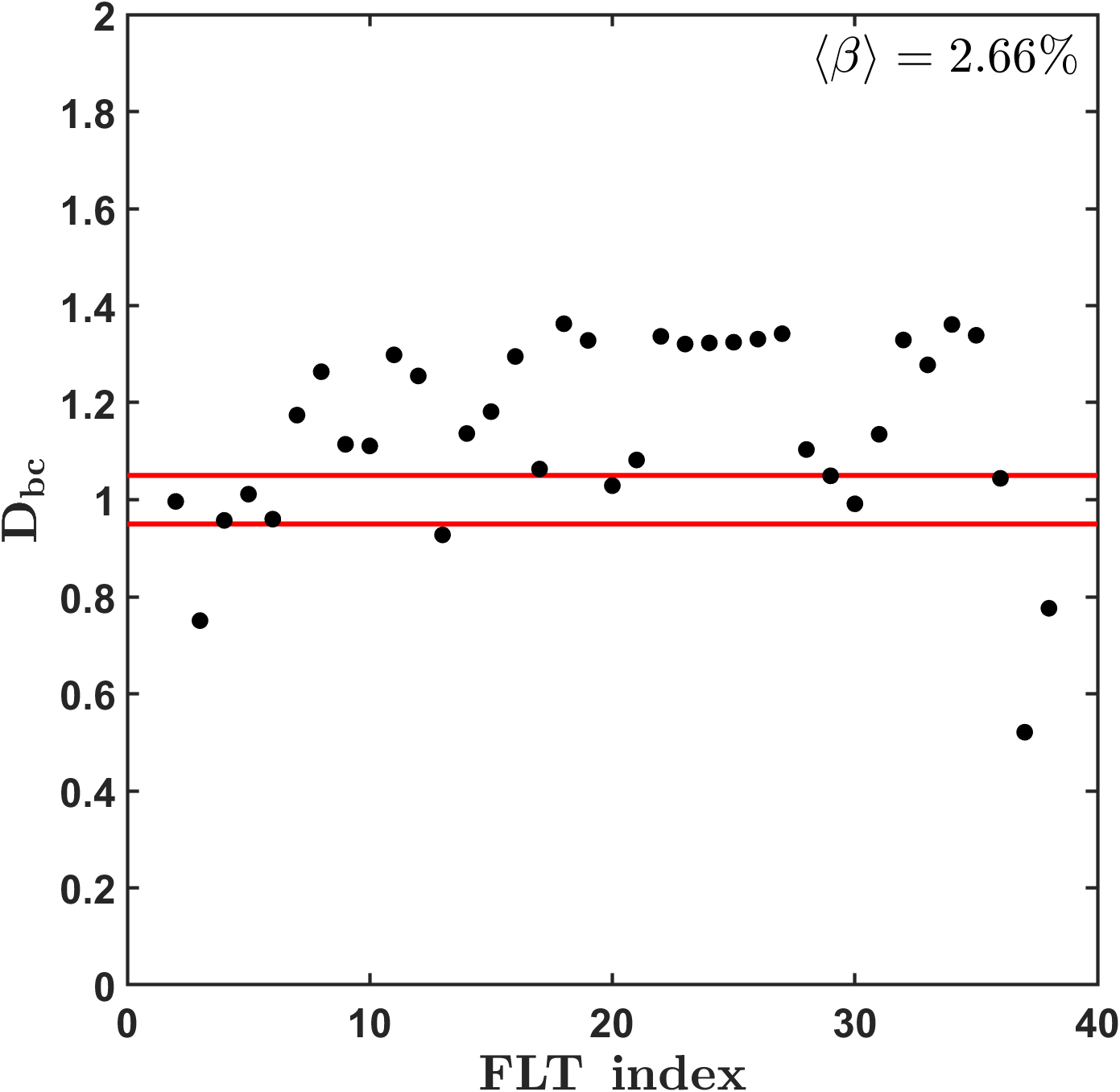}} \qquad
    \subfloat{\includegraphics[width=0.35\linewidth]{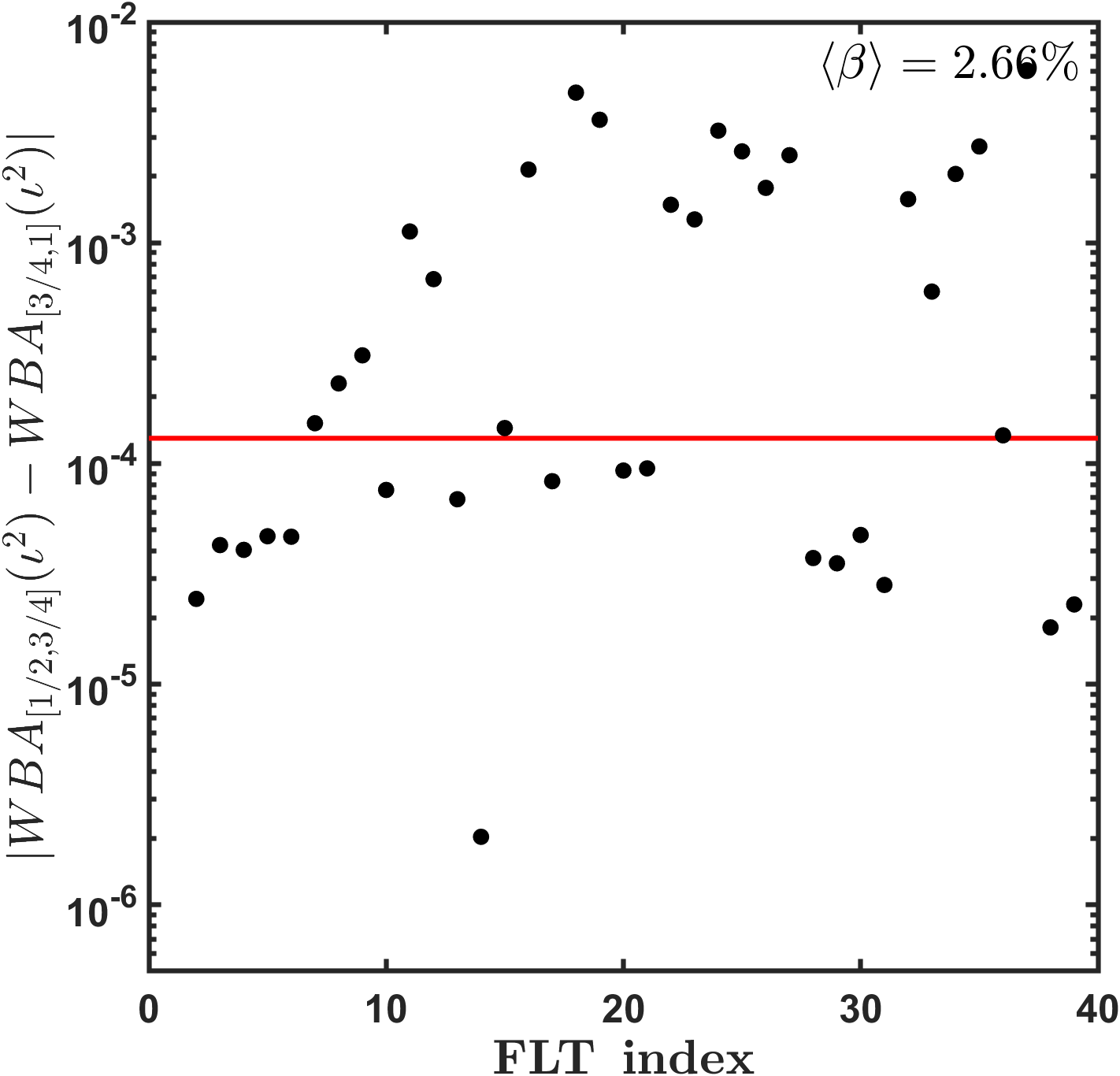}}
    \caption{Box-counting dimension (left column) and Weighted Birkhoff average (right column) for net-current-free finite-$\beta$ equilibria in the standard configuration with various $\langle\beta\rangle$. 
             The double red lines in the $\text{D}_{bc}$-plots indicate the range $[0.95,1.05]$, and the single red lines in the WBA-plots indicate the value $1.3\times10^{-4}$. 
             The abscissa represents the index of field line tracing from the magnetic axis to the plasma boundary.}
    \label{fig:BC_WBAs}
\end{figure}

\begin{figure}[htbp]
    % \vspace{-7em}
    \captionsetup[subfloat]{position=top,labelformat=empty}
    \centering
    \subfloat[$\text{D}_{bc}$]{\includegraphics[width=0.35\linewidth]{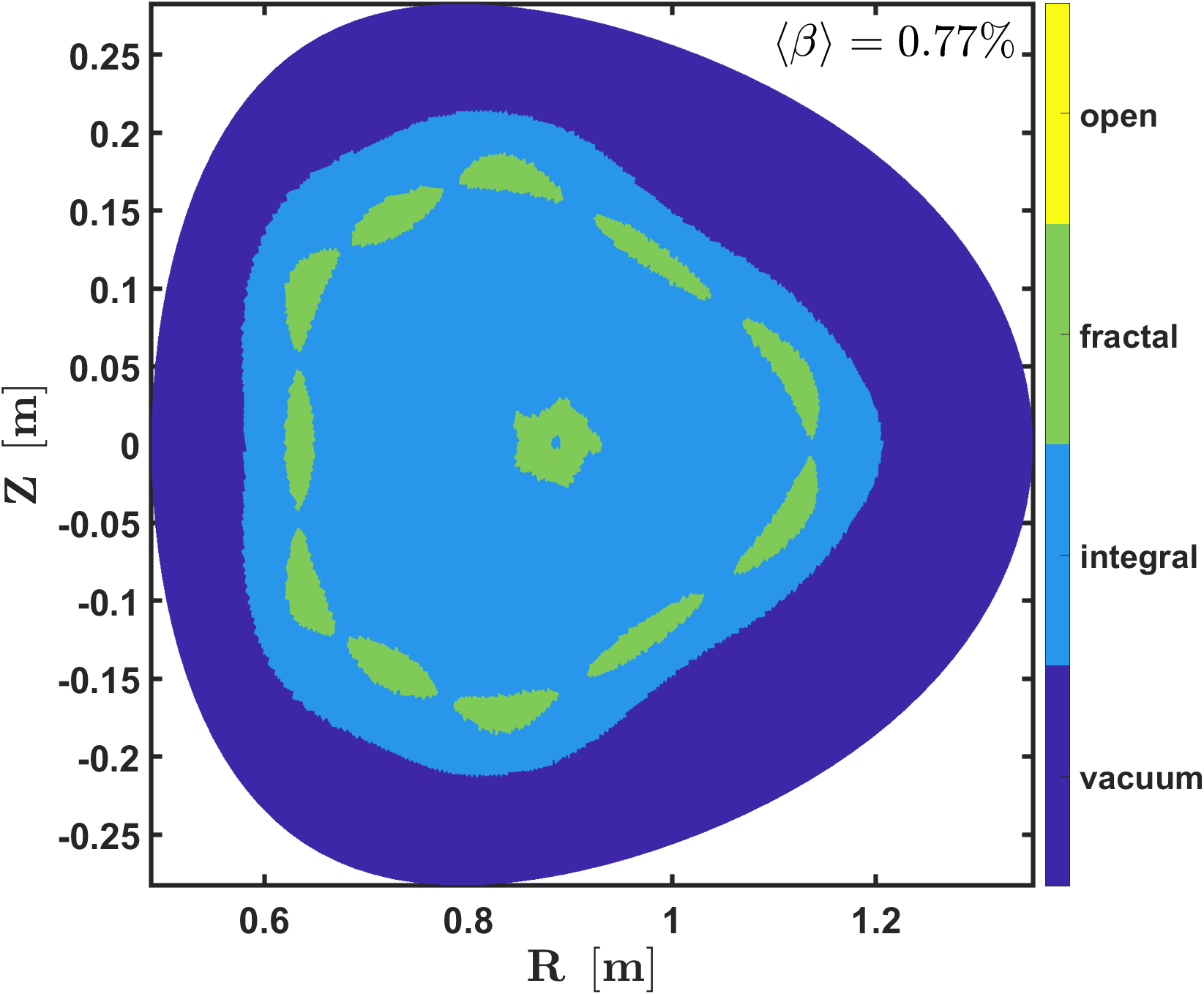}} \qquad
    \subfloat[WBA]{\includegraphics[width=0.35\linewidth]{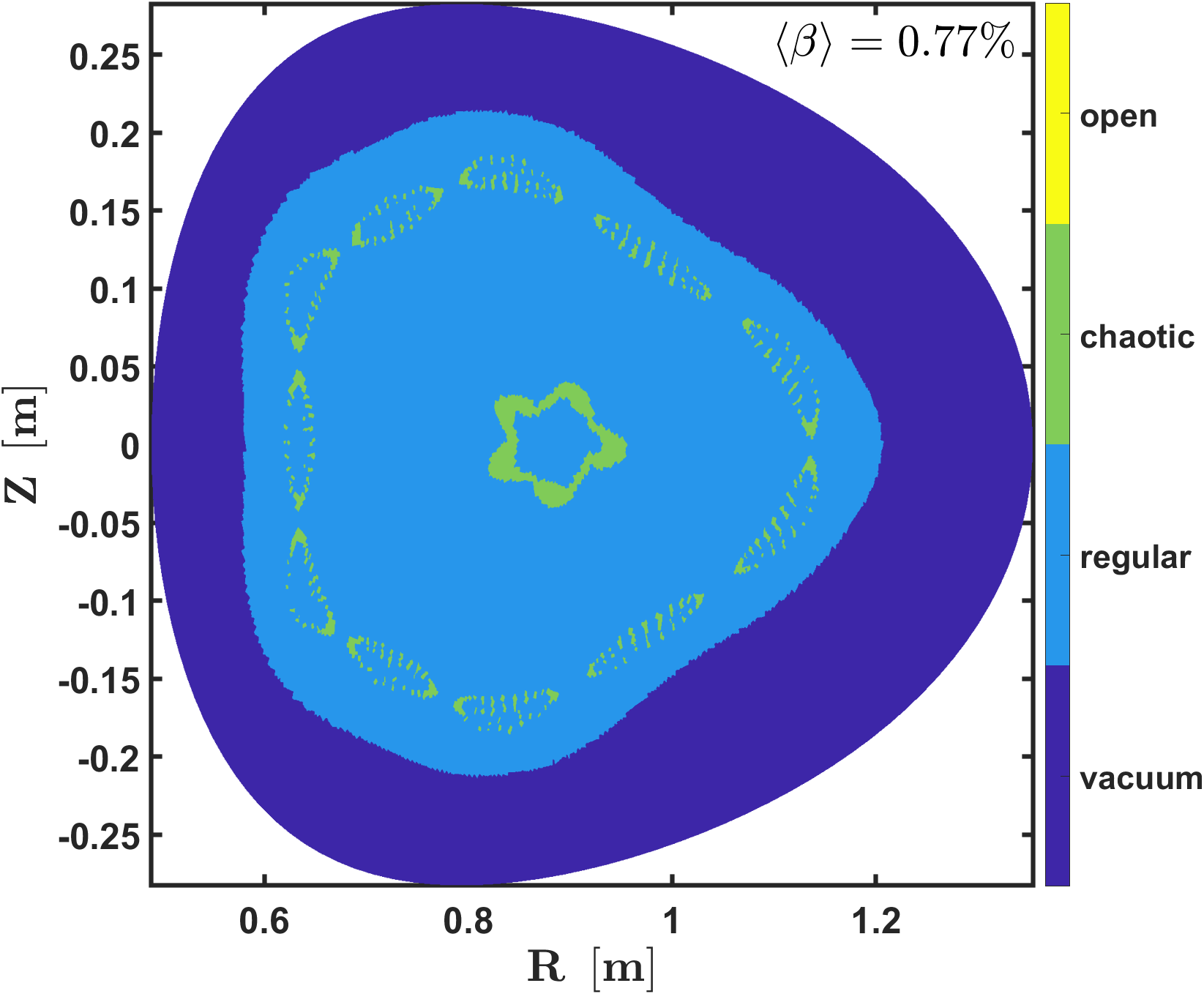}}\\ \vspace{-0.75em}
    \subfloat{\includegraphics[width=0.35\linewidth]{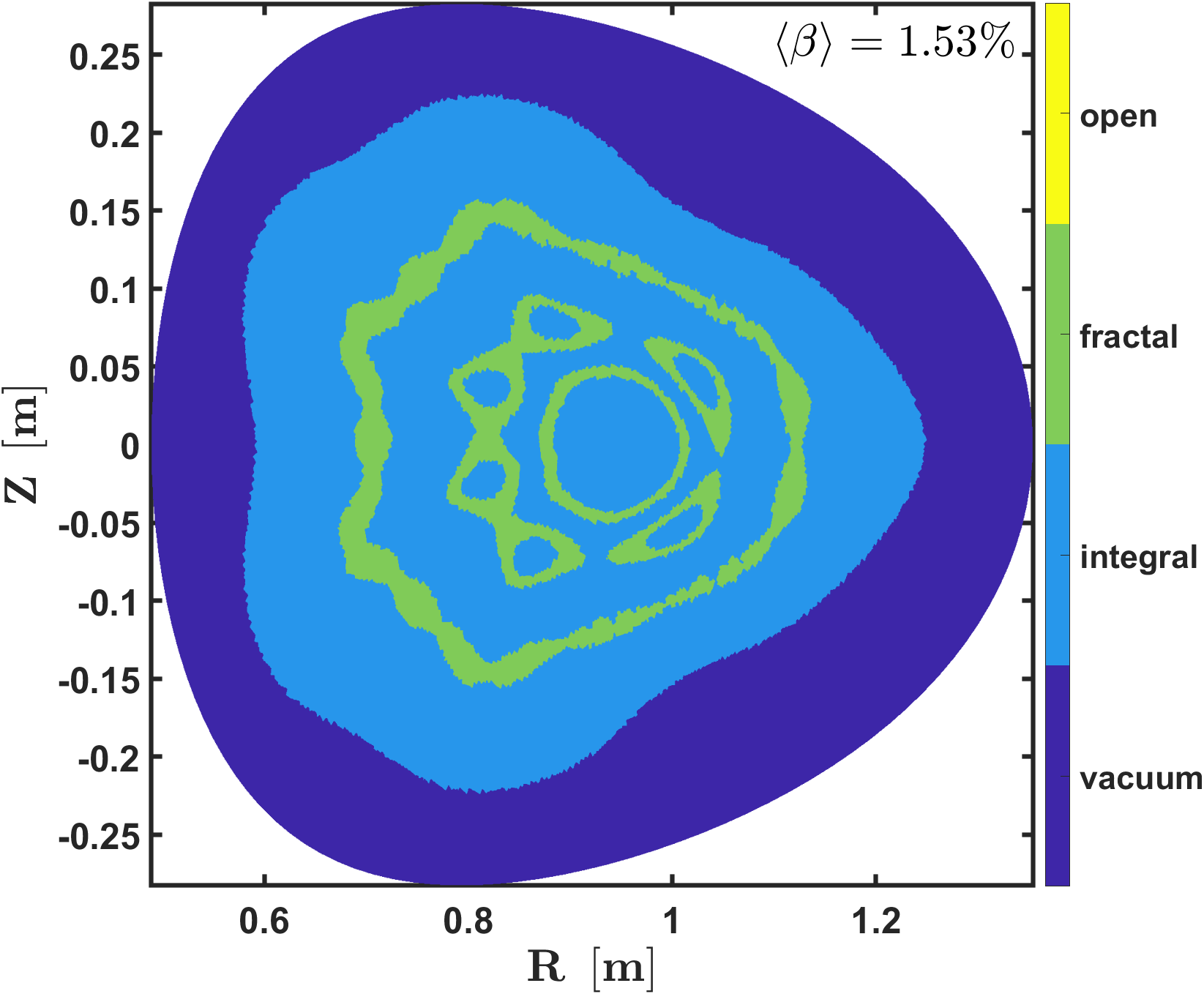}} \qquad
    \subfloat{\includegraphics[width=0.35\linewidth]{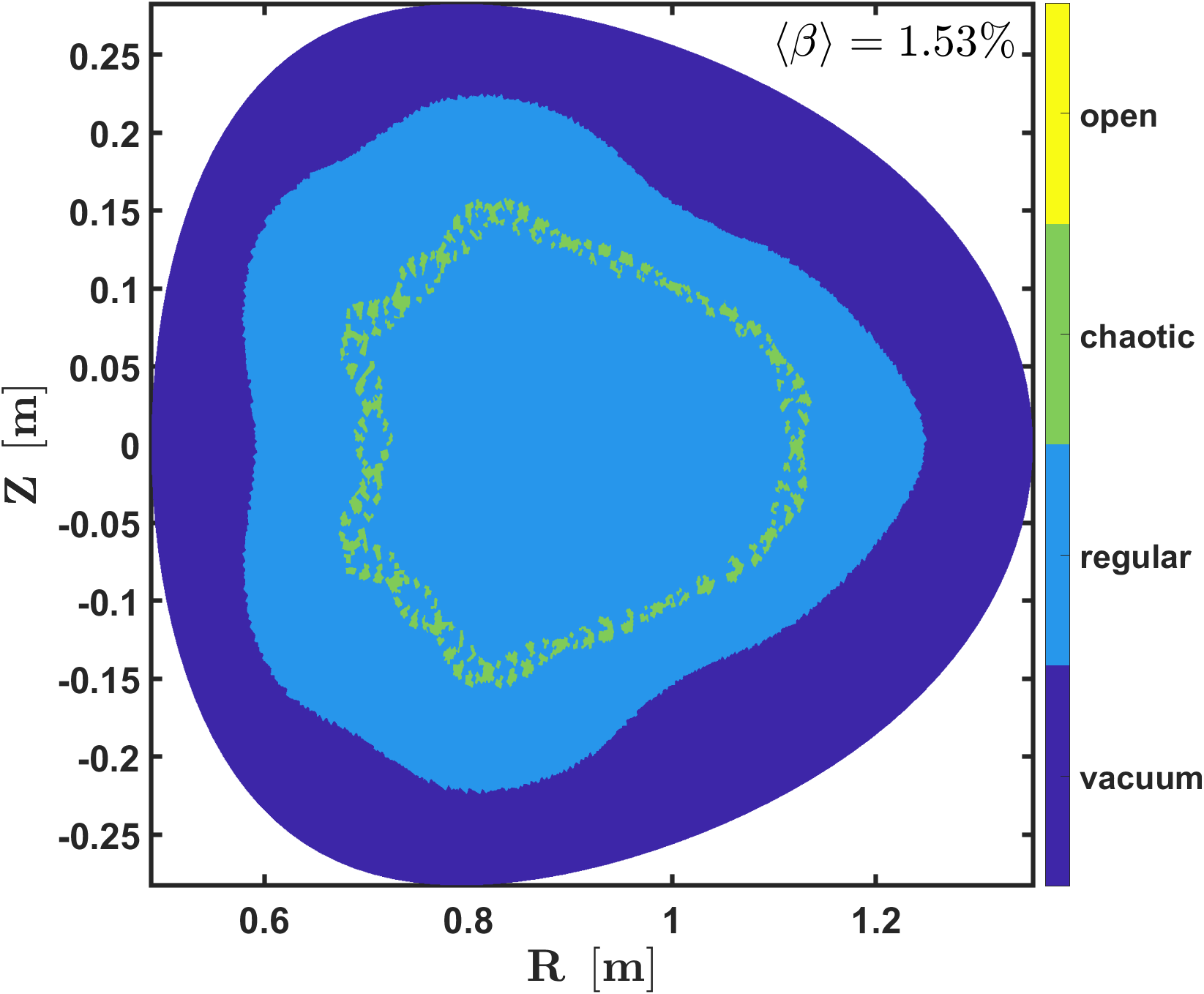}}\\ \vspace{-0.75em}
    \subfloat{\includegraphics[width=0.35\linewidth]{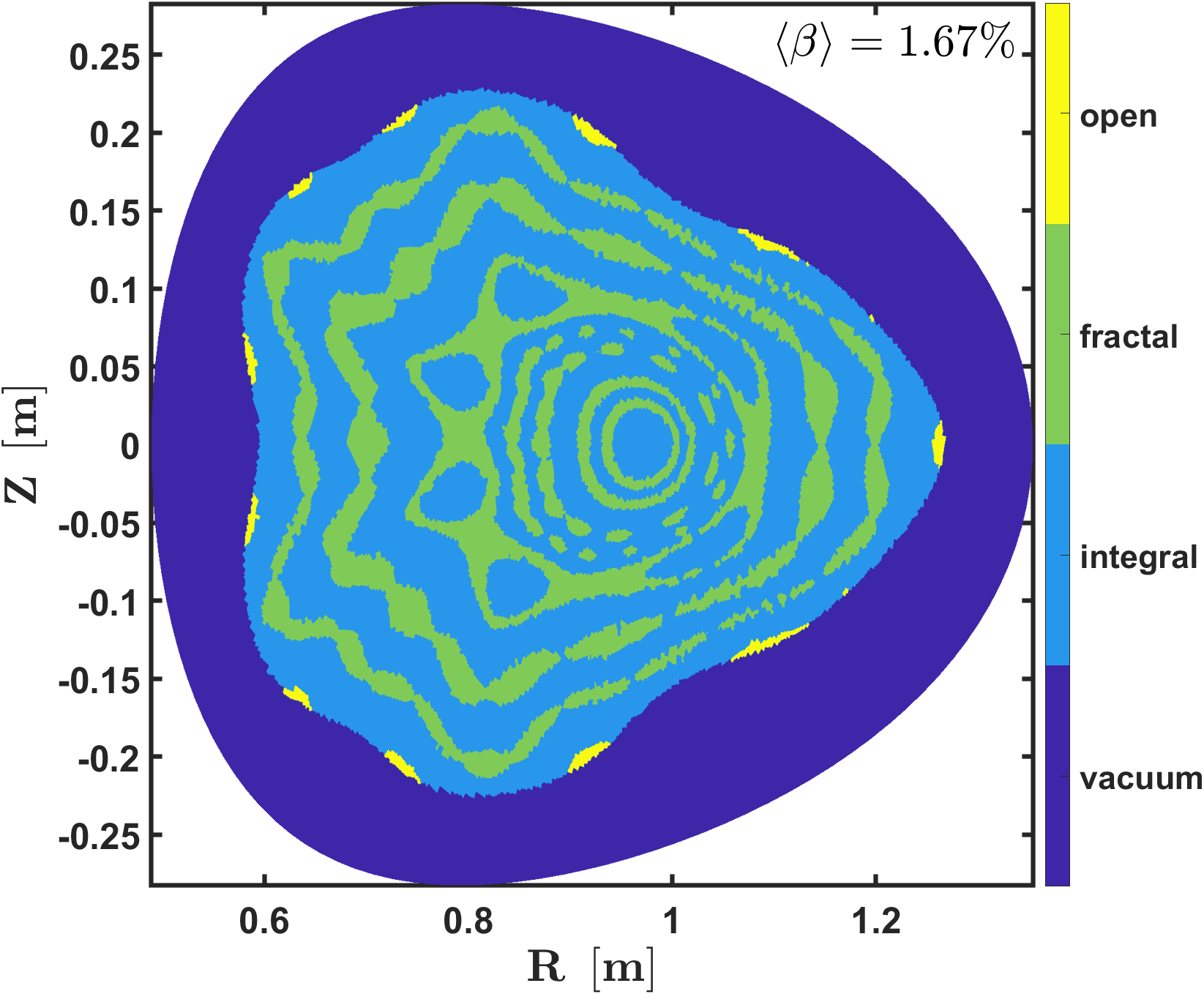}} \qquad
    \subfloat{\includegraphics[width=0.35\linewidth]{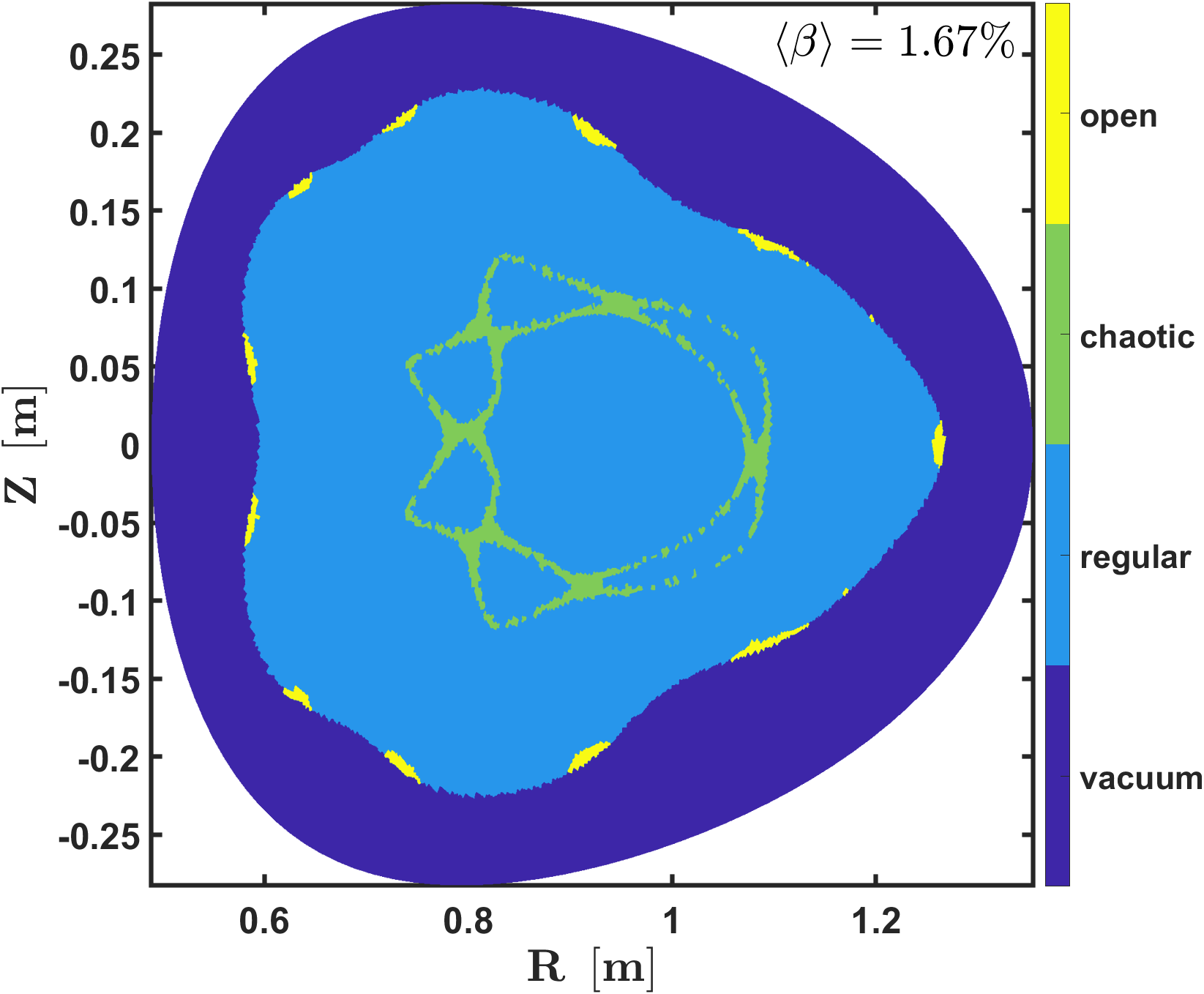}}\\ \vspace{-0.75em}
    \subfloat{\includegraphics[width=0.35\linewidth]{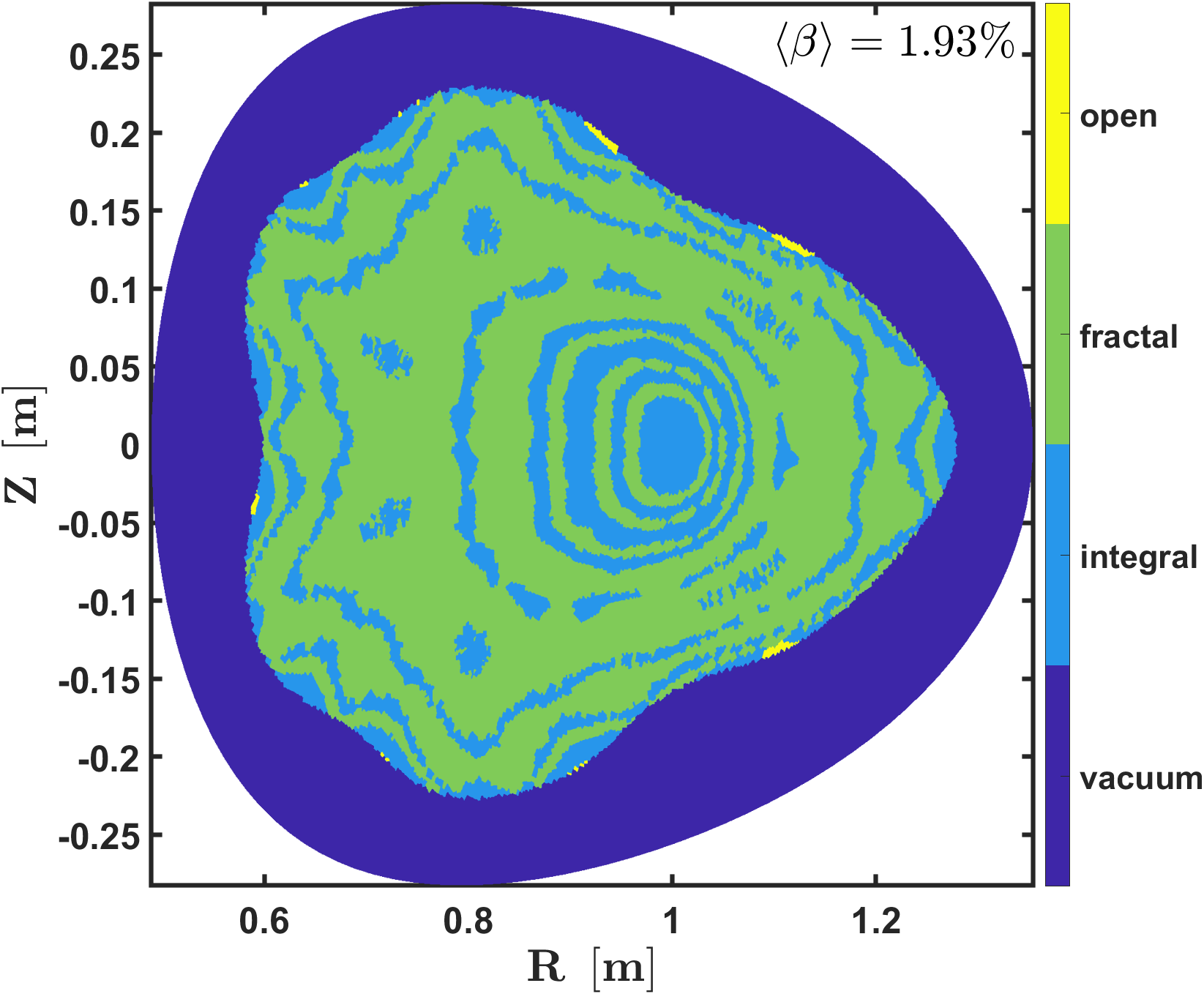}} \qquad
    \subfloat{\includegraphics[width=0.35\linewidth]{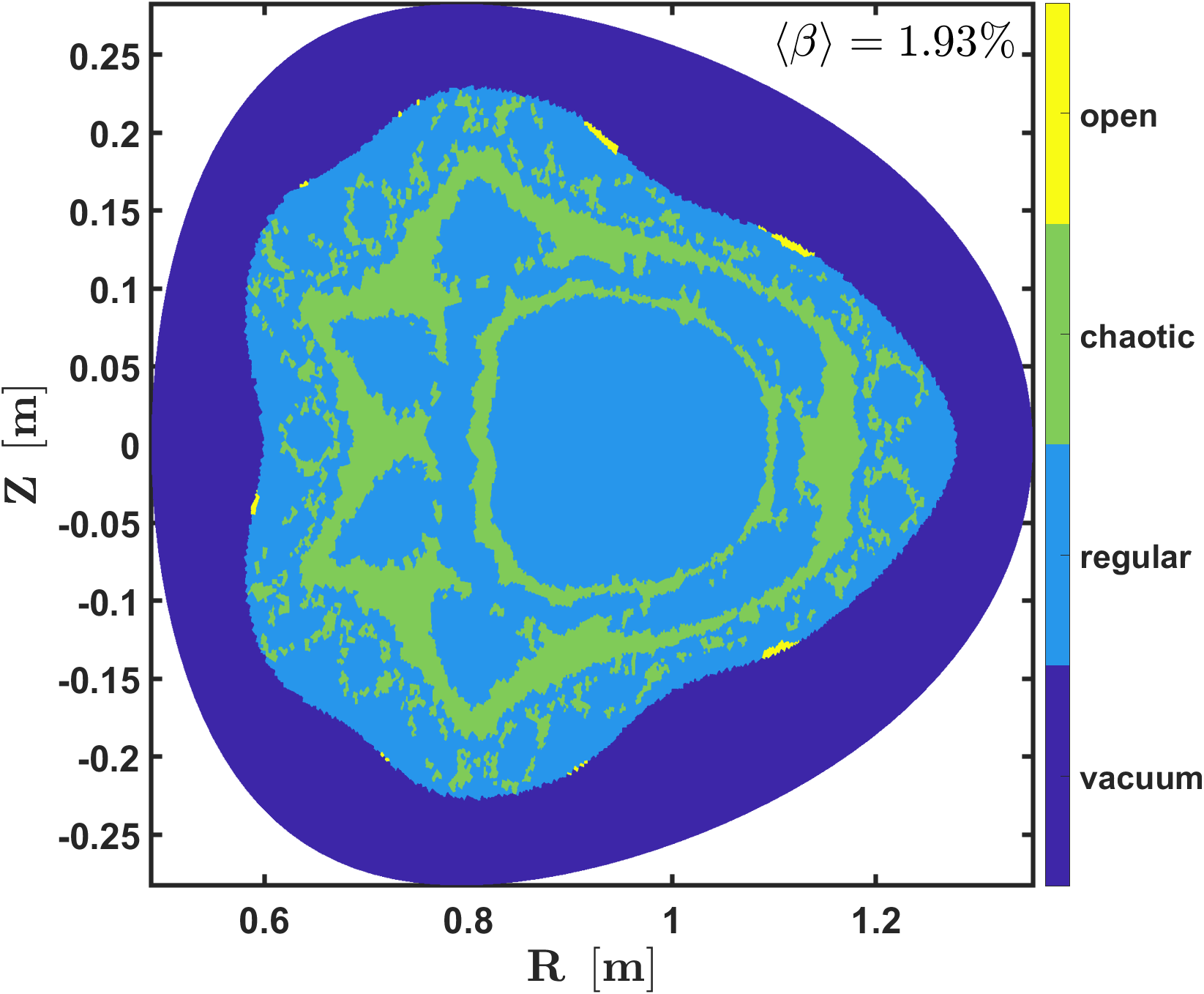}}
    \caption{Contours of the box-counting dimension (left column) and the weighted Birkhoff average (right column) for net-current-free finite-$\beta$ equilibria in the standard configuration with various $\langle\beta\rangle$. 
             The filled contours are drawn at the cross section $\phi=\pi/2$.}
    \label{fig:BC_WBA_contours}
\end{figure}

\begin{figure}[htbp]
    \centering
    \subfloat{\includegraphics[width=1\linewidth]{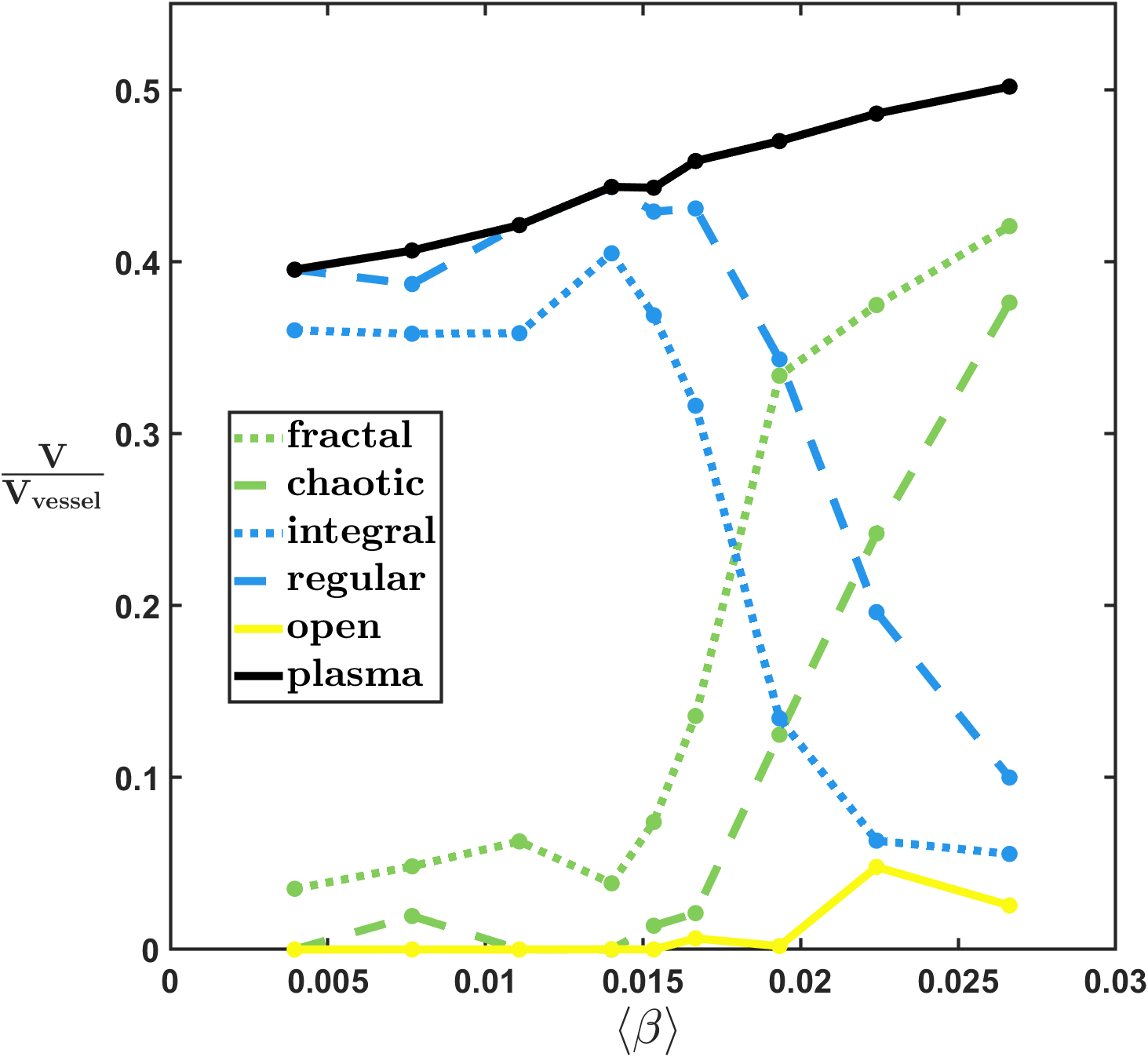}}
    \caption{Effective subvolumes divided by the box-counting dimension (dotted lines) or the weighted Birkhoff average (dashed lines)
             for net-current-free finite-$\beta$ equilibria in the standard configuration.}
    \label{fig:beta_vs_BC_WBA}
\end{figure}

\Cref{fig:BC_WBAs} shows the box-counting dimension $\text{D}_{bc}$ and the WBA for equilibria in the standard configuration. 
We remark that accurate calculation of the two measures require ultrahigh resolutions for 
field line tracing, and especially, equilibrium solutions. Thereby, the criteria are slightly relaxed for the limited computational resource. 
In the $\text{D}_{bc}$ calculation, a range between $0.95$ and $1.05$, bounded by two red lines, is prescribed to denote nested flux surfaces. 
While in the WBA calculation, the regular and the chaotic regions are separated by the value $1.3\times10^{-4}$, namely, 
the field lines with the WBA below the red line represent the regular regions. 
Both metrics signify that stochastic field lines emerge due to the finite-$\beta$ effect, 
and extend to nearly the entire plasma region for cases with high $\langle\beta\rangle$. 
To better quantify the volume of chaos, we divide the total computational domain into several subdomains with distinctive properties. 
The left column in \Cref{fig:BC_WBA_contours} illustrates the vacuum, integral-dimension, fractal-dimension, and open-field-line subregions, respectively. 
Note that all non-trivial regions are counted within the plasma region defined by the ratio $p/p_\text{max}\geq 10^{-2}$. 
The connection length is used to identify whether a field line is open or not \cite{Suzuki_2020_2}, 
and the integral-dimension and the fractal-dimension regions are identified using their box-counting values. 
Technically, field line tracing cannot cover the full computational area in the case where edge islands are embedded in open field lines 
and located outside the last closed flux surface (LCFS), 
and the spectral clustering method \cite{Luxburg_2007} is used to distinguish the mesh points. 
For low-$\langle\beta\rangle$ equilibria, there are a few thin layers of island chains, 
and the rest of the plasma region is filled with good flux surfaces. 
However, as $\langle\beta\rangle$ increases, the integral-dimension regions shrink significantly. 
The equilibrium with $\langle\beta\rangle=1.67\%$ exhibits a clearly stratified structure and 
the presence of a small open-field-line region adjacent to the plasma-vacuum separatrix. 
When $\langle\beta\rangle$ rises to about $2\%$, the fraction of the integral-dimension regions drops to an extremely low level. 
The right column in \Cref{fig:BC_WBA_contours} portrays the chaotic and the regular subregions.  
A transient state with slightly chaotic region caused by the interaction of the near-axis $n/m=2/5$ and the peripheral $n/m=4/11$ islands 
emerges at $\langle\beta\rangle=0.77\%$, and soon fades away as $\langle\beta\rangle$ continues to increase. 
The overlap related to internal $n/m=2/6$ island first appears at $\langle\beta\rangle=1.53\%$. Subsequently, 
the stochasticity in field lines near the $n/m=2/6$ islands can be obviously observed at $\langle\beta\rangle=1.67\%$ and $1.93\%$, 
and the impact of island overlap becomes much more severe for equilibria with higher $\langle\beta\rangle$. 
The general trend of topological variation can be quantitatively indicated by changes in their volumes normalized to the total plasma volume. 
\Cref{fig:beta_vs_BC_WBA} suggests that the non-nested-flux-surface regions, 
i.e., the fractal-dimension or the chaotic regions, start to surge at $\langle\beta\rangle\approx1.5\%$. 

\begin{figure}[htbp]
    % \vspace{-7em}
    \centering
    \subfloat{\includegraphics[width=0.35\linewidth]{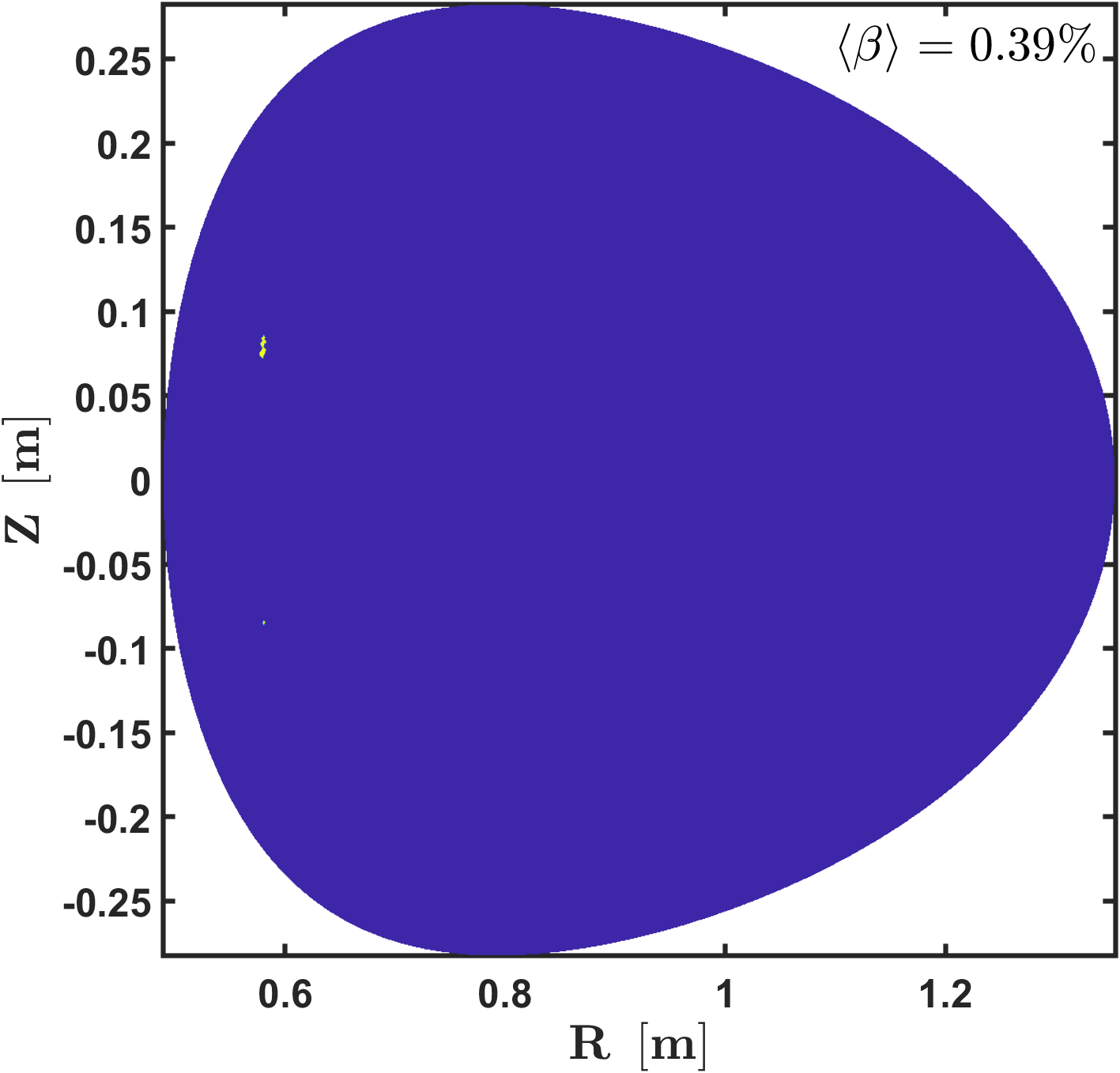}}
    \subfloat{\includegraphics[width=0.35\linewidth]{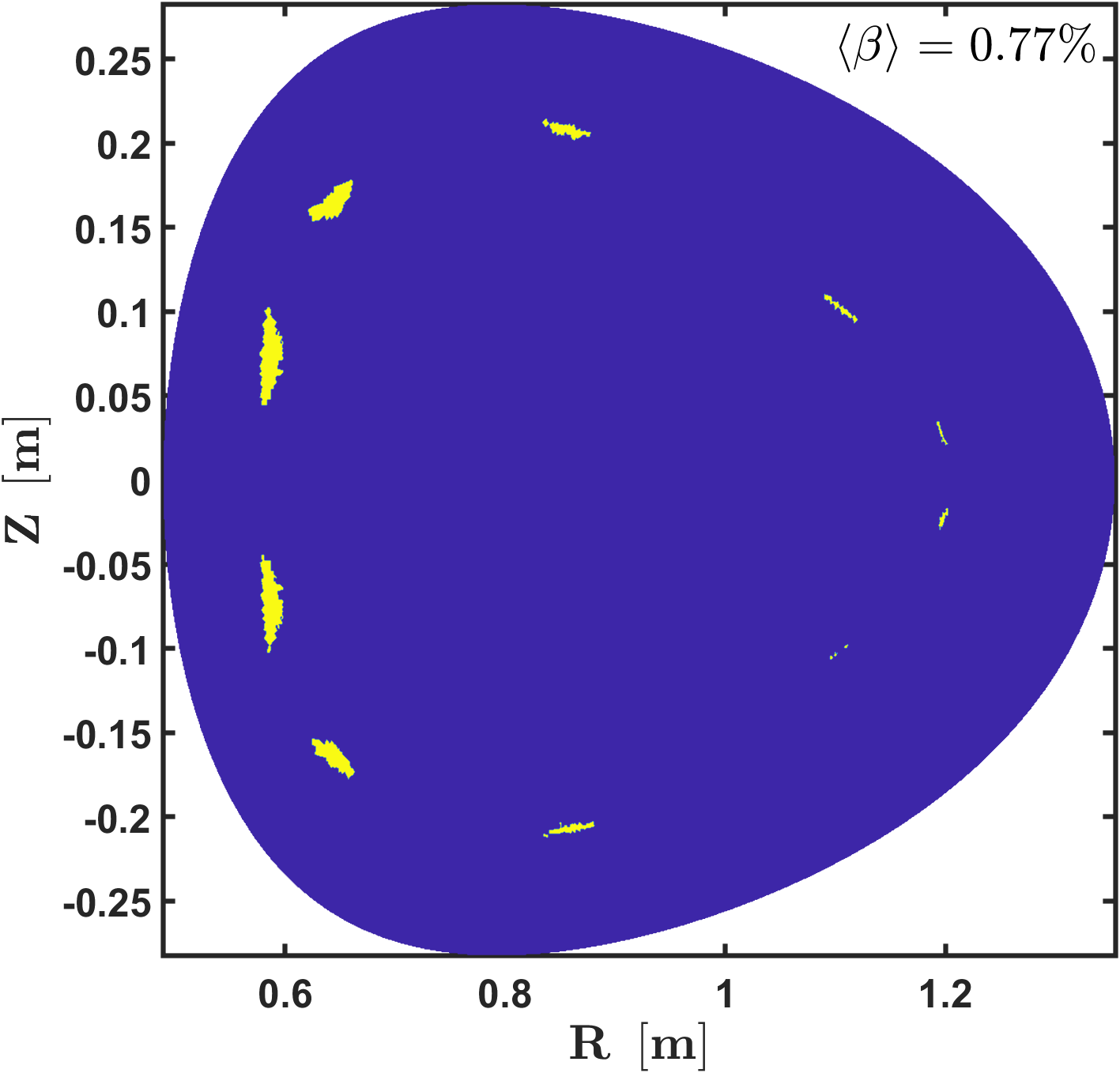}}
    \subfloat{\includegraphics[width=0.35\linewidth]{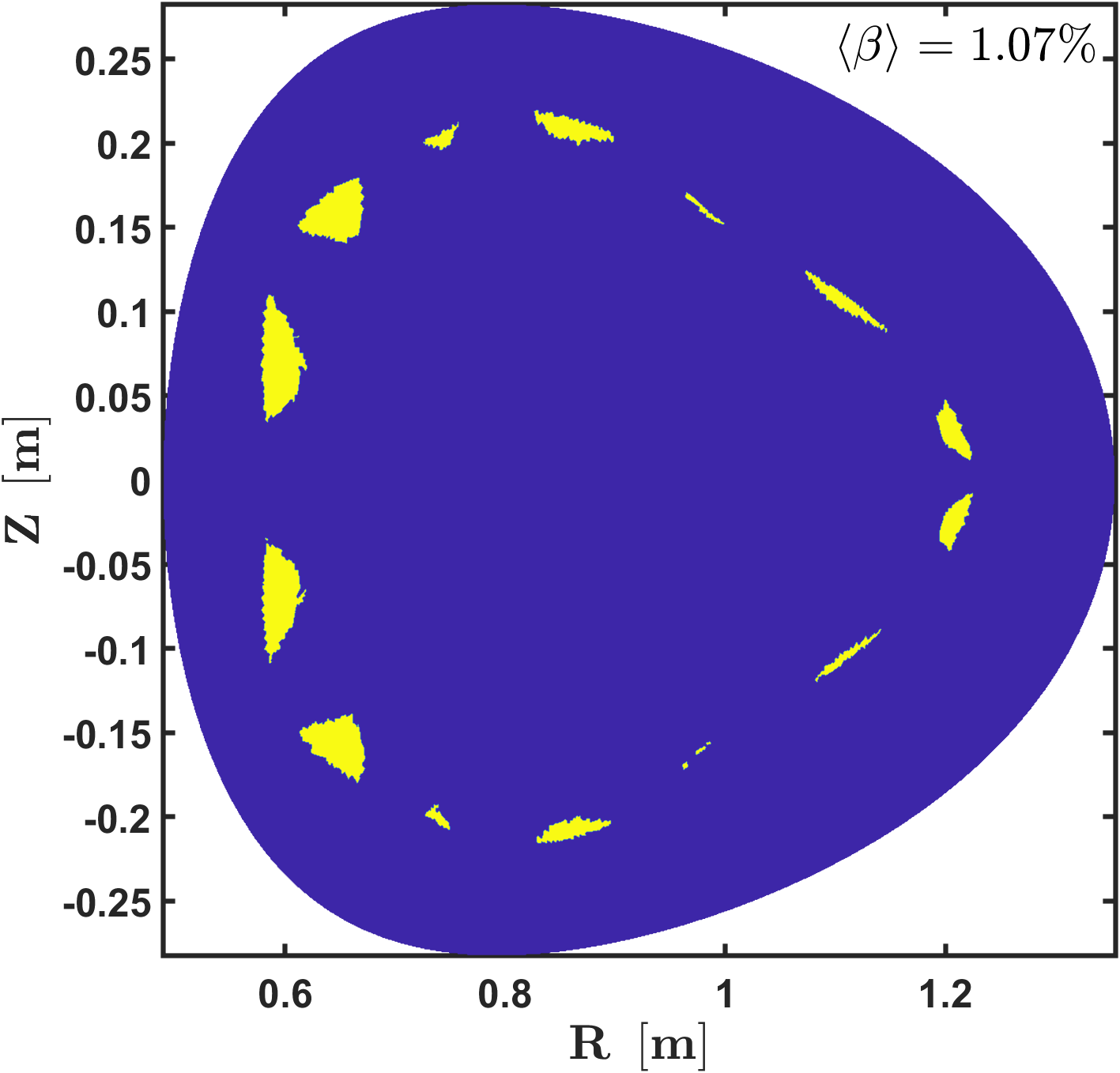}}\\ \vspace{-0.75em}
    \subfloat{\includegraphics[width=0.35\linewidth]{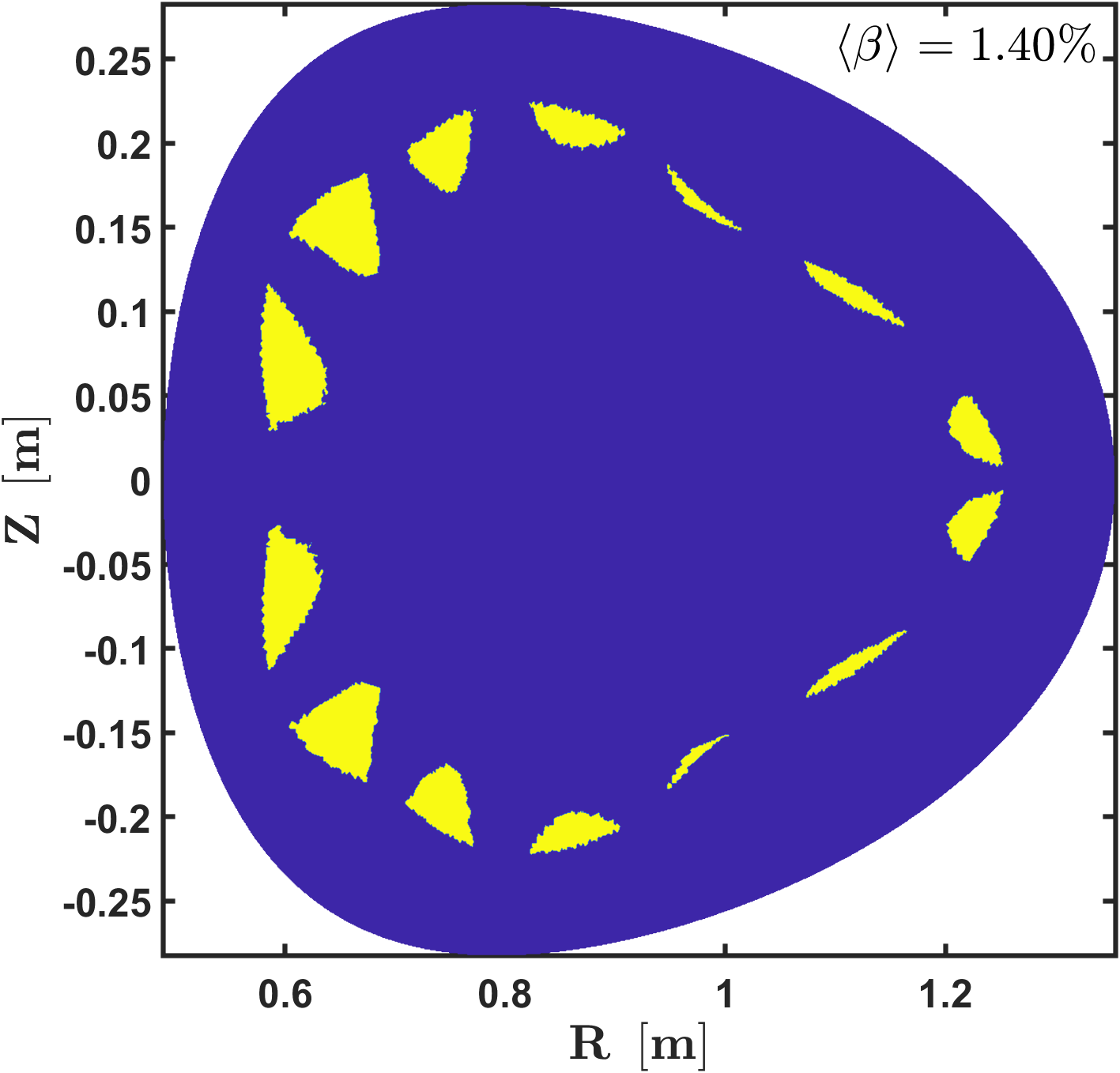}}
    \subfloat{\includegraphics[width=0.35\linewidth]{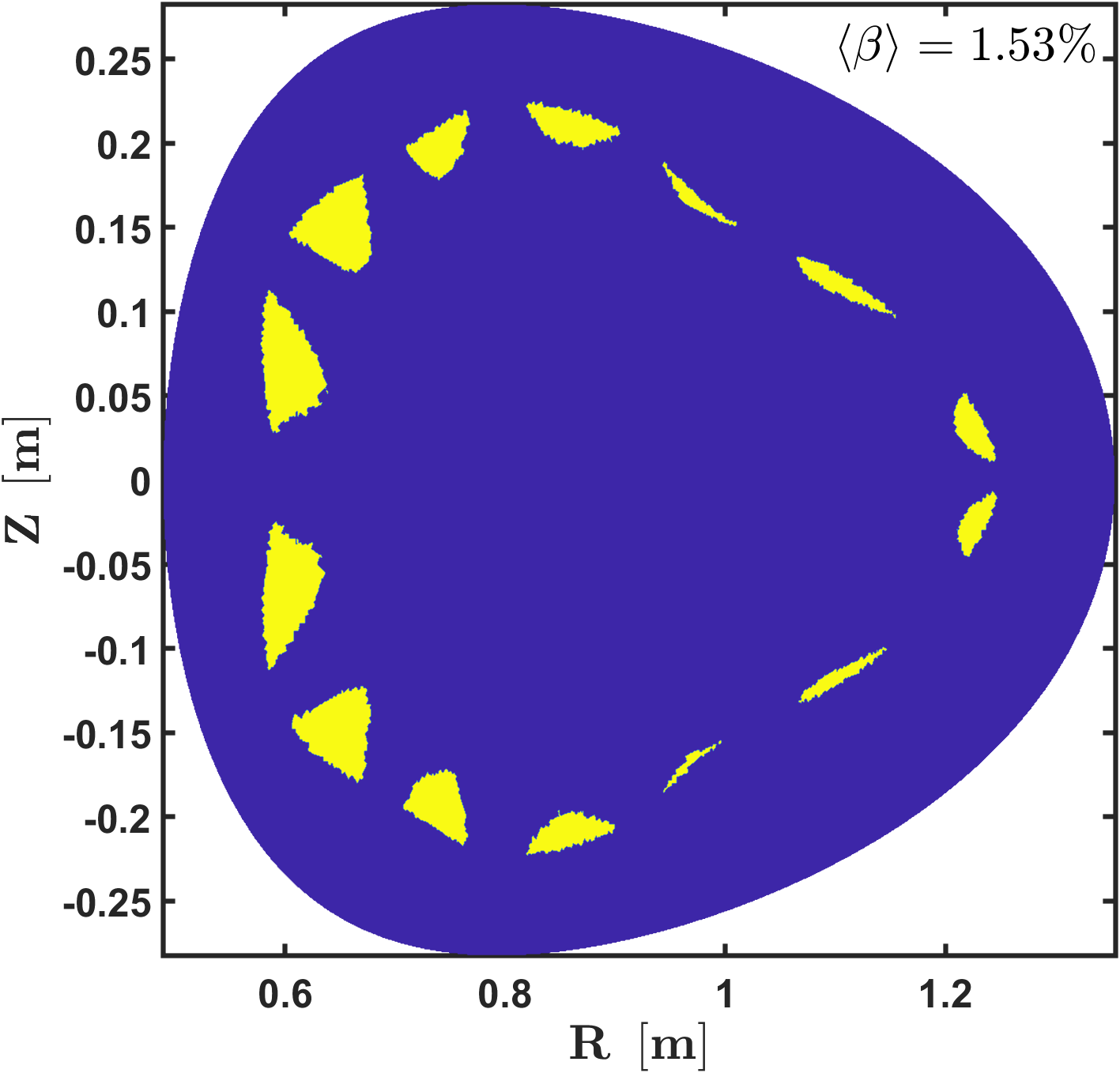}}
    \subfloat{\includegraphics[width=0.35\linewidth]{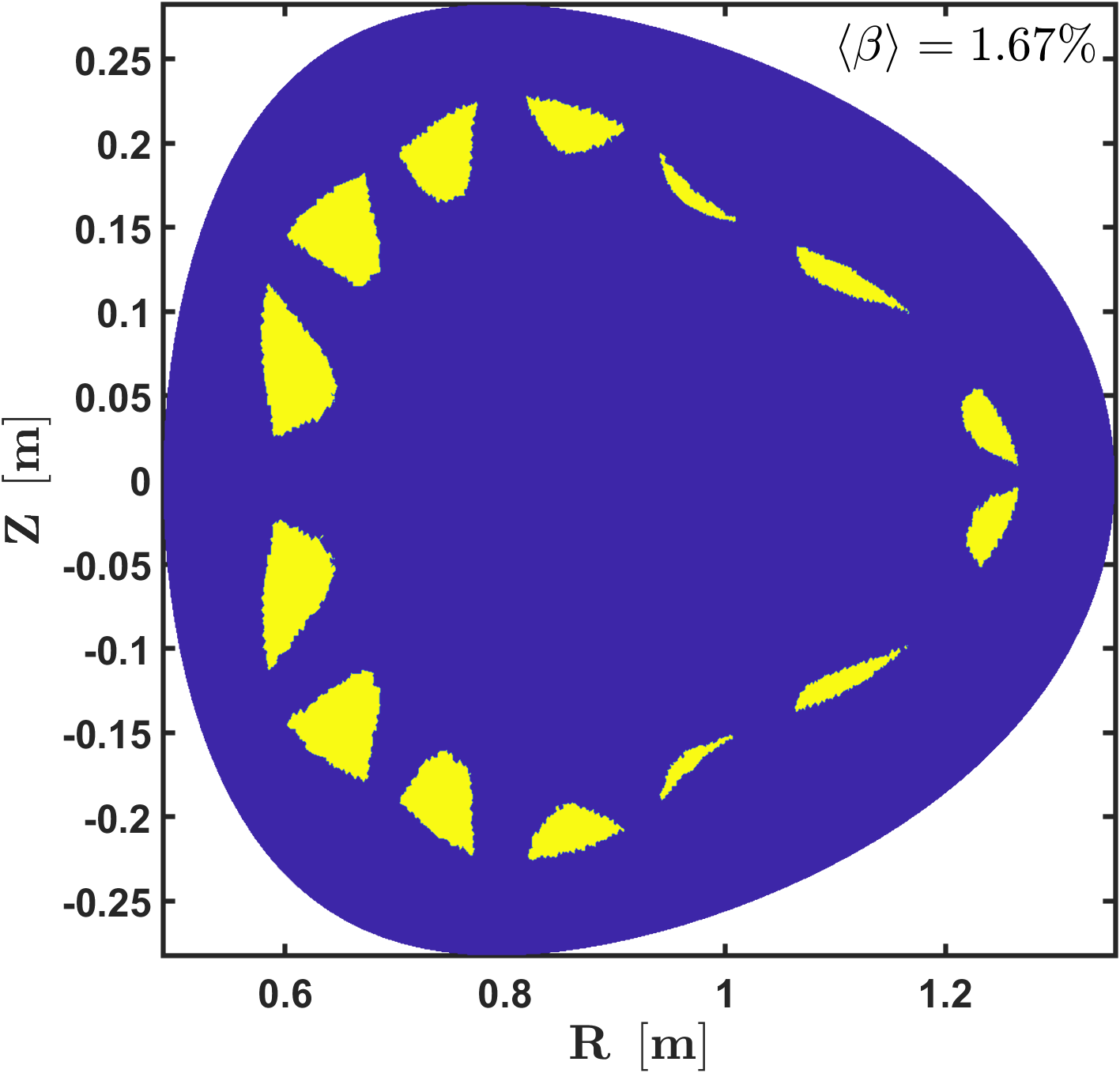}}\\ \vspace{-0.75em}
    \subfloat{\includegraphics[width=0.35\linewidth]{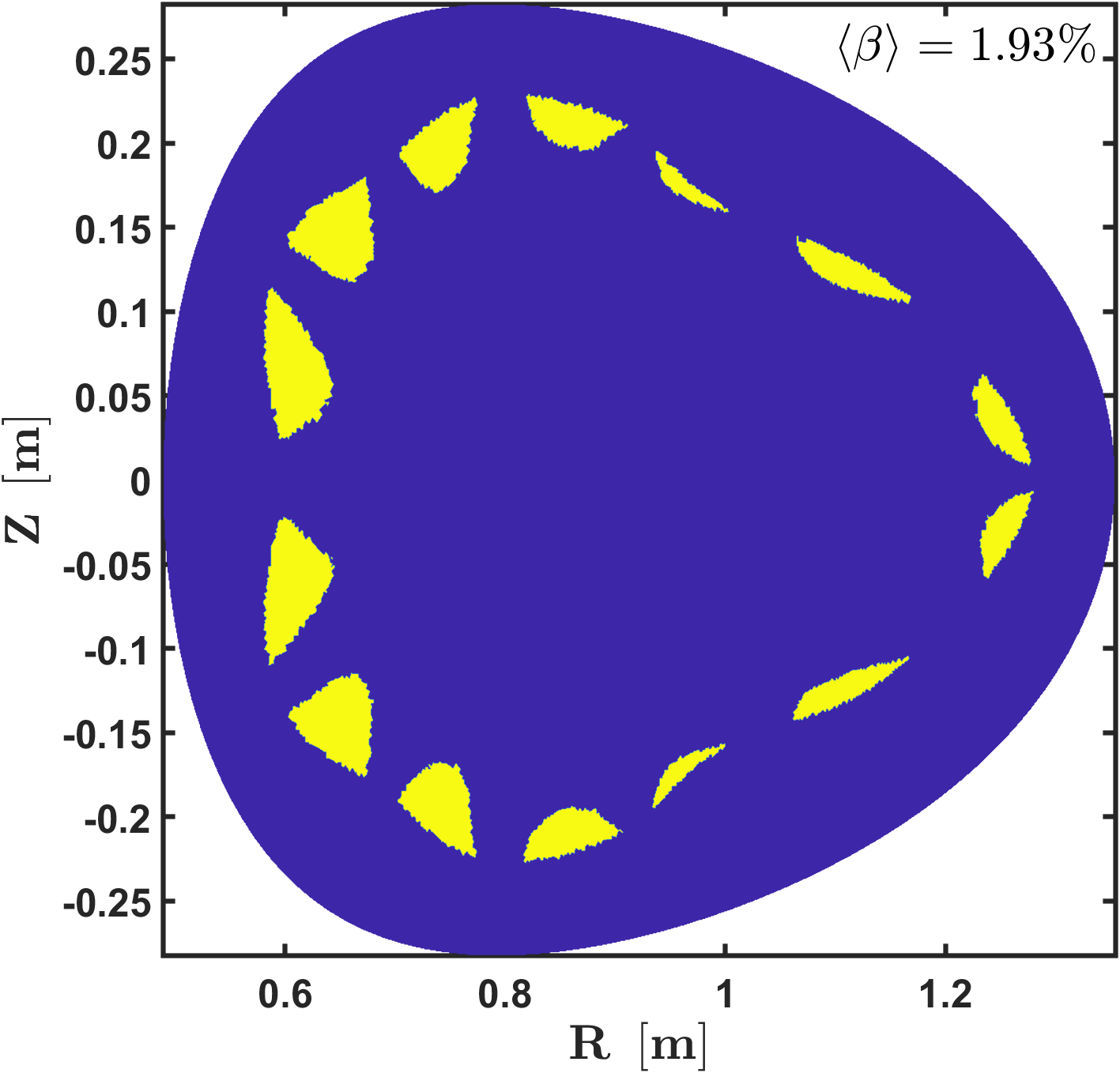}}
    \subfloat{\includegraphics[width=0.35\linewidth]{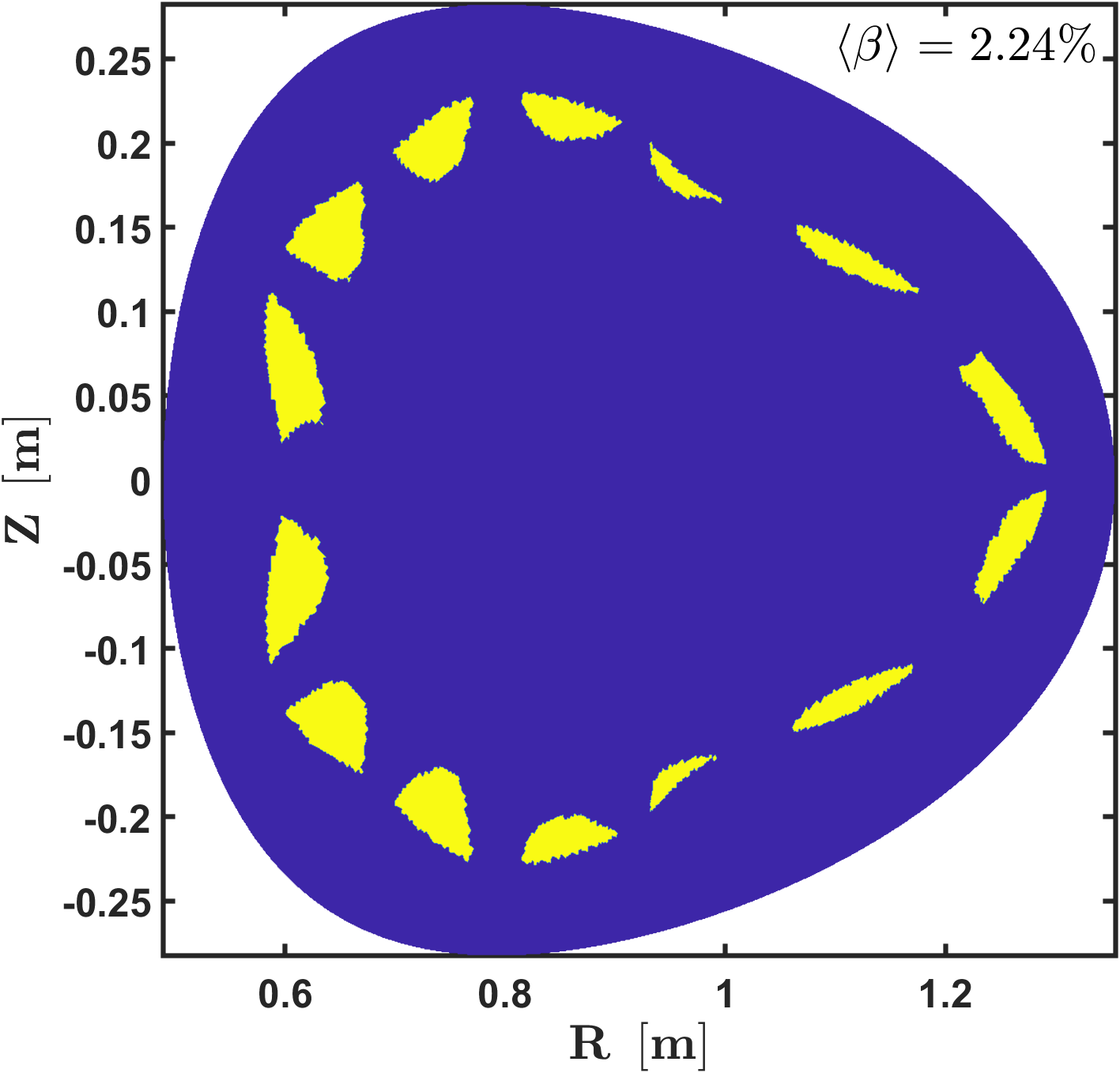}}
    \subfloat{\includegraphics[width=0.35\linewidth]{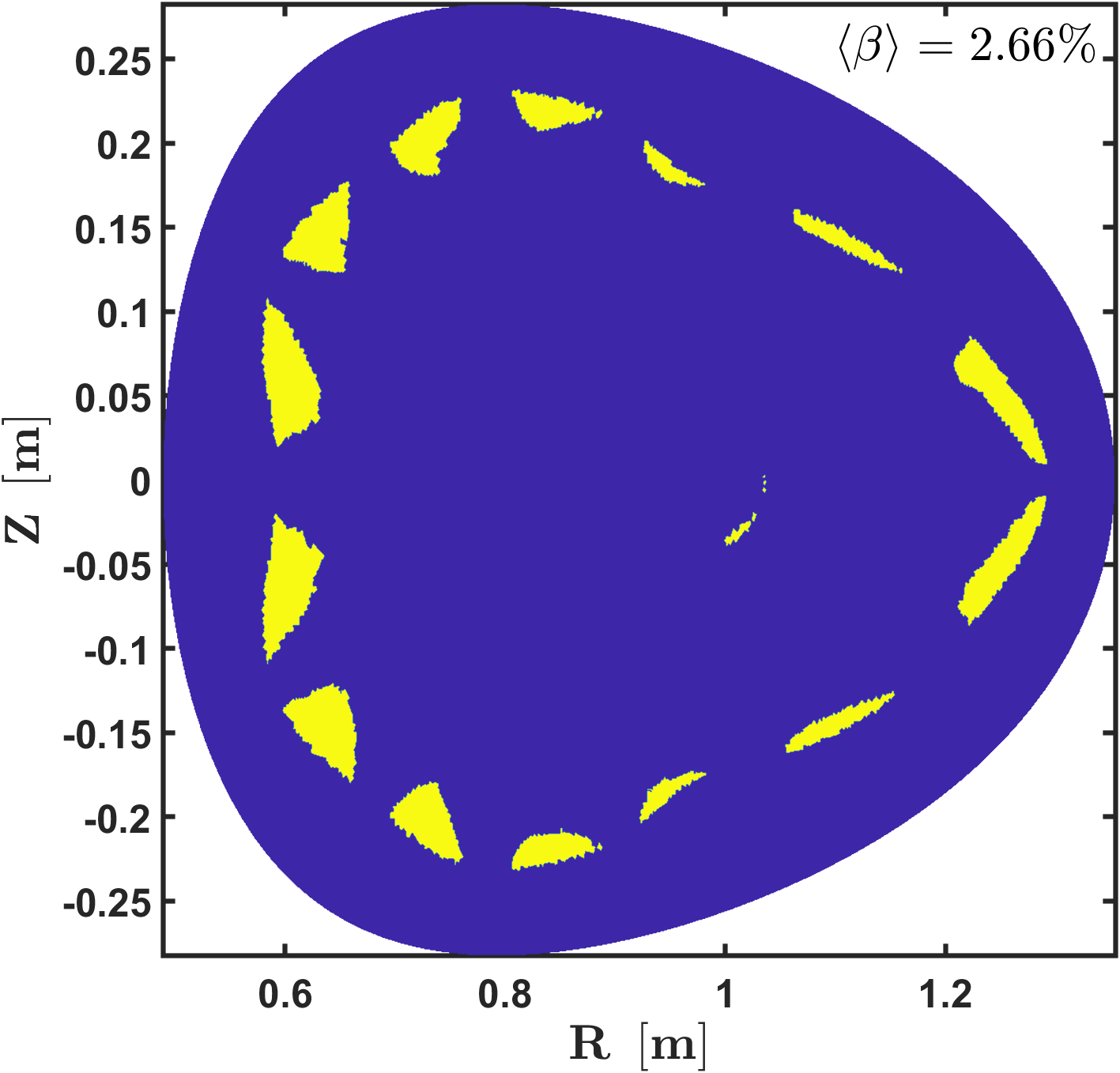}}
    \caption{Contours of the parallel diffusion for net-current-free finite-$\beta$ equilibria in the standard configuration with various $\langle\beta\rangle$. 
             The yellow-marked areas denote the $V_\text{PD}$ subregions.} The filled contours are drawn at the cross section $\phi=\pi/2$.
    \label{fig:VPD_contours}
\end{figure}

Unlike previous numerical studies in LHD \cite{Suzuki_2020, Suzuki_2020_2, Civit_2025} and W7-AS \cite{Zarnstorff_2004, Reiman_2007}, 
which indicate that a stochastic-field region forms first at the edge and then penetrates into the core as the pressure increases, 
the net-current-free equilibrium calculations for CFQS manifest that an overlap of internal main islands causes the 
localized disruption of nested flux surfaces. Nevertheless, we still assess the edge effect via the transport metrics $V_\text{PD}$.  
We set $\kappa_{\parallel}/\kappa_{\perp}=10^{5}$ in order to render the lowest-$\langle\beta\rangle$ equilibrium 
with virtually no $V_\text{PD}$, so that the grid error is minimized. 
\Cref{fig:VPD_contours} highlights the yellow-marked subregions characterized by effective parallel diffusion. 
Peripheral marked regions expand with increasing $\langle\beta\rangle$ in low-$\langle\beta\rangle$ cases, 
and nearly saturate in high-$\langle\beta\rangle$ cases. 
Moreover, the marked regions reside near the plasma-vacuum separatrix, and do not spread to or emerge in the internal area, 
indicating that the edge effect is not the primary reason for the flux surface breaking. 
The calculation of $V_\text{PD}$ further shows that, in addition to its gradual increase, the maximum $V_\text{PD}$ remains below $4.5\%$ across all equilibria. 
It is no surprise that internal chaotic regions do not reflect on $V_\text{PD}$. 
There was a similar finding for M3D-C1 fixed-plasma-boundary simulations on W7-X high-$\beta$ plasmas 
where internal stochastic field lines occupy a large part of the plasma region while $V_\text{PD}<4\%$ \cite{Zhou_2024}. 

\subsection{Bootstrap-current-carrying equilibria}\label{sec:standard_bcc}

In quasi-symmetric stellarators with nested flux surfaces, 
the bootstrap current distribution may be approximated by the Sauter-Redl-Landreman formulation \cite{Landreman_2022}. 
For simplicity, we employ the reduced form as $\langle \bm{J} \cdot \bm{B} \rangle = \alpha pp'$ to avoid assumptions on the separate profiles of the ion and electron, 
or calculations of other geometric information. The coefficient $\alpha$ is set to obey the scaling law, 
which is revealed by previous computations on CFQS using VMEC-BOOTSJ \cite{Wang_2021}, -SFINCS \cite{Liu_2023}, and -MONTS \cite{Yang_2025}. 
This law demonstrates a linear relationship between the total bootstrap current $I_{bs}$ and $\langle\beta\rangle$. 
In numerical implementations, we take the maximum value of pressure as the surrogate of $\langle\beta\rangle$, 
and replace the toroidal flux with the pressure isosurface \cite{Suzuki_2020}. 

\begin{figure}[htbp]
    % \vspace{-7em}
    \centering
    \subfloat{\includegraphics[width=0.35\linewidth]{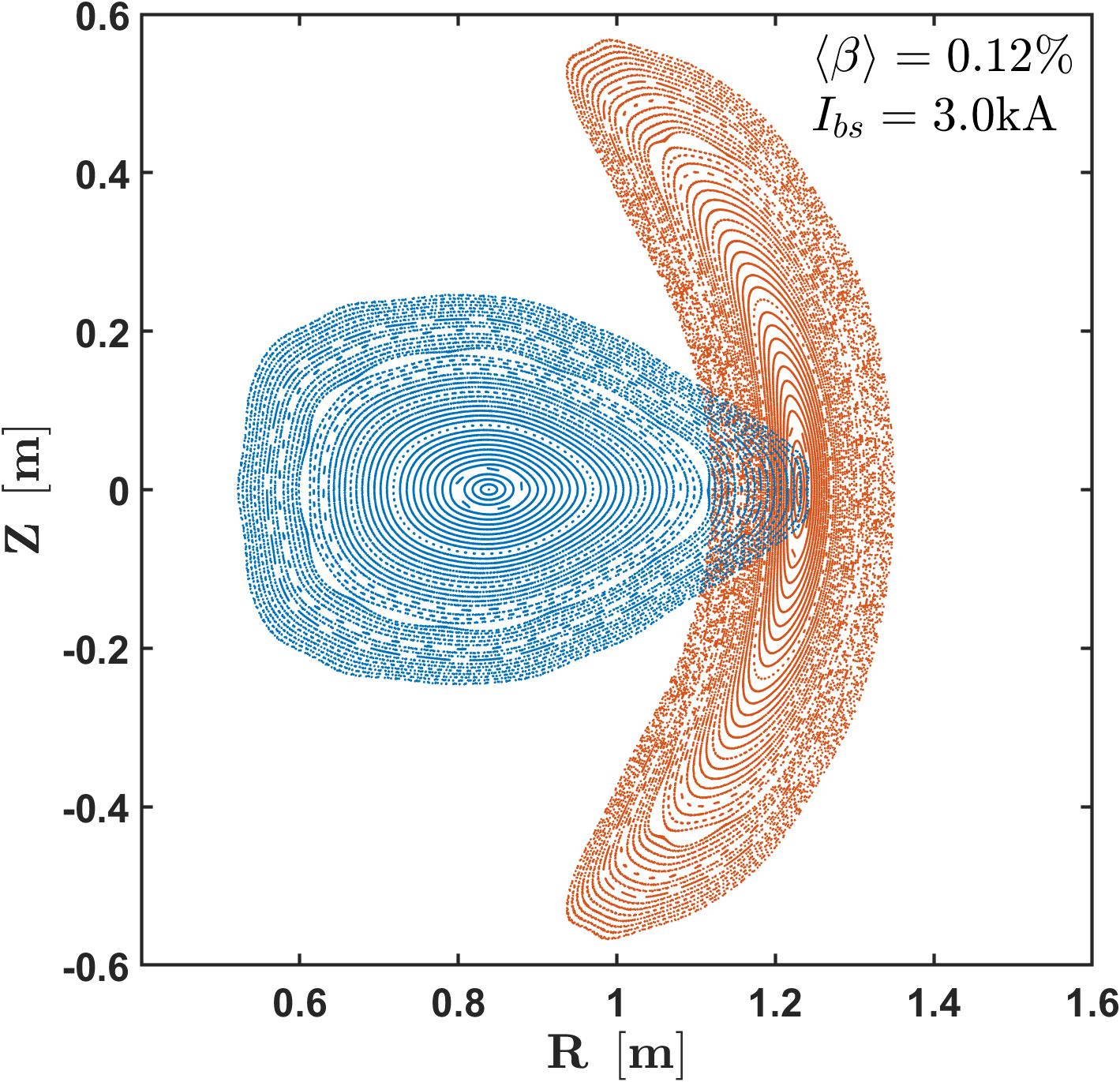}}
    \subfloat{\includegraphics[width=0.35\linewidth]{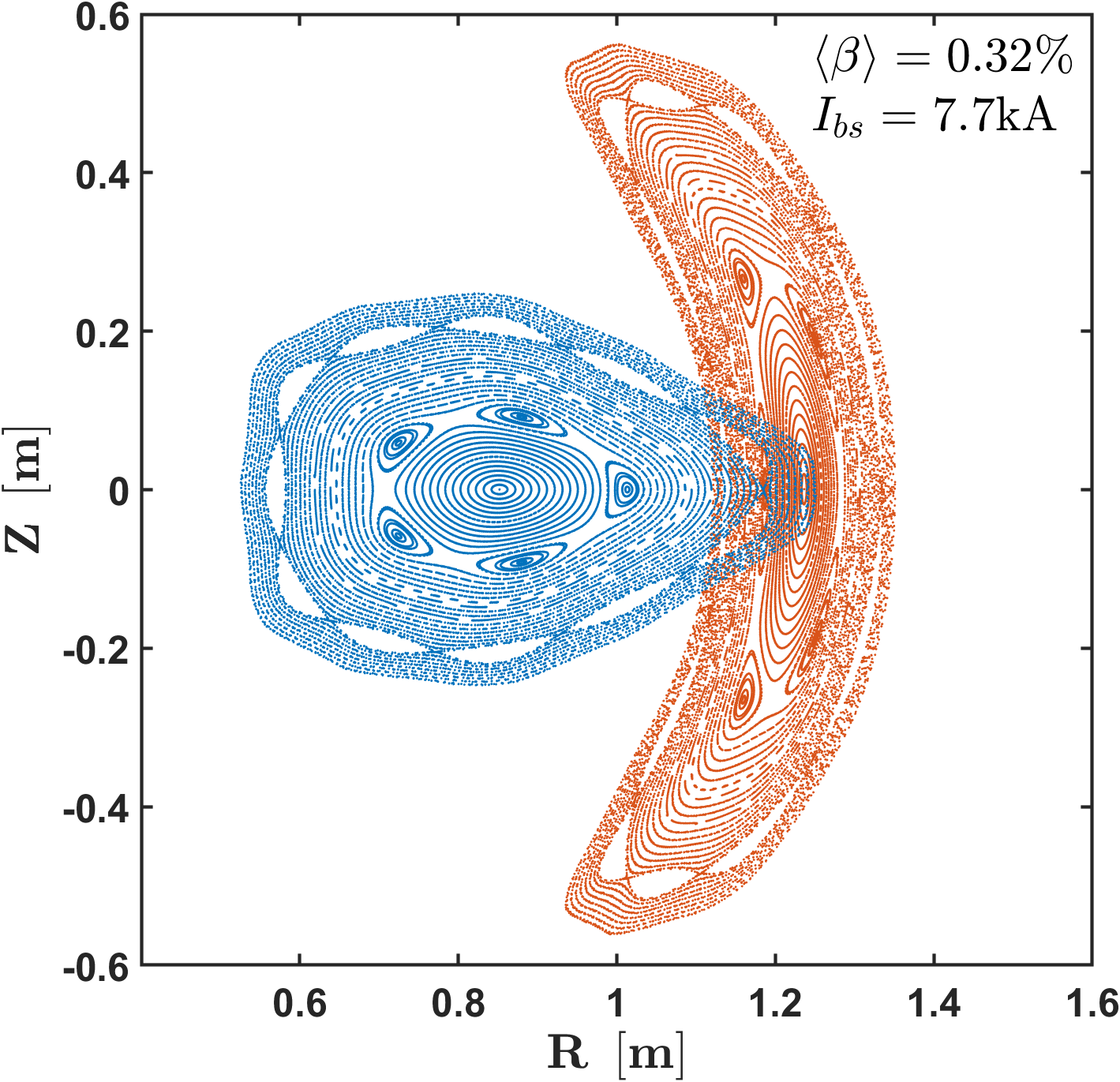}}
    \subfloat{\includegraphics[width=0.35\linewidth]{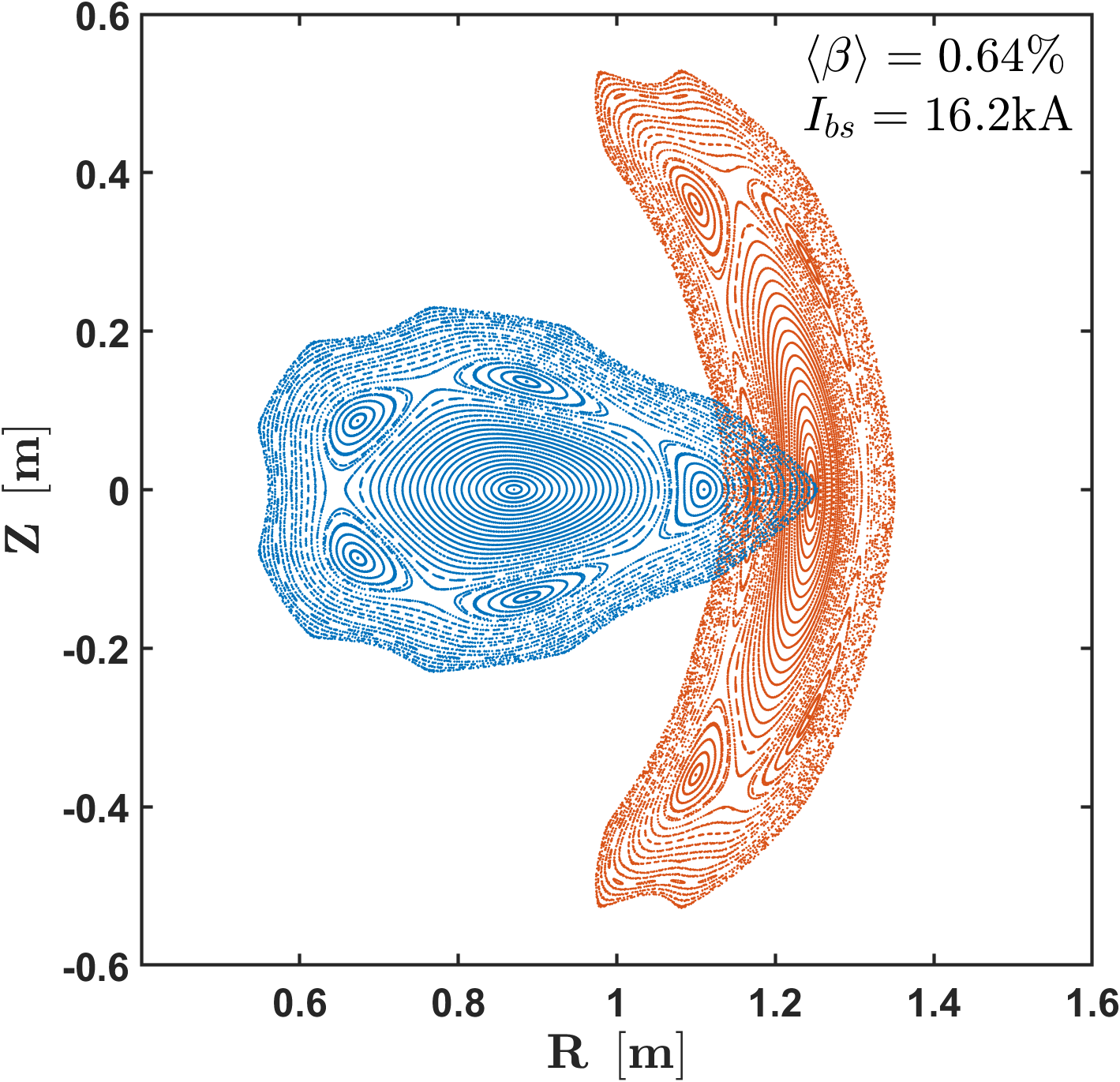}}\\ \vspace{-0.75em}
    \subfloat{\includegraphics[width=0.35\linewidth]{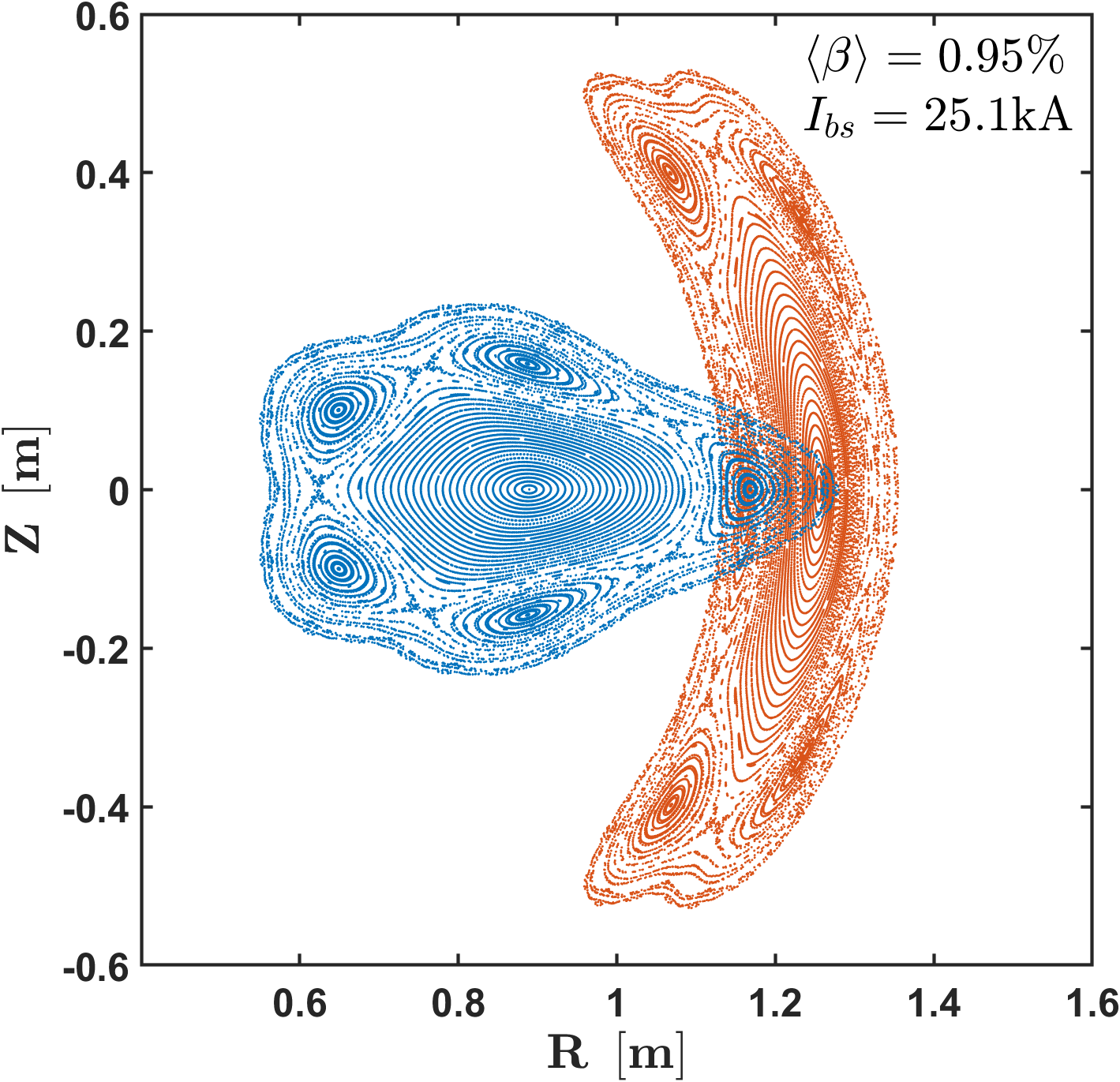}}
    \subfloat{\includegraphics[width=0.35\linewidth]{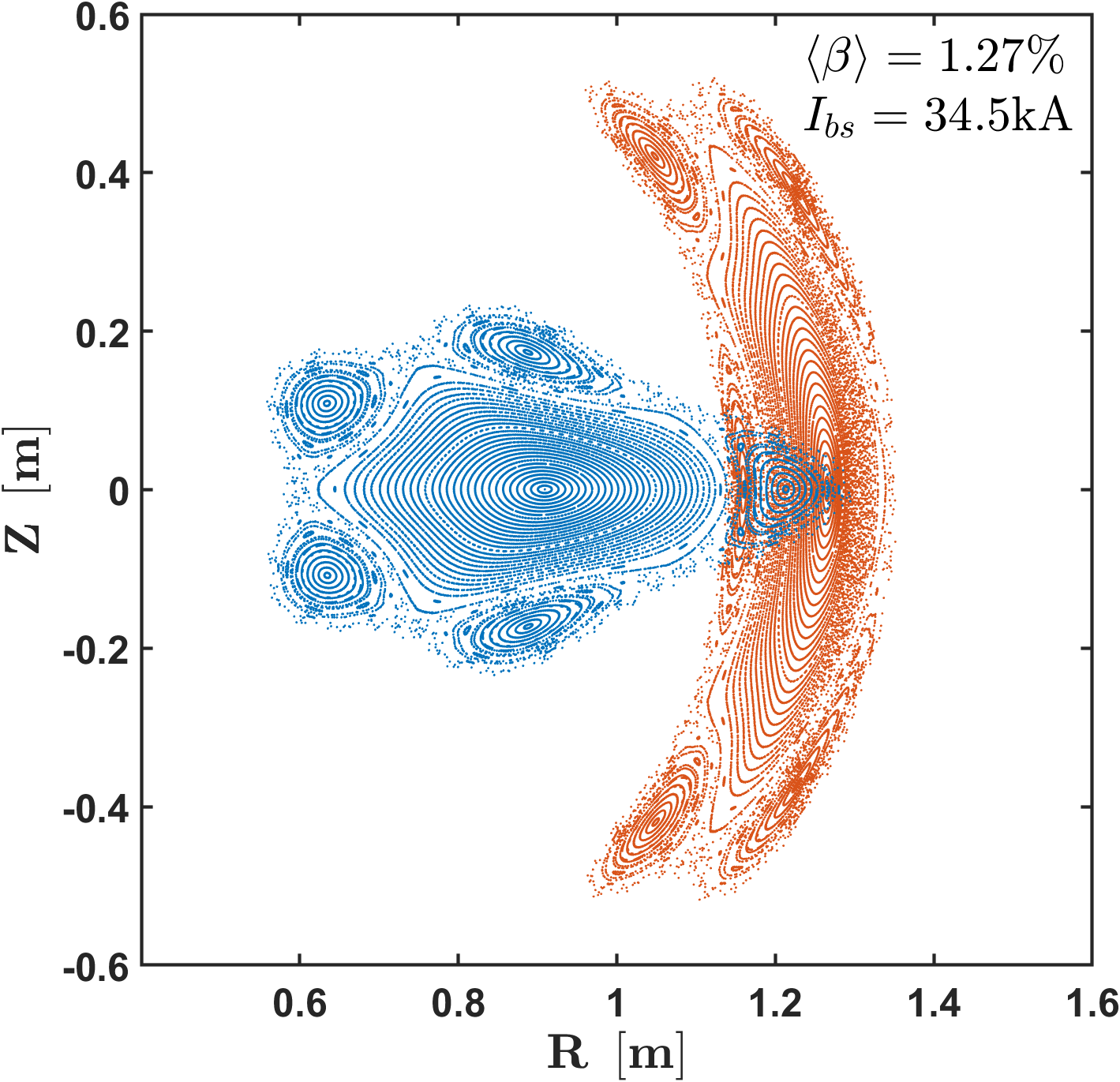}}
    \subfloat{\includegraphics[width=0.35\linewidth]{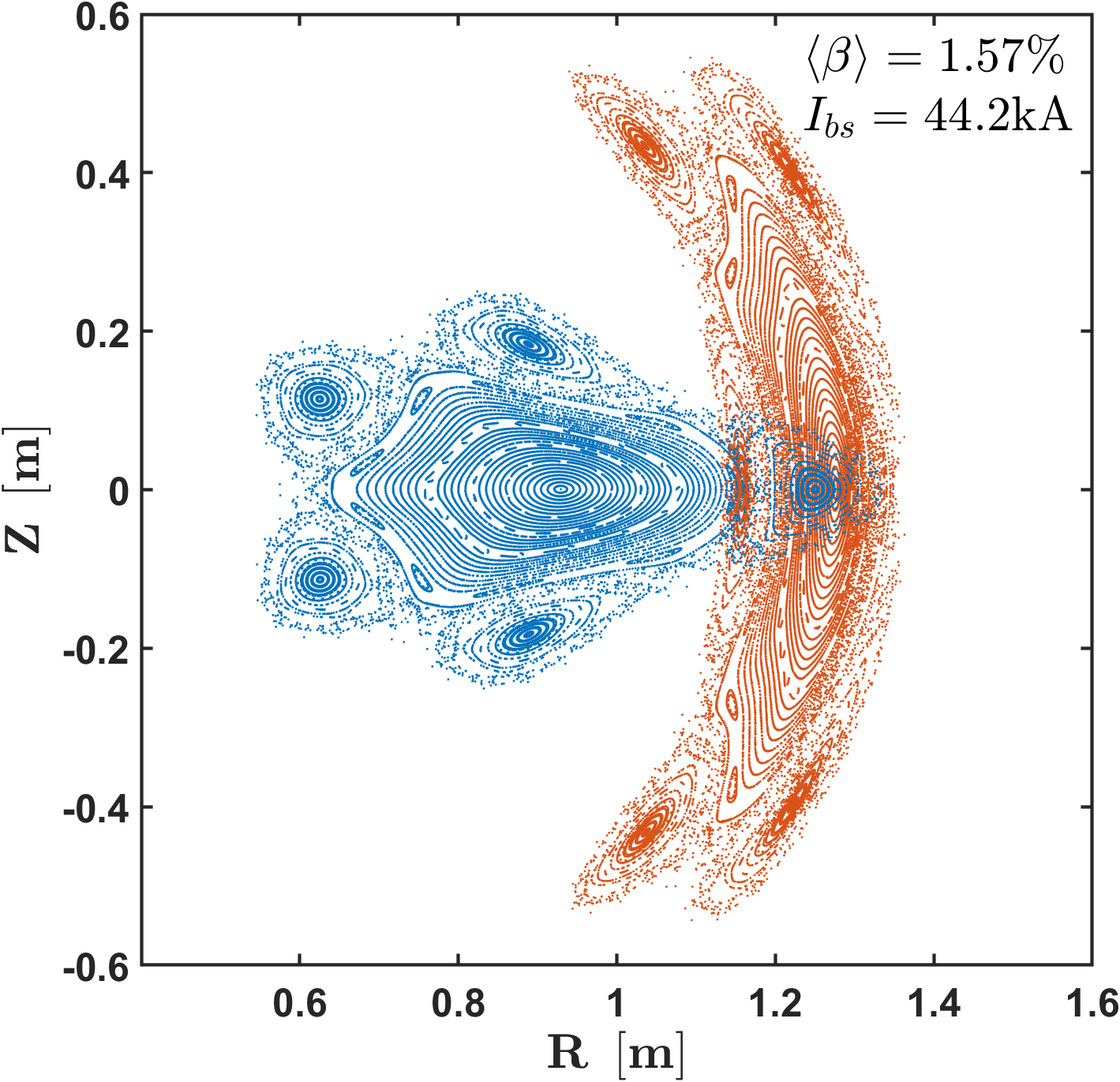}}\\ \vspace{-0.75em}
    \subfloat{\includegraphics[width=0.35\linewidth]{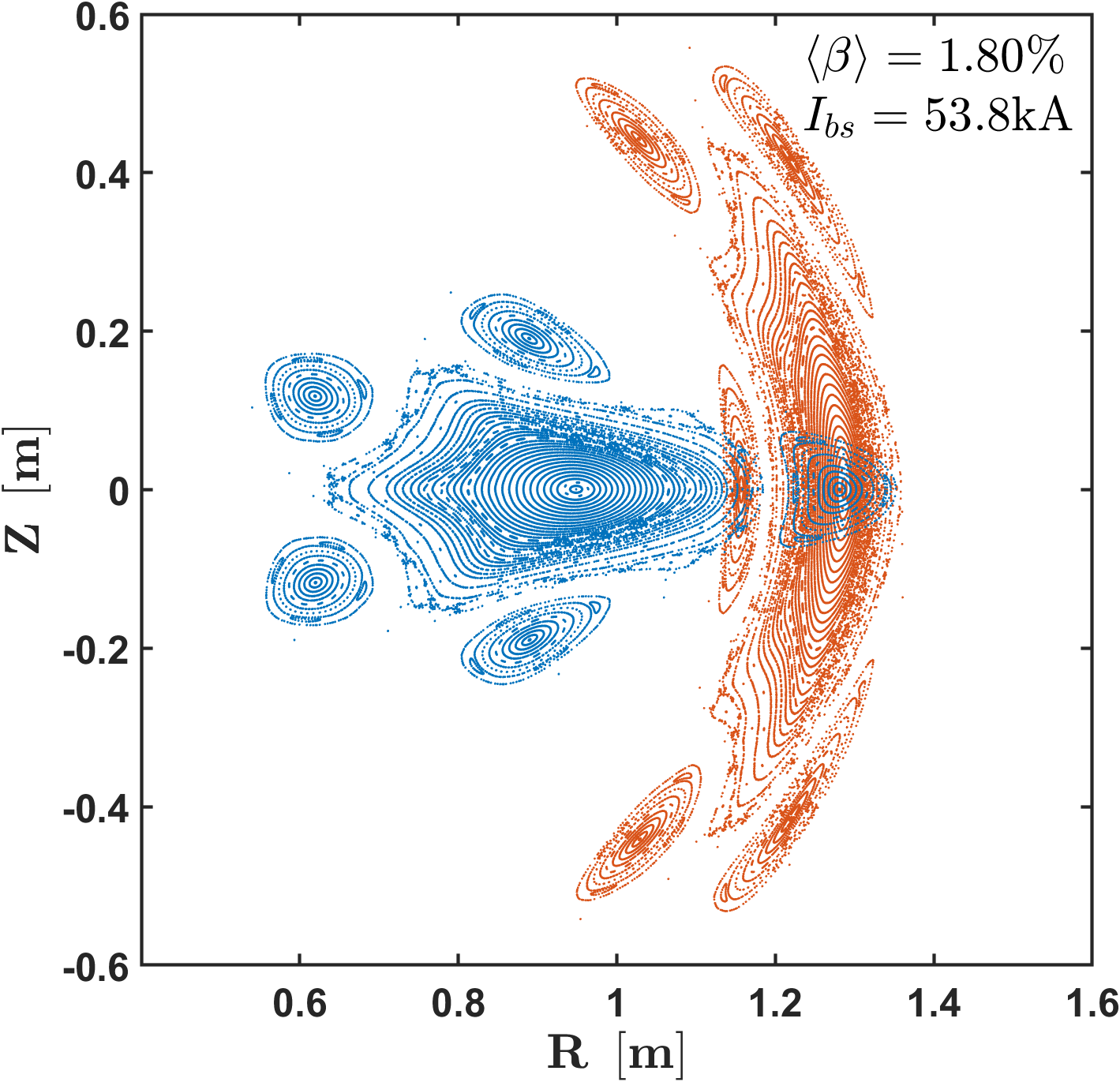}}
    \subfloat{\includegraphics[width=0.35\linewidth]{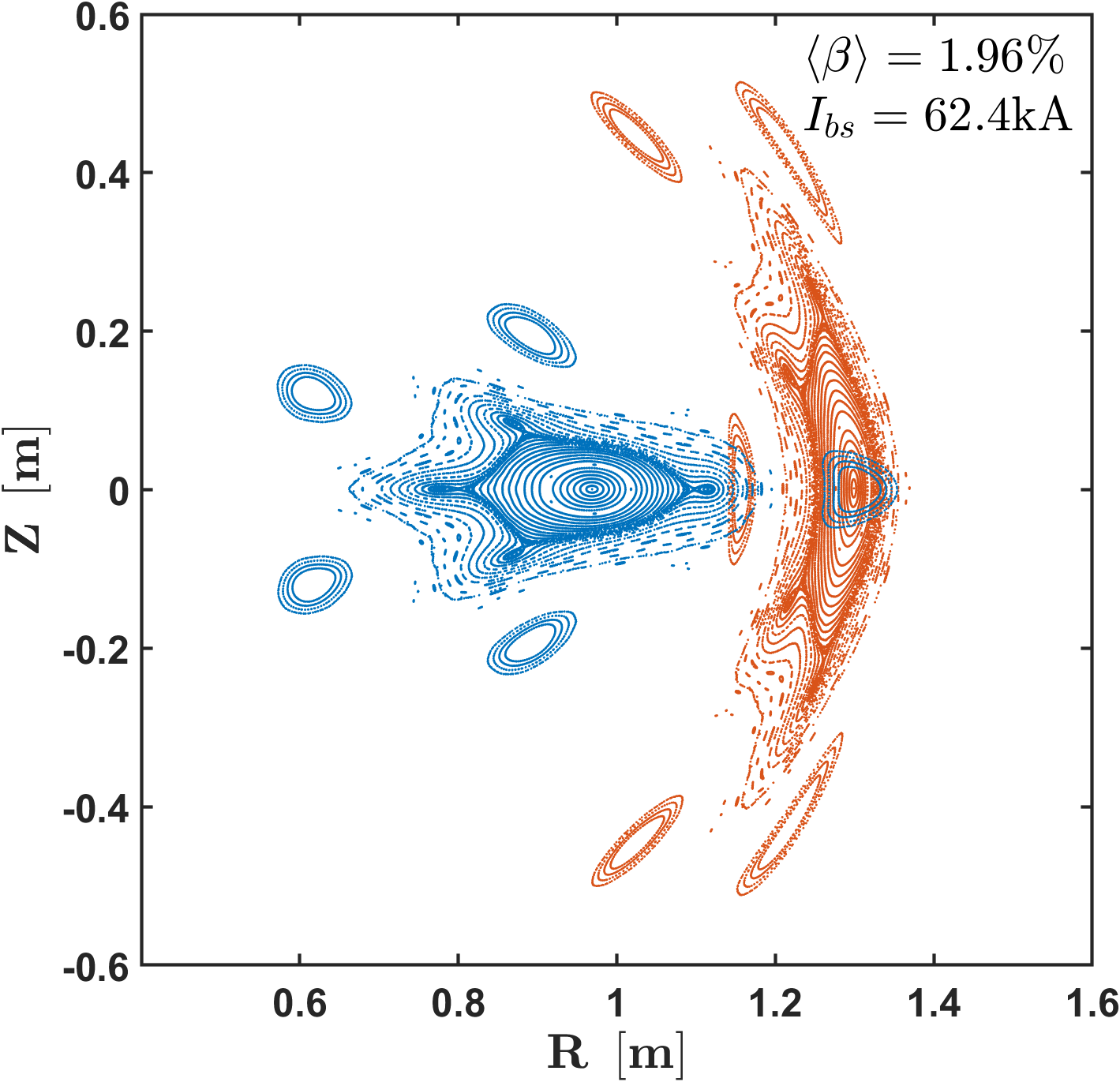}}
    \subfloat{\includegraphics[width=0.35\linewidth]{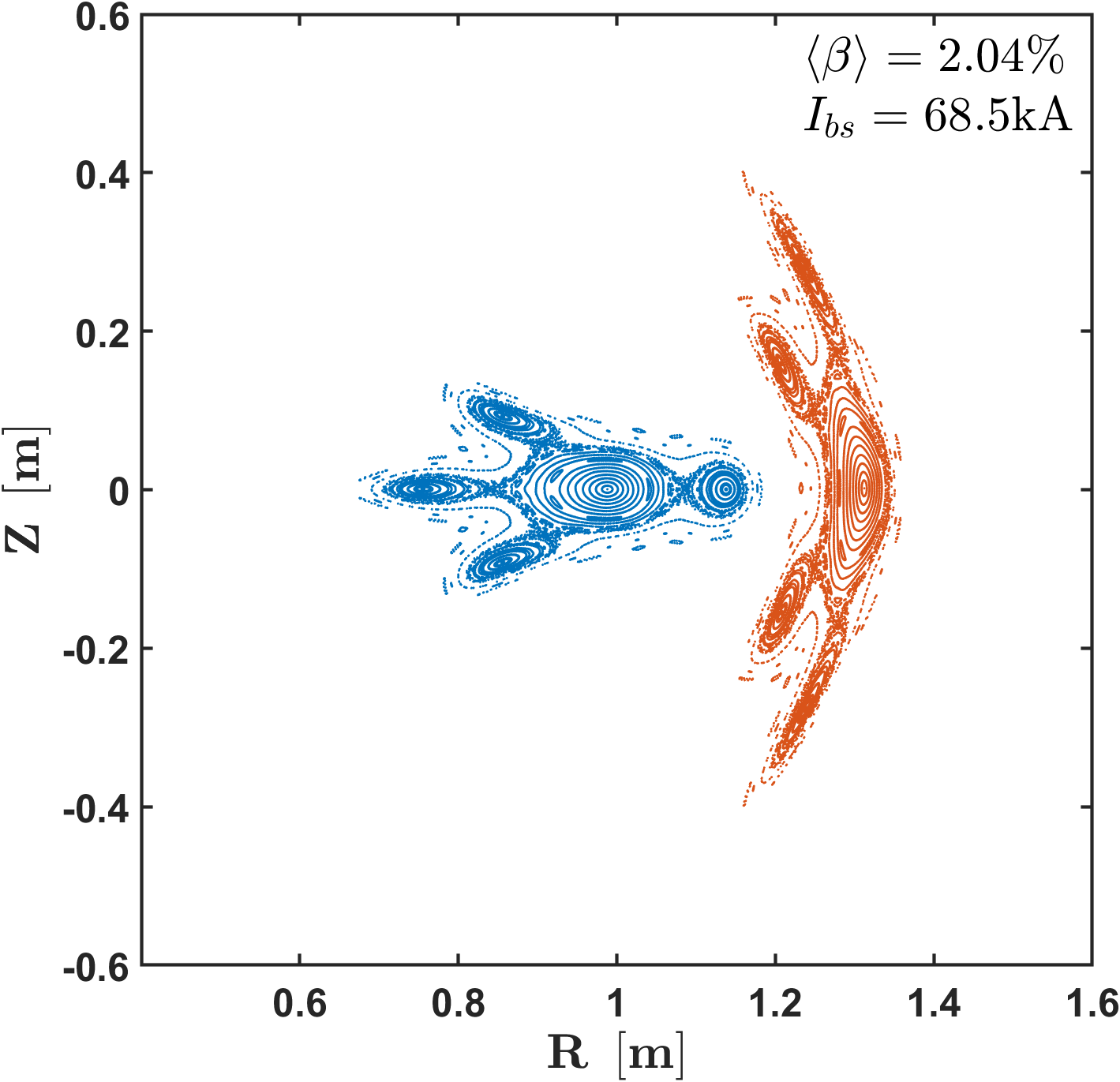}}
    \caption{Poincaré plots of the bootstrap-current-carrying finite-$\beta$ equilibria in the standard configuration at two cross sections 
             $\phi=0\:\text{(red)}\:\text{and}\:\pi/2\:\text{(blue)}$ with various $\langle\beta\rangle$.}
    \label{fig:poincare_QA_bootstrap}
\end{figure}

\begin{figure}[htbp]
    \centering
    \subfloat{\includegraphics[width=1\linewidth]{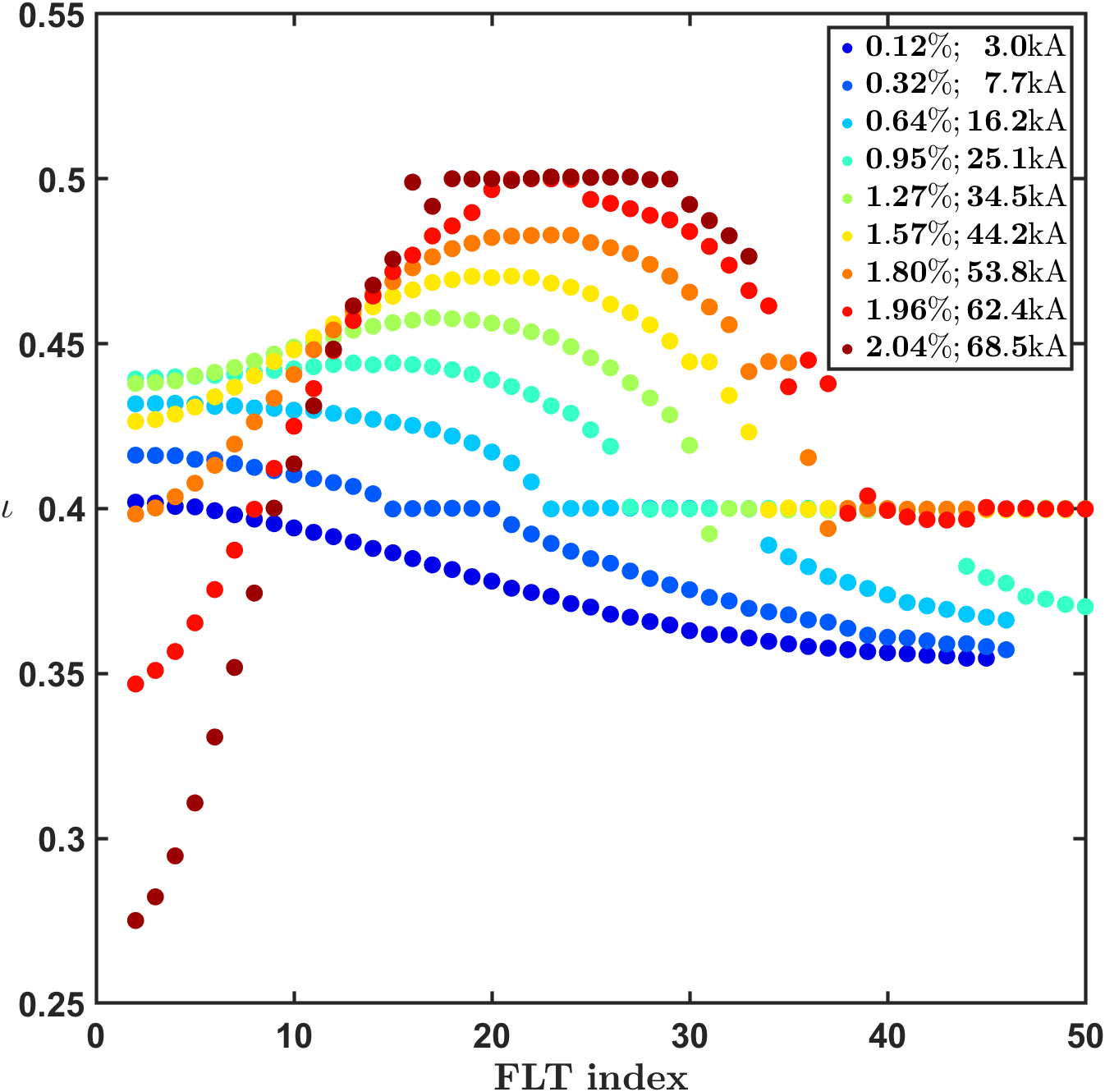}}
    \caption{Profiles of the rotational transform $\iota$ with respect to the equilibria in \Cref{fig:poincare_QA_bootstrap}. 
             The abscissa represents the index of field line tracing from the magnetic axis to the plasma boundary.}
    \label{fig:beta_vs_iota_bootstrap}
\end{figure}

\Cref{fig:poincare_QA_bootstrap} shows the Poincaré plots for the equilibria with bootstrap currents. Clearly, 
the overall trend of magnetic topology variation differs from the net-current-free cases. 
Internal low-order $n/m=2/5$ islands emerge at low-$\beta$ equilibria, however, without triggering a significant island overlap 
that may induce a large chaotic region. The equilibria with $\langle\beta\rangle=0.32\%$ and $0.64\%$ are dominated by nested flux surfaces, 
and possess a spontaneously formed divertor magnetic configuration with $n/m=2/5$ islands. 
For $\langle\beta\rangle<1.5\%$, the $n/m=2/5$ islands progressively move outward and saturate in size. 
The island overlap between $n/m=2/5$ and $n/m=4/9$ thus happens at $\langle\beta\rangle=1.57\%$. As $\langle\beta\rangle$ further increases, 
the LCFS shrinks, $n/m=2/4$ islands dominate the $n/m=2/5$ ones. The profiles of rotational transform plotted in \Cref{fig:beta_vs_iota_bootstrap} 
show that $\iota$ in the middle of the plasma region keeps increasing and is finally flattened to $0.5$. In addition, the edge $\iota$ remains $0.4$ 
in high-$\beta$ equilibria, which also indicates an evident modification of edge magnetic topology due to the effect of bootstrap currents. 

\begin{figure}[htbp]
    % \vspace{-7em}
    \captionsetup[subfloat]{position=top,labelformat=empty}
    \centering
    \subfloat[$\text{D}_{bc}$]{\includegraphics[width=0.35\linewidth]{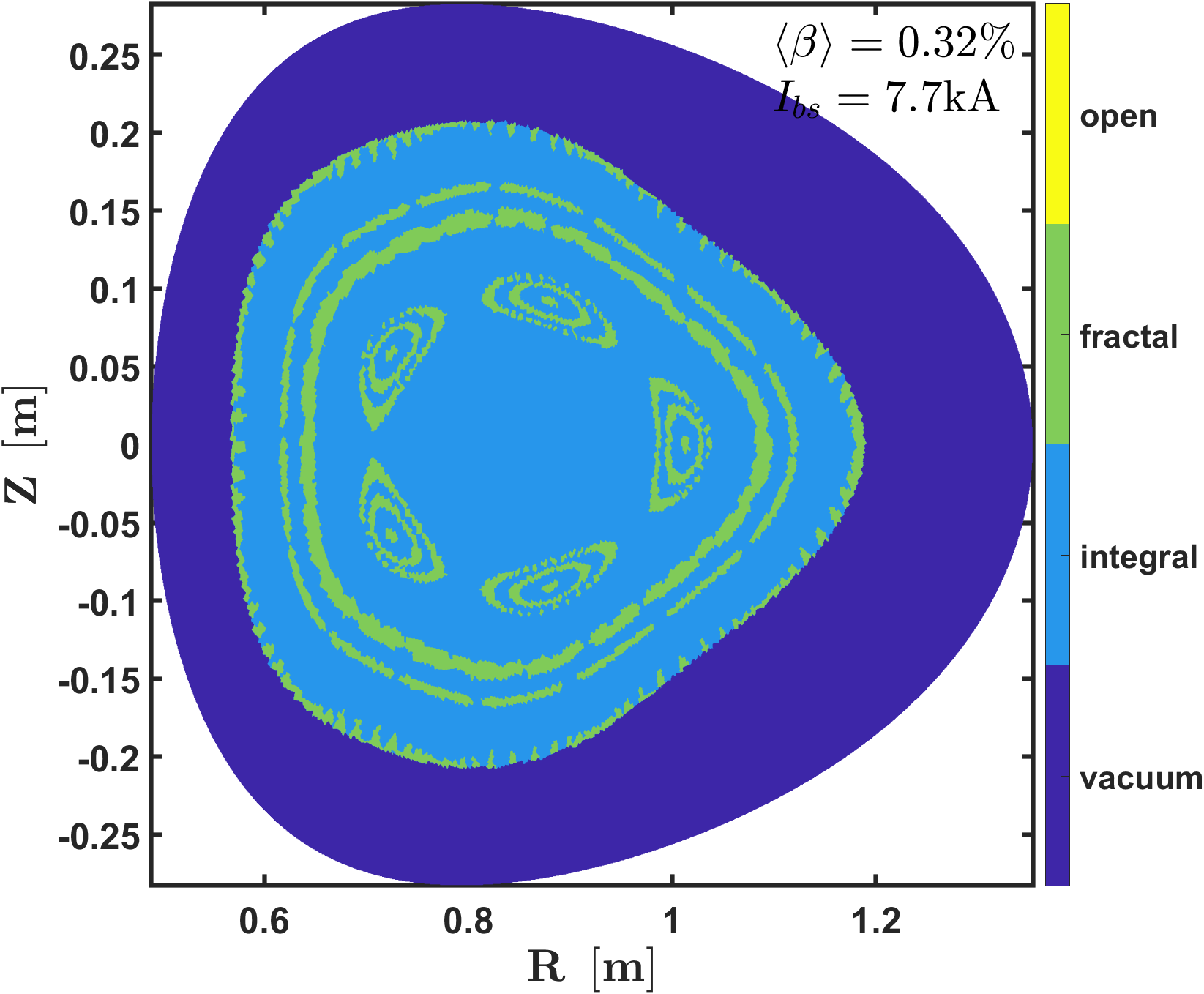}} \qquad
    \subfloat[WBA]{\includegraphics[width=0.35\linewidth]{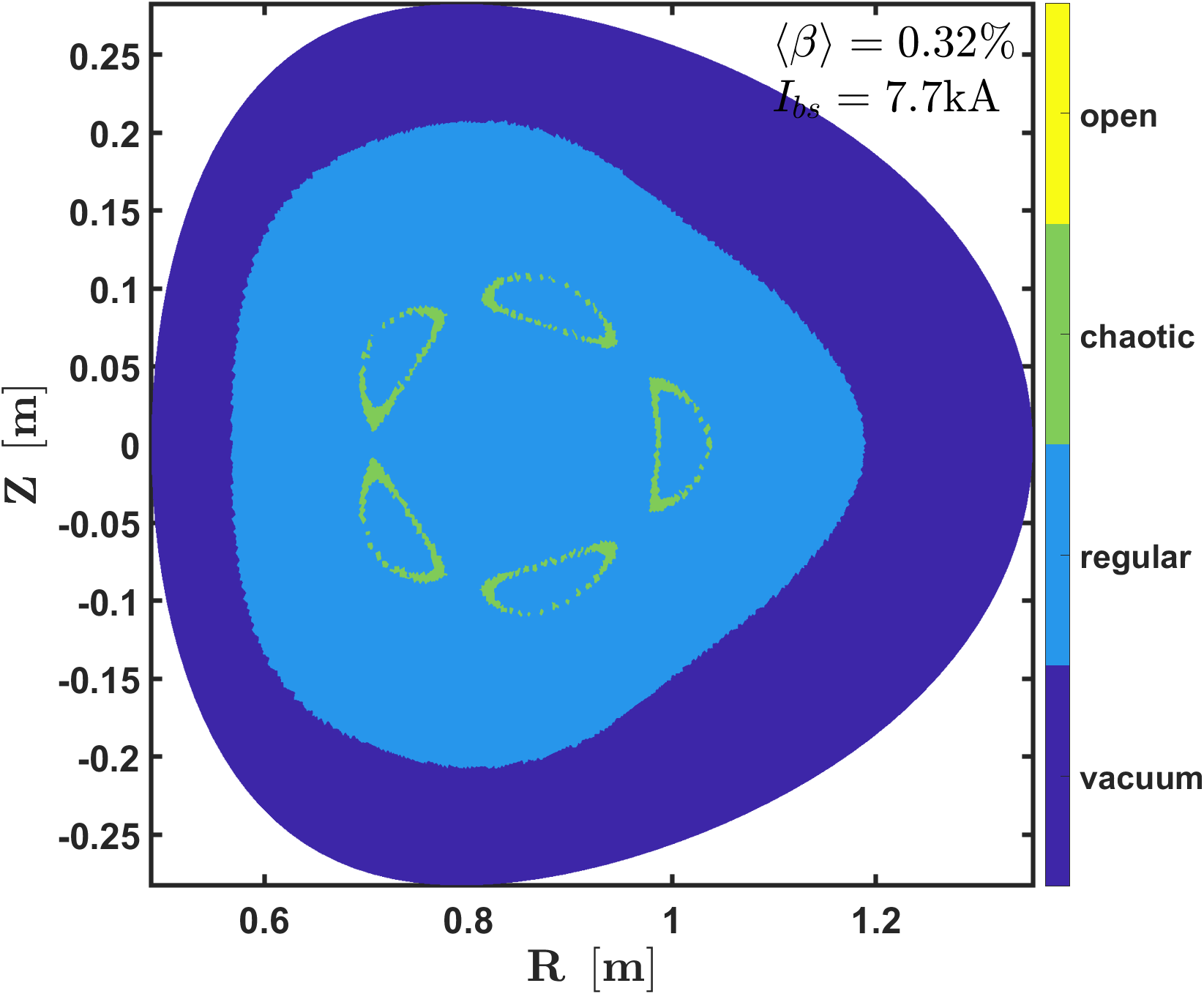}}\\ \vspace{-0.75em}
    \subfloat{\includegraphics[width=0.35\linewidth]{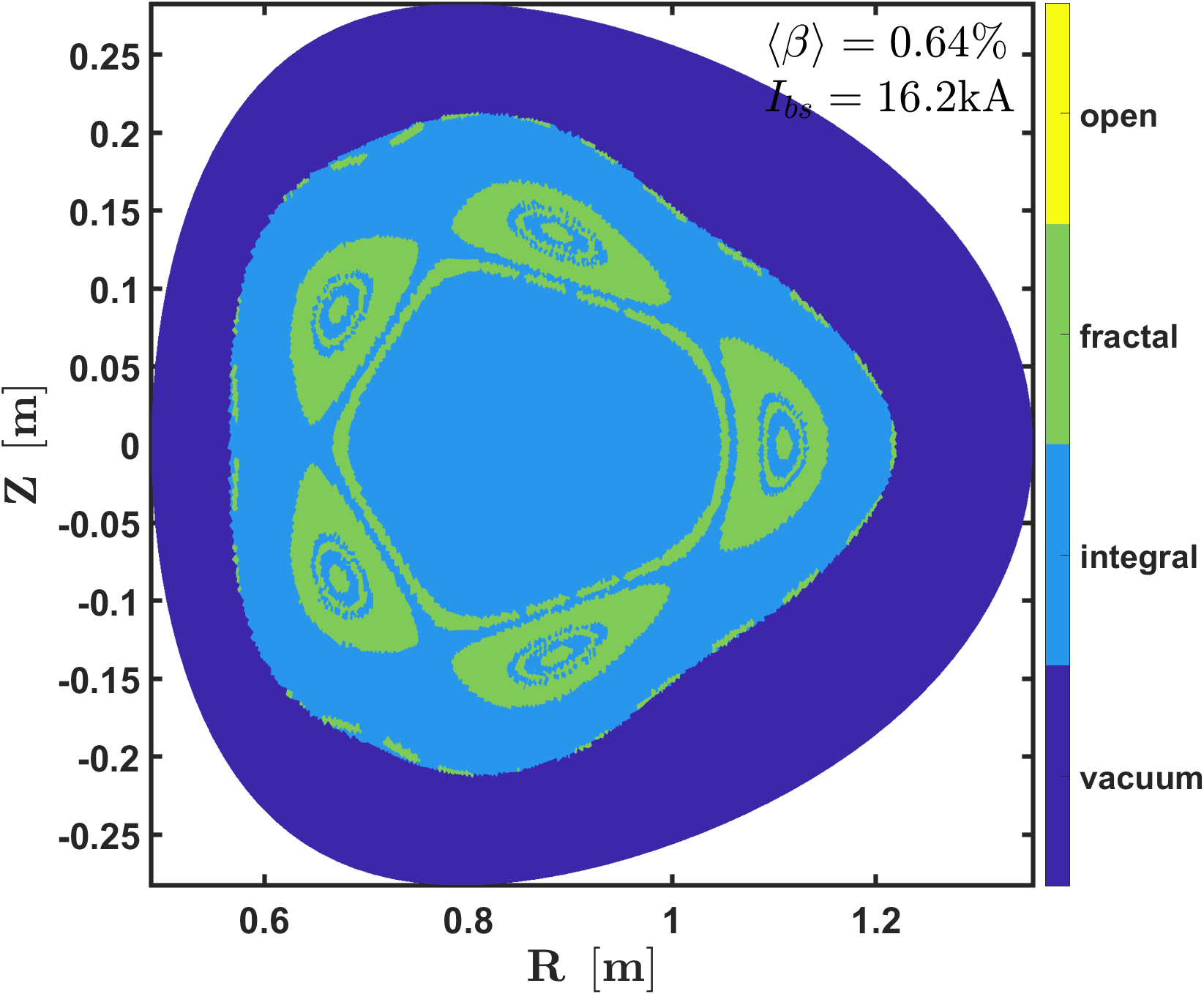}} \qquad
    \subfloat{\includegraphics[width=0.35\linewidth]{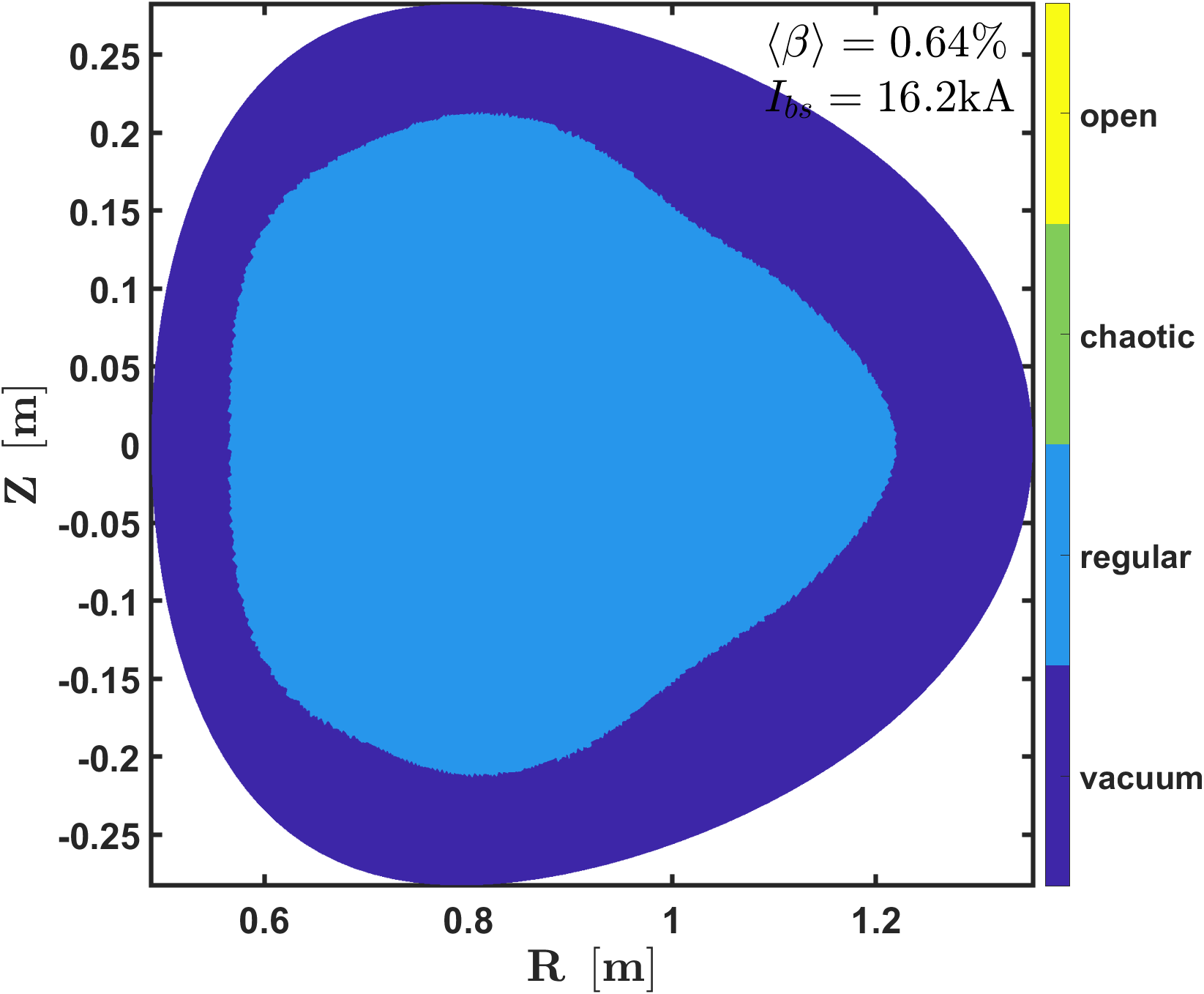}}\\ \vspace{-0.75em}
    \subfloat{\includegraphics[width=0.35\linewidth]{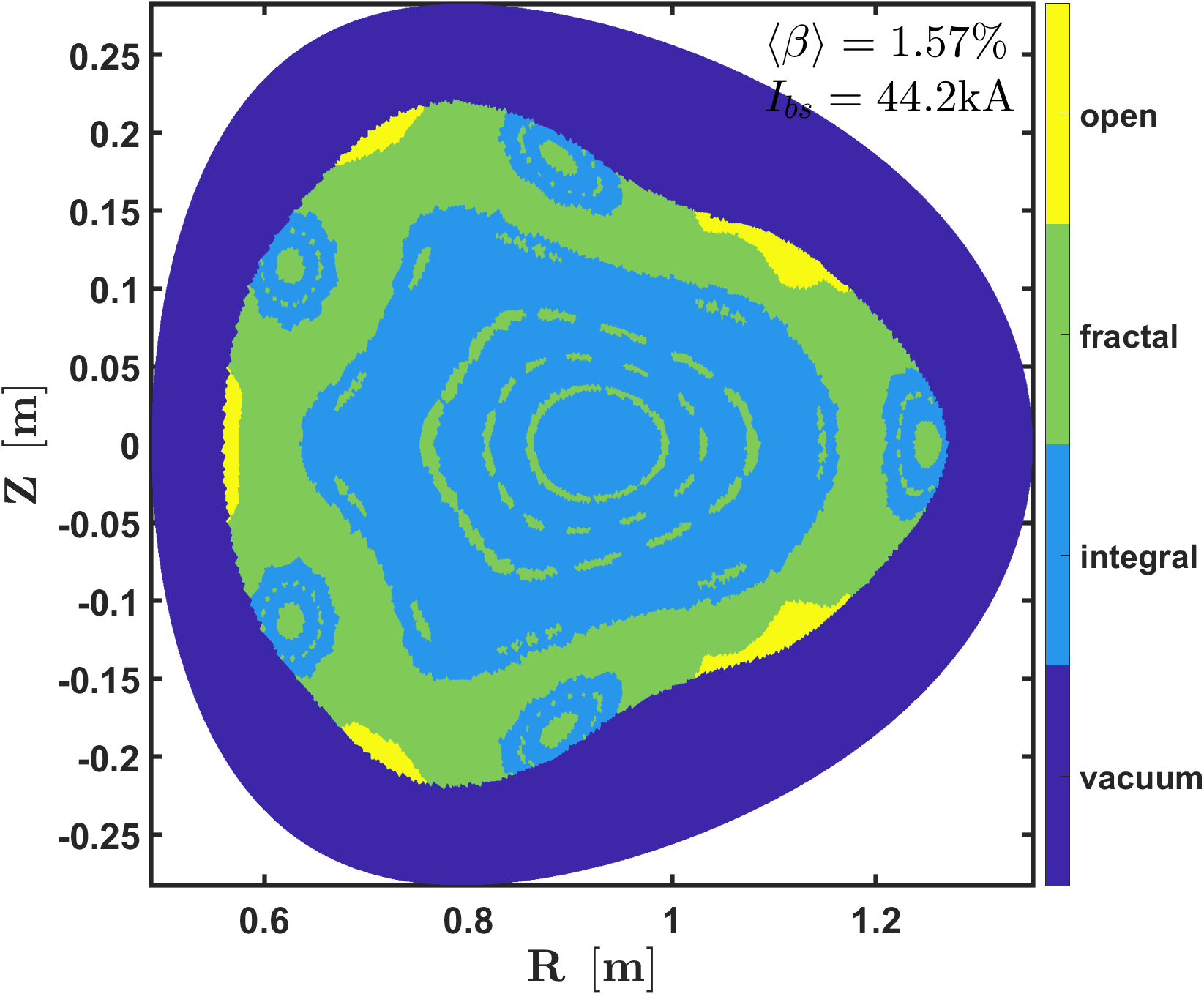}} \qquad
    \subfloat{\includegraphics[width=0.35\linewidth]{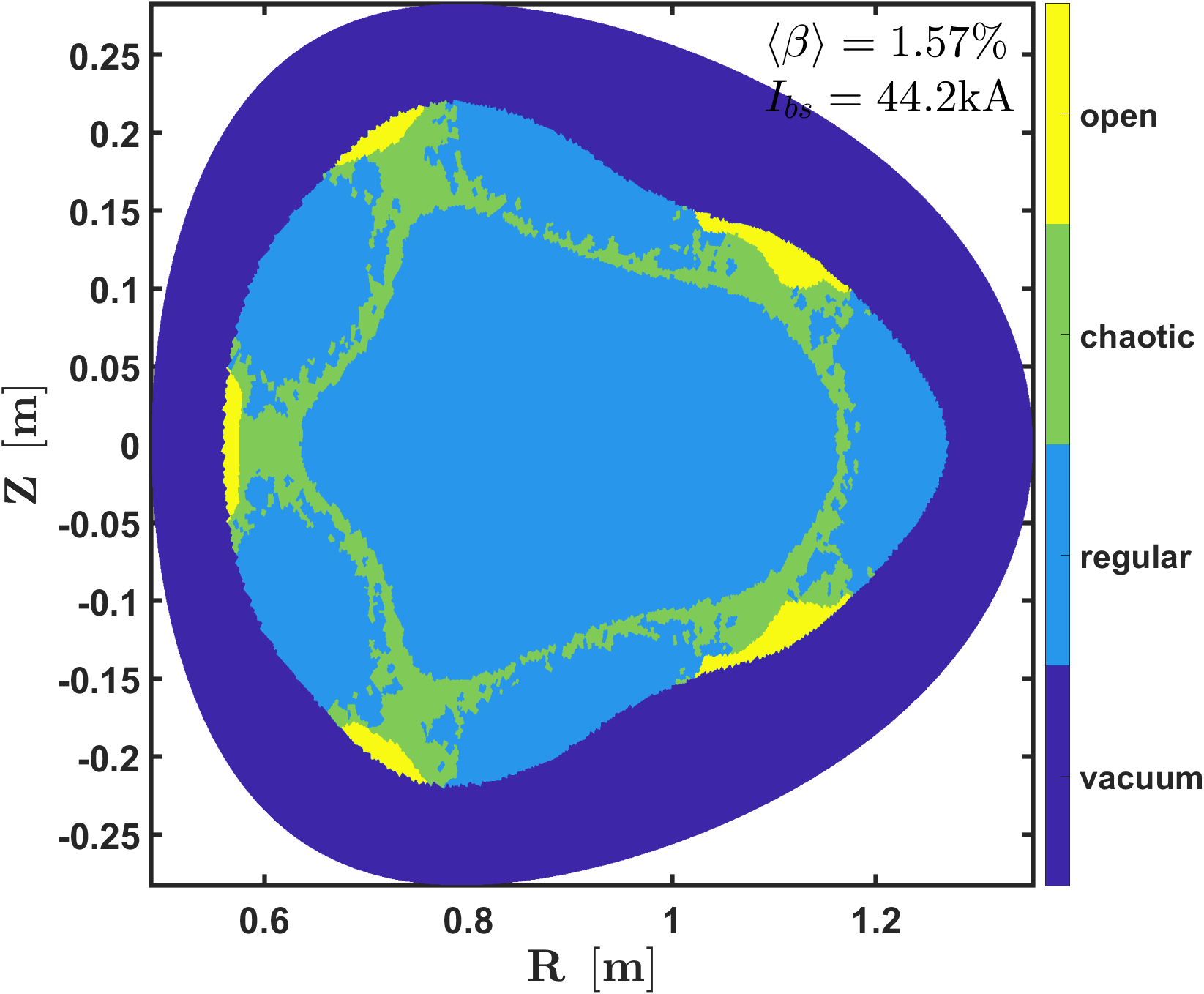}}\\ \vspace{-0.75em}
    \subfloat{\includegraphics[width=0.35\linewidth]{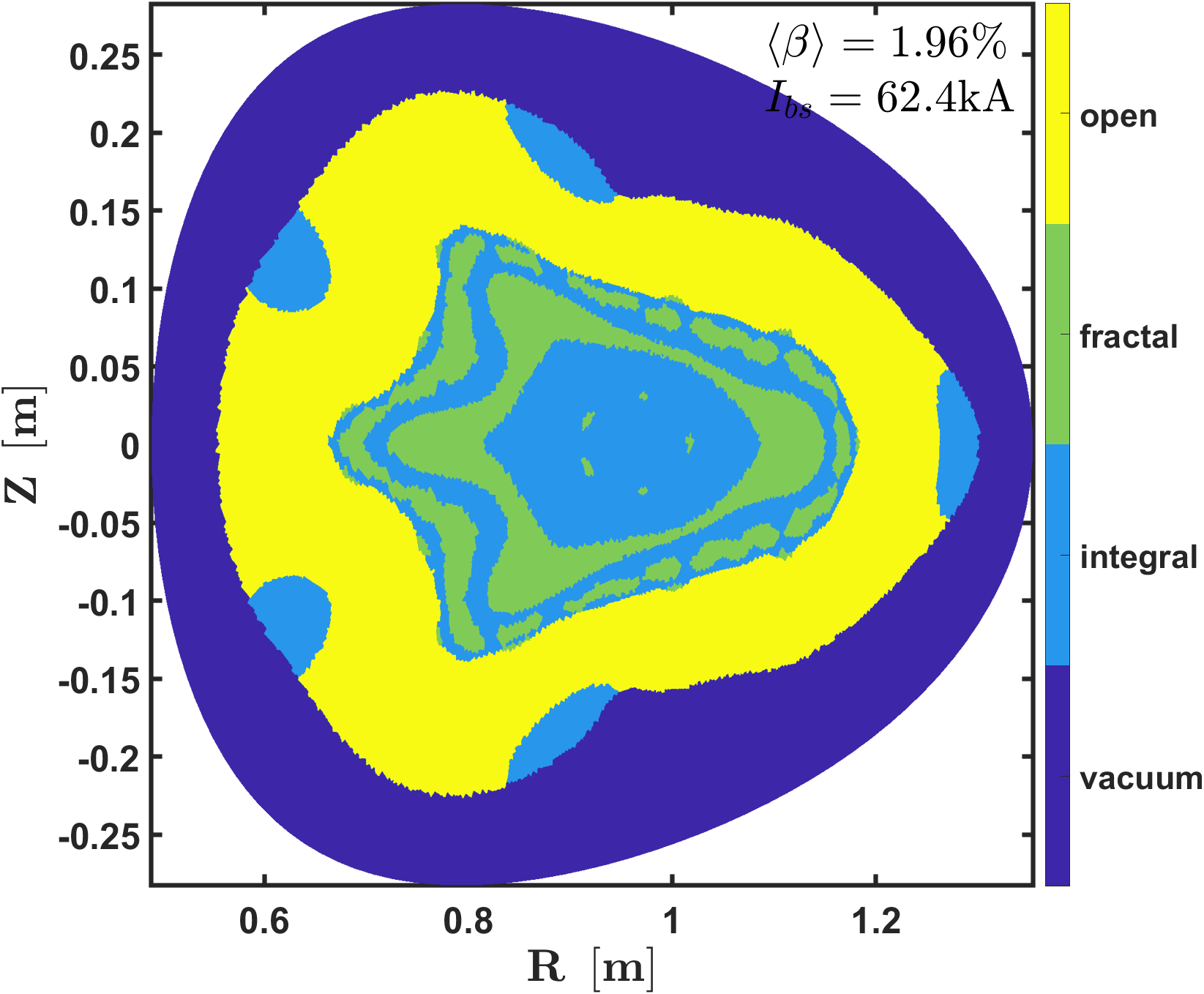}} \qquad
    \subfloat{\includegraphics[width=0.35\linewidth]{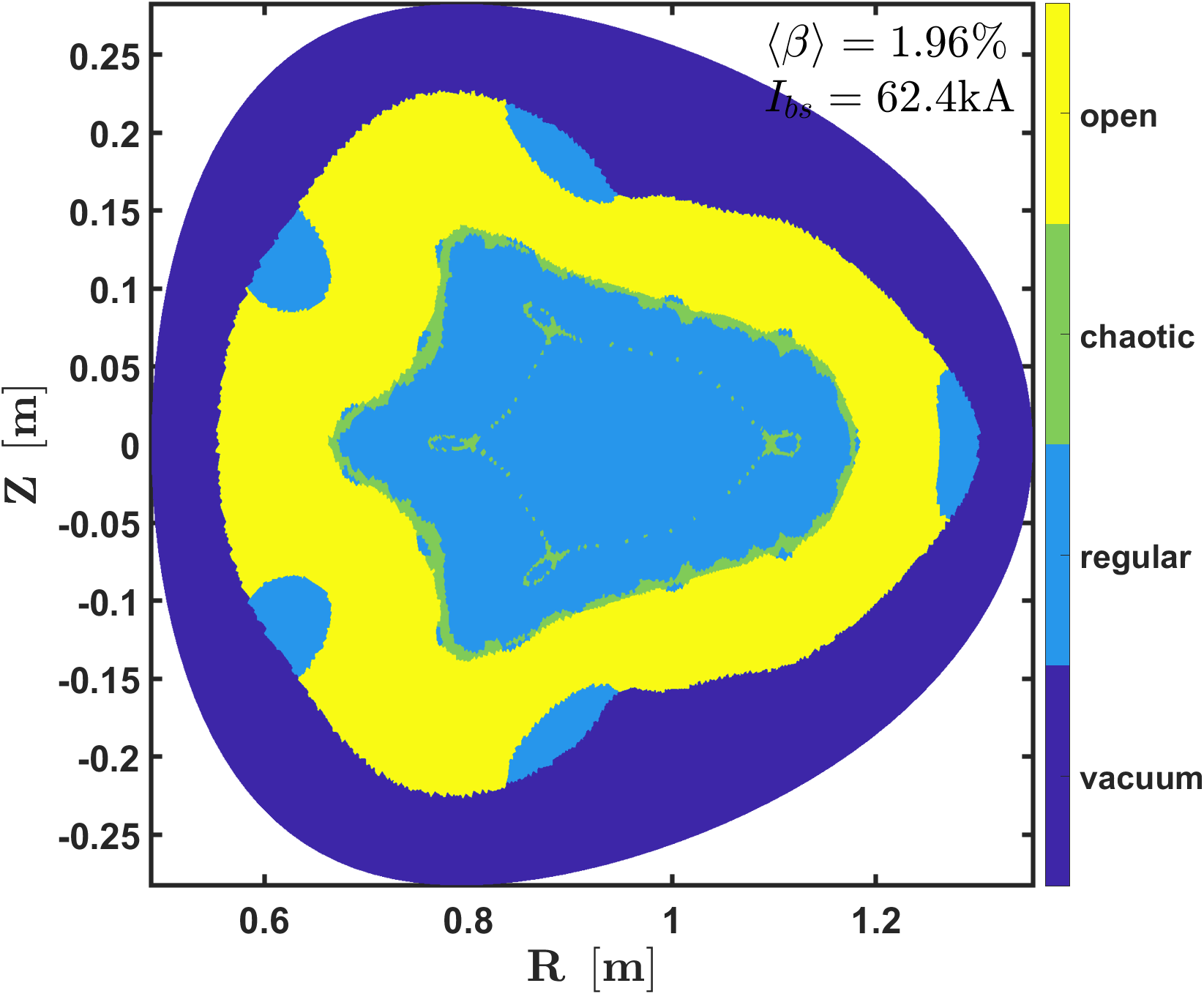}}
    \caption{Contours of the box-counting dimension (left column) and the weighted Birkhoff average (right column) for bootstrap-current-carrying finite-$\beta$ equilibria in the standard configuration with various $\langle\beta\rangle$. 
             The filled contours are drawn at the cross section $\phi=\pi/2$.}
    \label{fig:BC_WBA_contours_bootstrap}
\end{figure}

\begin{figure}[htbp]
    \centering
    \subfloat{\includegraphics[width=1\linewidth]{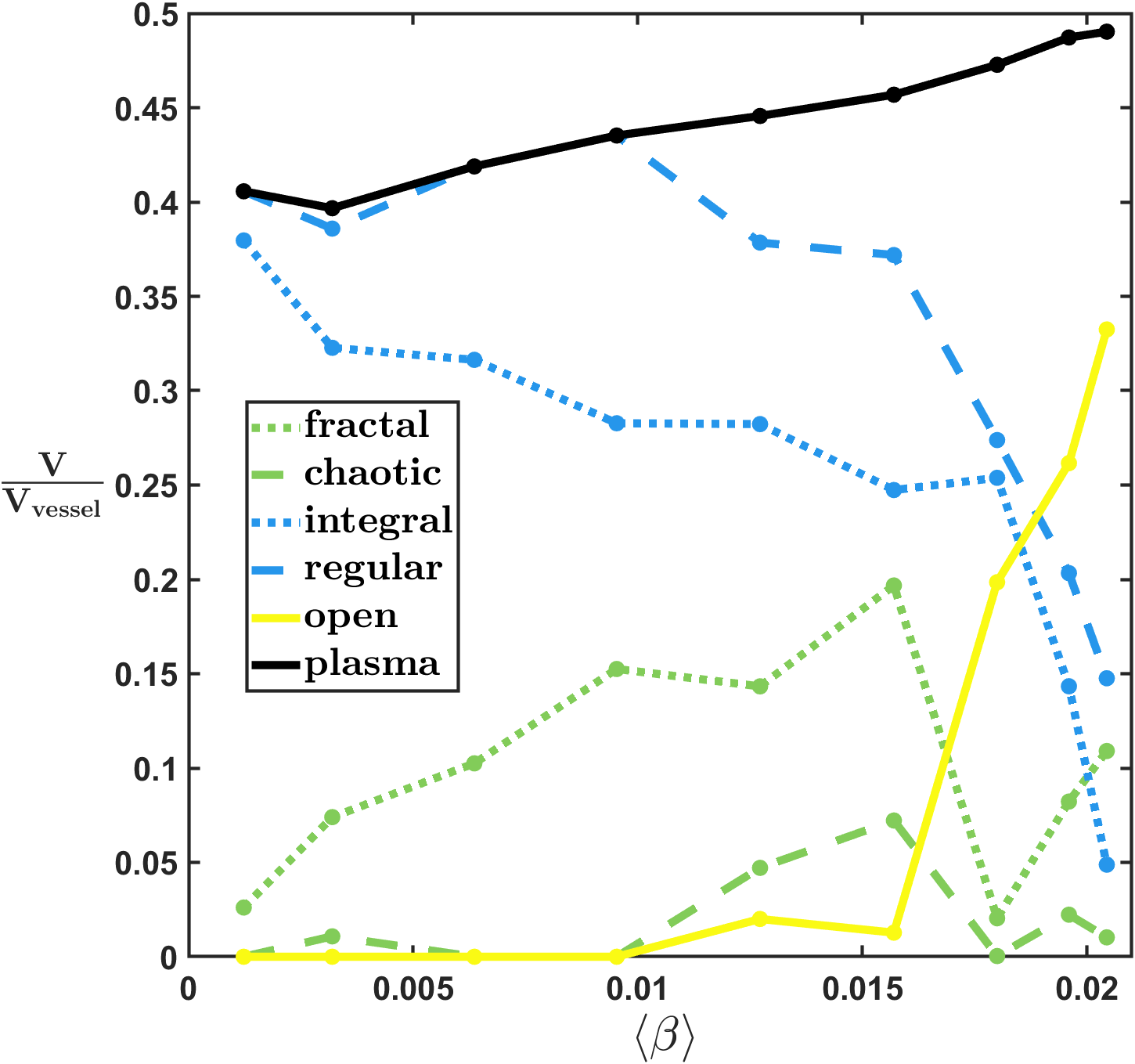}}
    \caption{Effective subvolumes divided by the box-counting dimension (dotted lines) or the weighted Birkhoff average (dashed lines)
             for the bootstrap-current-carrying finite-$\beta$ equilibria in the standard configuration.}
    \label{fig:beta_vs_BC_WBA_bootstrap}
\end{figure}

The contours of the box-counting dimension and the WBA for the equilibria with bootstrap currents are shown in 
\Cref{fig:BC_WBA_contours_bootstrap}. Apparently, in the equilibrium with $\langle\beta\rangle=1.57\%$, 
the fractal/chaotic regions adjoin the open-field-line regions in the area between the edge islands and the LCFS. 
Hence, the perturbed field lines outside the LCFS tend to become open with increasing $\beta$. 
The open-field-line regions, instead of the fractal/chaotic regions, take place of the integral/regular regions in high-$\beta$ equilibria, 
which is significantly different from the net-current-free cases. 
\Cref{fig:beta_vs_BC_WBA_bootstrap} further illustrates the variation of the volumes of various subregions with $\langle\beta\rangle$. 
It is found that the rapid growth of the open-field-line subvolume, corresponding to the plunge of the integral/regular subvolume, 
starts at $\langle\beta\rangle\approx1.5\%$, which implies that the equilibrium $\beta$ limit of the bootstrap-current-carrying standard configuration is reached. 

\begin{figure}[htbp]
    % \vspace{-7em}
    \centering
    \subfloat{\includegraphics[width=0.35\linewidth]{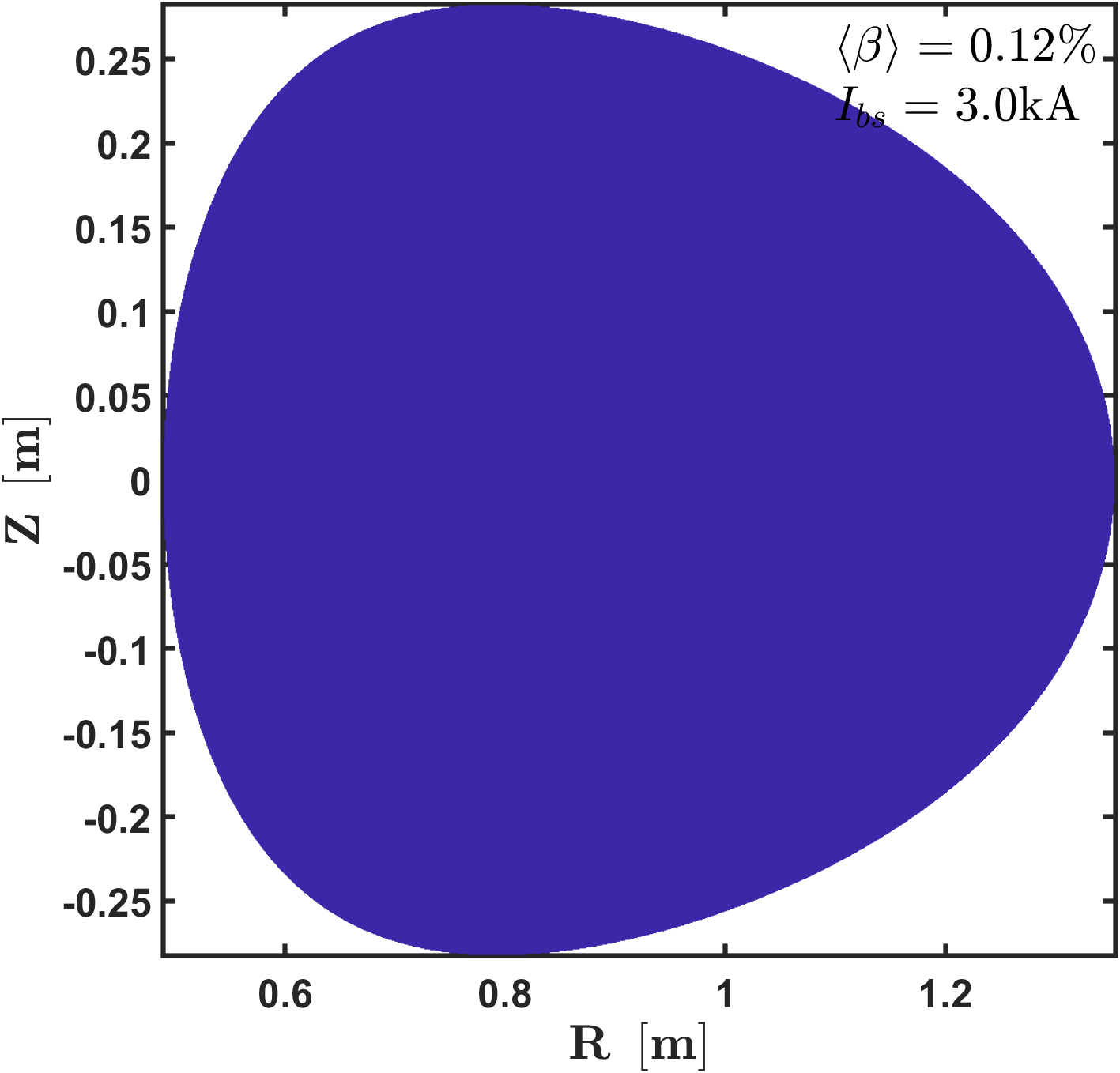}}
    \subfloat{\includegraphics[width=0.35\linewidth]{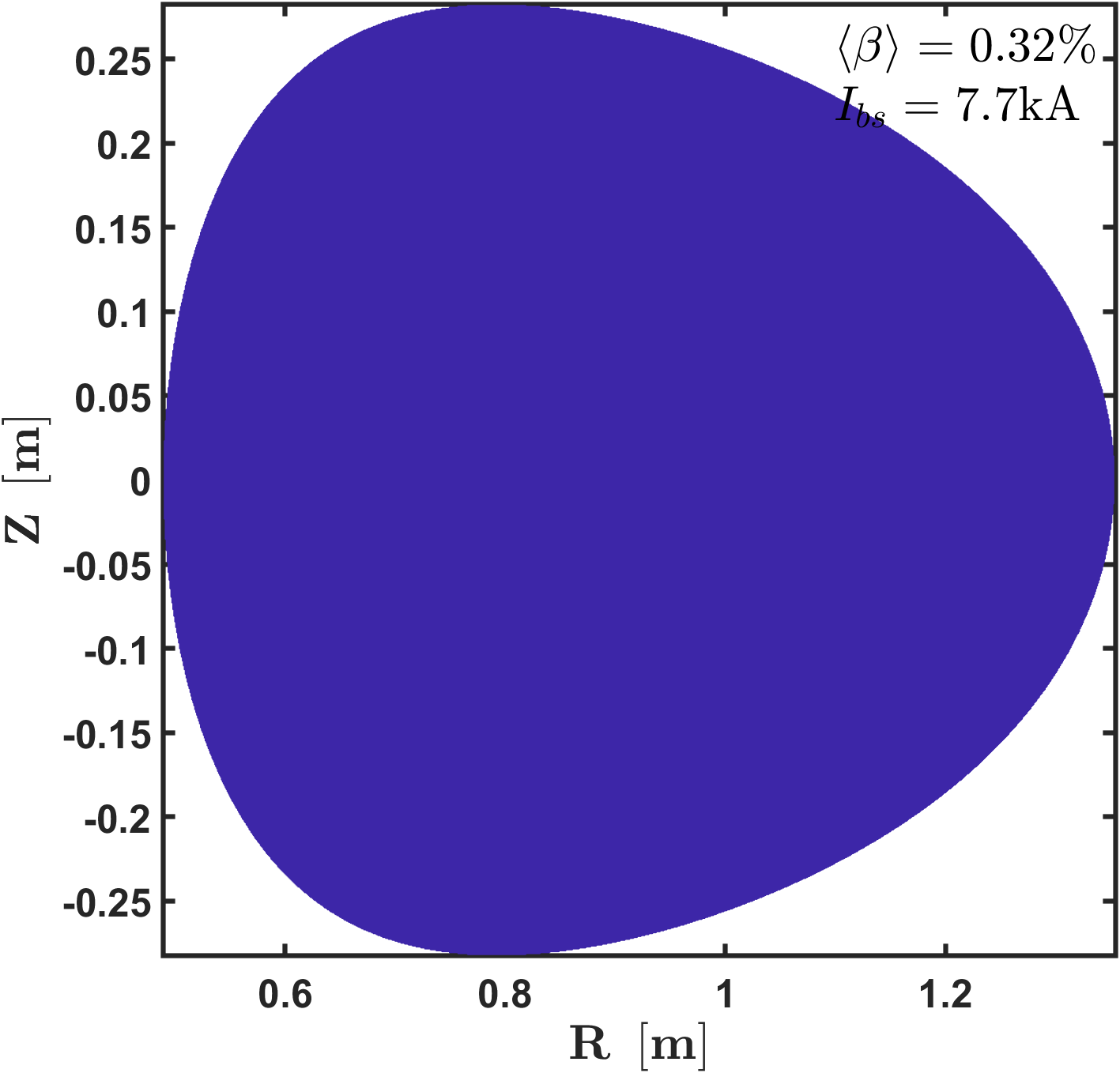}}
    \subfloat{\includegraphics[width=0.35\linewidth]{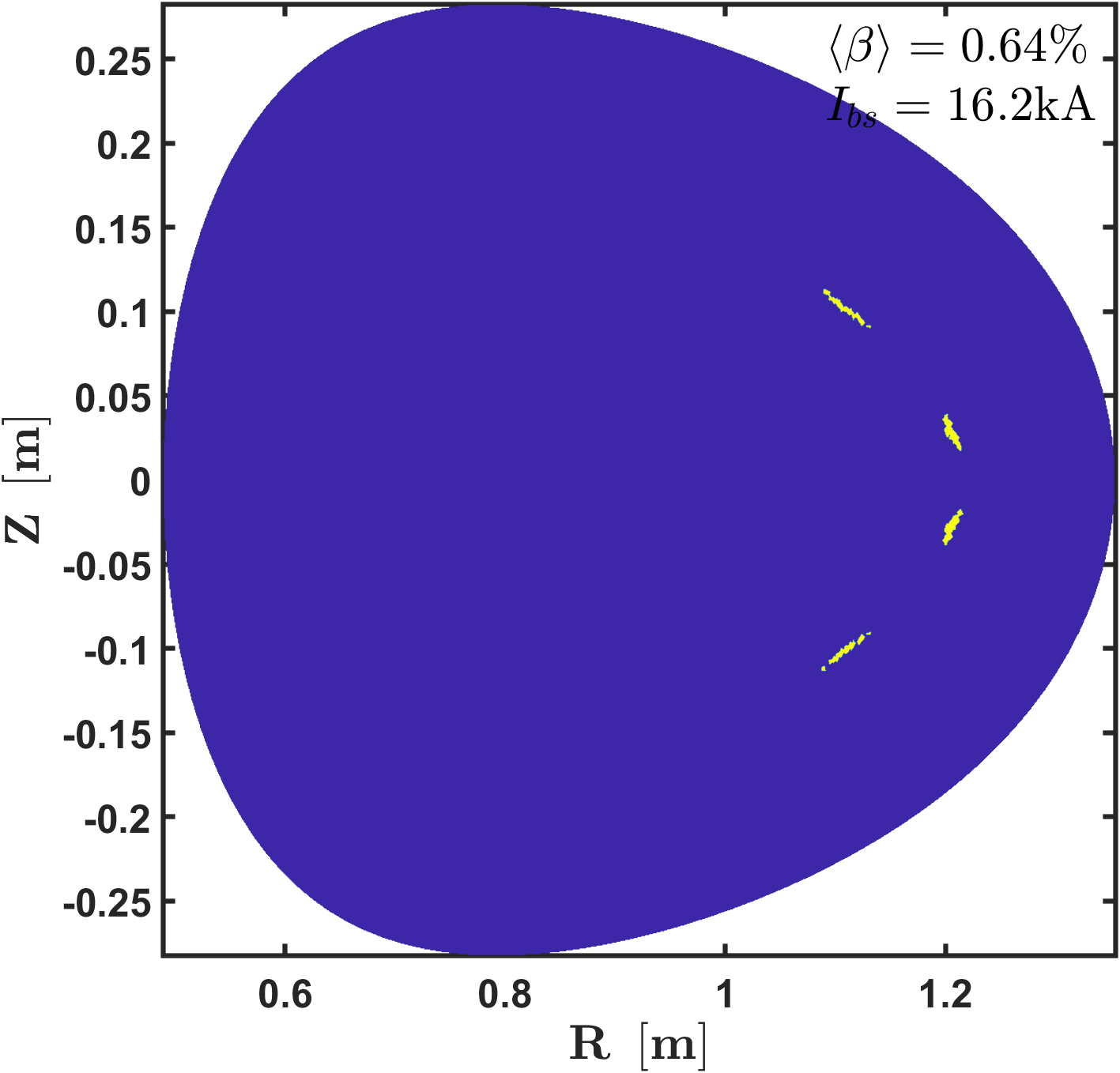}}\\ \vspace{-0.75em}
    \subfloat{\includegraphics[width=0.35\linewidth]{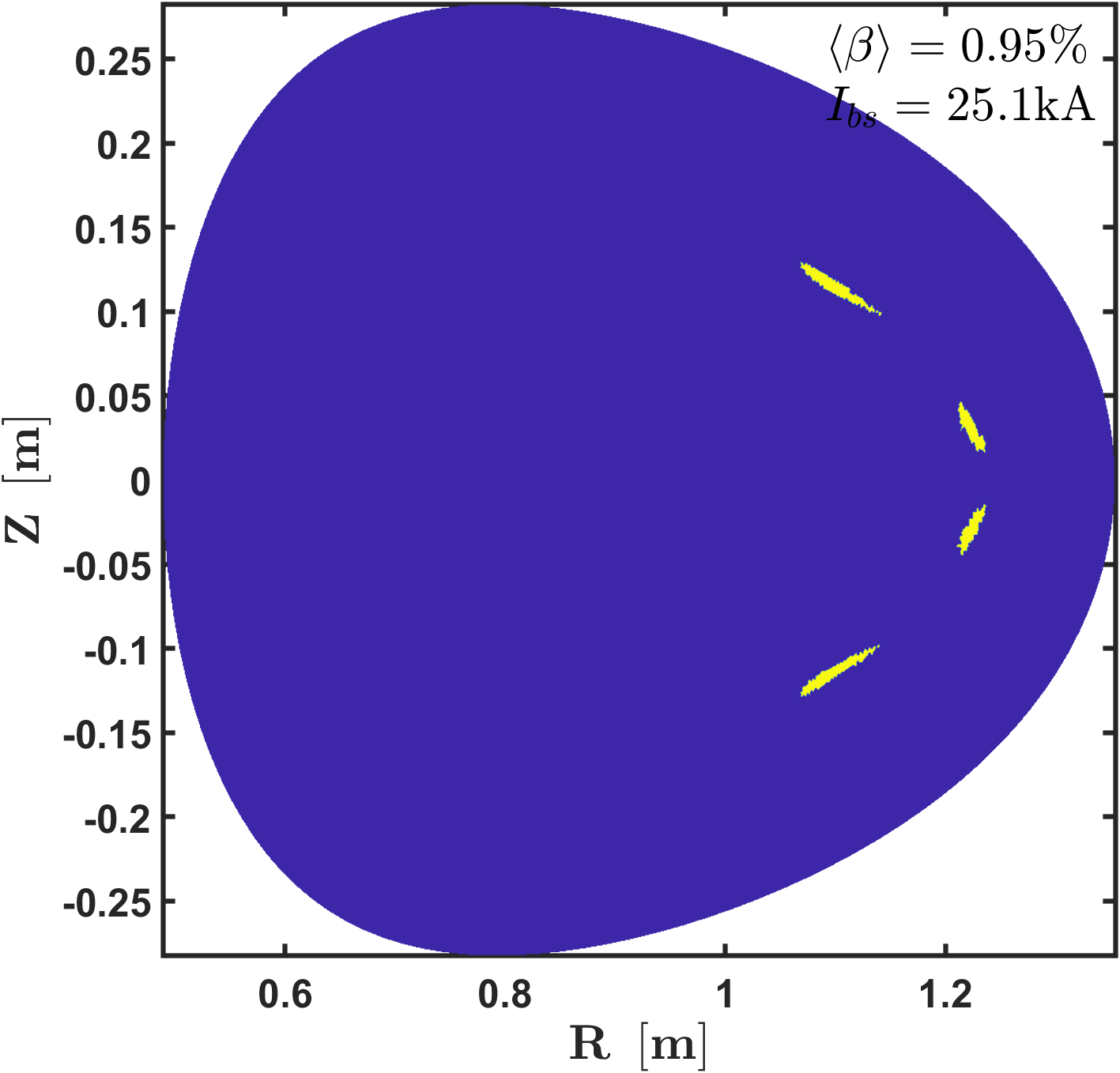}}
    \subfloat{\includegraphics[width=0.35\linewidth]{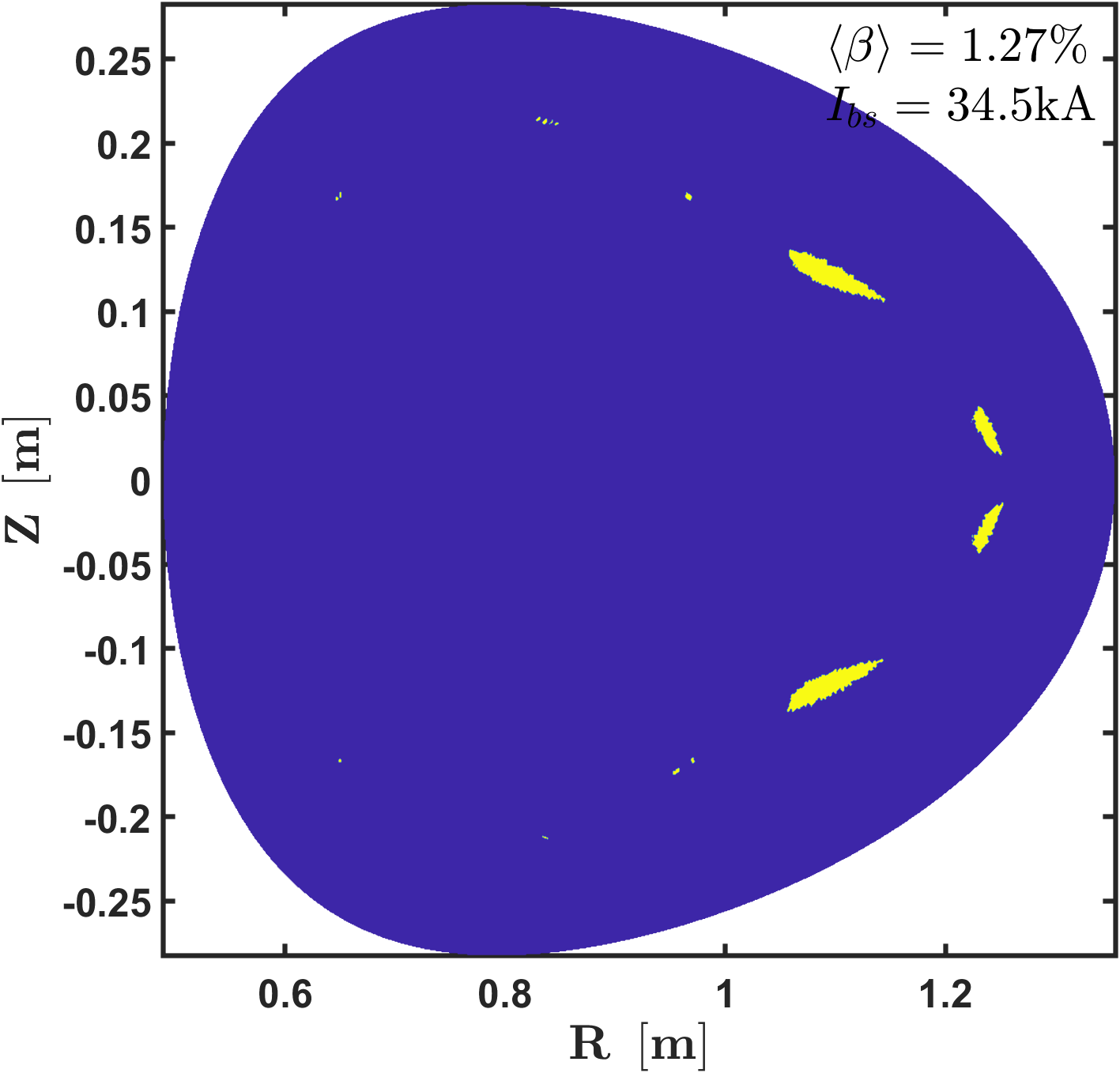}}
    \subfloat{\includegraphics[width=0.35\linewidth]{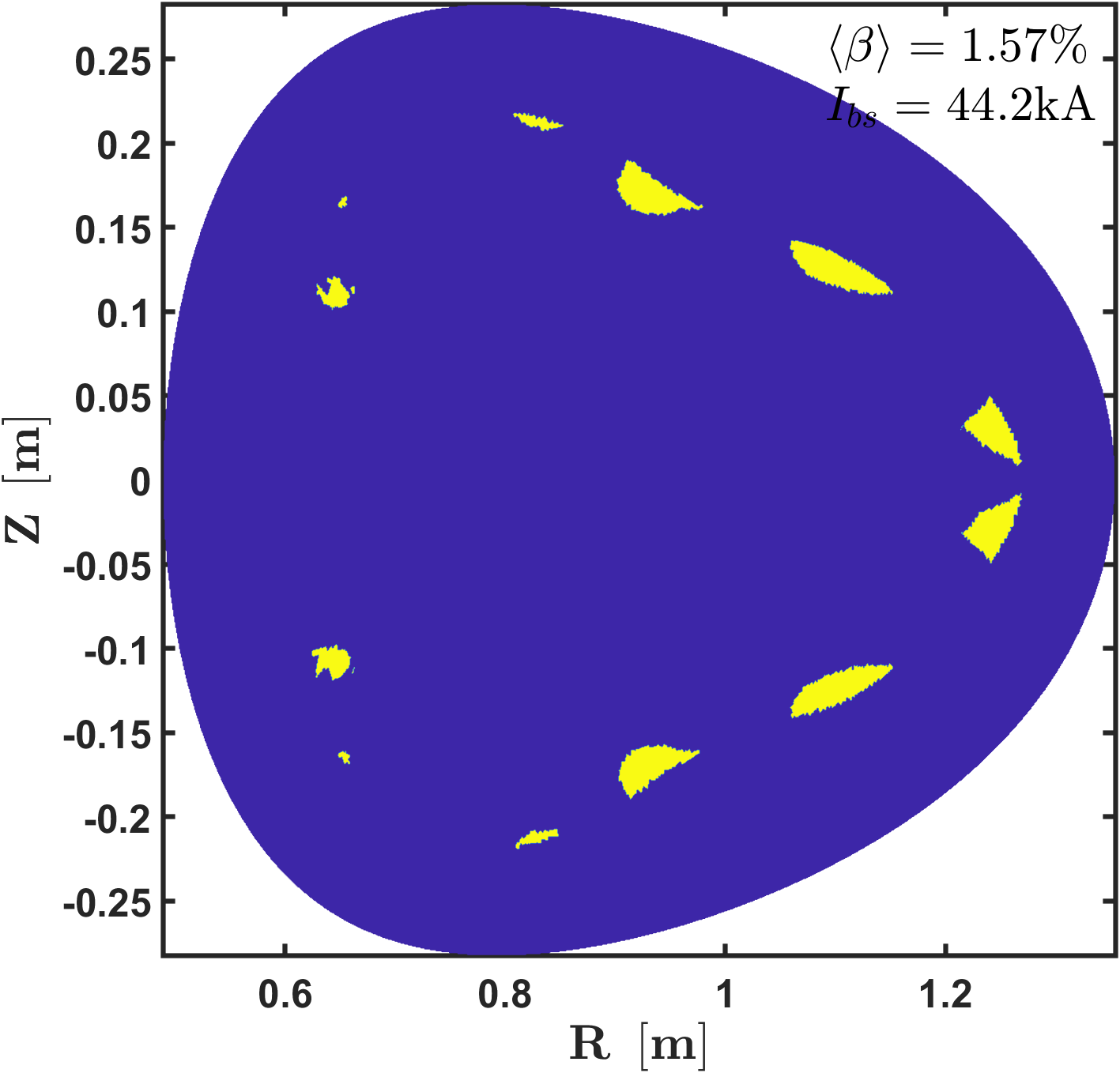}}\\ \vspace{-0.75em}
    \subfloat{\includegraphics[width=0.35\linewidth]{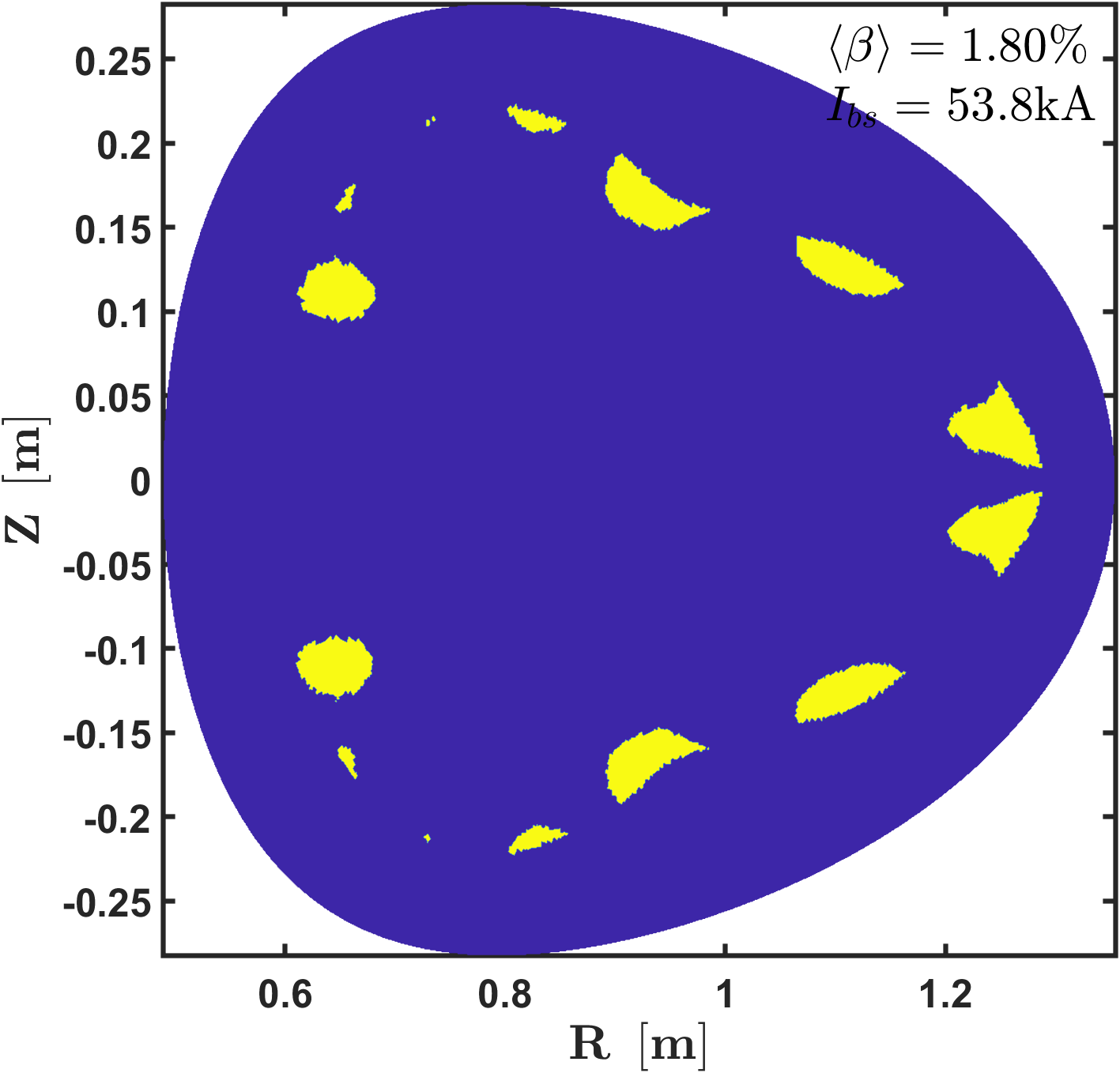}}
    \subfloat{\includegraphics[width=0.35\linewidth]{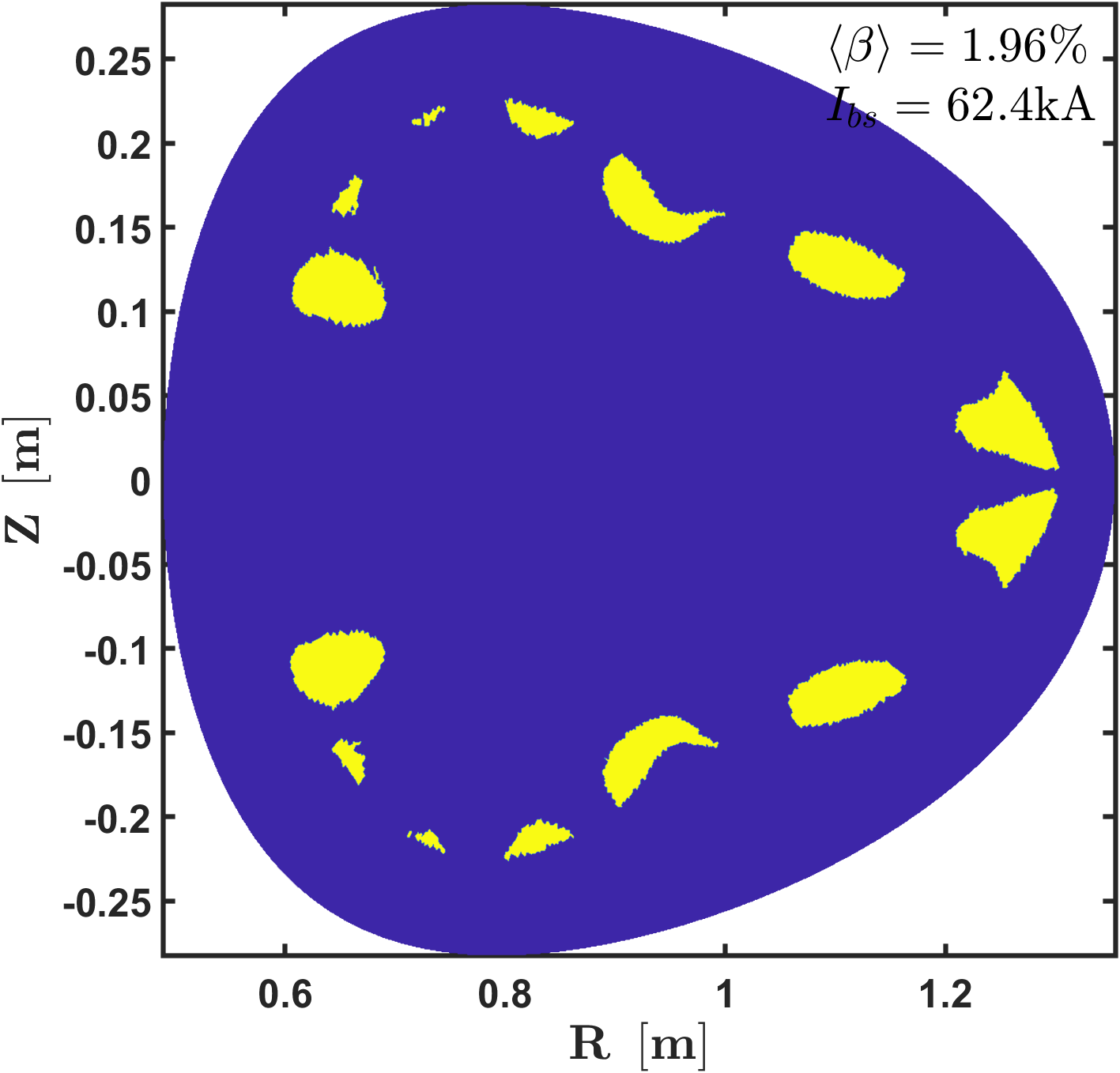}}
    \subfloat{\includegraphics[width=0.35\linewidth]{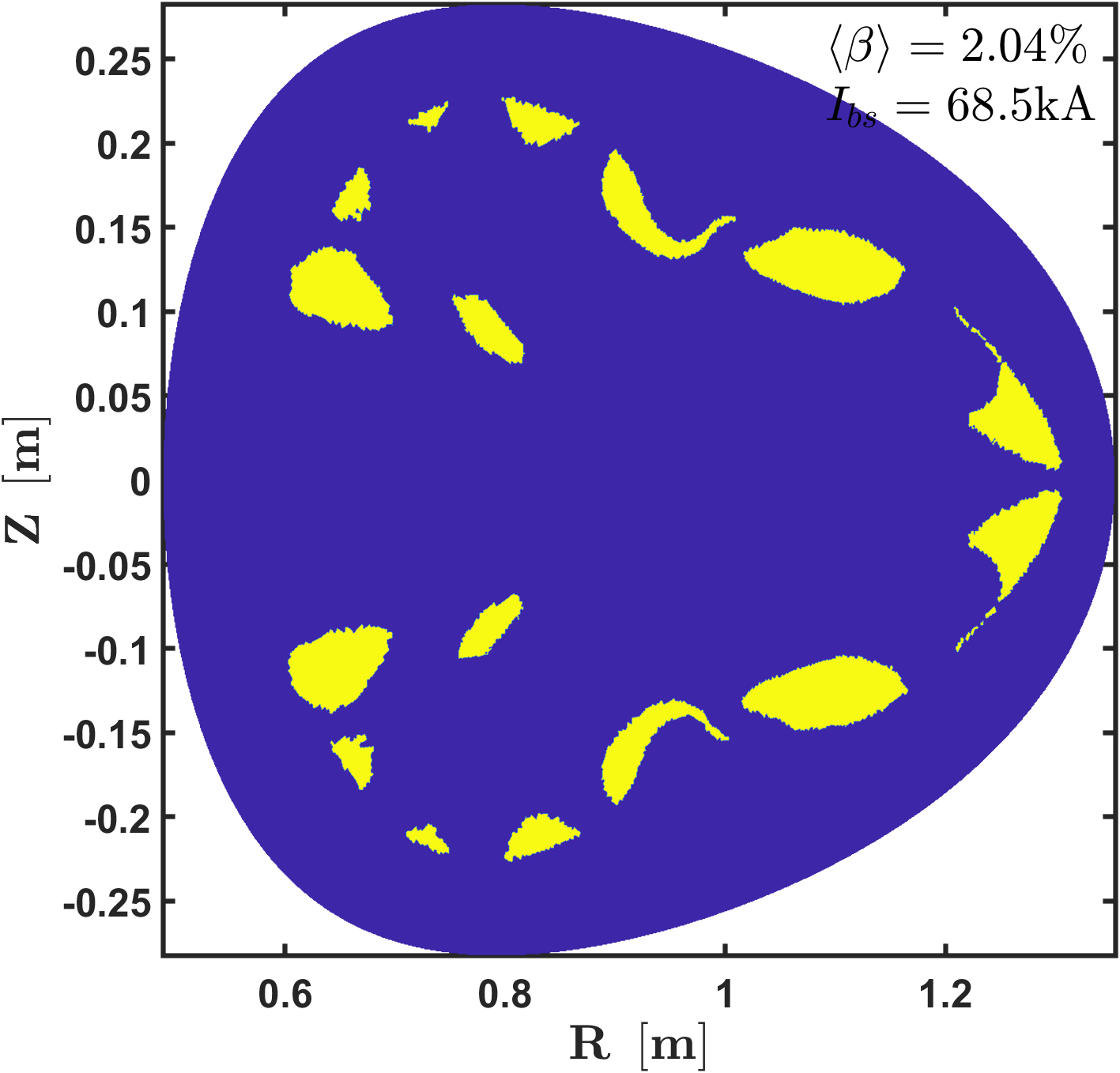}}
    \caption{Contours of the parallel diffusion for bootstrap-current-carrying finite-$\beta$ equilibria in the standard configuration with various $\langle\beta\rangle$. 
             The yellow-marked areas denote the $V_\text{PD}$ subregions. The filled contours are drawn at the cross section $\phi=\pi/2$.}
    \label{fig:VPD_contours_bootstrap}
\end{figure}

\Cref{fig:VPD_contours_bootstrap} draws the contours of effective parallel diffusion. Although the marked regions still do not expand into the internal area, 
they are more spacious than those in net-current-free cases, which is apparently revealed by their volume variation. 
$V_\text{PD}$ quickly rises from merely $1\%$ at $\langle\beta\rangle\approx1.0\%$, and then, reaches about $15\%$ but does not saturate at $\langle\beta\rangle\approx2.0\%$. 
Large $V_\text{PD}$ indicates that the connection length is too short to maintain the condition $\nabla_{\parallel}p=0$ in the periphery. In a word, 
the edge effect is more severe in the bootstrap-current-carrying equilibria than in the net-current-free ones, which is probably due to the fact that
the bootstrap currents can significantly modify the edge magnetic topology.

\section{Equilibrium \texorpdfstring{$\beta$}{beta} limits of magnetic island configurations}\label{sec:island}

The vacuum fields of two island configurations are produced with MC-TFCs and MC-PFCs, respectively, 
The adjustment by TFCs raises $\iota$ to pass through $1/4$ for $n/m=2/5$ islands, 
while the other by PFCs lowers $\iota$ across $1/3$ for $n/m=2/6$ islands.
The coil currents and the magnetic topology of vacuum fields are described in Reference \cite{Liu_2025}. 

\begin{figure}[htbp]
    % \vspace{-7em}
    \centering
    \subfloat{\includegraphics[width=0.35\linewidth]{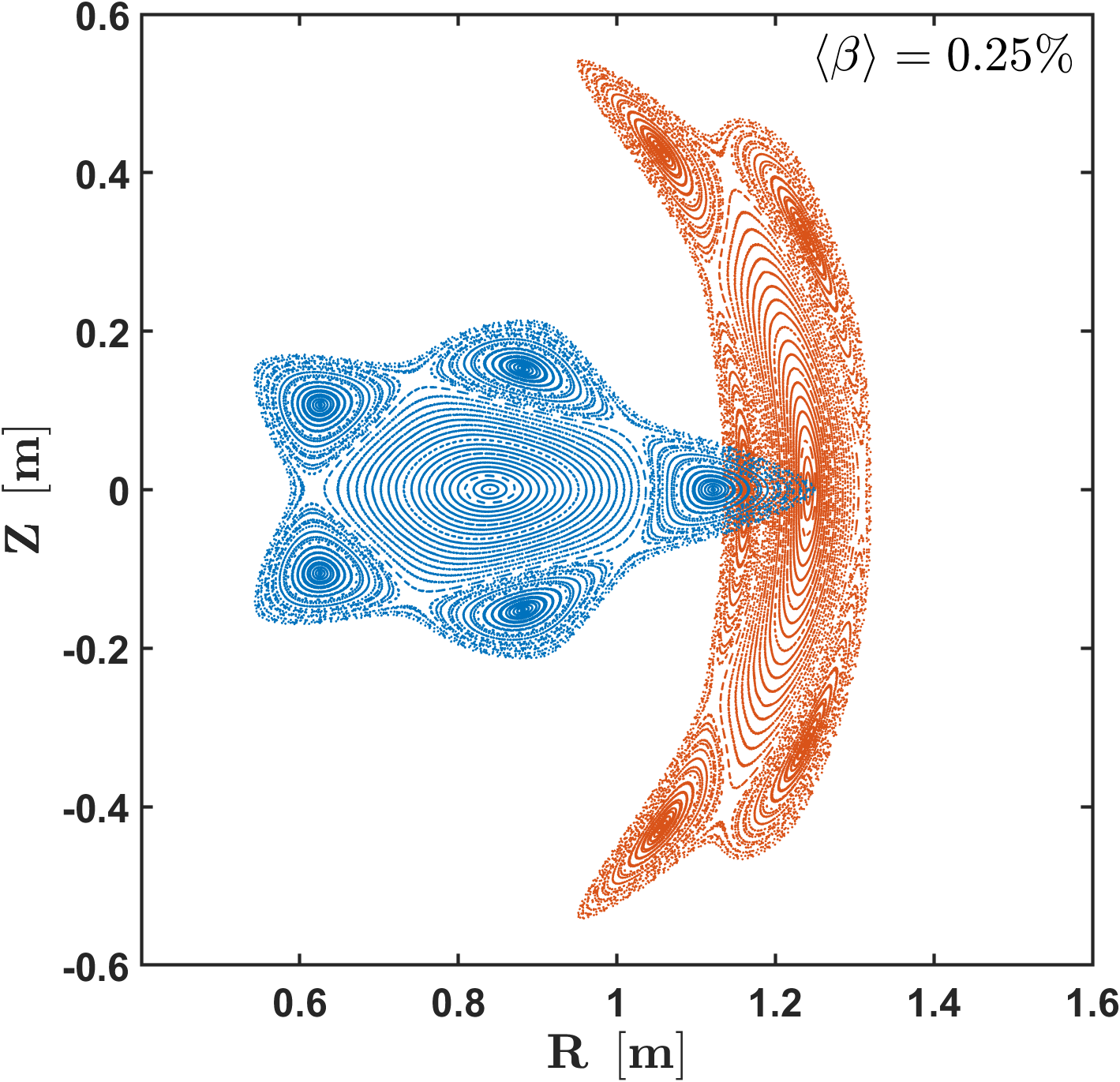}}
    \subfloat{\includegraphics[width=0.35\linewidth]{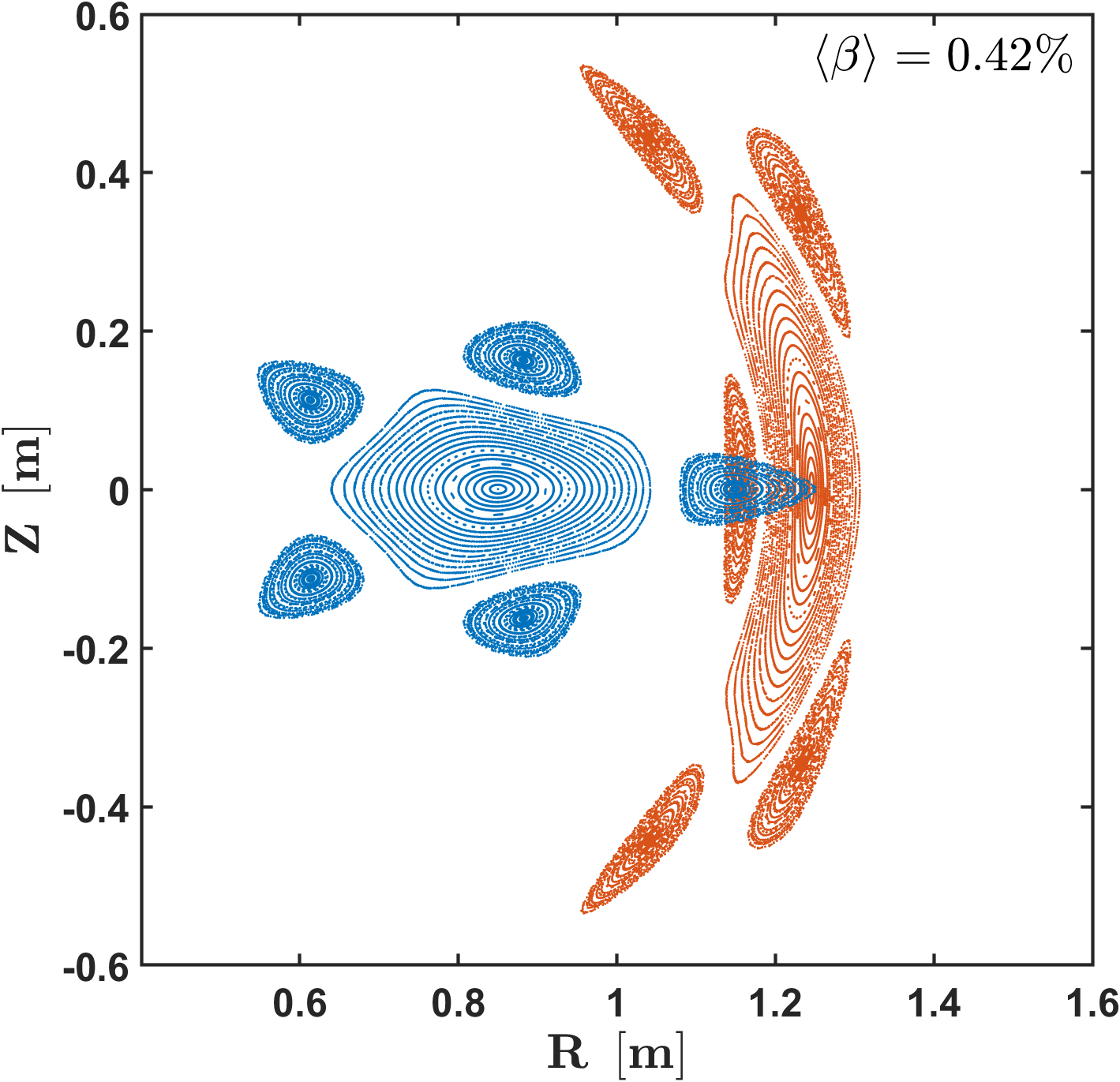}}
    \subfloat{\includegraphics[width=0.35\linewidth]{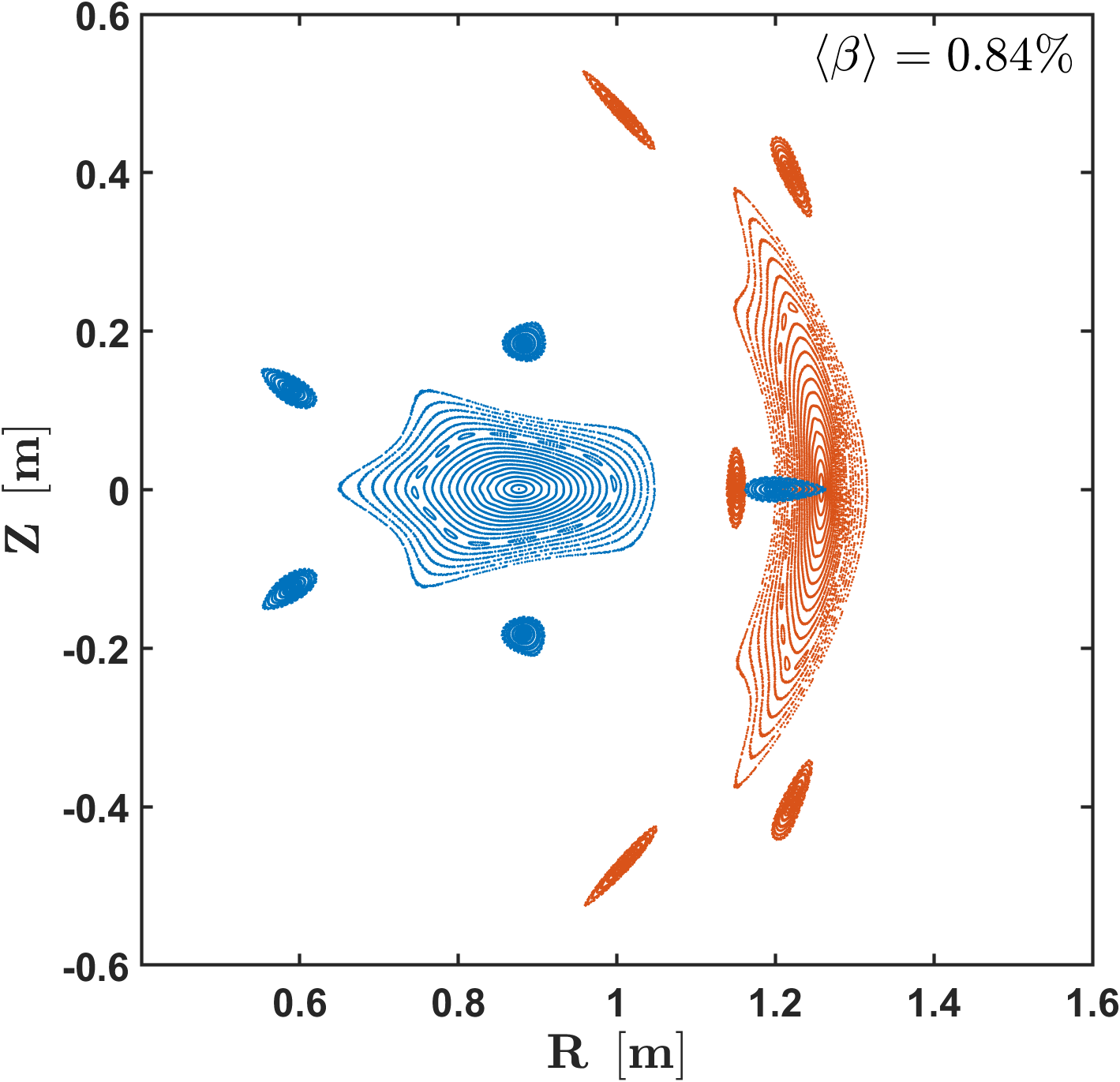}}\\ \vspace{-0.75em}
    \subfloat{\includegraphics[width=0.35\linewidth]{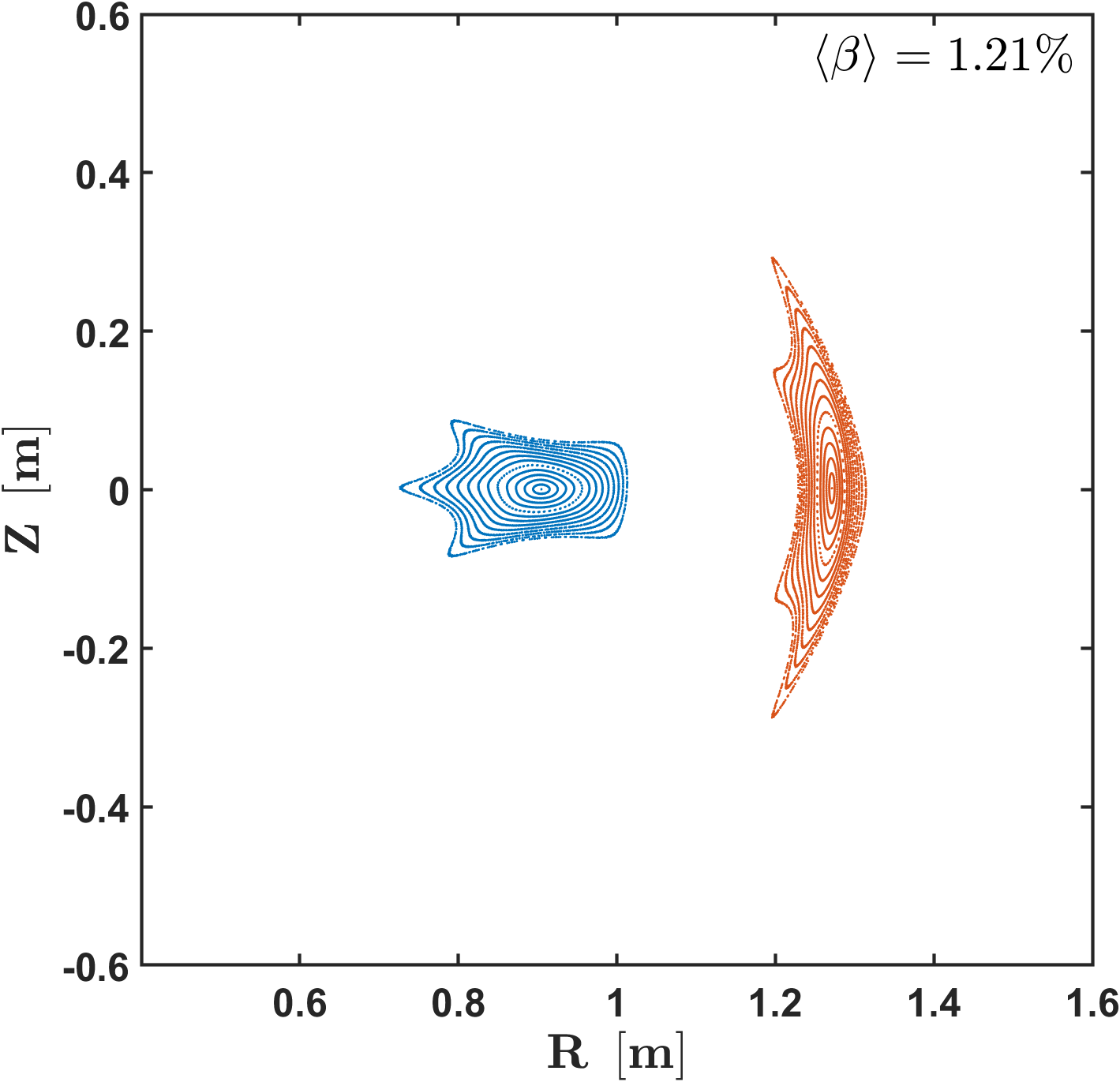}}
    \subfloat{\includegraphics[width=0.35\linewidth]{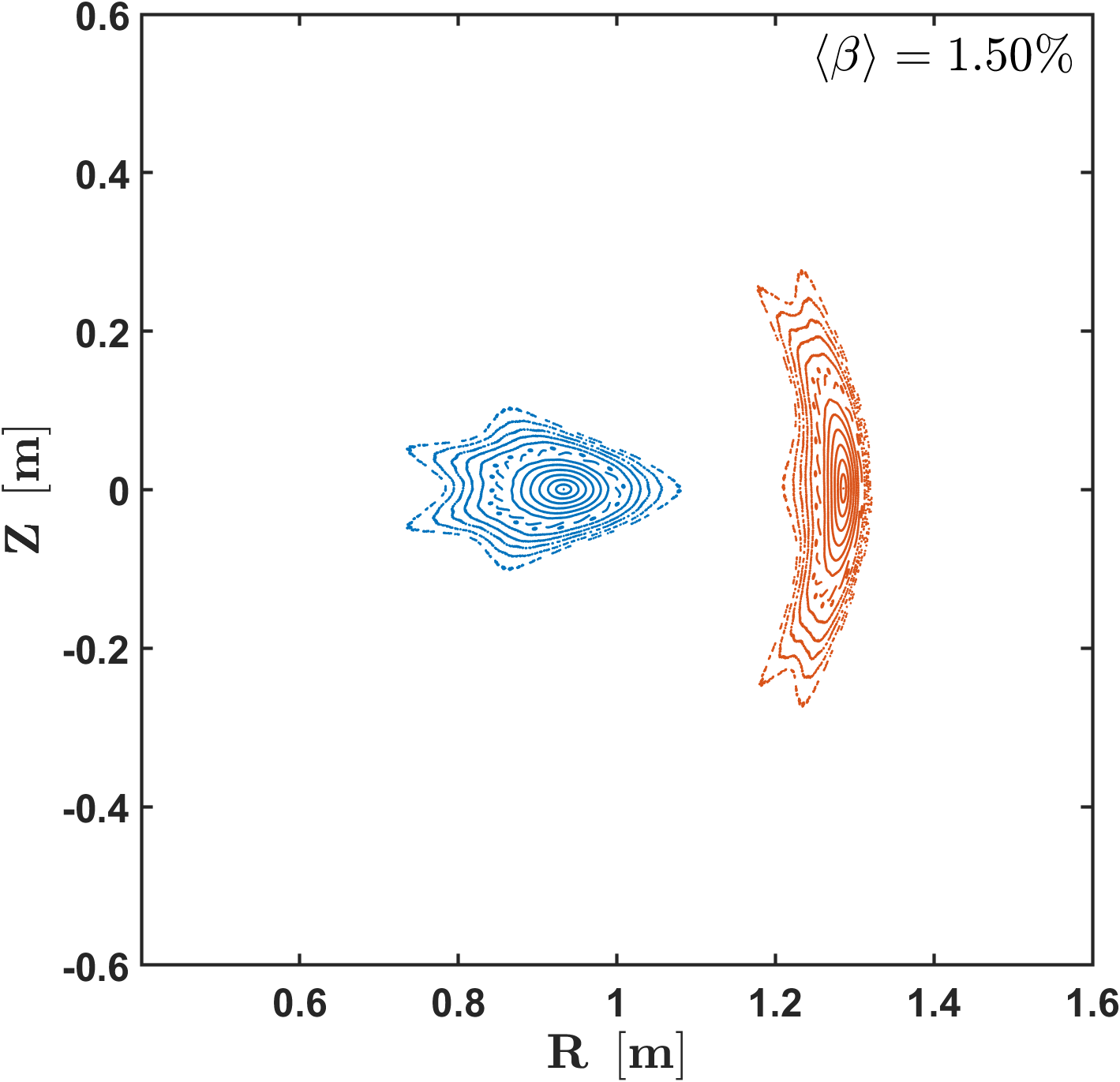}}
    \subfloat{\includegraphics[width=0.35\linewidth]{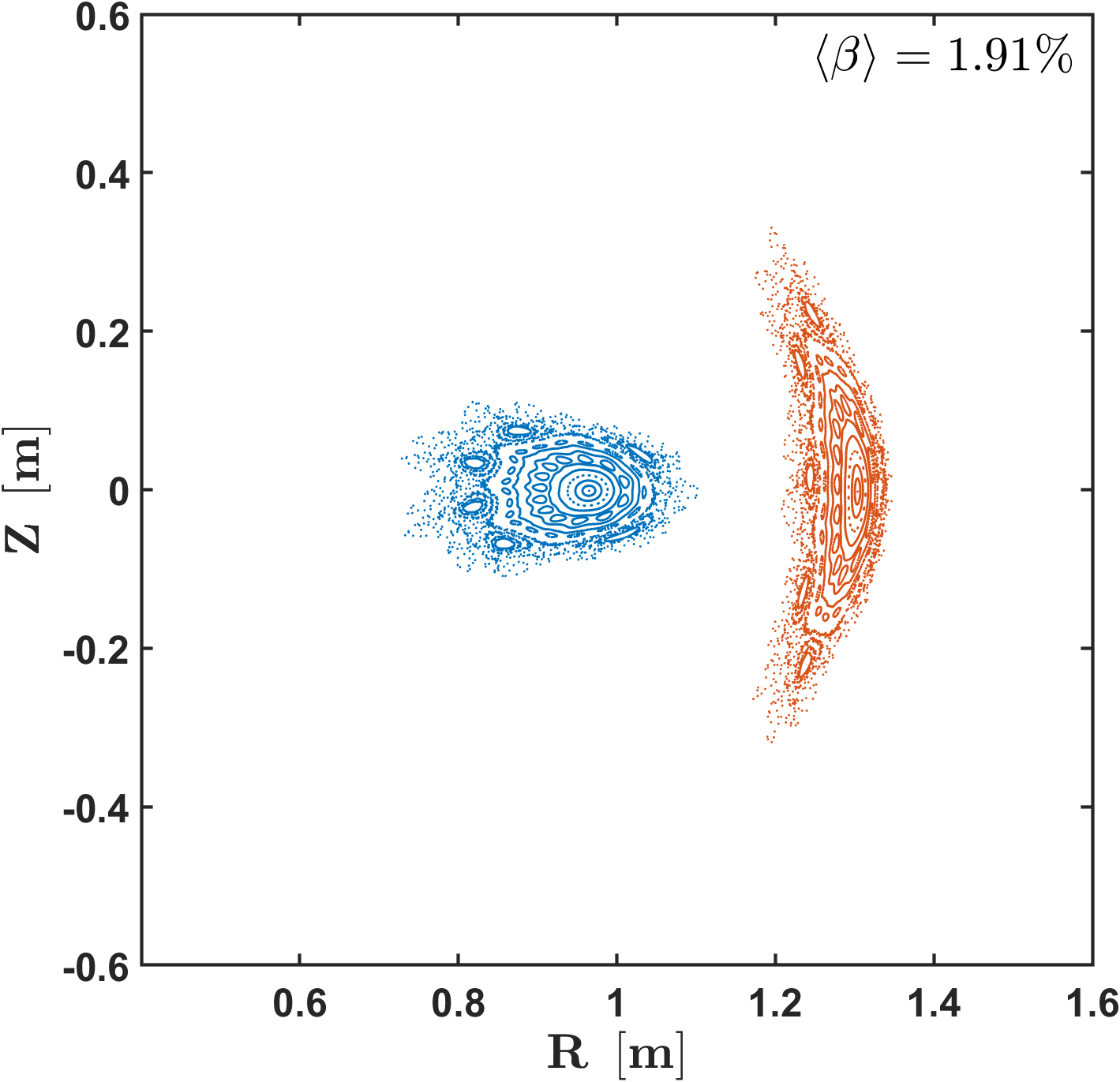}}\\ \vspace{-0.75em}
    \subfloat{\includegraphics[width=0.35\linewidth]{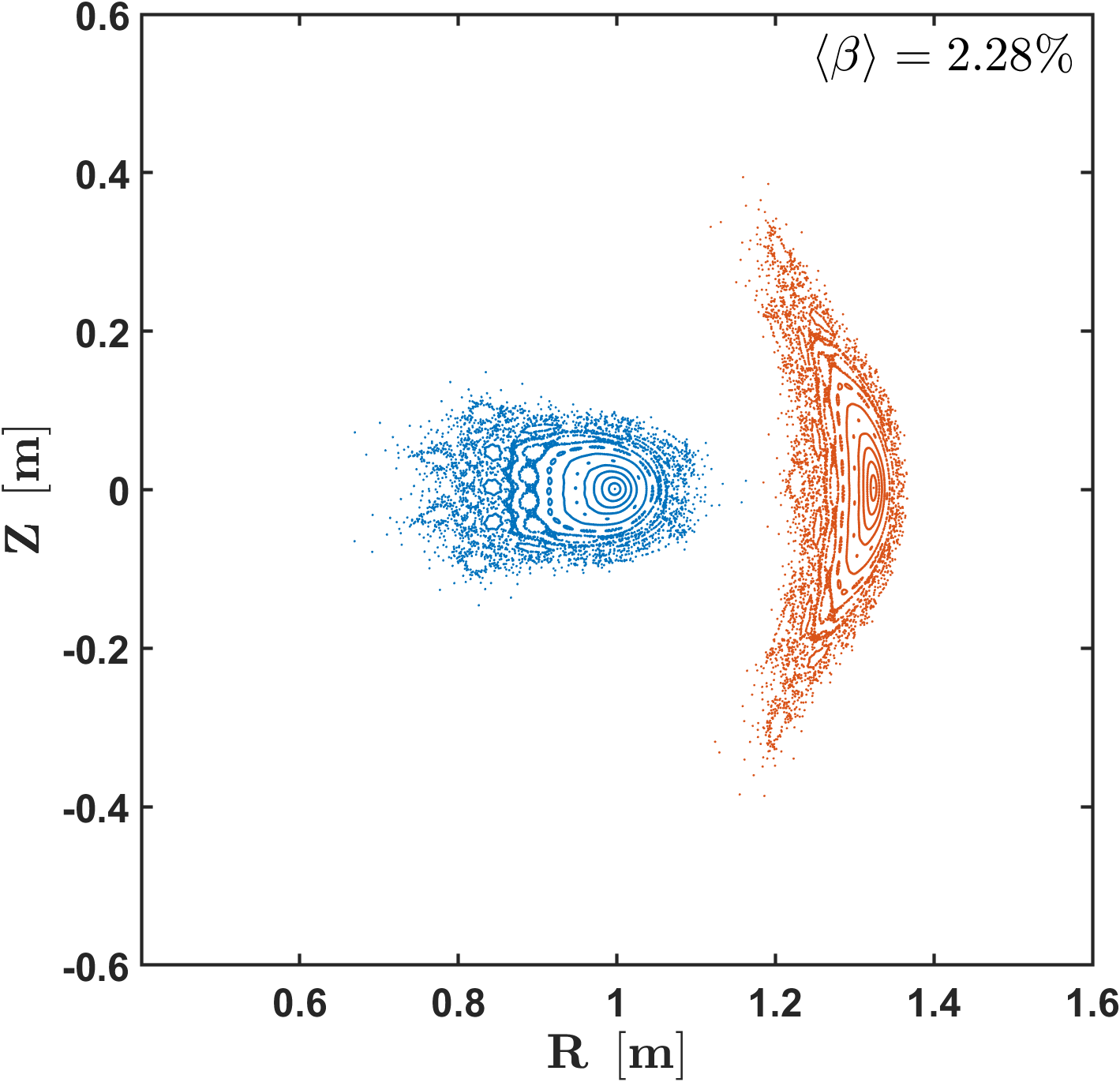}}
    \subfloat{\includegraphics[width=0.35\linewidth]{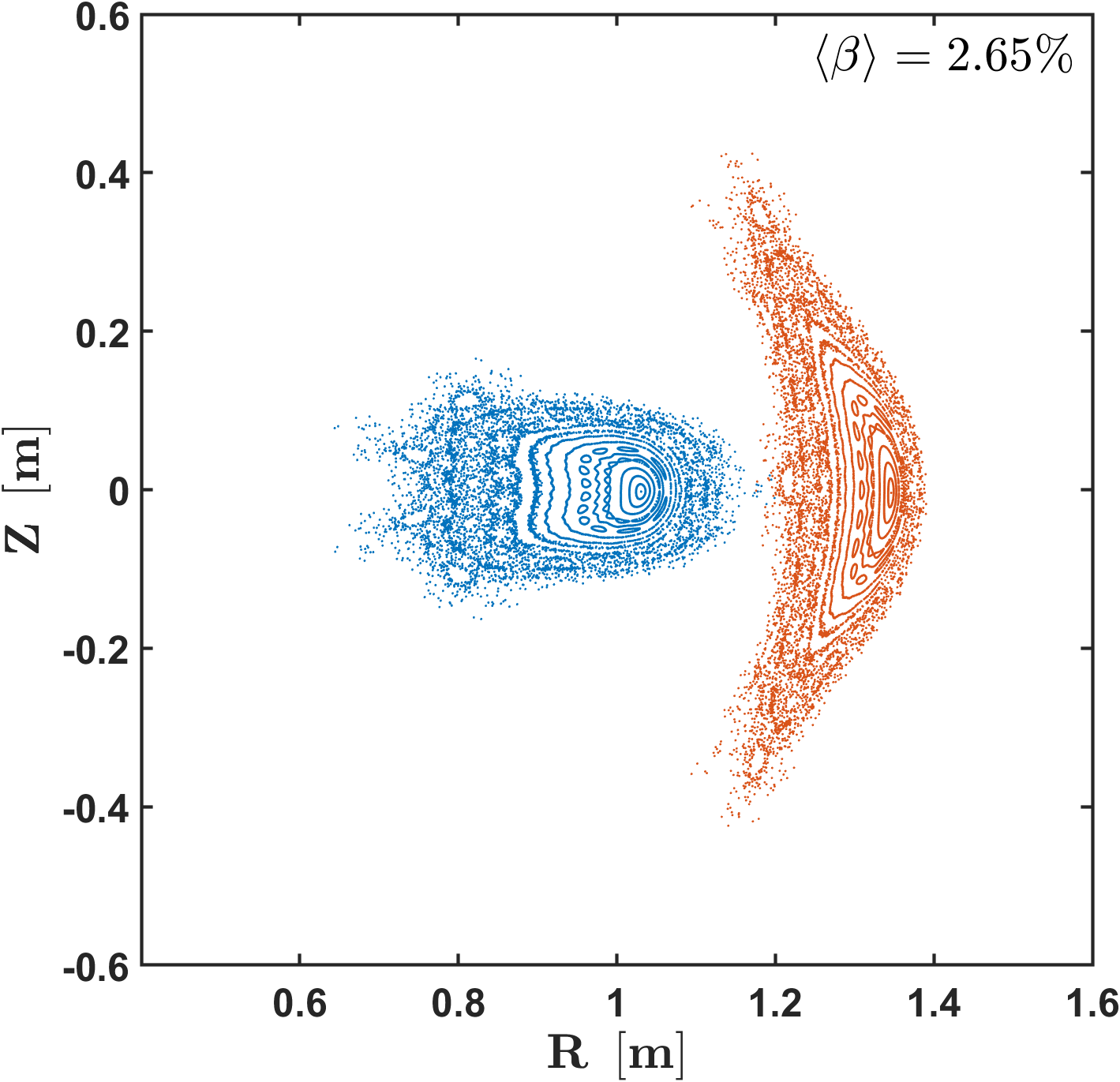}}
    \subfloat{\includegraphics[width=0.35\linewidth]{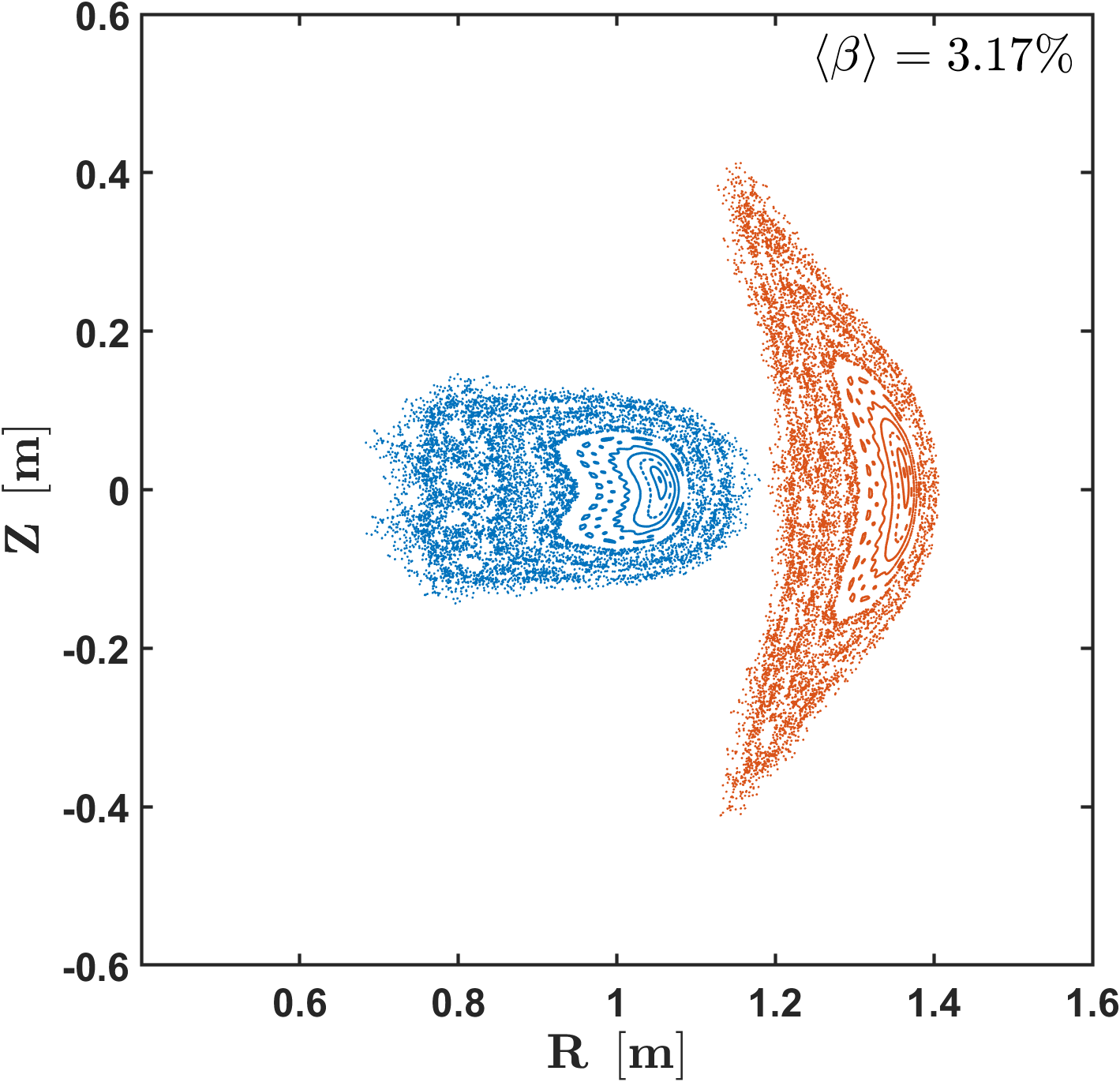}}
    \caption{Poincaré plots of the finite-$\beta$ equilibria in the $n/m=2/5$ island configuration at two cross sections 
             $\phi=0\:\text{(red)}\:\text{and}\:\pi/2\:\text{(blue)}$ with various $\langle\beta\rangle$.}
    \label{fig:poincare_island_n2m5}
\end{figure}

\begin{figure}[htbp]
    % \vspace{-7em}
    \centering
    \subfloat{\includegraphics[width=0.35\linewidth]{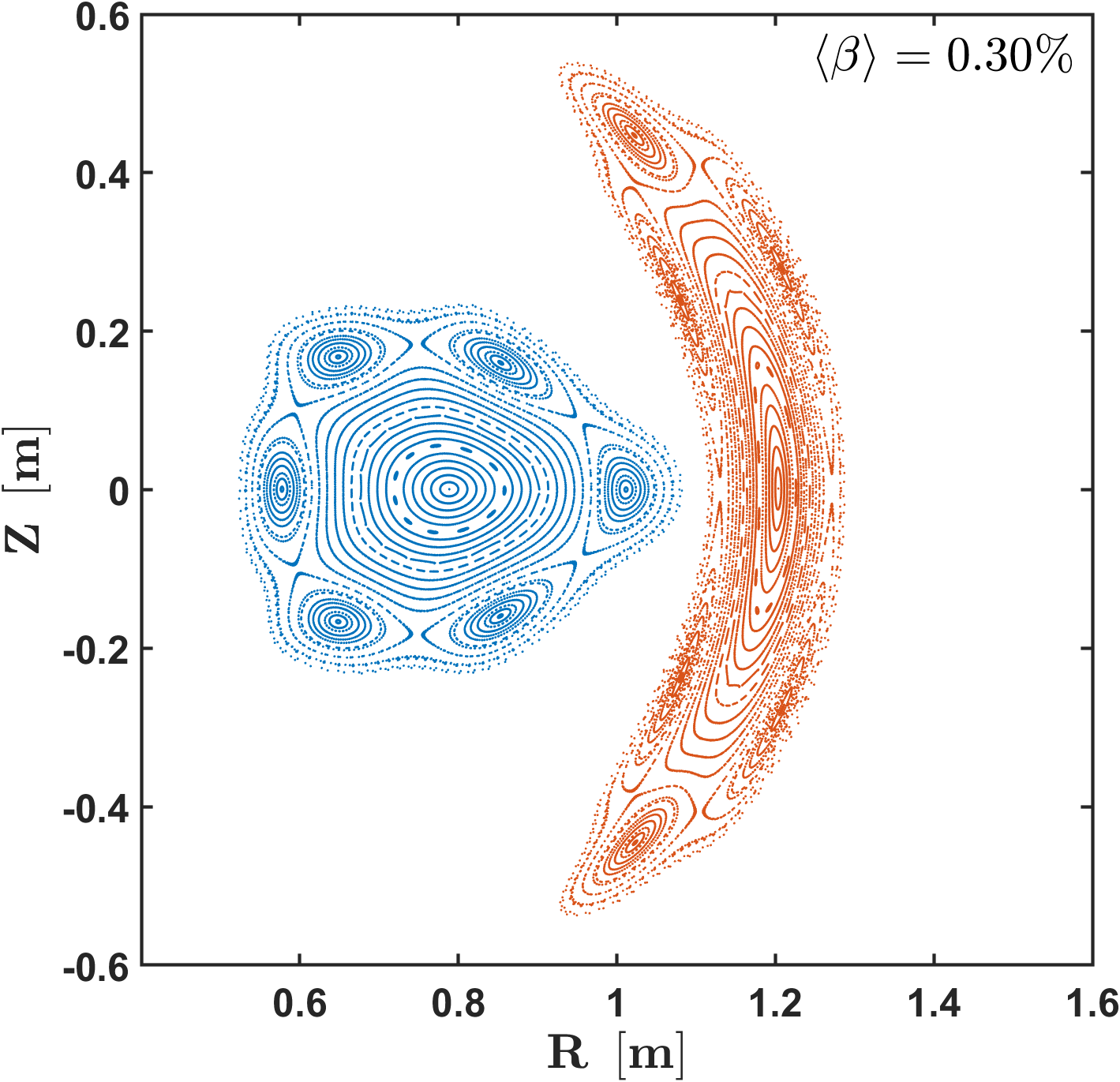}}
    \subfloat{\includegraphics[width=0.35\linewidth]{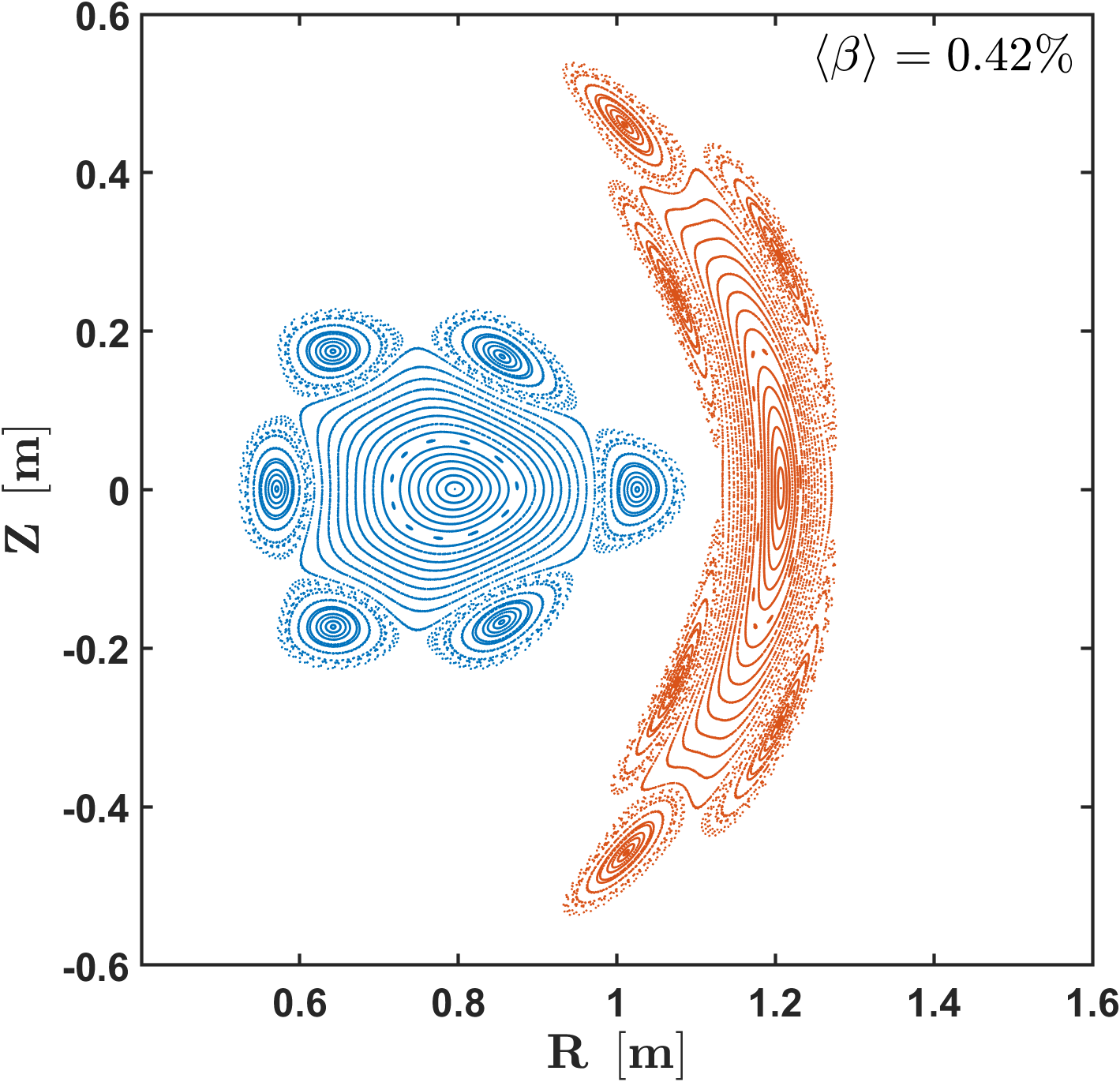}}
    \subfloat{\includegraphics[width=0.35\linewidth]{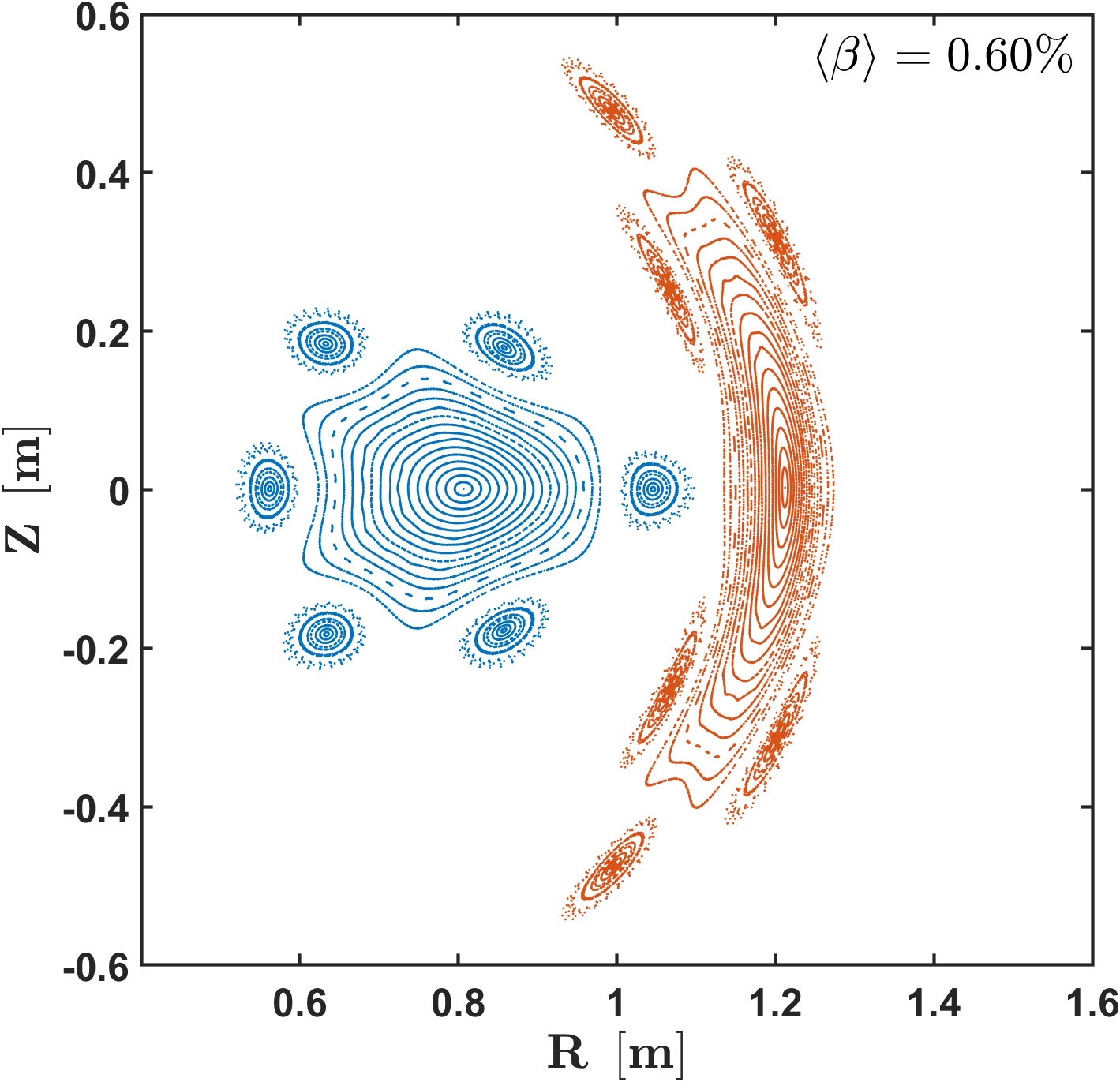}}\\ \vspace{-0.75em}
    \subfloat{\includegraphics[width=0.35\linewidth]{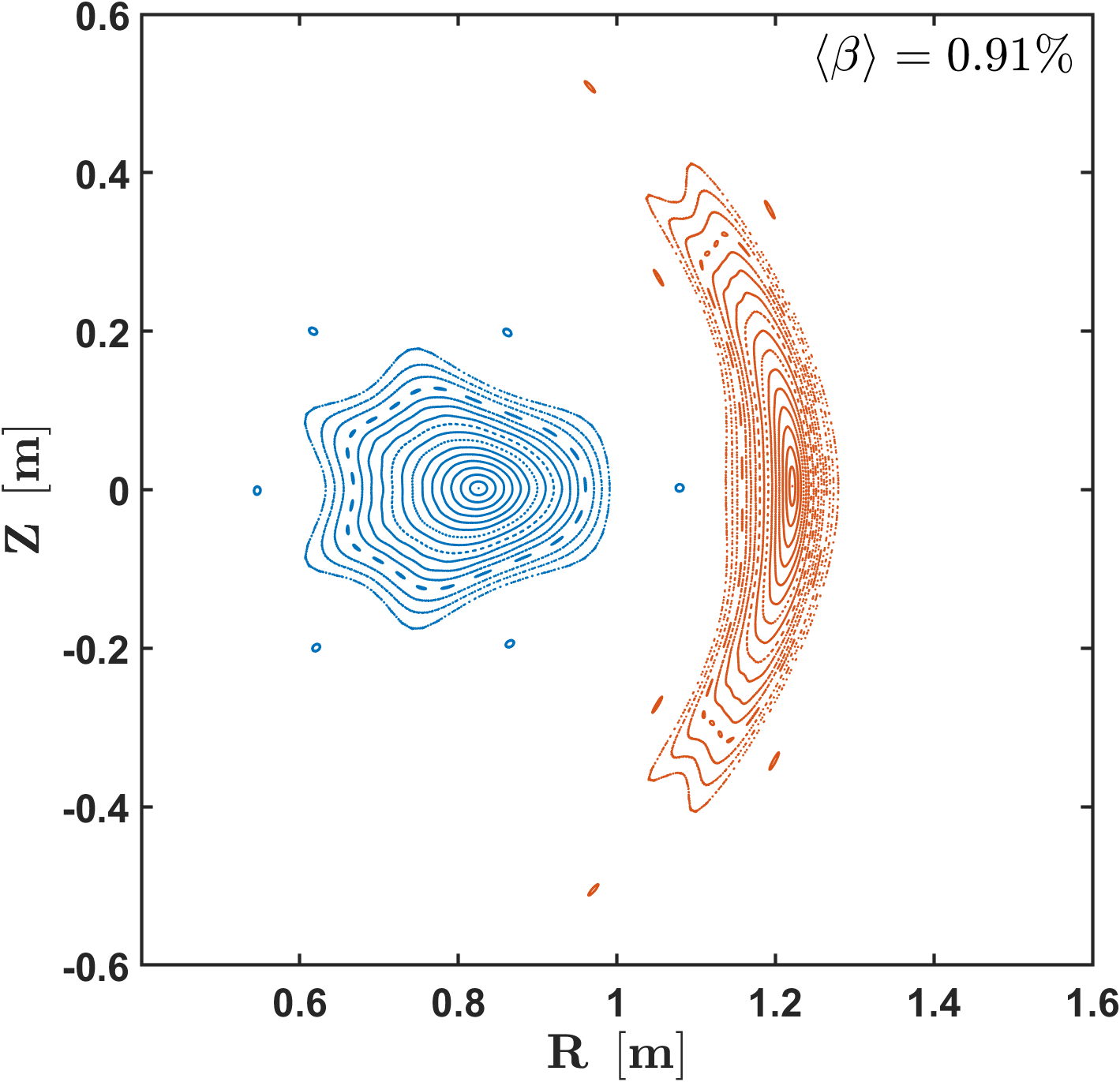}}
    \subfloat{\includegraphics[width=0.35\linewidth]{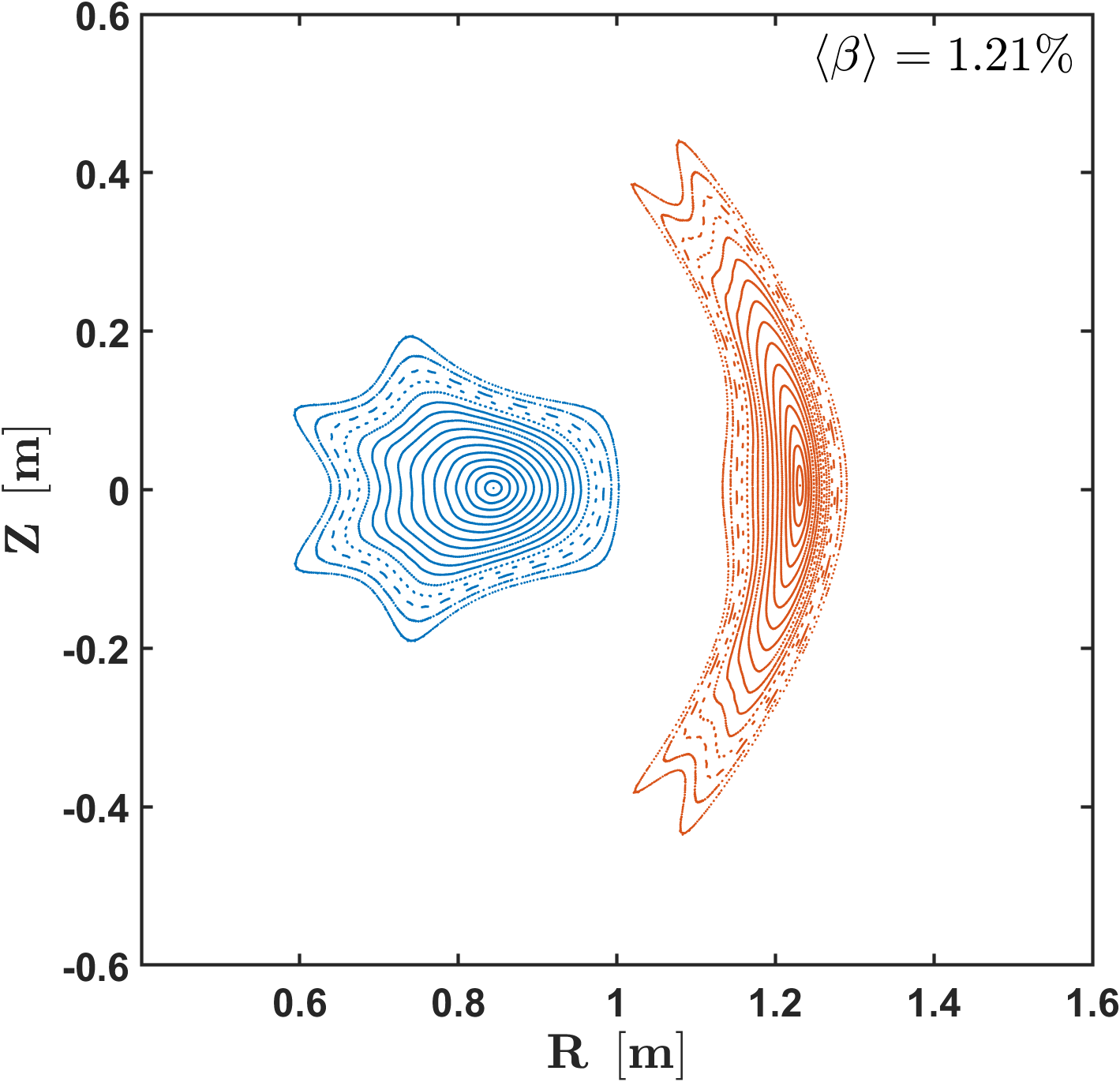}}
    \subfloat{\includegraphics[width=0.35\linewidth]{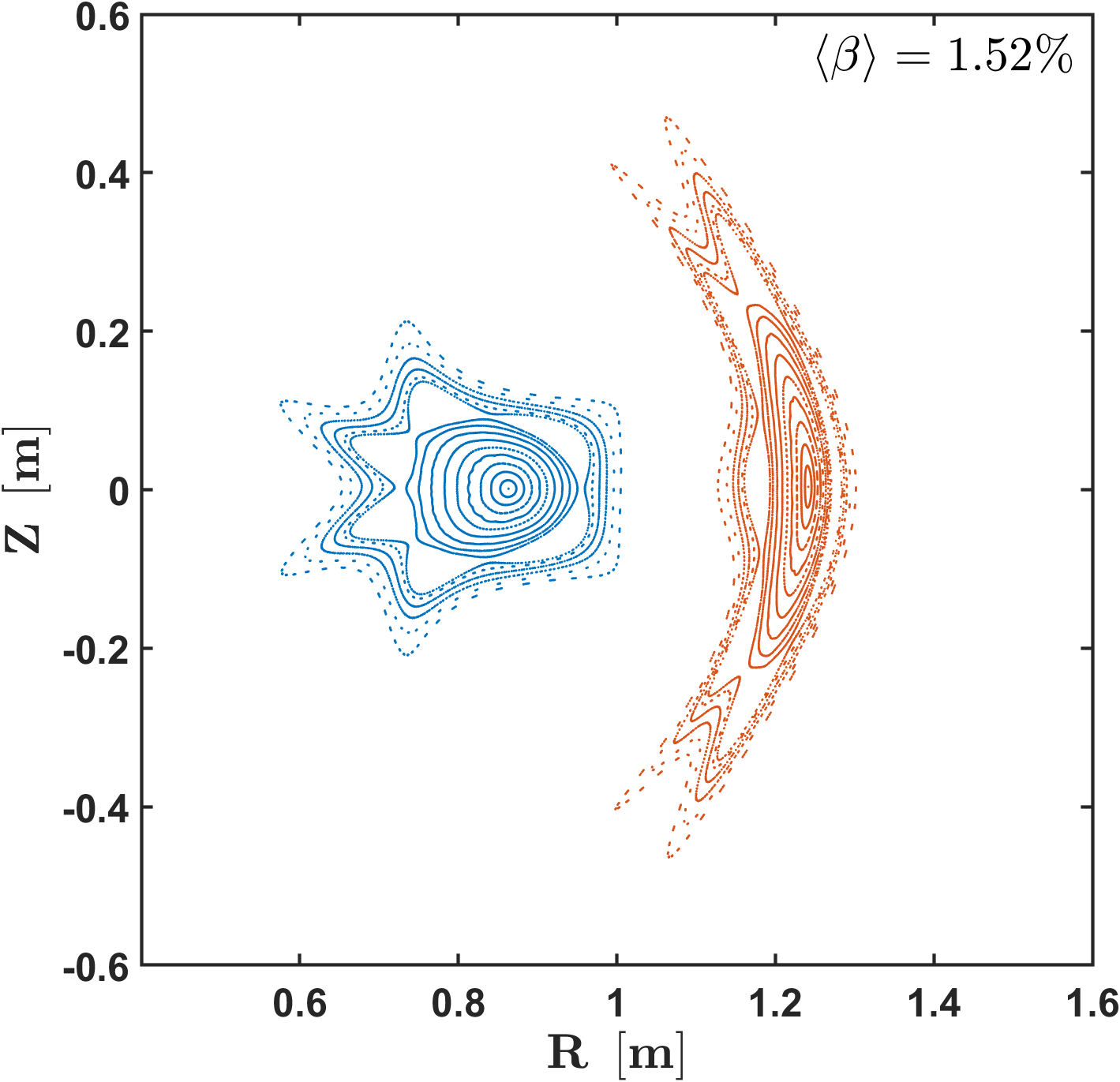}}\\ \vspace{-0.75em}
    \subfloat{\includegraphics[width=0.35\linewidth]{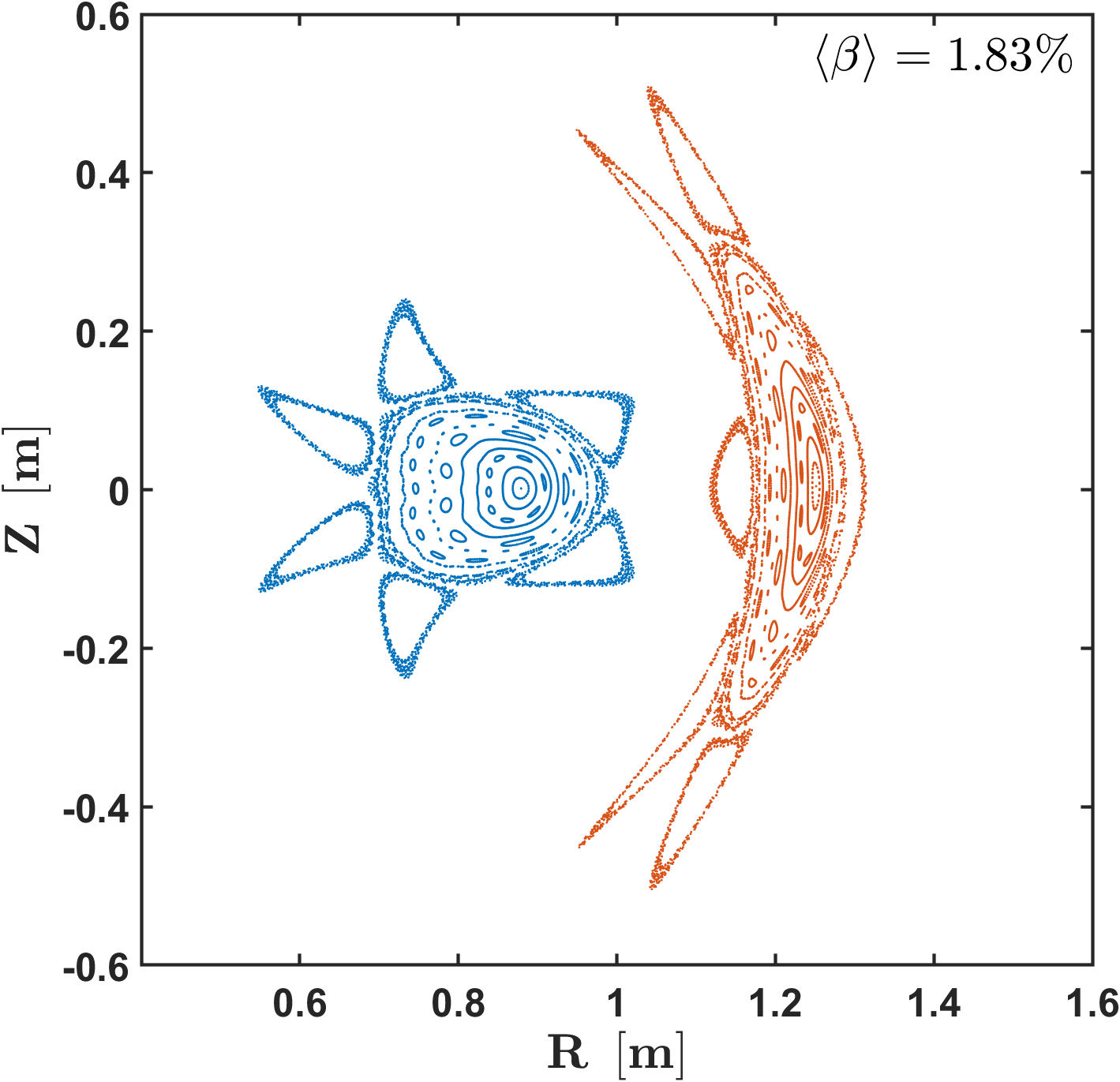}}
    \subfloat{\includegraphics[width=0.35\linewidth]{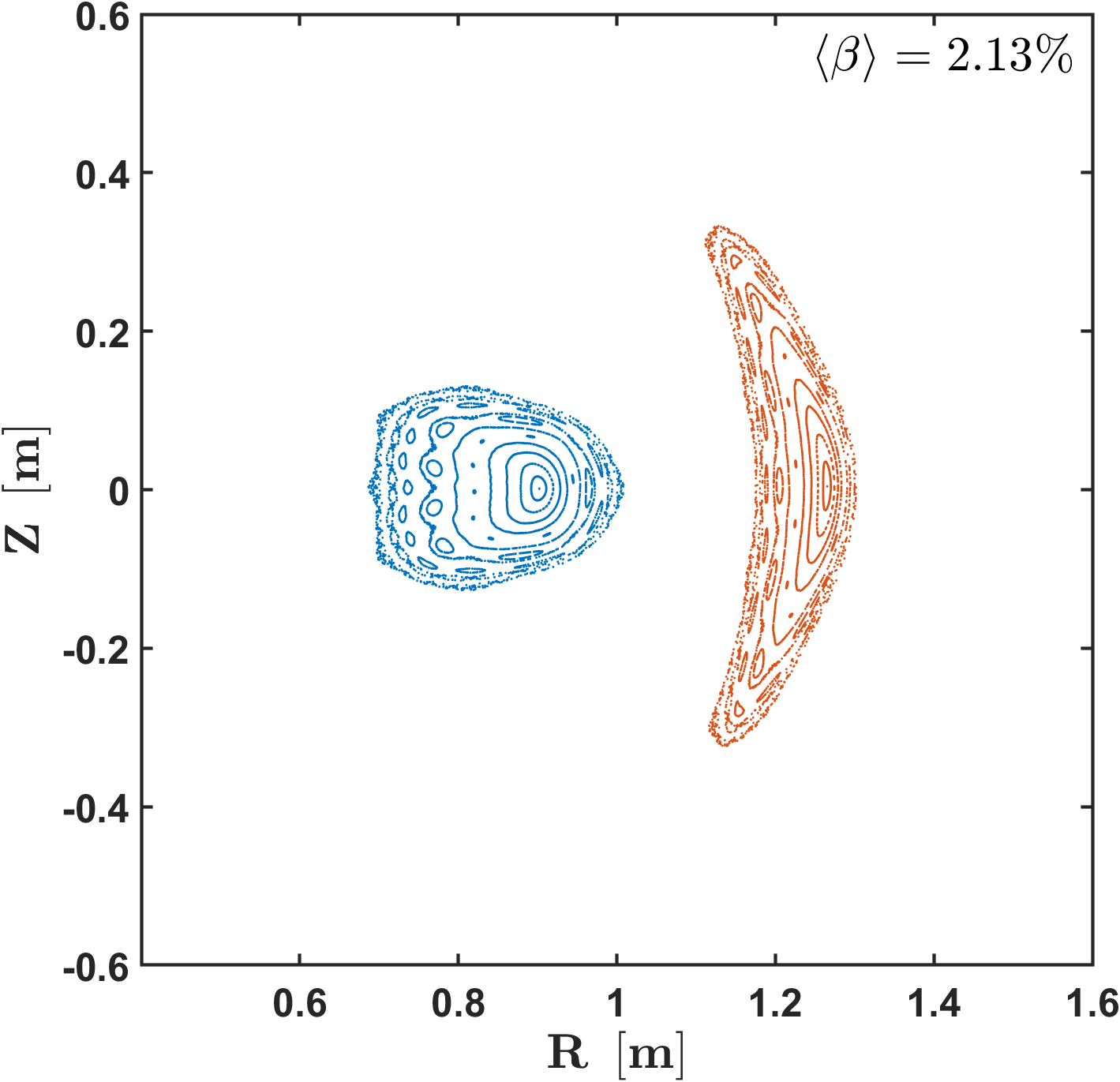}}
    \subfloat{\includegraphics[width=0.35\linewidth]{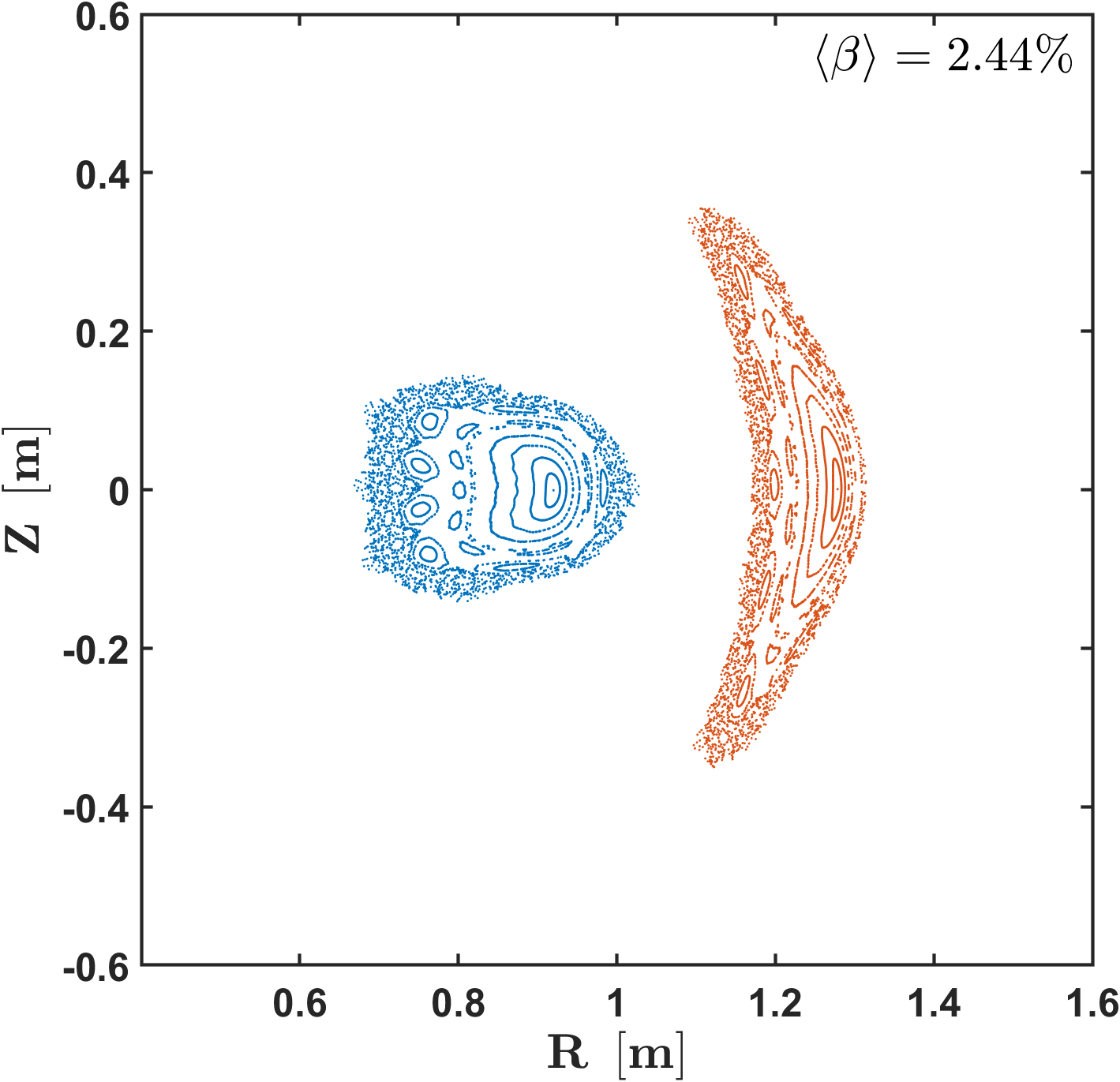}}
    \caption{Poincaré plots of the finite-$\beta$ equilibria in the $n/m=2/6$ island configuration at two cross sections 
             $\phi=0\:\text{(red)}\:\text{and}\:\pi/2\:\text{(blue)}$ with various $\langle\beta\rangle$.}
    \label{fig:poincare_island_n2m6}
\end{figure}

\begin{figure}[htbp]
    % \vspace{-7em}
    \captionsetup[subfloat]{position=top,labelformat=empty}
    \centering
    \subfloat[$n/m=2/5$]{\includegraphics[width=0.35\linewidth]{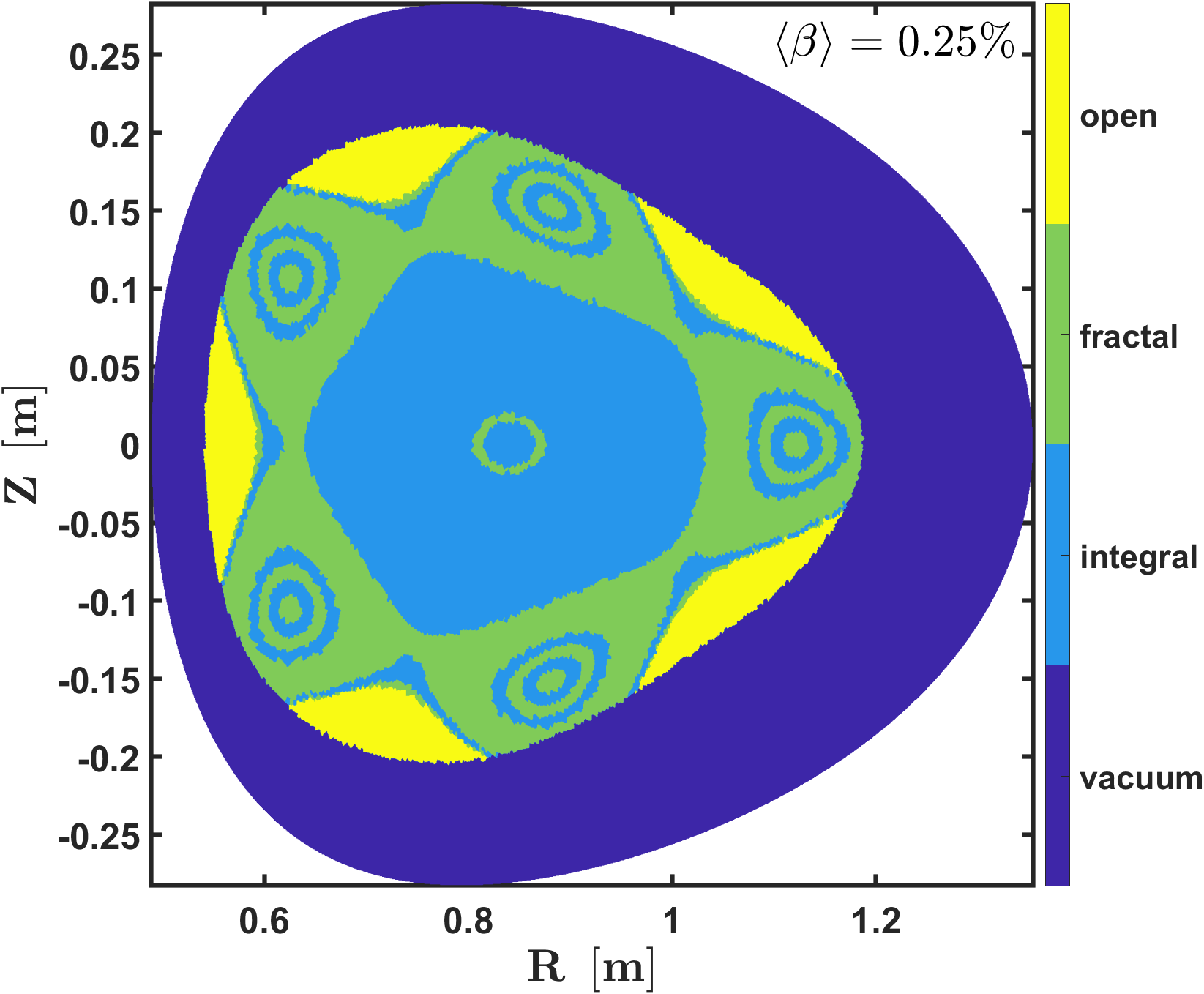}} \qquad
    \subfloat[$n/m=2/6$]{\includegraphics[width=0.35\linewidth]{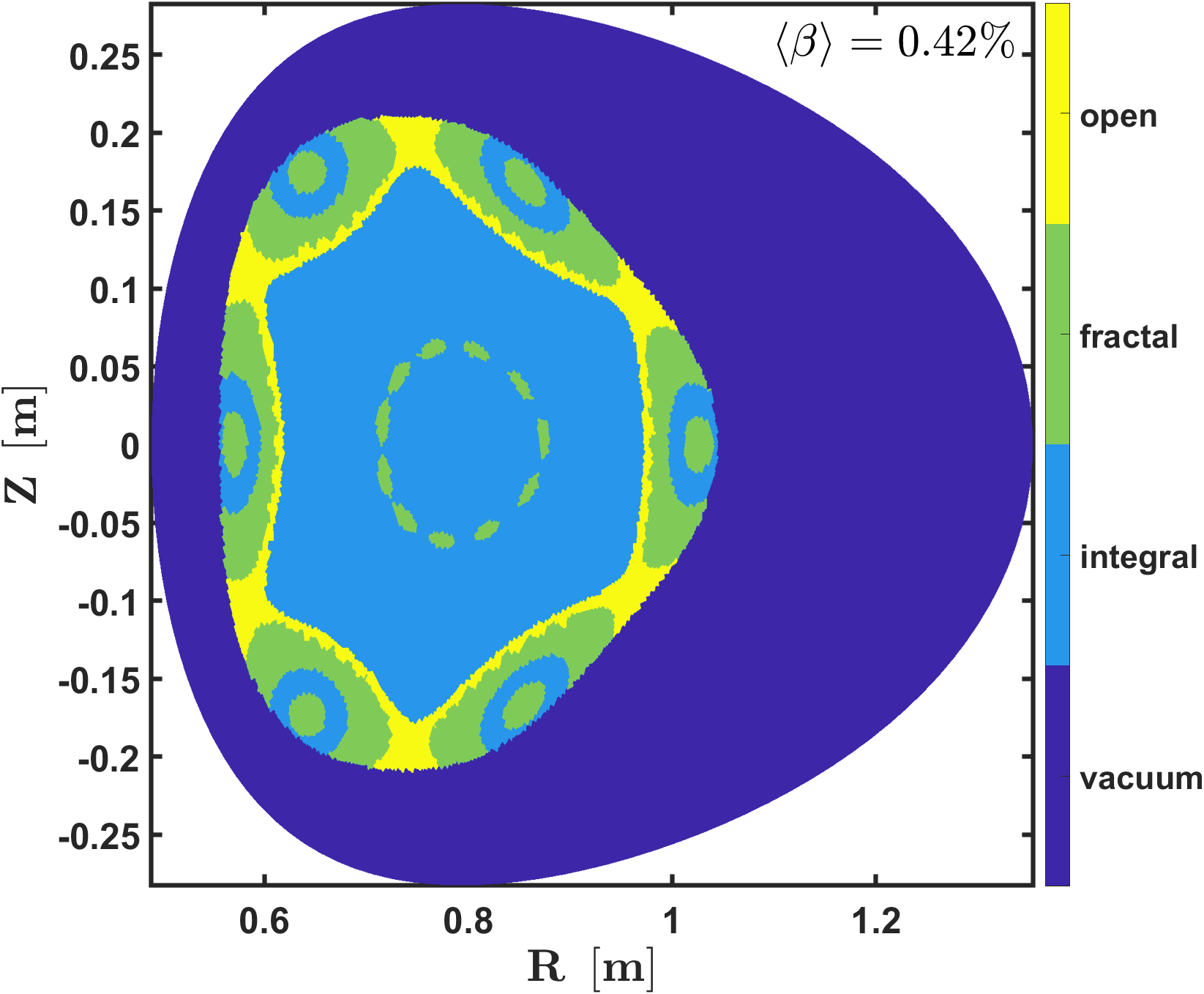}}\\ \vspace{-0.75em}
    \subfloat{\includegraphics[width=0.35\linewidth]{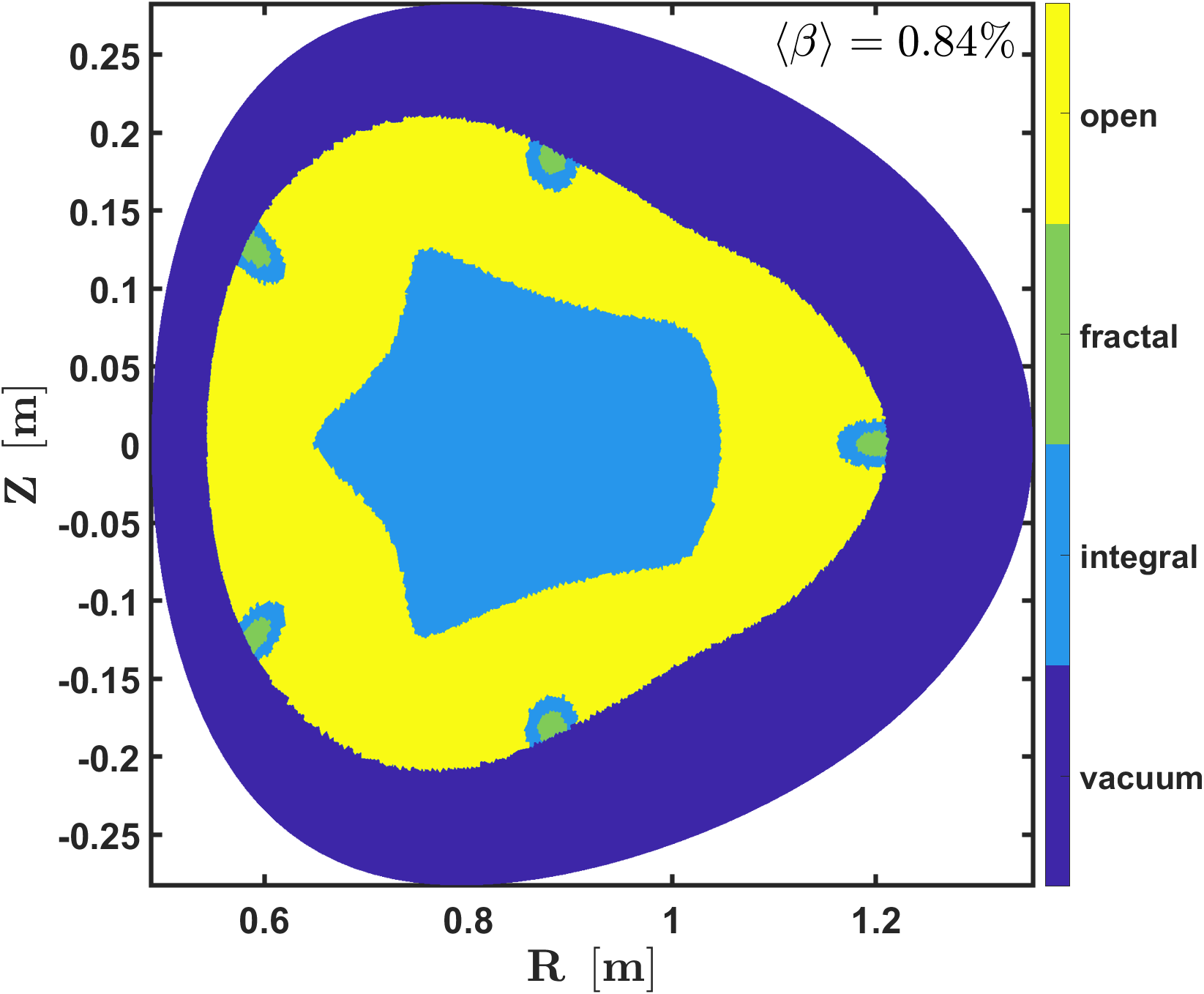}} \qquad
    \subfloat{\includegraphics[width=0.35\linewidth]{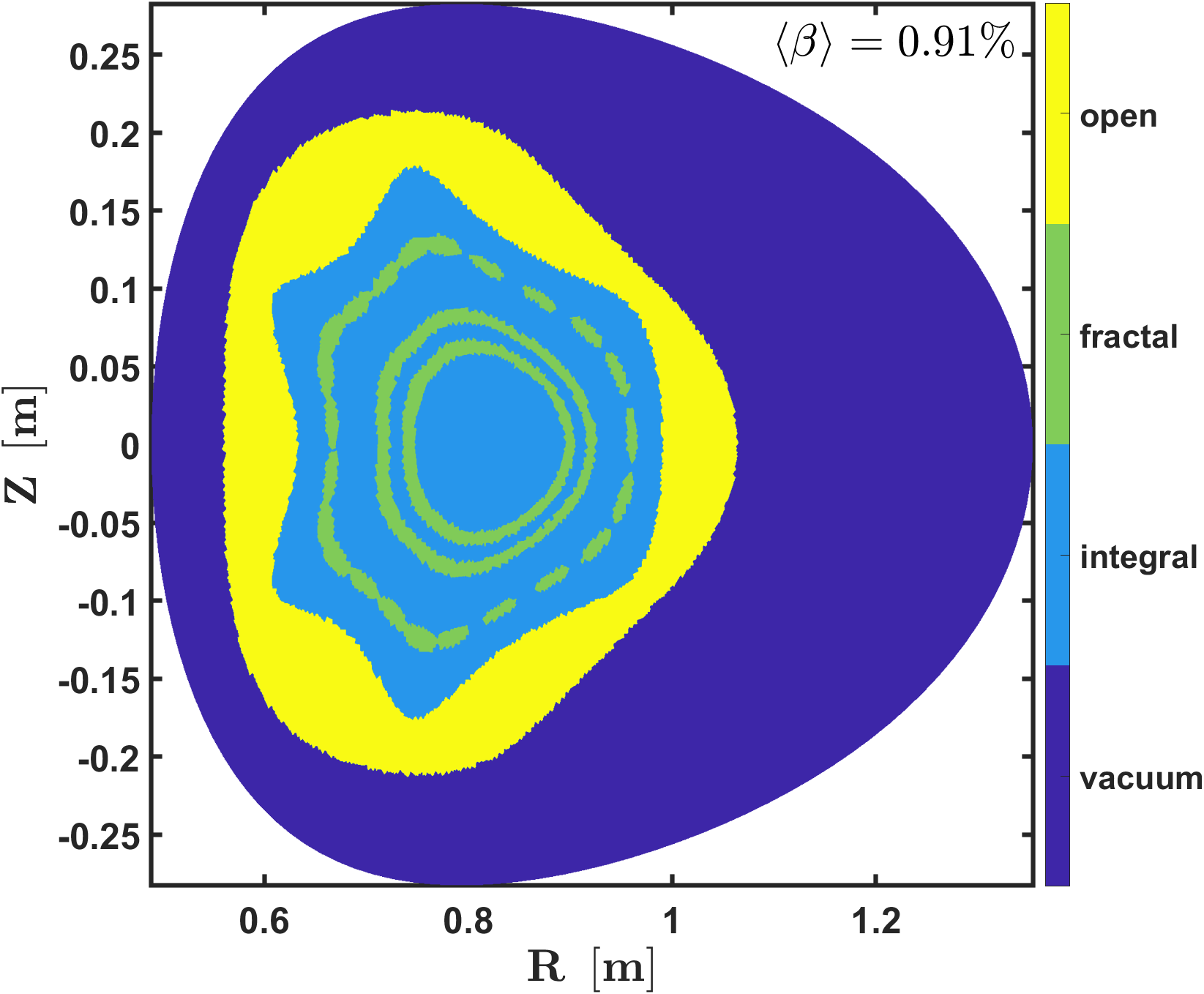}}\\ \vspace{-0.75em}
    \subfloat{\includegraphics[width=0.35\linewidth]{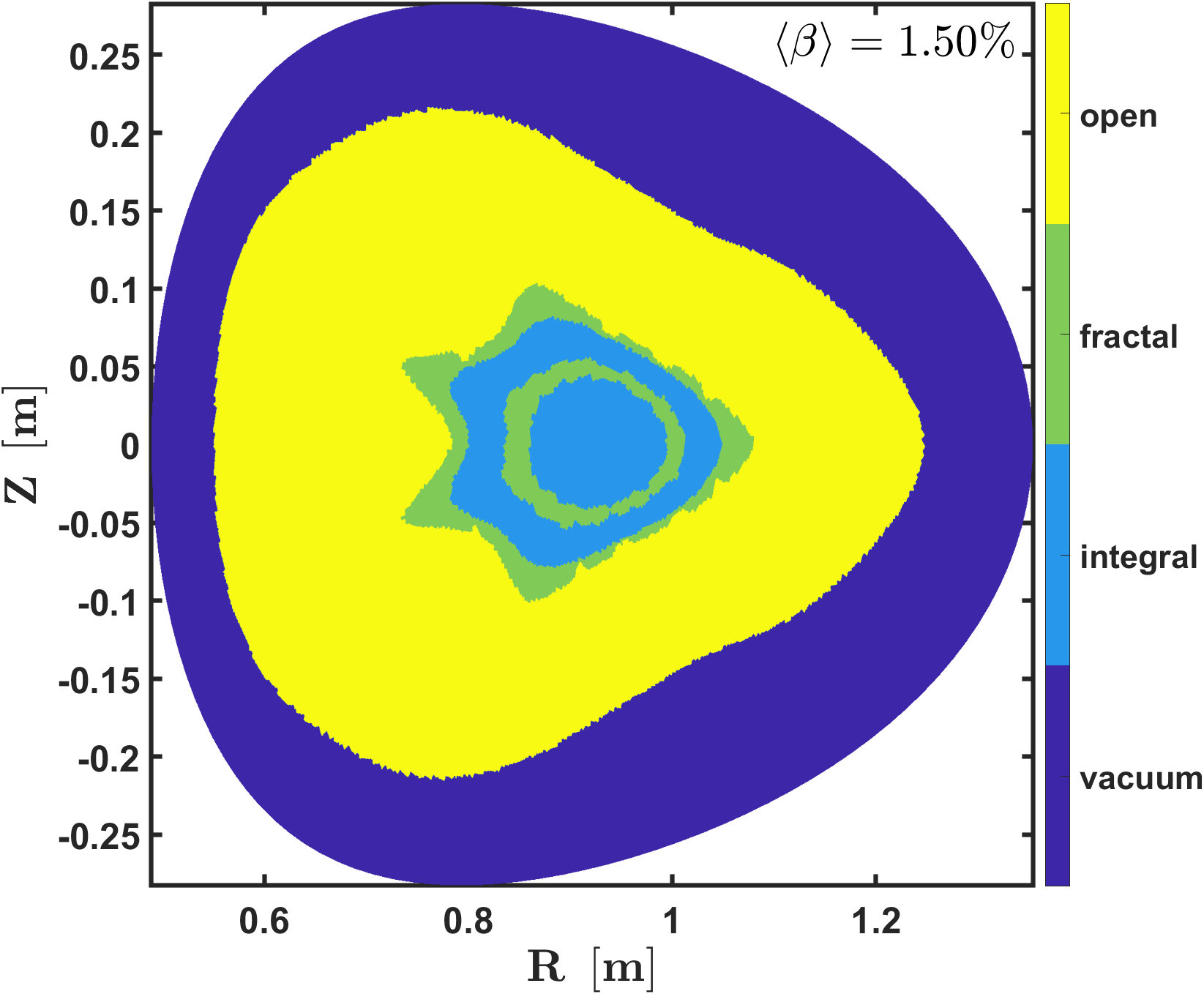}} \qquad
    \subfloat{\includegraphics[width=0.35\linewidth]{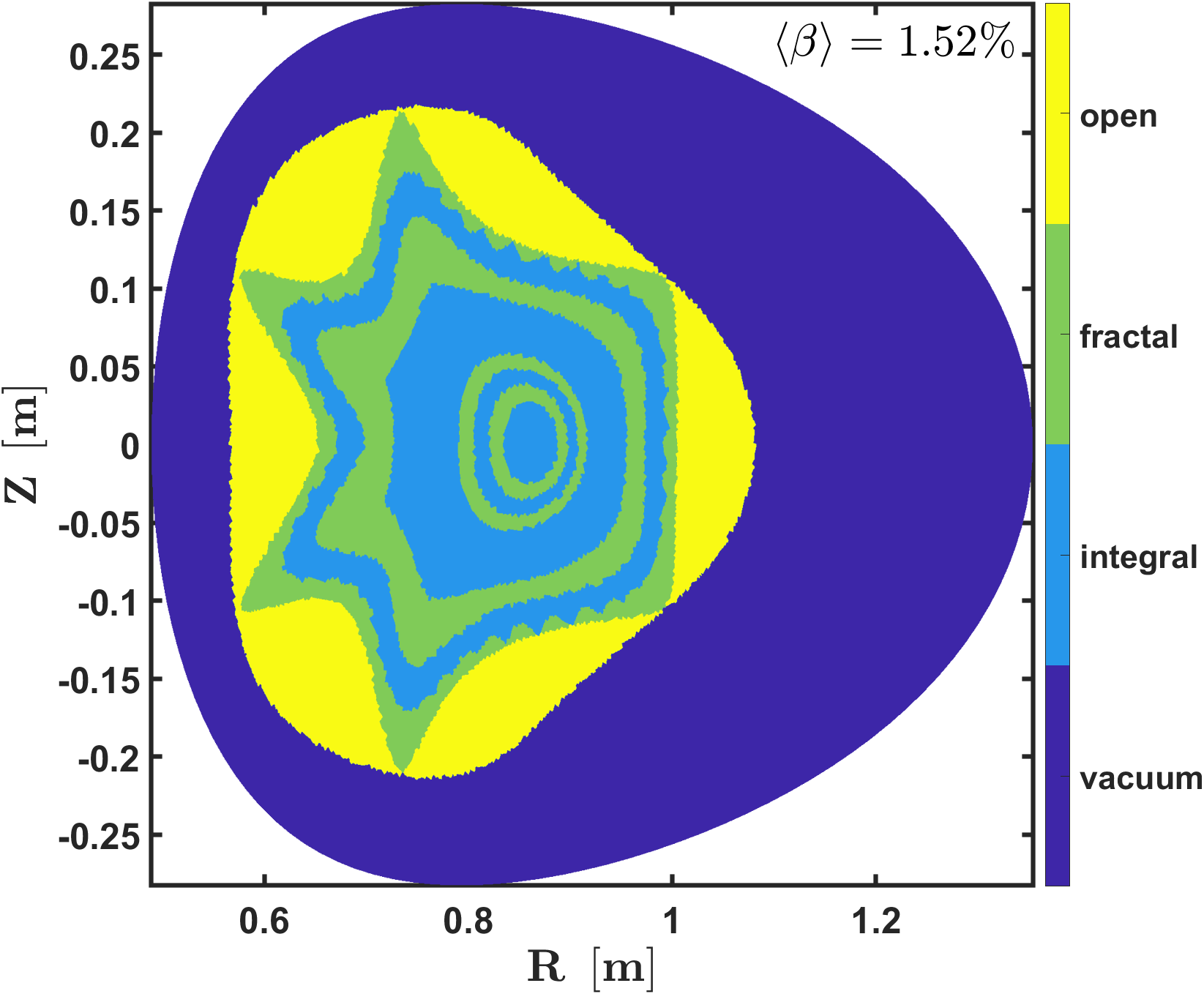}}\\ \vspace{-0.75em}
    \subfloat{\includegraphics[width=0.35\linewidth]{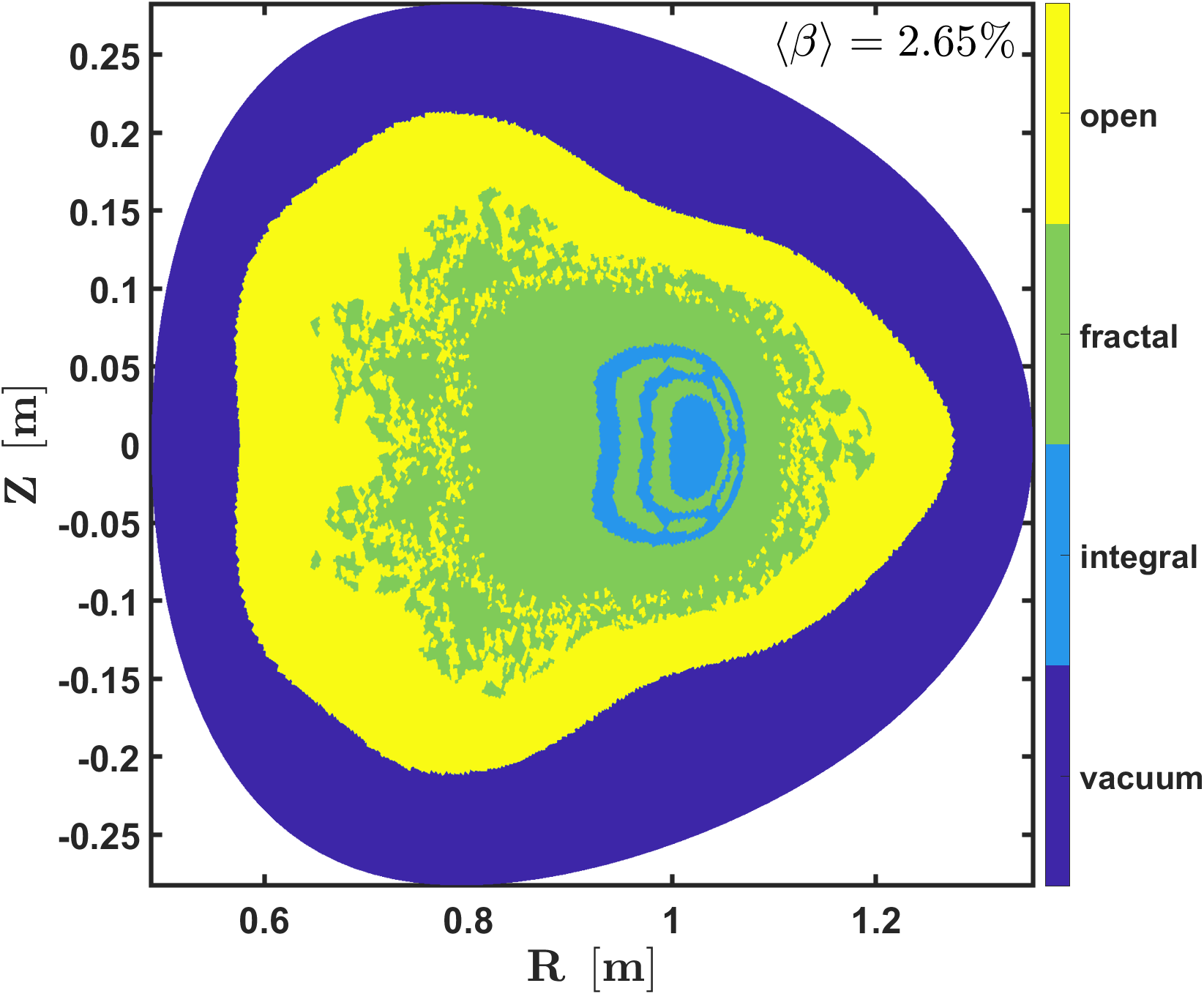}} \qquad
    \subfloat{\includegraphics[width=0.35\linewidth]{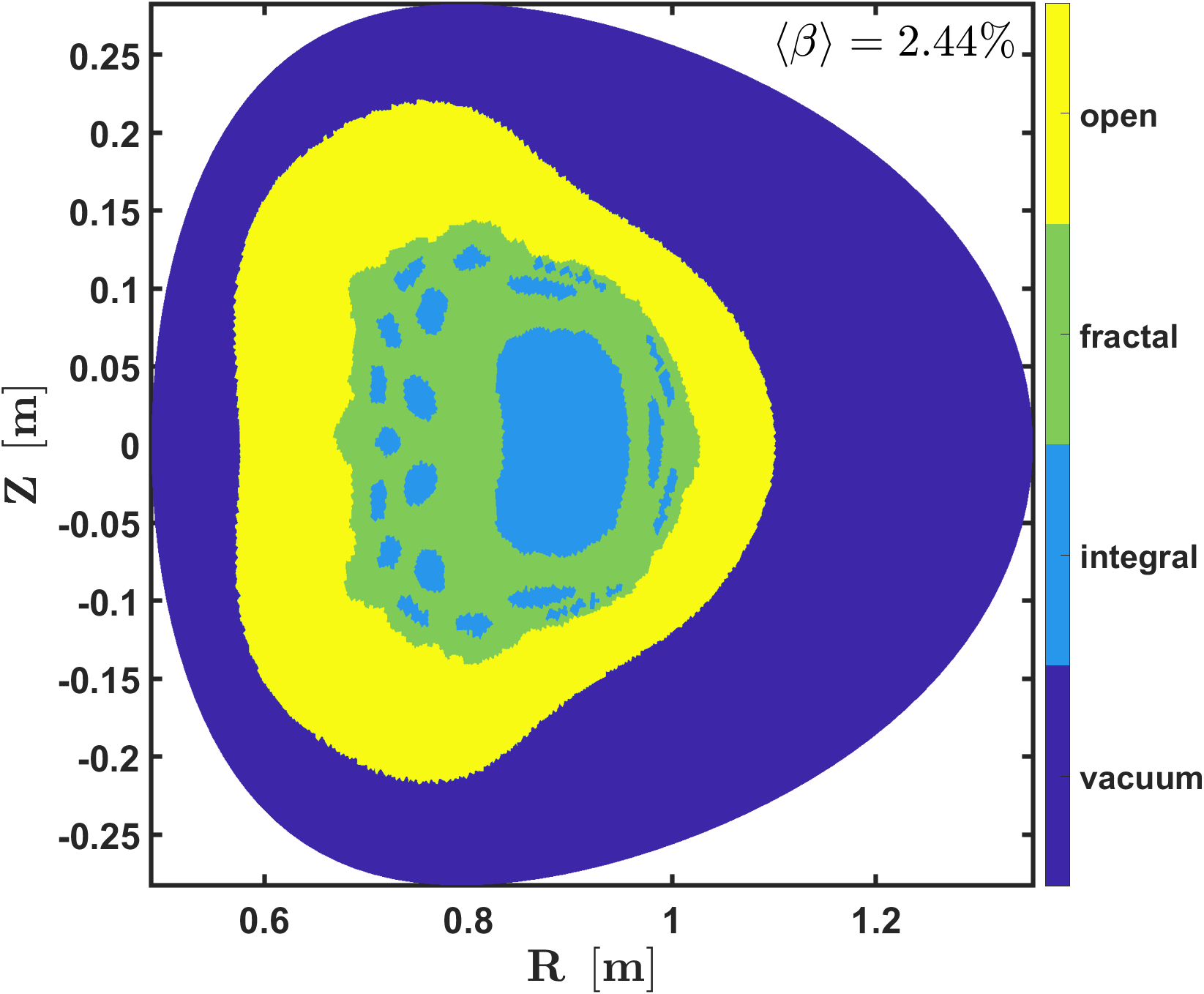}}
    \caption{Contours of the box-counting dimension for finite-$\beta$ equilibria in the $n/m=2/5$ (left column) and the $n/m=2/6$ (right column) island configurations with various $\langle\beta\rangle$. 
             The filled contours are drawn at the cross section $\phi=\pi/2$.}
    \label{fig:BC_contours_n2m5_n2m6}
\end{figure}

\begin{figure}[htbp]
    % \vspace{-7em}
    \captionsetup[subfloat]{position=top,labelformat=empty}
    \centering
    \subfloat[$n/m=2/5$]{\includegraphics[width=0.35\linewidth]{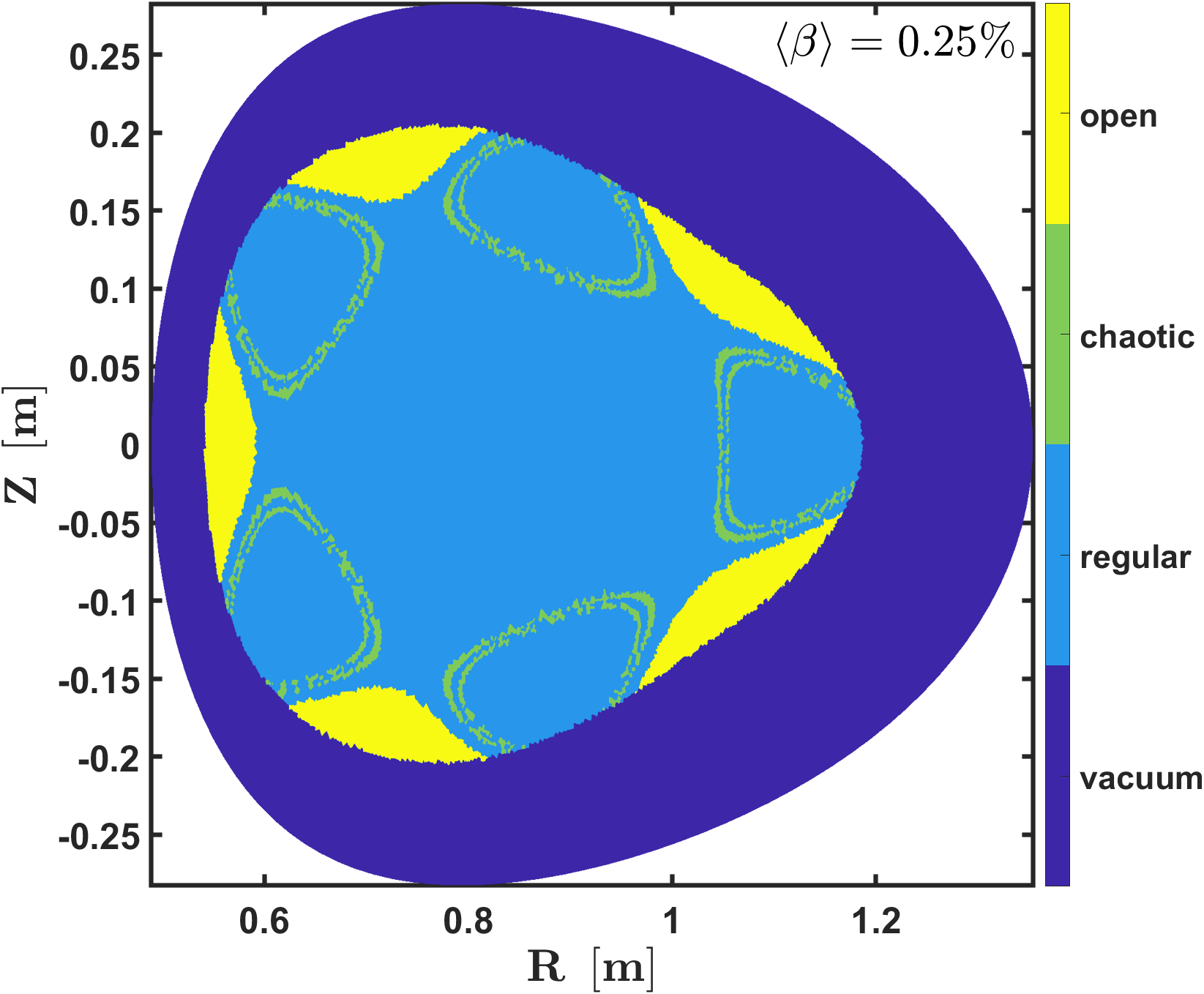}} \qquad
    \subfloat[$n/m=2/6$]{\includegraphics[width=0.35\linewidth]{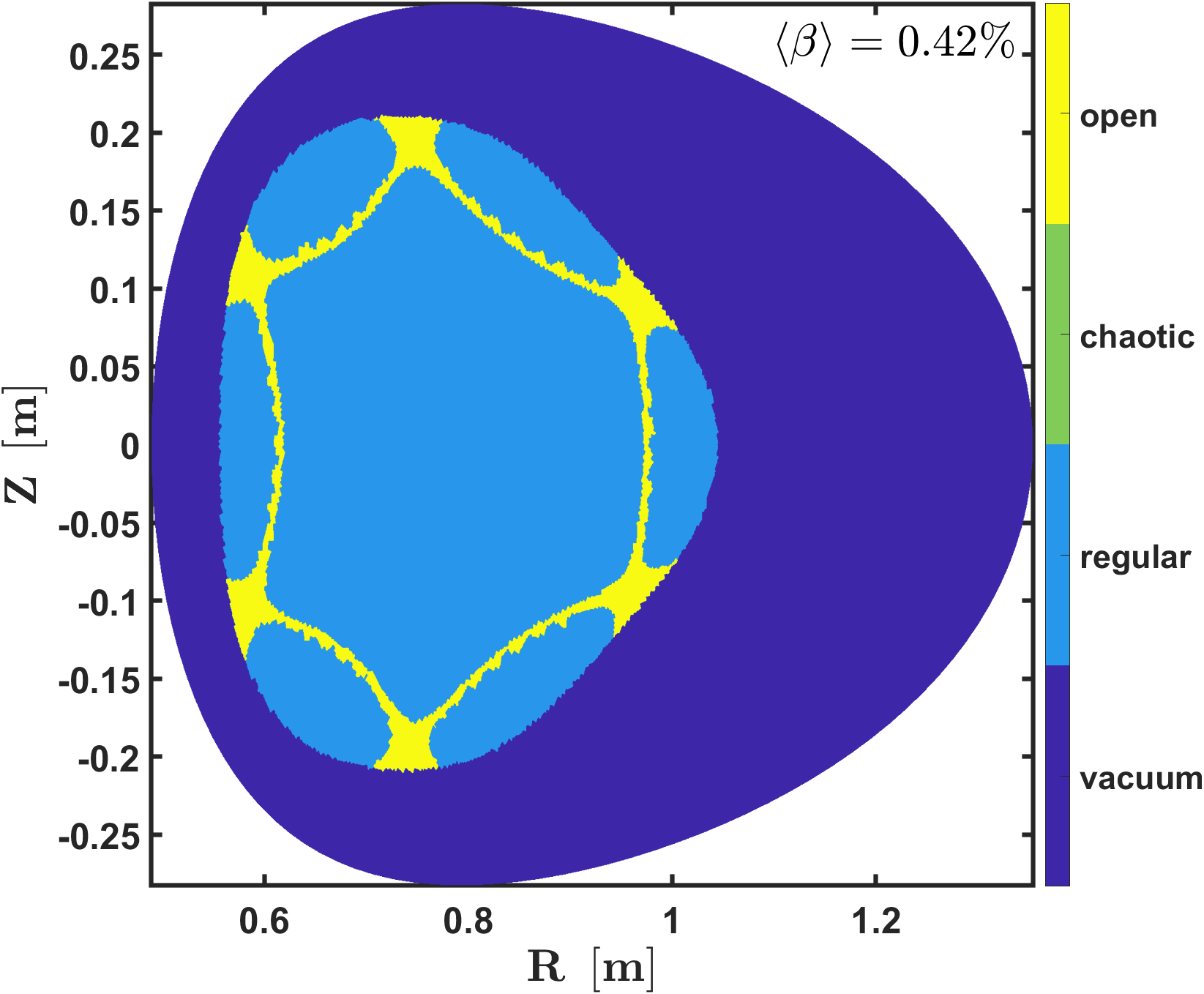}}\\ \vspace{-0.75em}
    \subfloat{\includegraphics[width=0.35\linewidth]{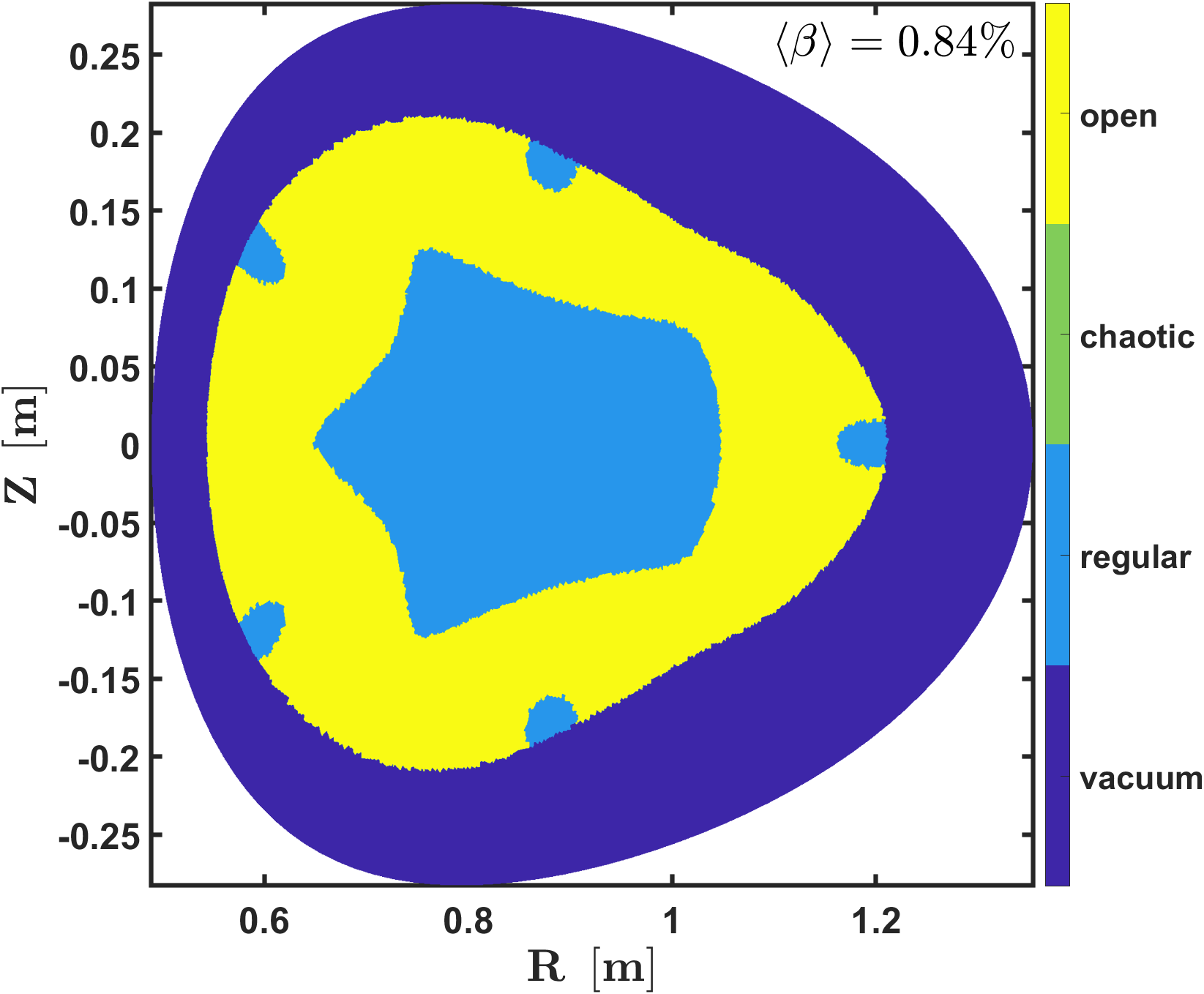}} \qquad
    \subfloat{\includegraphics[width=0.35\linewidth]{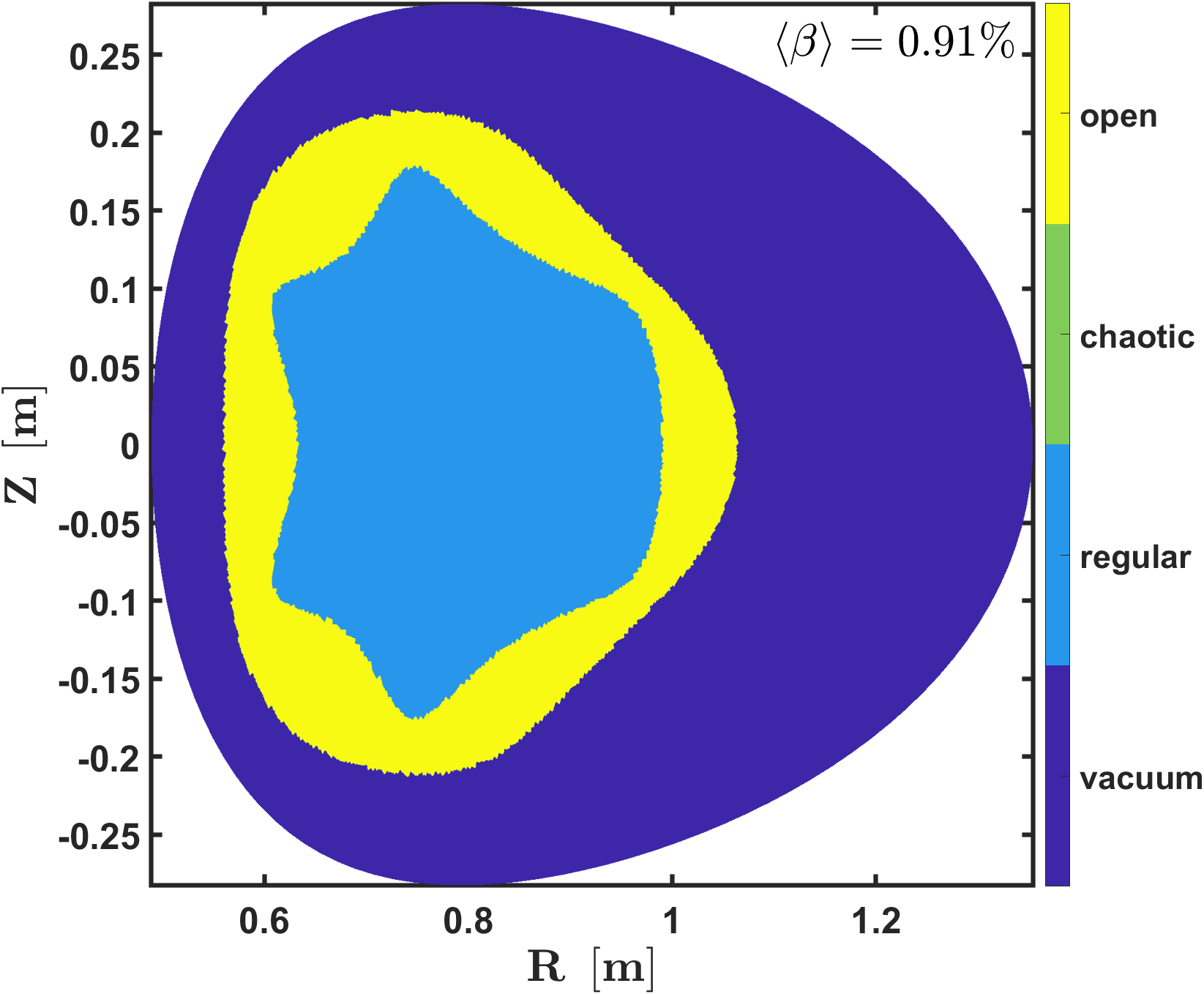}}\\ \vspace{-0.75em}
    \subfloat{\includegraphics[width=0.35\linewidth]{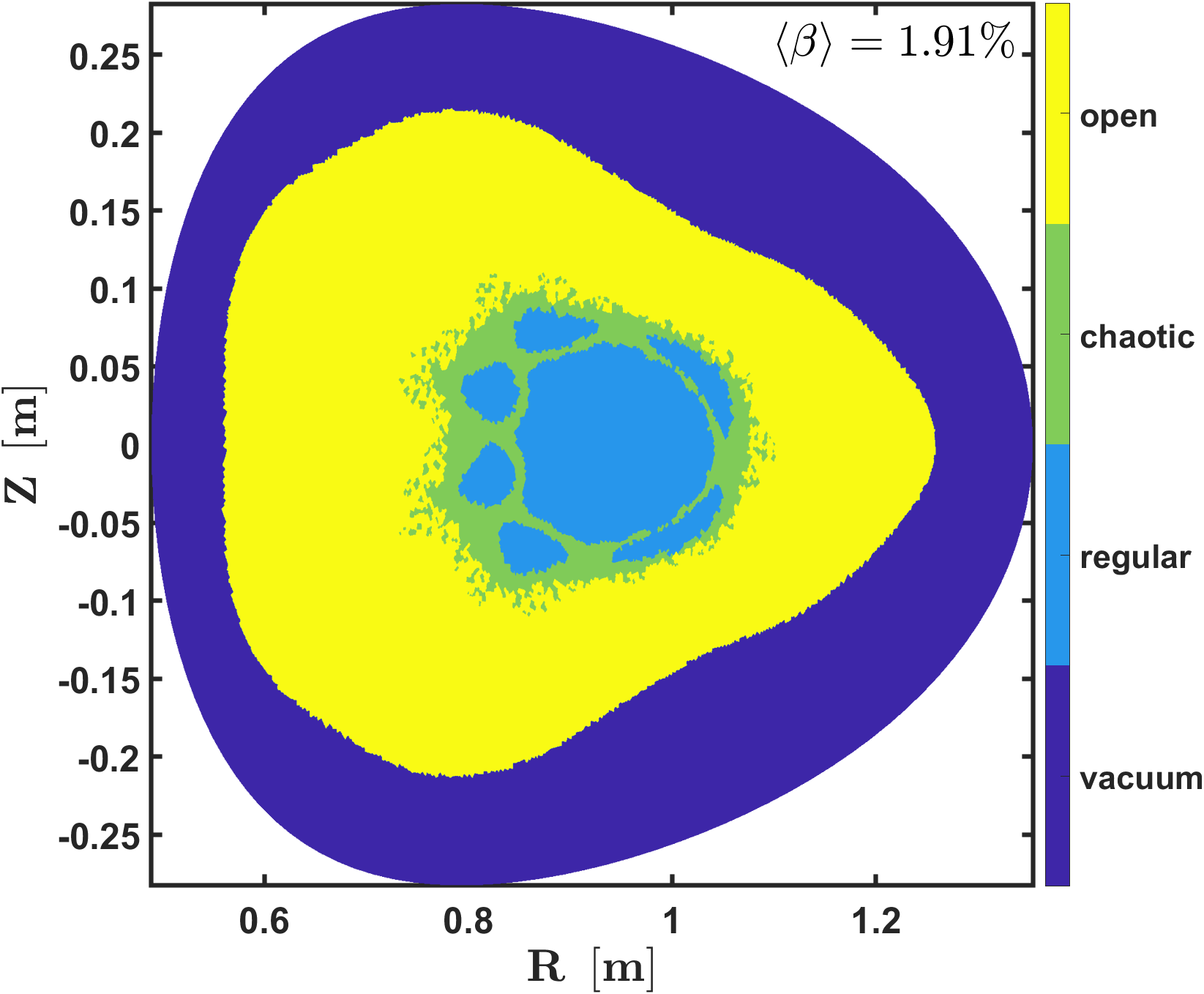}} \qquad
    \subfloat{\includegraphics[width=0.35\linewidth]{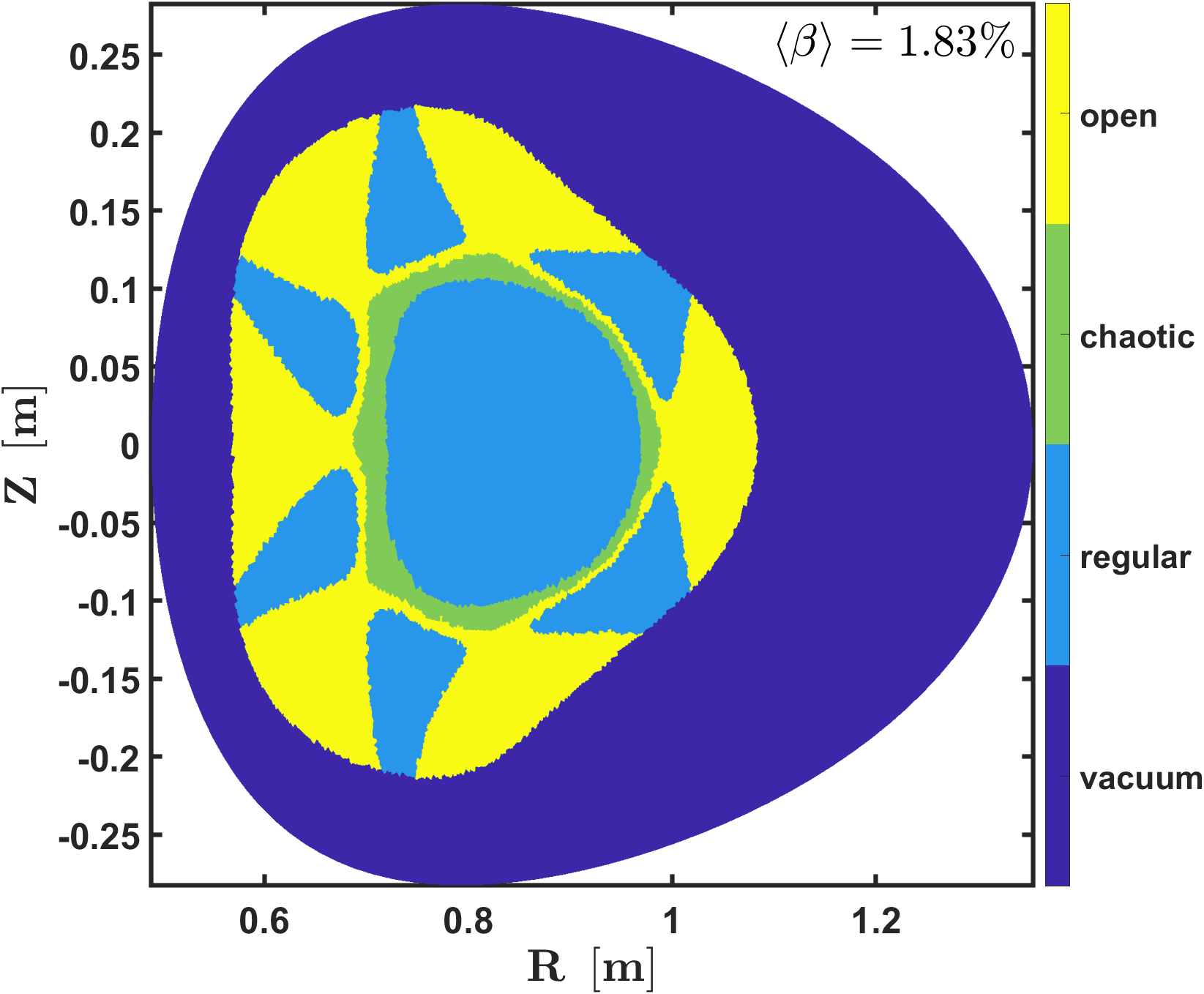}}\\ \vspace{-0.75em}
    \subfloat{\includegraphics[width=0.35\linewidth]{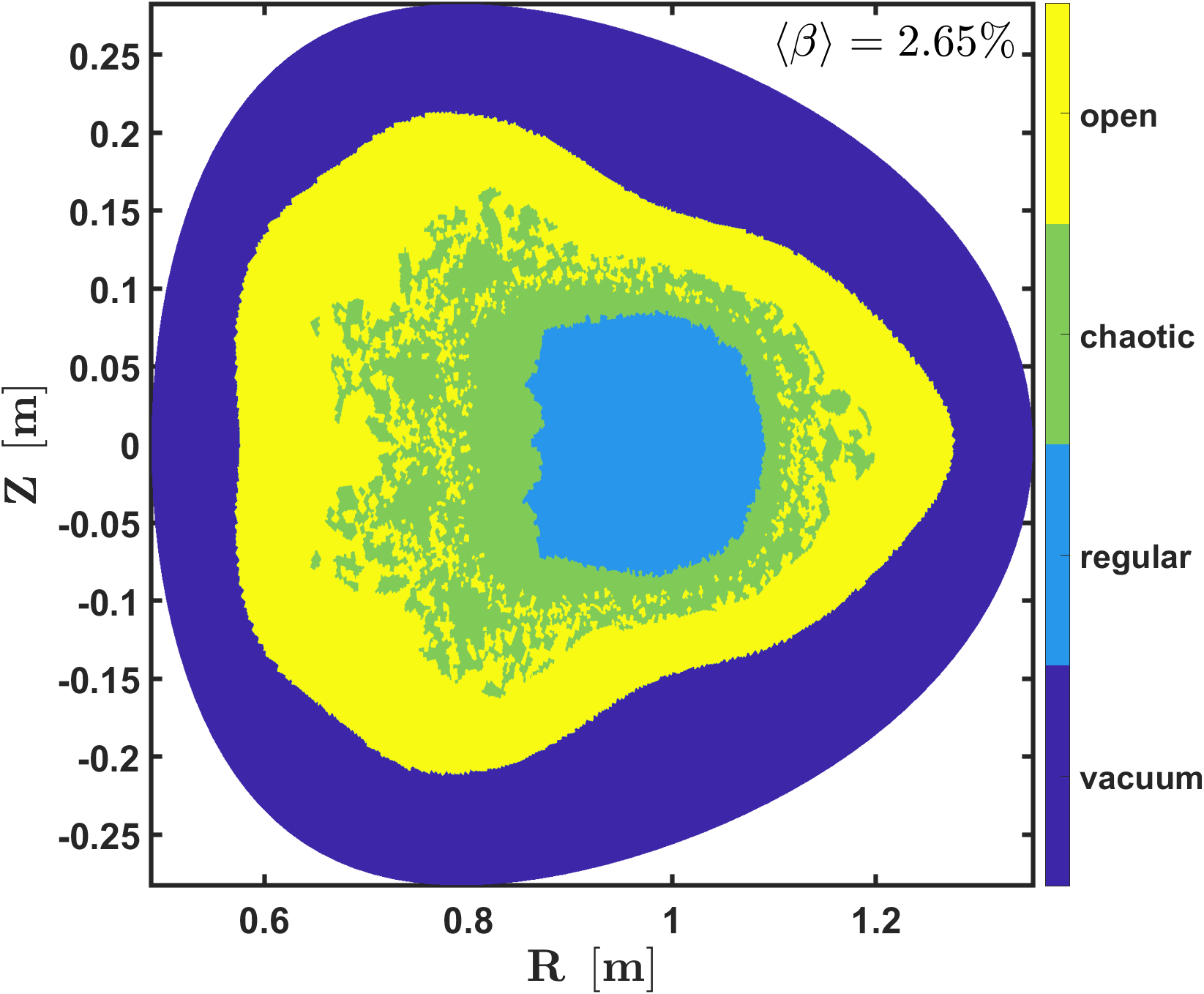}} \qquad
    \subfloat{\includegraphics[width=0.35\linewidth]{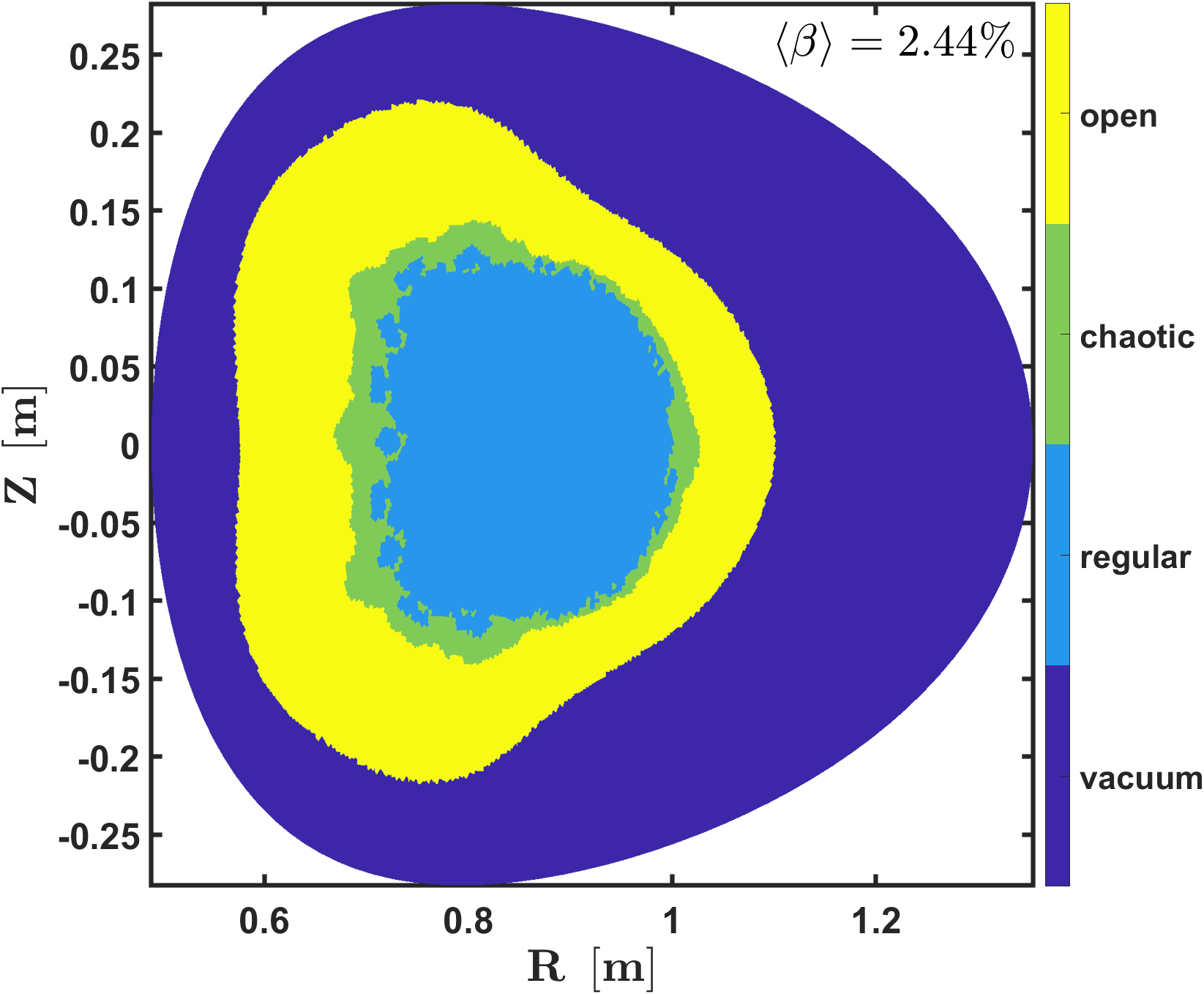}}
    \caption{Contours of the weighted Birkhoff average for finite-$\beta$ equilibria in the $n/m=2/5$ (left column) and $n/m=2/6$ (right column) island configurations with various $\langle\beta\rangle$. 
             The filled contours are drawn at the cross section $\phi=\pi/2$.}
    \label{fig:WBA_contours_n2m5_n2m6}
\end{figure}

\begin{figure}[htbp]
    \captionsetup[subfloat]{position=top,labelformat=empty}
    \centering
    \subfloat[$n/m=2/5$]{\includegraphics[width=0.6\linewidth]{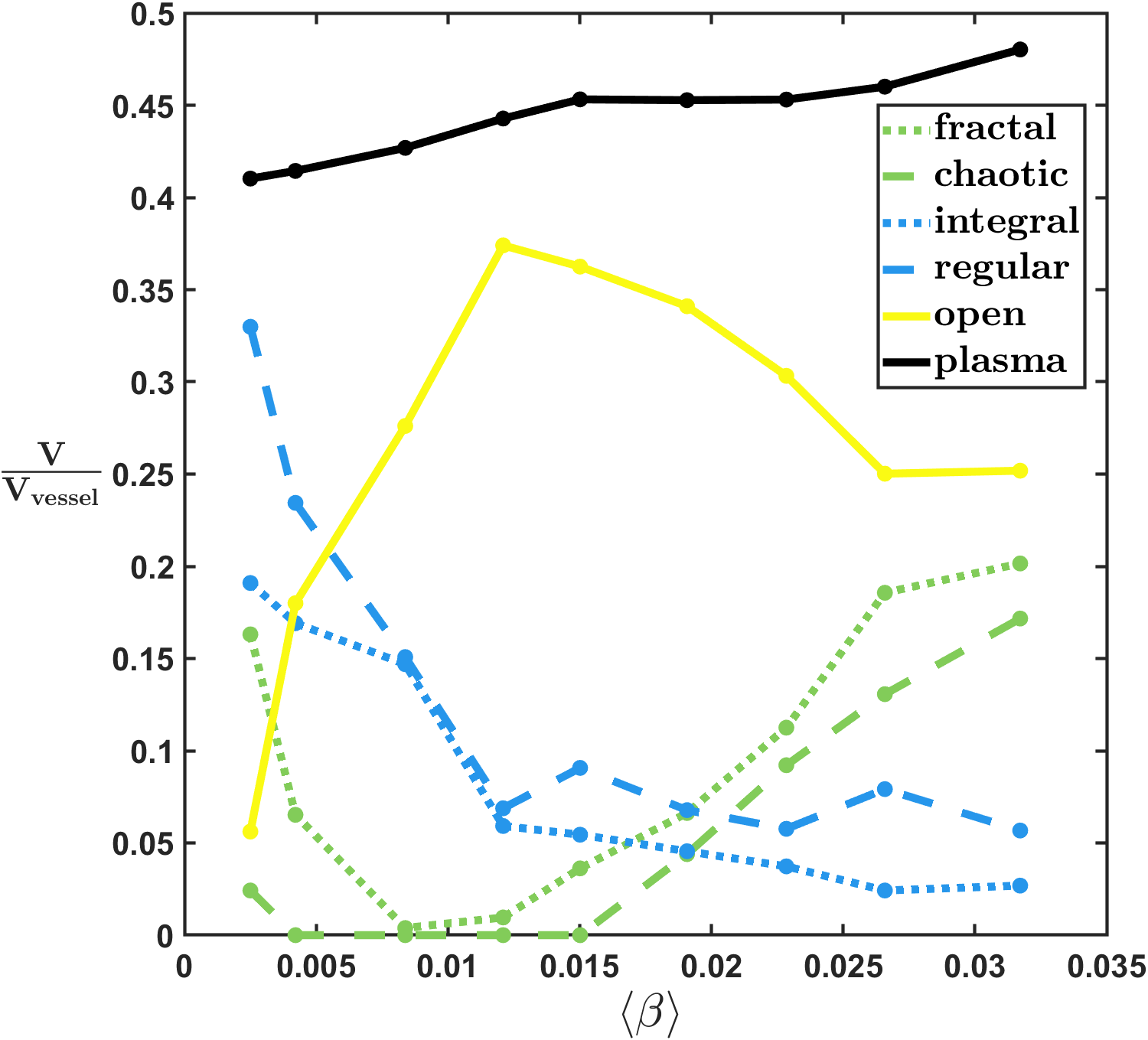}}\\ \vspace{+0.75em}
    \subfloat[$n/m=2/6$]{\includegraphics[width=0.6\linewidth]{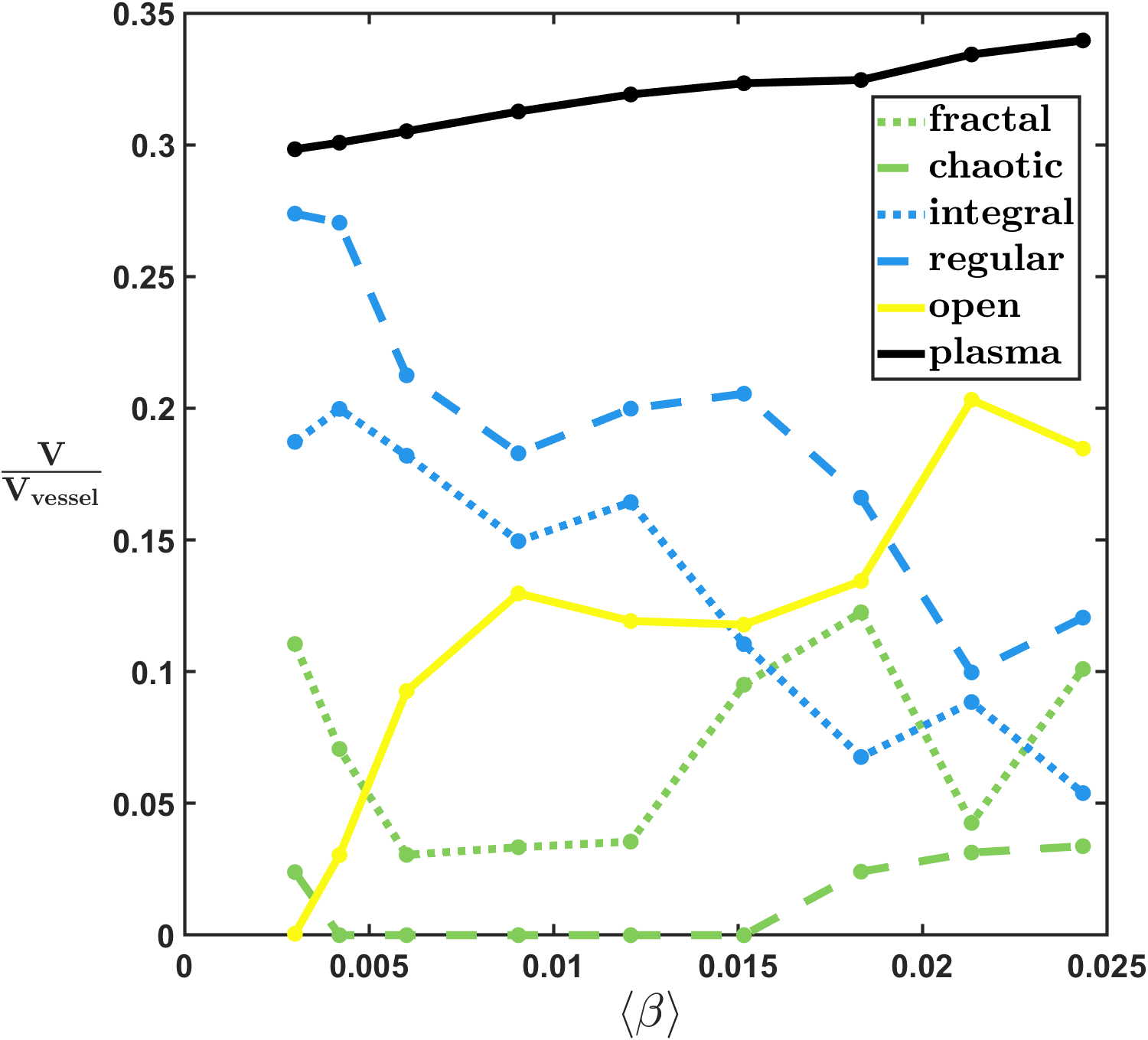}}
    \caption{Effective subvolumes divided by the box-counting dimension (dotted lines) or the weighted Birkhoff average (dashed lines)
             for finite-$\beta$ equilibria in the $n/m=2/5$ and the $n/m=2/6$ island configurations.}
    \label{fig:beta_vs_BC_WBA_n2m5_n2m6}
\end{figure}

Discussions in \Cref{sec:standard_bcc} suggest that the transition of field lines takes place between the LCFS and edge islands, thereby, 
it is worthwhile to evaluate the robustness and resilience of edge islands first. 
In the low $\langle\beta\rangle$ regime, the finite-$\beta$ effect does not substantially alter the shape of the plasma region in the net-current-free standard configuration 
as shown in \Cref{fig:poincare_QA}. However, for island configurations, the edge islands 
may transform into open field lines when the LCFS diminishes, even prior to the overlap onset of inner islands. 
\Cref{fig:poincare_island_n2m5} and \Cref{fig:poincare_island_n2m6} display the corresponding Poincaré plots. 
It is clear that an obvious gap between the edge islands and the LCFS appears even at $\langle\beta\rangle=0.42\%$ for the $n/m=2/5$ island configuration, 
and at $\langle\beta\rangle=0.60\%$ for the $n/m=2/6$ island configuration. 
These results suggest that the candidate island divertor configuration remains viable at low $\langle\beta\rangle$ if the bootstrap currents are absent. 
On the other hand, internal field line stochastization in both island configurations follows the pattern comparable to the net-current-free standard equilibria, 
as shown by the Poincaré plots for $\langle\beta\rangle=1.91\%$ (Config. $n/m=2/5$) and $1.83\%$ (Config. $n/m=2/6$) where 
the overlap between internal $n/m=2/6$ islands with other higher-order island chains results in chaotic field line behavior. 
This outcome is probably attributable to the fact that TFC and PFC currents may be insufficient 
to produce an influence greater than that from pressure-driven currents near the axis. \Cref{fig:BC_contours_n2m5_n2m6} reveals 
the distinctions between the subregions characterized by the box-counting dimension for the two island configurations. 
In the $n/m=2/5$ island equilibria, the open-field-line regions have already taken over the bulk of the plasma region before the onset of chaos, 
and the mixture area of the open-field-line and the chaotic regions is evidently visible in high $\langle\beta\rangle$ cases. 
On the contrary, for the $n/m=2/6$ island configuration, 
the LCFS gradually shrinks and open-field-line regions continue to expand as $\langle\beta\rangle$ increases. 
\Cref{fig:WBA_contours_n2m5_n2m6} shows the contours of subregions divided by the WBA, 
from which we may draw the identical conclusions. The consistent trend in the fractal-integral and the chaotic-regular variation is illustrated in 
\Cref{fig:beta_vs_BC_WBA_n2m5_n2m6} via the effective volume. 
In terms of the criterion, the equilibrium $\beta$ limits of the island configurations should be much lower than those of the standard configuration. 
Despite that, the nested flux surfaces near the core persist at $\langle\beta\rangle\approx1.5\%$. Accordingly, 
if we only focus on the islands and the associated stochasticity close to the core region, 
the $\beta$ limits are inferred to be close to the standard configuration. 

\begin{figure}[htbp]
    % \vspace{-7em}
    \captionsetup[subfloat]{position=top,labelformat=empty}
    \centering
    \subfloat[$n/m=2/5$]{\includegraphics[width=0.3\linewidth]{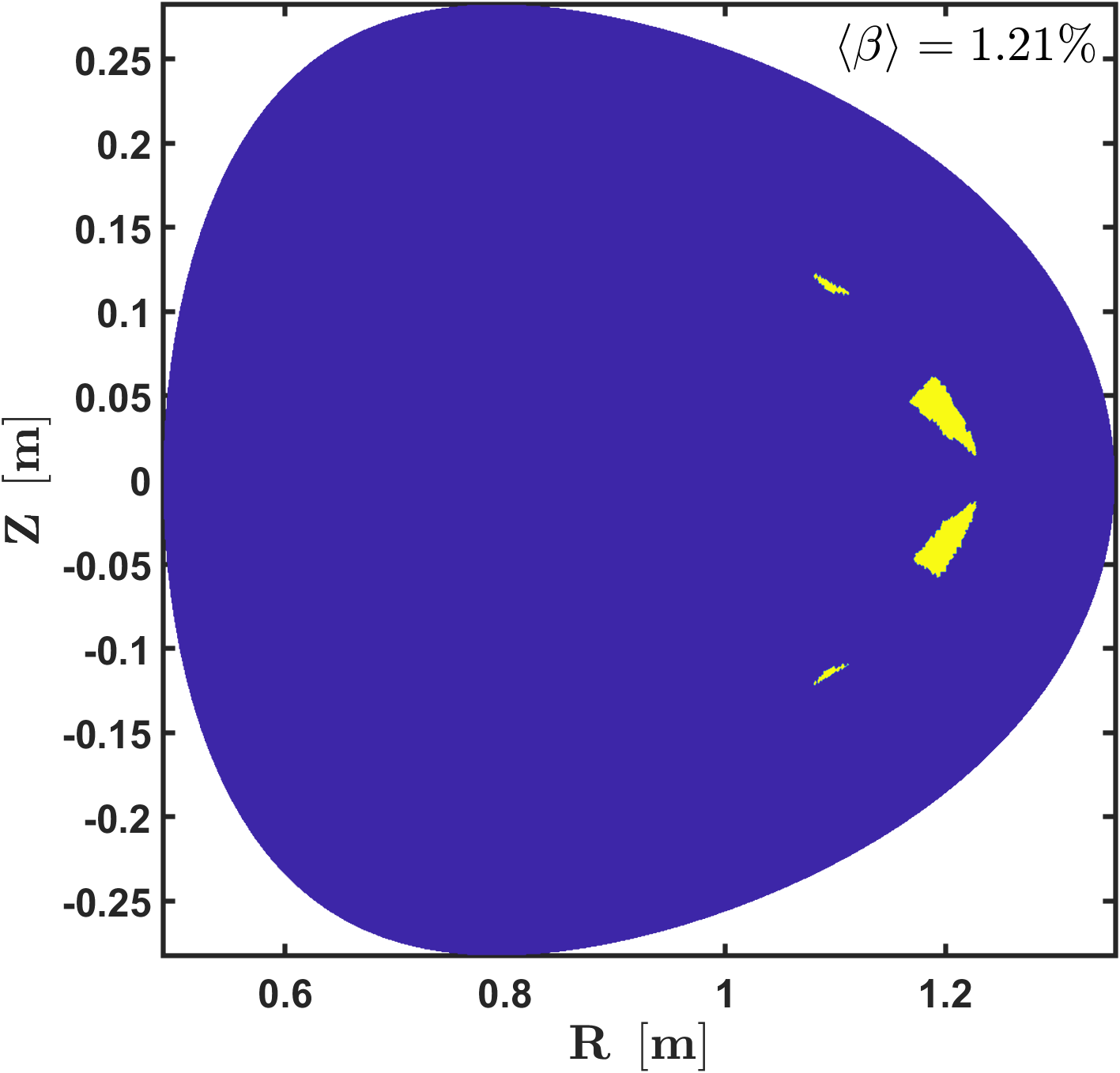}} \qquad
    \subfloat[$n/m=2/6$]{\includegraphics[width=0.3\linewidth]{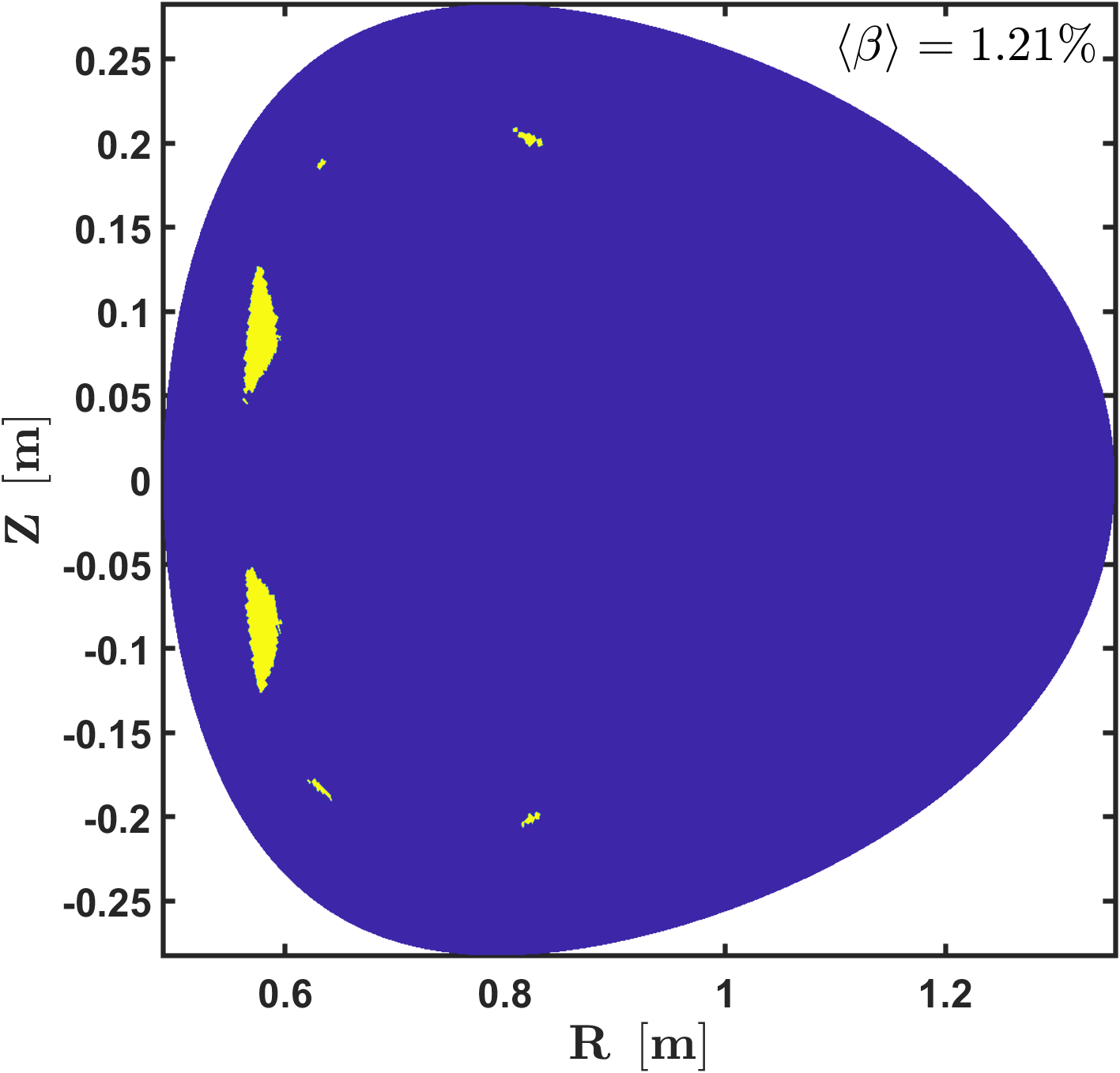}}\\ \vspace{-0.75em}
    \subfloat{\includegraphics[width=0.3\linewidth]{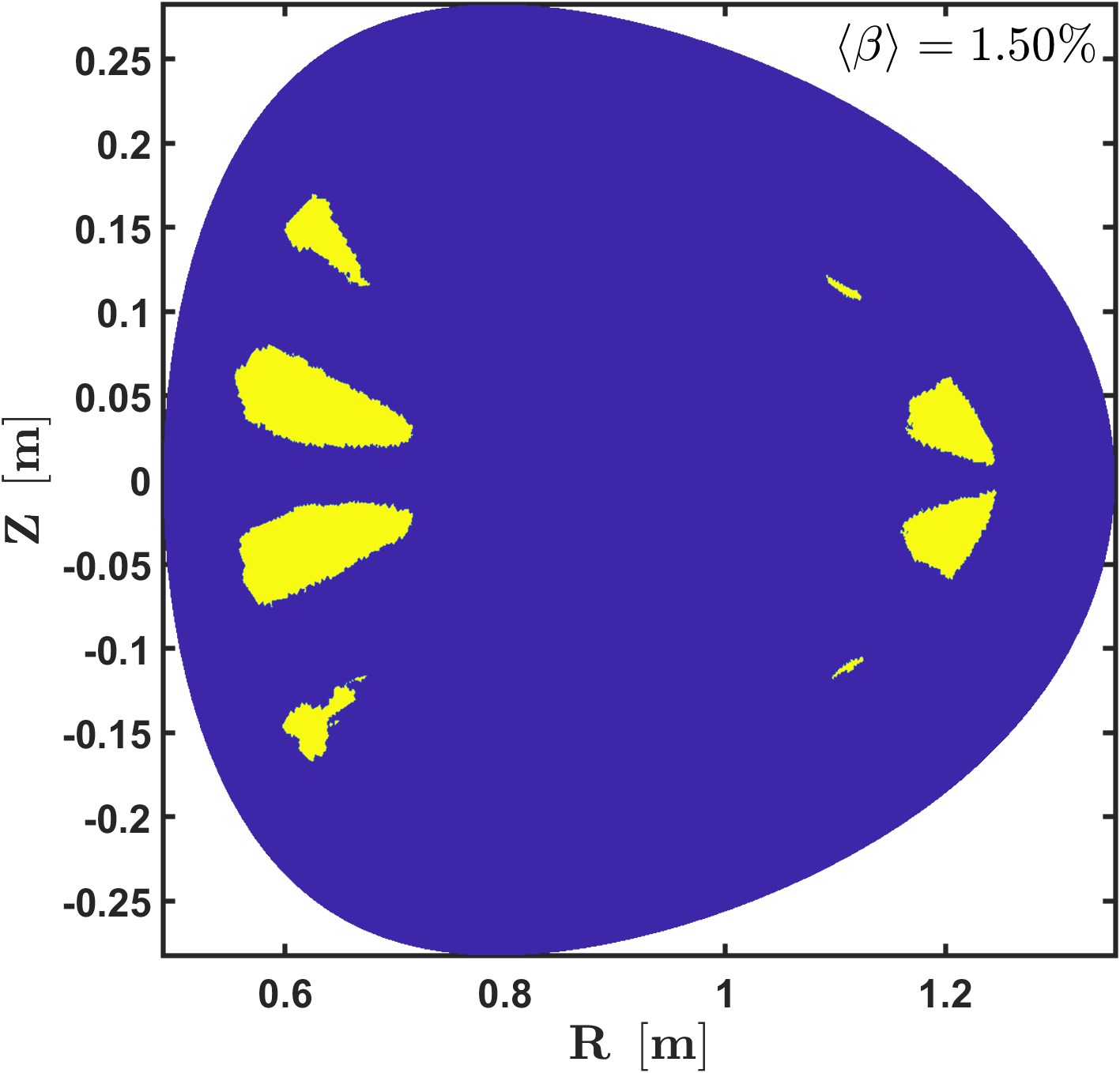}} \qquad
    \subfloat{\includegraphics[width=0.3\linewidth]{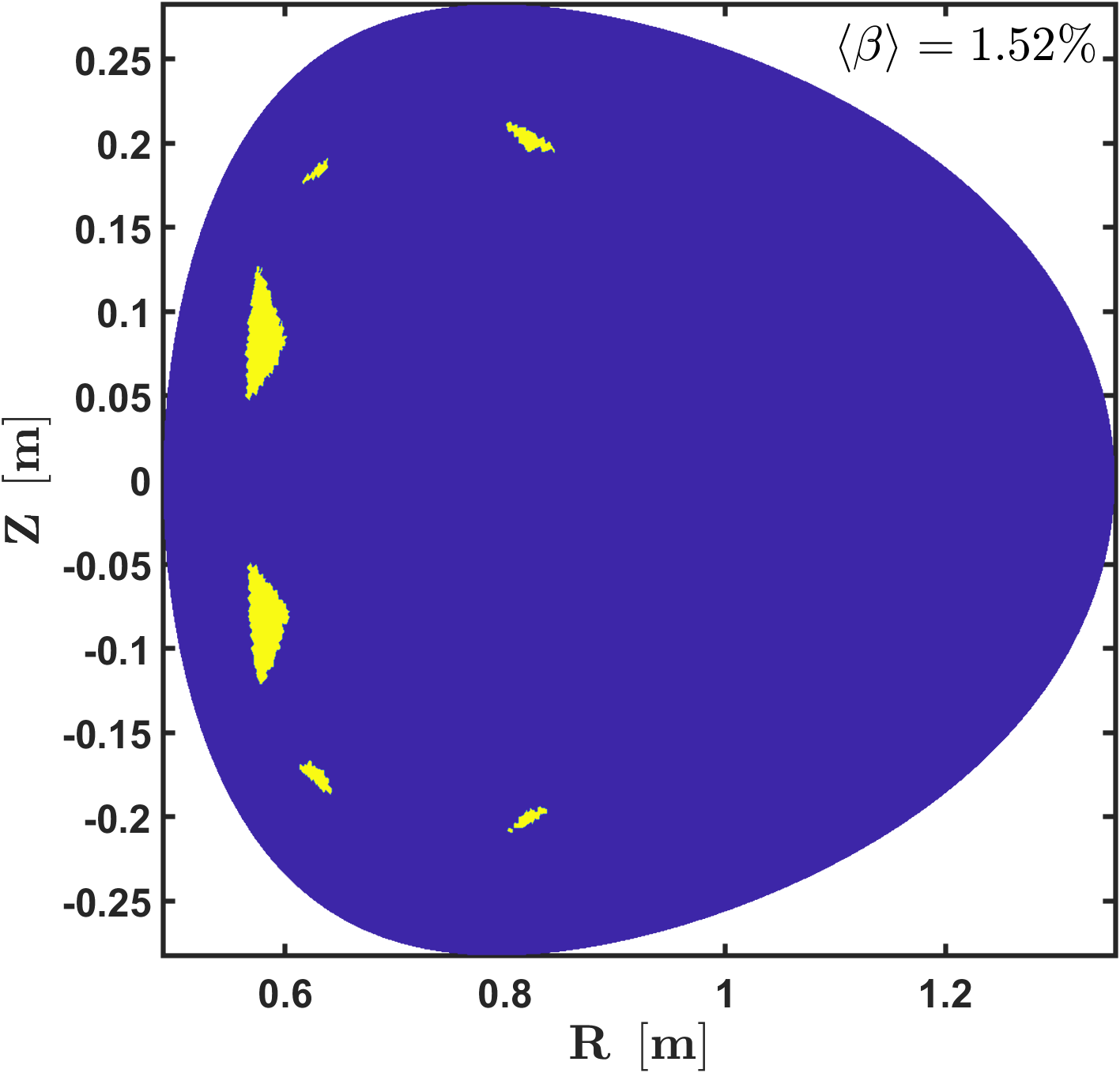}}\\ \vspace{-0.75em}
    \subfloat{\includegraphics[width=0.3\linewidth]{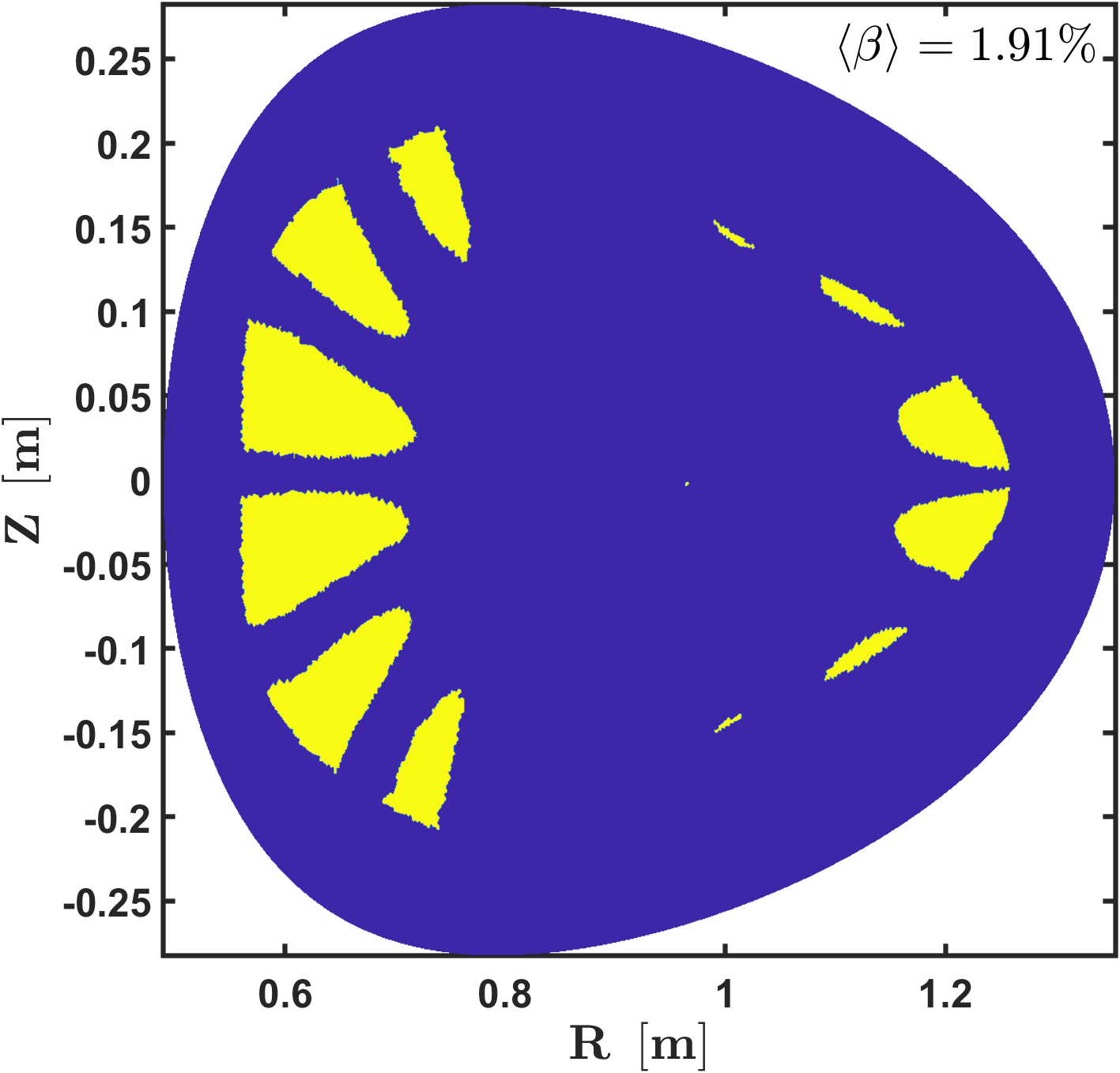}} \qquad
    \subfloat{\includegraphics[width=0.3\linewidth]{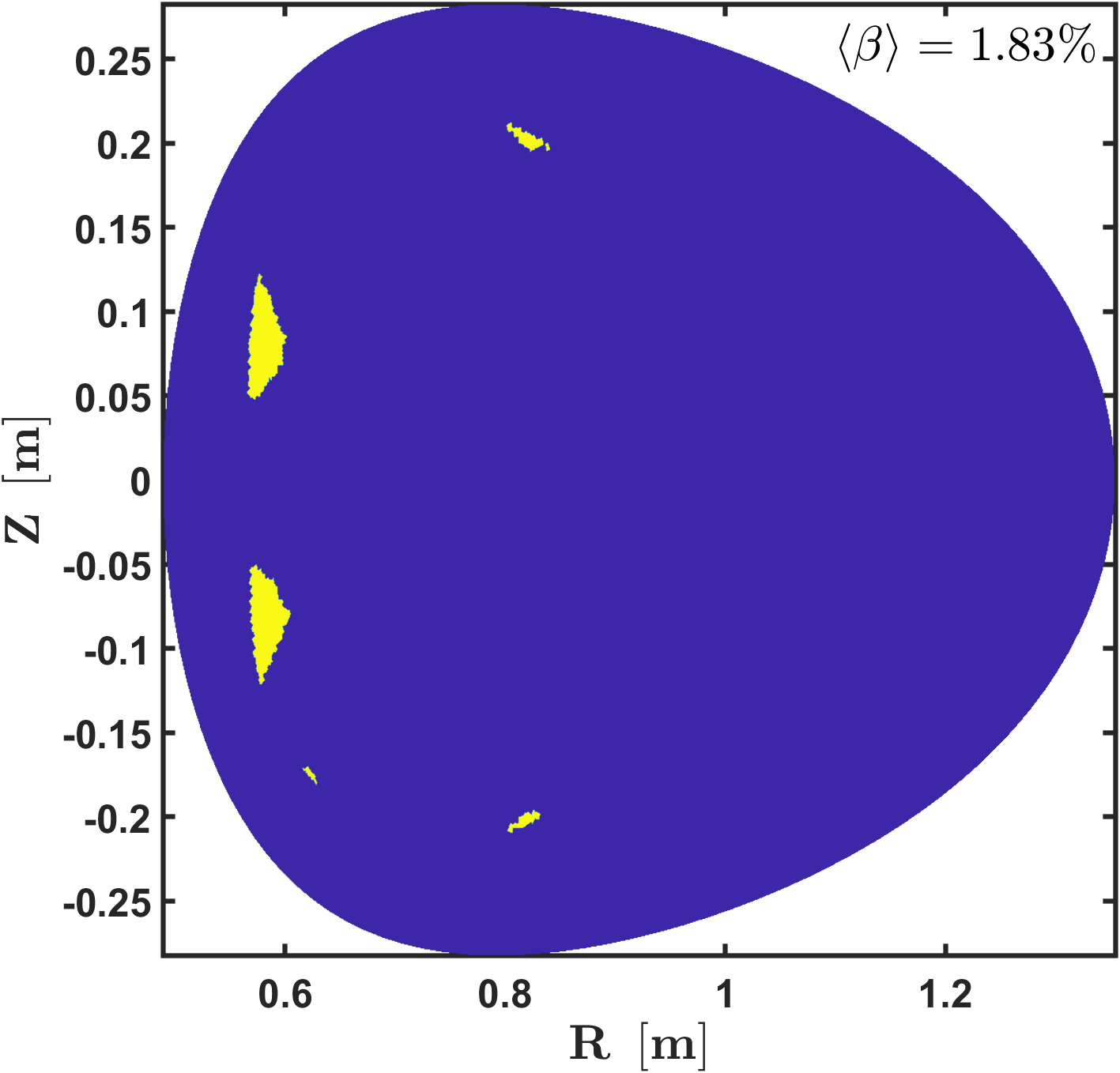}}\\ \vspace{-0.75em}
    \subfloat{\includegraphics[width=0.3\linewidth]{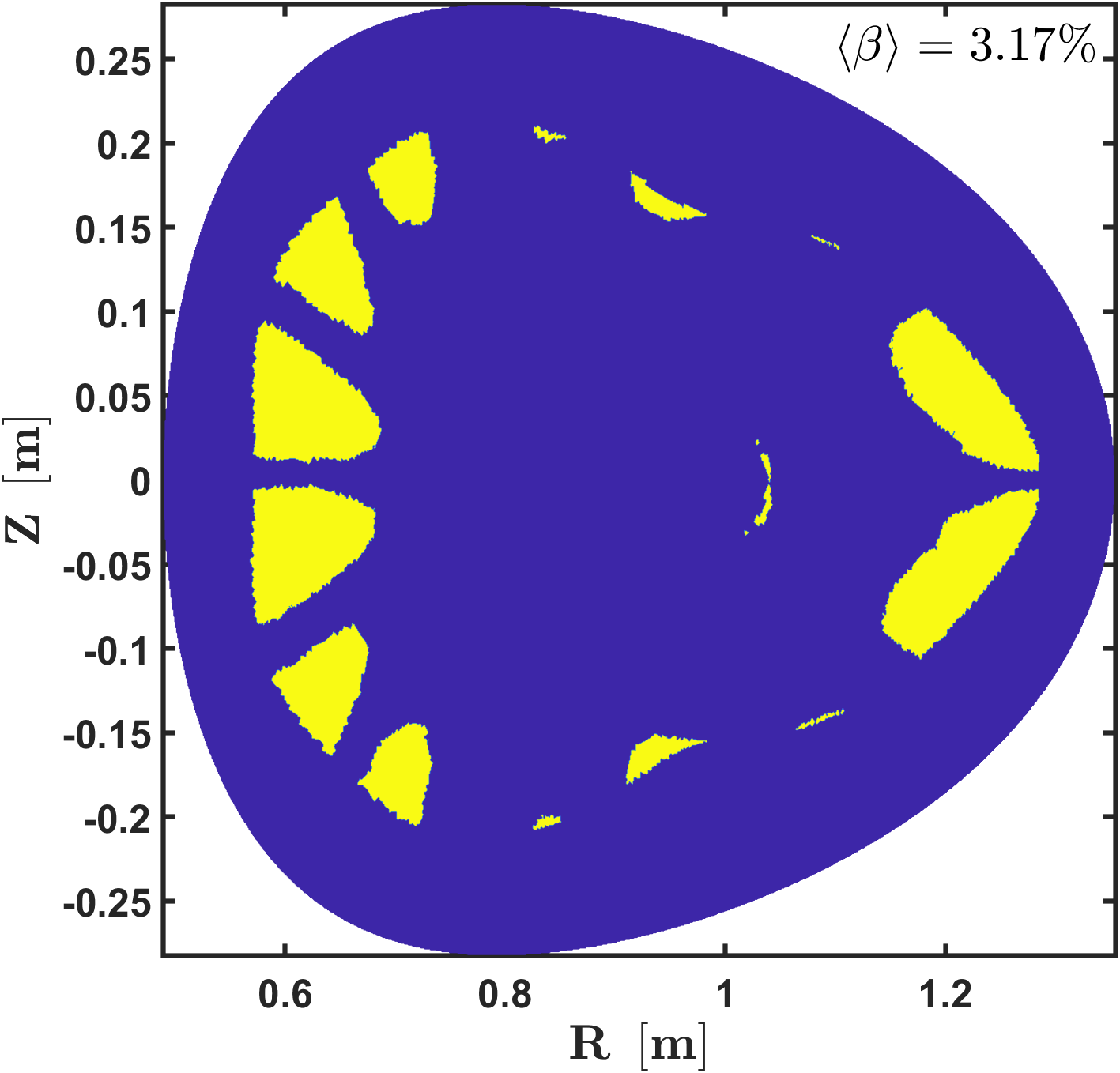}} \qquad
    \subfloat{\includegraphics[width=0.3\linewidth]{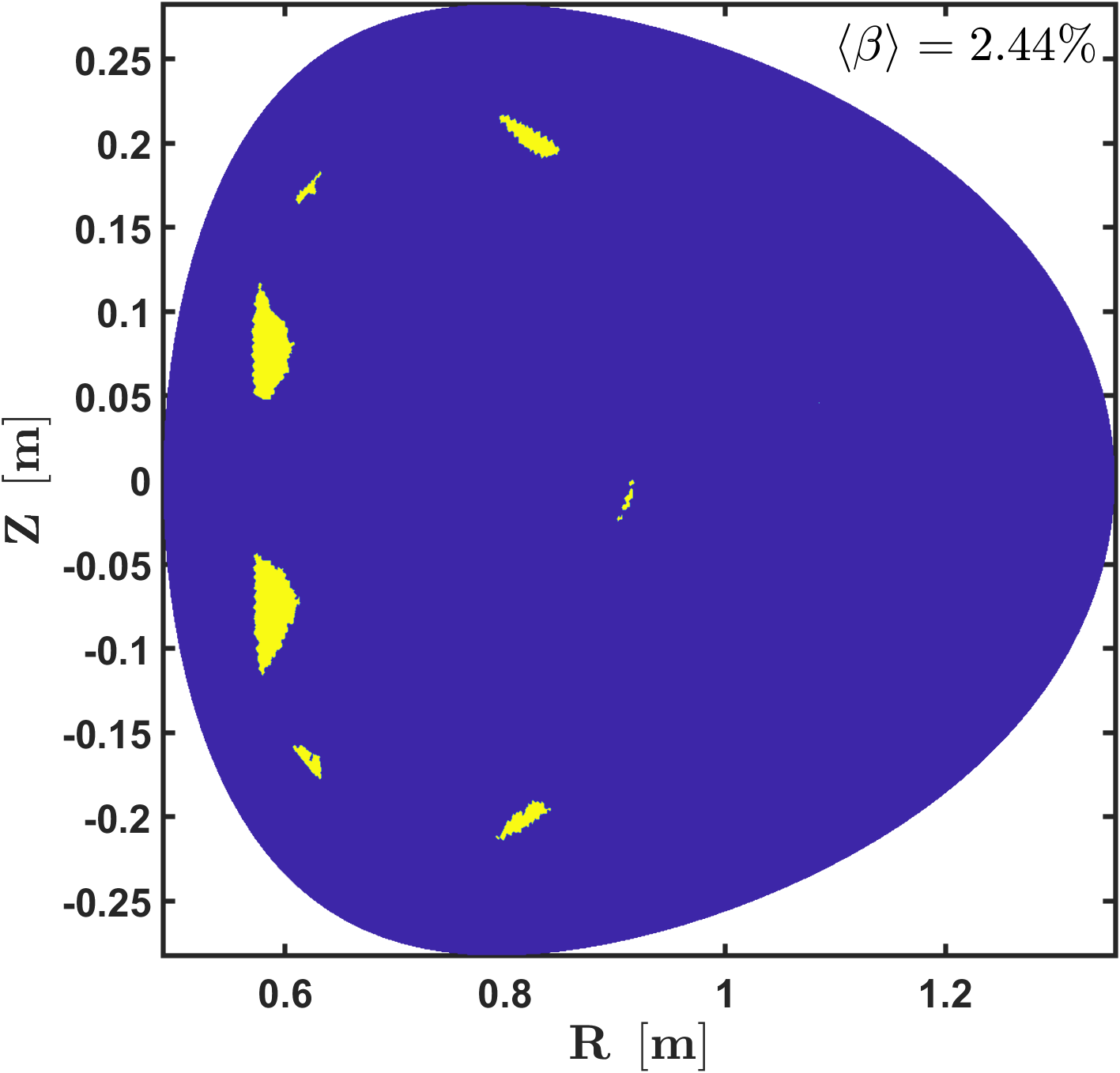}}
    \caption{Contours of the by parallel diffusion for finite-$\beta$ equilibria in the $n/m=2/5$ (left column) and the $n/m=2/6$ (right column) island configurations with various $\langle\beta\rangle$. 
             The yellow-marked areas denote the $V_\text{PD}$ subregions. The filled contours are drawn at the cross section $\phi=\pi/2$.}
    \label{fig:VPD_contours_n2m5_n2m6}
\end{figure}

\Cref{fig:VPD_contours_n2m5_n2m6} shows the contours of effective parallel diffusion for the island configurations. 
In the cases of the $n/m=2/5$ island configuration, 
the area of marked region manifests a rapid increase with $\langle\beta\rangle$ from $1.21\%$ to $1.91\%$, and then saturates. 
The marked region links the plasma-vacuum separatrix and the LCFS at $\langle\beta\rangle=1.50\%$, which agrees with the mixture between 
open field lines and chaotic regions shown in the left columns of \Cref{fig:BC_contours_n2m5_n2m6}. 
Moreover, the behavior of the parallel diffusion area becomes significantly different in the $n/m=2/6$ island configuration. 
The marked region occupies only a small portion of the plasma region in all cases, showing that the connection length from 
the plasma-vacuum separatrix to the LCFS is sufficiently long to satisfy the condition $\bm{B}\cdot\nabla p = 0$. 
The calculation of the effective volume subsequently shows that  
the $V_\text{PD}$ value surges to $8.5\%$ and saturates at $\langle\beta\rangle=1.50\%$ for the $n/m=2/5$ island configuration, 
and remains at a low level near $1\%$ for $n/m=2/6$. 

\section{Summary and future work}\label{sec:summary}

Our calculations indicate that, in the standard CFQS configuration, the primary cause of the disruption of nested flux surfaces is 
the critical overlap between the high-order island chains and the internal major islands that correspond to $n/m = 2/6$ in the net-current-free case 
and $n/m = 2/5$ in the bootstrap-current-carrying case. In addition, the bootstrap current can alter the edge magnetic topology, 
making field lines outside the LCFS more likely to open and connect with the vessel boundary, in contrast to the net-current-free case. 
It is predicted that the equilibrium $\beta$ limits for both types of equilibria in the standard configuration are around $1.5\%$. 
Moreover, we have also investigated two divertor island configurations of the net-current-free equilibria. 
We observe that the edge islands are embedded in open field lines even at low $\langle\beta\rangle$, which is unfavorable for resilient divertor operations. 
As $\beta$ increases, the LCFS shrinks and the edge islands are replaced by open field lines, 
which is more severe in Config. $n/m=2/5$ than in Config. $n/m=2/6$. Fortunately, 
the inner flux surfaces can persist at $\langle\beta\rangle\approx1.5\%$ for both island configurations, 
which is comparable to the standard configuration. 

In this paper, the bootstrap current is not treated self-consistently, 
although such currents would be important in a quasi-axisymmetric stellarator \cite{Okamura_2004}.
This choice is motivated by the difficulty of determining its distribution from solving the drift-kinetic equations in the absence of nested flux surfaces, 
In future work, we will test and validate several approaches, such as an analytical model for quasi-symmetric stellarator \cite{Saxena_2025}, in order 
to incorporate bootstrap and other external currents more self-consistently into NTEC. 

\section*{Acknowledgments}
\addcontentsline{toc}{section}{Acknowledgments}
The first author thanks the fluid engineering group in Hiroshima University for its hospitality during a research visit. 
This work is supported by the Ministry of Education under Grant No. ZYGXONJSKYCXNLZCXM-P5, 
the National MCF Energy R\&D Program of China under Grant No. 2019YFE03050004, 
the U.S. Department of Energy Grant No. DE-FG02-86ER53218, 
the Hubei International Science and Technology Cooperation Project under Grant No. 2022EHB003, 
the Science and Technology Plan Project in Sichuan Province of China under Grant No. 24NSFJQ0201, 
and the Graduate Short-Term Study Abroad Program in HUST. 
The computing work in this paper is supported by the Public Service Platform of High Performance Computing by Network and Computing Center of HUST.

\section*{Appendix A}\label{App:A}
\begin{figure}[htbp]
    % \vspace{-7em}
    \centering
    \subfloat{\includegraphics[width=0.65\linewidth]{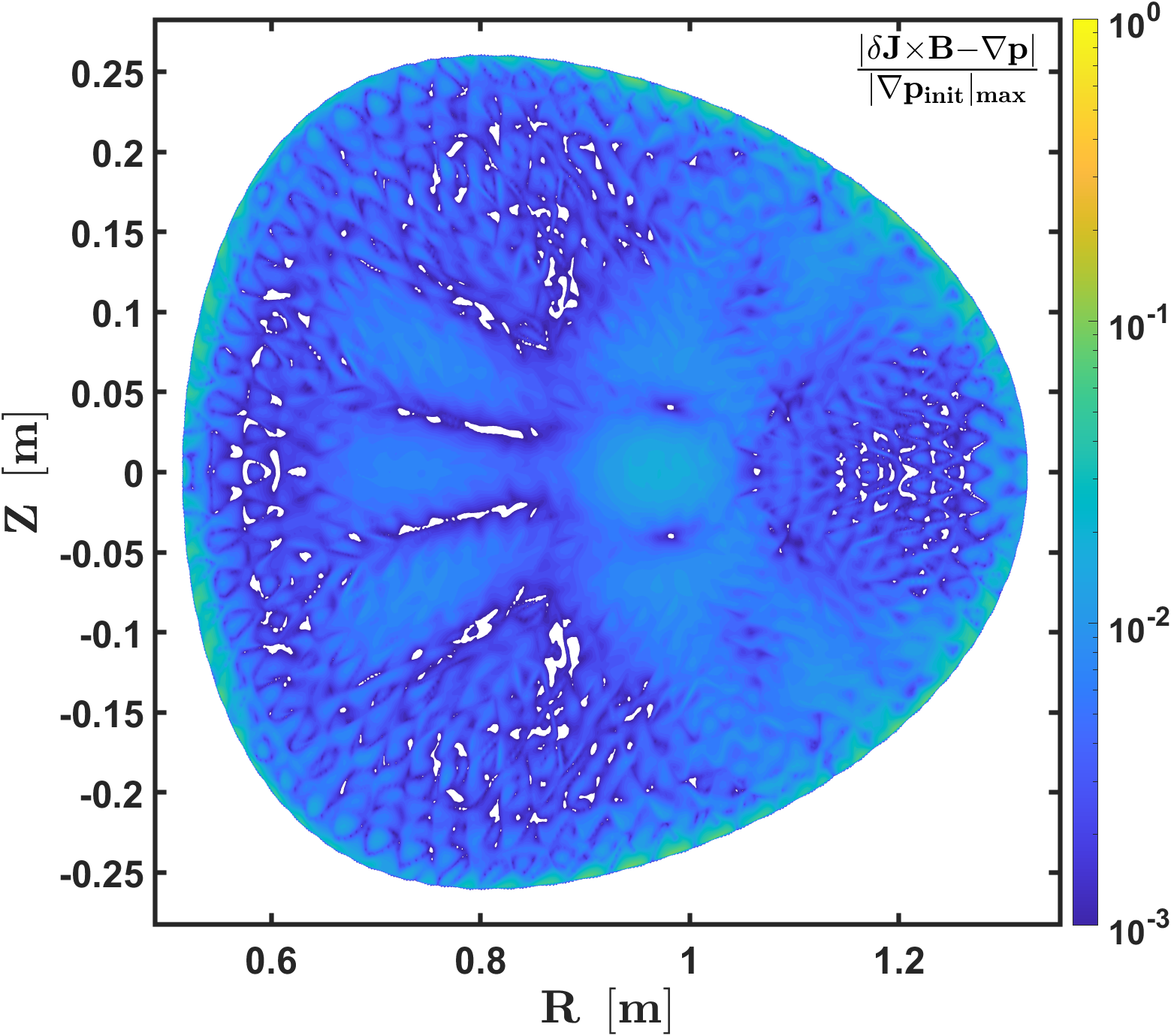}}\\ \vspace{-0.75em}
    \subfloat{\includegraphics[width=0.65\linewidth]{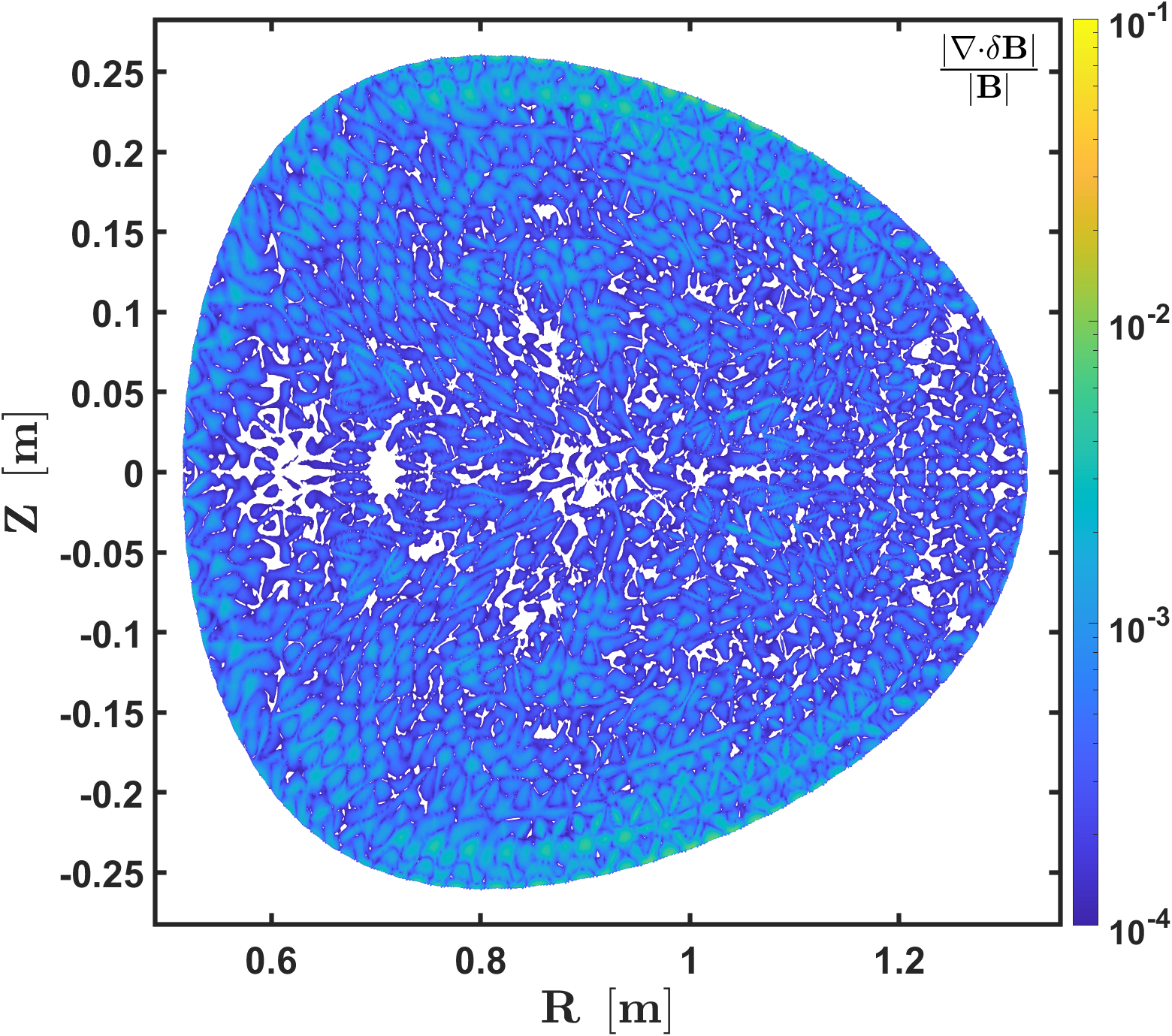}}
    \caption{Contours of (upper:) MHD force residues and (lower:) magnetic divergence error of plasma-induced magnetic fields 
             for the bootstrap-current-carrying equilibrium at $\langle\beta\rangle=2.04\%$ in the standard configuration. 
             The internal blank areas indicate that the residue is less than the minimum in the colorbar.}
    \label{fig:eq_error}
\end{figure}

\addcontentsline{toc}{section}{Appendix A}
We use $1102$ quintic triangles in the poloidal plane and $25$ Fourier modes corresponding to $[0,2,4,6,\dots,24]$ in the toroidal direction. 
The mesh convergence study is performed by solving the vector Poisson equation to obtain vacuum fields with various parameters. 
We have scanned the grid parameters up to $[N_\text{triangles}=1958,N_\text{modes}=29]$, and found that the grid parameters used in the paper are sufficient to numerically resolve the magnetic topology. 
Regarding the finite-$\beta$ equilibrium convergence properties, 
we require that the normalized residual force $|\delta\bm{J} \times \bm{B} - \nabla p| / |\nabla p_{\text{init}}|_{\text{max}}$ drops below $1\%$ on average at the specified grid resolution, 
and the fictitious force $F_{\text{fict}}=|(\nabla \cdot \delta \bm{B})\bm{B}/\mu_{0}|$ 
reduces to be comparable with or smaller than the residual force $F_{\text{res}}=|\delta\bm{J} \times \bm{B} - \nabla p|$ in the final steady state, 
where $\delta$ indicates the $\beta$-induced variation. For illustration, 
we take the bootstrap-current-carrying case with the maximum $\langle\beta\rangle$ used in standard configuration calculations as an example. 
\Cref{fig:eq_error} plots the contours of MHD force residues and magnetic divergence errors at the cross section $\phi=\pi/2$. 
The averaged normalized errors of MHD force and magnetic divergence are less than $10^{-2}$ and $10^{-3}$, respectively, 
which indicates that $\langle F_{\text{res}} \rangle / \langle F_{\text{fict}} \rangle \sim O(1)$ for 
$\langle p \rangle / (r_\text{eff} \langle B \rangle) \sim O(10^{5})$ in this case. 

The initial state for finite-$\beta$ equilibrium calculations is obtained by adding a pressure field to the vacuum field. 
The pressure profile evolves under the no-slip boundary condition, and is initially prescribed as 
\begin{equation}\label{eq:pres_init}
p=\begin{cases}
p_{0}(1-3\rho_{N}^3+2\rho_{N}^4) & \text{if}\:\rho_{N}\leq 0.6\\
0 & \text{if}\:\rho_{N}>0.6
\end{cases}, 
\end{equation}
where $\rho_{N}$ is the normalized radial label in the NTEC computational coordinates. During the iteration, we keep the maximum pressure constant. 

% refs
\bibliographystyle{unsrt}
\bibliography{refs}
\clearpage

% figs
\clearpage

\clearpage

\clearpage

\clearpage

\clearpage

\clearpage

% \begin{figure}[htbp]
%     \centering
%     \subfloat{\includegraphics[width=1\linewidth]{figs/beta_vs_VPD.png}}
%     \caption{Effective volume of parallel diffusion for net-current-free finite-$\beta$ equilibria in the standard configuration.}
%     \label{fig:beta_vs_VPD}
% \end{figure}
% \clearpage

%% figs_bootstrap

\clearpage

\clearpage

\clearpage

\clearpage

\clearpage

% \begin{figure}[htbp]
%     \centering
%     \subfloat{\includegraphics[width=1\linewidth]{figs/beta_vs_VPD_bootstrap.png}}
%     \caption{Effective volume of parallel diffusion for bootstrap-current-carrying finite-$\beta$ equilibria in the standard configuration.}
%     \label{fig:beta_vs_VPD_bootstrap}
% \end{figure}
% \clearpage

%% figs_islands

% \begin{figure}[htbp]
%     % \vspace{-7em}
%     \centering
%     \subfloat{\includegraphics[width=0.65\linewidth]{figs/cfqs_L12_N25_p0_00k_n2m5_poincare.png}}\\ \vspace{-0.75em}
%     \subfloat{\includegraphics[width=0.65\linewidth]{figs/cfqs_L12_N25_p0_00k_n2m6_poincare.png}}
%     \caption{Poincaré plots of vacuum fields in the island configurations at two cross sections 
%              $\phi=0\:\text{(red)}\:\text{and}\:\pi/2\:\text{(blue)}$.}
%     \label{fig:poincare_islands_vac}
% \end{figure}
% \clearpage

\clearpage

\clearpage

\clearpage

\clearpage

\clearpage

% \begin{figure}[htbp]
%     \centering
%     \subfloat{\includegraphics[width=1\linewidth]{figs/beta_vs_BC_WBA_n2m5.png}}
%     \caption{Effective subvolumes divided by box-counting dimension (dotted lines) or weighted Birkhoff average (dashed lines)
%              for finite-$\beta$ equilibria in the $n/m=2/5$ island configuration.}
%     \label{fig:beta_vs_BC_WBA_n2m5}
% \end{figure}
% \clearpage

% \begin{figure}[htbp]
%     \centering
%     \subfloat{\includegraphics[width=1\linewidth]{figs/beta_vs_BC_WBA_n2m6.png}}
%     \caption{Effective subvolumes divided by box-counting dimension (dotted lines) or weighted Birkhoff average (dashed lines)
%              for finite-$\beta$ equilibria in the $n/m=2/6$ island configuration.}
%     \label{fig:beta_vs_BC_WBA_n2m6}
% \end{figure}
% \clearpage

\clearpage

\clearpage

% \begin{figure}[htbp]
%     \centering
%     \subfloat{\includegraphics[width=1\linewidth]{figs/cfqs_L12_N25_p0_40k_force_init.png}}
%     \caption{Contours of initial pressure distribution at the cross section $\phi=\pi/2$. 
%              The internal blank areas indicate that the residue is less than the minimum in the colorbar.}
%     \label{fig:pres_init}
% \end{figure}
% \clearpage

\end{document}